\documentclass[final,3p,times,fleqn]{elsarticle}
\usepackage{mathtools,booktabs}
\usepackage{bm}
\usepackage{autobreak}
\usepackage{amsmath}
\usepackage{amsthm}
\usepackage{gensymb}
\usepackage{subfigure}
\usepackage{epstopdf}
\usepackage{color}
\usepackage{hyperref}

\biboptions{sort&compress}
\begin{document}
	\begin{frontmatter}
		\title{Numerical Simulation of Power-Law Fluid Flow in a Trapezoidal Cavity using the Incompressible Finite-Difference Lattice Boltzmann Method}
		\author[a]{Xinmeng Chen}
		\author[a,b,c]{Zhenhua Chai }
		\author[d]{Yong Zhao}
		\author[a,b,c]{Baochang Shi \corref{cor1}}
		\ead{shibc@hust.edu.cn}
		\address[a]{School of Mathematics and Statistics, Huazhong University of Science and Technology, Wuhan 430074, China}
		\address[b]{Institute of Interdisciplinary Research for Mathematics and Applied Science, Huazhong University of Science and Technology, Wuhan 430074, China}
		\address[c]{Hubei Key Laboratory of Engineering Modeling and Scientific Computing, Huazhong University of Science and Technology, Wuhan 430074, China}
		\address[d]{School of Mathematics and Statistics, Changsha University of Science and Technology, Changsha 410114, Hunan, China}
		\cortext[cor1]{Corresponding author.}
		\begin{abstract}
			In this paper, a numerical investigation of power-law fluid flow in the trapezoidal cavity has been conducted by incompressible finite-difference lattice Boltzmann method (IFDLBM).
			By designing the equilibrium distribution function, the Navier-Stokes equations (NSEs) can be recovered exactly.
			Through the coordinate transformation method, the body-fitted grid in physical
			region is transformed into a uniform grid in computational region. The effect of Reynolds ($Re$) number, the
			power-law index $n$ and the vertical angle $\theta$ on the trapezoidal cavity are investigated.
			According to the numerical results, we come to some conclusions.
			For low $Re$ number $Re=100$, it can be found that the behavior of power-law fluid flow becomes more complicated with the increase
			of n. And as vertical angle $\theta$ decreases, the flow becomes smooth and
			the number of vortices decreases. For high Re numbers, the flow development
			becomes more complex, the number and strength of vortices increase. If the Reynolds number
			increases further, the power-law fluid will changes from steady flow to periodic flow and then to
			turbulent flow. For the steady flow, the lager the $\theta$, the more complicated the vortices. And the
			critical Re number from steady to periodic state decreases with the decrease of power-law index $n$.
		\end{abstract}
	\end{frontmatter}
	\section{Introduction}
	
	Over the last couple of decades, tremendous amount of research has been carried out in solving NSEs, such as finite difference method \cite{JWT2013finite}, finite element method \cite{CJ2009Element}, finite volume method \cite{RJL2002FiniteV}.
	In particular, the lattice Boltzmann method (LBM), as a novel alternative method, has been widely concerned now \cite{Benzi1992LB,Chen1998LB,Succi2001LB,Aidum2010LB}.
	As a classic benchmark problem described by NSEs, the two-dimensional lid-driven flow in a square cavity has also been widely investigated \cite{2000lid,2018book,J2004lid,Chai2006lid}, including the study of high Reynolds number flow \cite{Zhou2004lid} and the three-dimensional lid-driven cavity flow \cite{He2004lid}. On this basis, the lid-driven flows of different cavities shapes were also simulated. In 2006, Patil et al. \cite{Patil2006} applied the lattice Boltzmann equation to simulate the lid-driven flow in a two-dimensional rectangular deep cavity. He studied several features of the flow, such as the location and strength of the primary vortex, and the corner-eddy dynamics.
	Then, Cheng et al. \cite{Cheng2006} investigated the vortex structure in a lid-driven rectangular cavity at different depth-to-width ratios and Reynolds numbers by a lattice Boltzmann method.
	Zhang et.al \cite{Zhang2010} used the lattice BGK model to simulate lid-driven flow in a two-dimensional trapezoidal cavity.
	In addition, Li et al. \cite{Li1996} presented an accurate and efficient calculations of the flow inside a triangular cavity for high Reynolds numbers.
	And Erturk et al. \cite{Ercan2007} studied the numerical solutions of 2-D steady incompressible flow in a driven skewed cavity.
	
	However, all above works only studied the Newtonian fluids. Non-newtonian fluids are widely observed in nature and industrial production, such as petroleum, food, geophysics, lubricants, chemistry, hydrogeology, to name but a few \cite{CR2010}. Unlike the Newtonian fluid, the relationship between shear stress and shear strain rate of non-Newtonian fluid is nonlinear. As a result, the non-Newtonian fluid will show shear thickening and shear variation characteristics. Due to the complicated constitutive equation of non-Newtonian fluid, it is a challenge to investigate the non-Newtonian fluid behavior by numerical methods. Recently, there are many efforts have been made to simulate non-Newtonian fluid flows through LBM in various computational geometries \cite{AE1993,Gz2002,GS200515,JB200616,MY200717,HH201120,W201121}, such as the Non-Newtonian flow through porous media \cite{AE1993}, the filling of expanding cavities by Bingham fluids \cite{Gz2002} and the non-Newtonian pseudo-plastic fluid in a micro-channel \cite{HH201120}.
	In addition,
	Gabbanelli et al. \cite{GS200515} studied the shear-thinning and shear-thickening fluids in parallel and reentrant geometries by LBM.
	Boy et al. \cite{JB200616} presented a second-order accurate LBM for the simulations of the power-law fluid in a two-dimensional rigid pipe flow.
	Yoshino et al. \cite{MY200717} developed a LBM to investigate the power-law model in a reentrant corner geometry and flows inside a
	three-dimensional porous structure.
	Mendu and Das \cite{SS201218} applied the LBM to study the power-law fluids inside a two-dimensional enclosure driven by the motion of the two facing lids.
	Psihogios et al. \cite{PJ200719} investigated the non-Newtonian shear-thinning fluid flow in three dimensional digitally reconstructed porous domain.
	Hamedi and Rahimian \cite{HH201120} simulated the power-law model for pseudo-plastic fluids in micro-channel by using LBM.
	Wang and Ho \cite{W201121} investigated the shear thinning non-Newtonian blood flows through LBM.
	Chai et al. used the multi-relaxation-time lattice Boltzmann method (MRT-LBM) to simulate the generalized Newtonian fluid flow.
	Qi et al. \cite{QZ2018} investigated the wake effect on the interaction between particle and power-law fluid flow by the parallel three-dimensional LBM.
	And they also investigated the interaction between fluid rheology and bed properties through LBM \cite{ZA2020}.
	
	For the Non-newtonian fluids in two-dimensional cavity, Li et al. \cite{Li2014} used the MRT-LBM to study power-law fluid flows in square cavity. Besides, MRT-LBM has been applied to simulate the power-law fluid in square enclosures with undulation in Ref. \cite{MB2020}. At present, the non-Newtonian fluid flow problem in trapezoidal cavity has not been investigated. Obviously, this problem will be more complex than that in the square cavity. The angle of trapezoid, the power-law index and the $Re$ number are the key factors affecting the flow. On the one hand, a more stable model is required for the simulation due to the characteristics of shear thickening and shear variation in Non-newtonian fluids. On the other hand, the curved boundary treatment method should be applied for the boundary of trapezoid.
	However, the stability of standard LBM is less than the finite difference LBM (FDLBM) \cite{Chen2021}, and it is also difficult to implement the body-fit grid in the trapezoidal cavity \cite{Chen2020}. Using the curved boundary treatment method means that the computational domain will be expanded into a rectangle, which will increase the computation.
	Based on the above problems, we find some works have been conducted on FDLBM to simulate complex flows in order to improve numerical scheme accuracy and geometric flexibility, including
	three-dimensional incompressible flows~\cite{ezzatneshan2019simulation}, two-phase liquid-vapor flows~\cite{hejranfar2015simulation}, natural
	convection in some special geometries~\cite{sai2019natural,Khakrah2019Numerical} and blood flow~\cite{Sakthivel2019blood}.
	Hence, the finite difference LBM (FDLBM) is more suitable to simulate the power-law fluid in the trapezoidal cavity.
	Compared to conventional numerical methods, one of the characteristics of IFDLBM is that the shear tensor
	can be computed locally without taking space derivatives of the velocity field \cite{SU2005,MY200717}.
	For transport phenomena in complex geometries, the LBM is more efficient than the finite difference method \cite{MY2004} and the finite volume method \cite{YY2004}.
	And the IFDLBM, as a mesoscopic numerical method, also possesses this characteristic.
	Besides, compared to the LBM, the FDLBM is more stable and the flow details of non-Newtonian fluids can
	be better captured even at high Re numbers through the FDLBM  \cite{Chen2020,Chen2021}.
	And the space-time is decoupled in FDLBM, it is convenient to use a body-fit mesh to simulate the trapezoidal cavity problem.

	In this paper, the incompressible FDLBM (IFDLBM) has been proposed as a core solver to simulate the power law flow in a two-dimensional trapezoidal cavity.
	The rest of the paper is organized as follows. The physical model and the governing equation are expressed in Sec. 2.
	In Sec. 3, we give the IFDLBM and list the calculational process. Through coordinate transformation, the general formula of body-fitting mesh transformation is given in Sec. 4.
	Then, the code validation and the grid independence testing are performed in Sec. 5. In Sec. 6, we showed the numerical results and discuss the fluid behavior.
	Finally, a brief summary was made in Sec. 7.
	
	\section{Physical model and governing equation}\label{GoverEqs}
	
	The IFDLBM for incompressible power-law flow in the trapezoidal cavity (TC) can be expressed as
	\begin{equation}\label{eq1.1}
		\nabla\cdot \bm u=0,
	\end{equation}
	\begin{equation}\label{eq1.2}
		\frac{\partial \bm u}{\partial t}+\nabla\cdot( \bm u\bm u)=-\nabla P+\nabla\cdot \bm \tau,
	\end{equation}
	where $\bm u$ is the fluid velocity, $P$ is the pressure. And $\bm\tau$ is the shear stress, which can be defined as
	\begin{equation}\label{eq1.3}
		\tau_{ij}=2\mu D_{ij},
	\end{equation}
	where
	\begin{equation}\label{eq1.3.1}
		D_{ij}=\frac{1}{2}(\nabla\bm u+\nabla \bm u^T),
	\end{equation}
	it indicates the rate of deformation tensor for the two dimensional Cartesian coordinate. $\mu$ is apparent viscosity which is given as
	\begin{equation}\label{eq1.4}
		\mu_\alpha=K(2D_{ij}D_{ij})^{\frac{n-1}{2}},
	\end{equation}
	\begin{equation}\label{eq1.4.1}
		D_{ij}=\frac{1}{2}(\nabla \bm u+\nabla \bm u^T),
	\end{equation}
	where $K$ is the consistency coefficient and $n$ is the power-law index.
	According to the value of $n$, the Power-law fluid can be divided into three different types of fluid. When $n<1$, it is the shear-thinning or pseudoplastic fluid. And when $n>1$, it is a shear-thickening or dilatant fluid. The case $n=1$ corresponds to the Newtonian fluid.
	
\begin{figure}[htbp]
		\centering
		\includegraphics[scale=0.65]{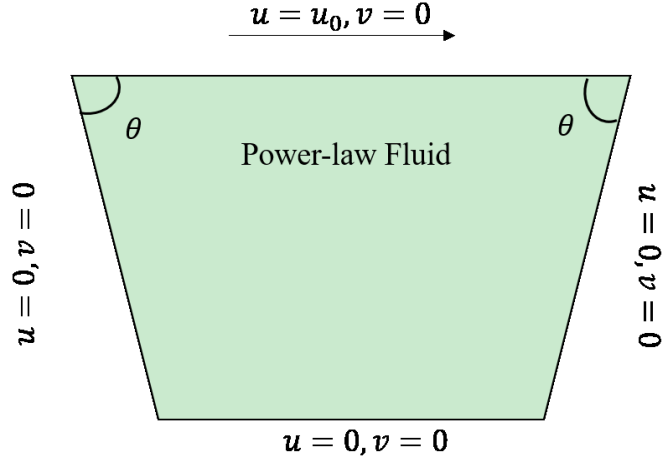}
		\caption{Geometry of the trapezoidal cavity} \label{fig:Lid}
	\end{figure}

	In the present work, we mainly consider three cases of isosceles trapezoids, namely $\theta=75^\circ,60^\circ,45^\circ$.
	The physical domain is shown in Fig. \ref{fig:Lid}.
	The boundary conditions of the problem are given as:
	\begin{eqnarray}\label{1.5}
		&Left\quad wall: u=v=0,\quad \\
		&Right\quad wall: u=v=0,\quad \\
		&Upper\quad wall: u=u_0,v=0,\quad for\quad (y=L,0\leq x \leq L),\\
		&Lower\quad wall: u=v=0,\quad for\quad (y=0,0\leq x \leq L).
	\end{eqnarray}

	\section{The incompressible finite-difference lattice Boltzmann method }\label{LBMs}
	
	In this section, the incompressible FDLBM will be presented where the collision term is discreted by BGK model \cite{Chen2020}.
	For the power-law fluid flow, we consider the BGK model combined with FDLBM.
	First, let's begin from the discrete velocity Boltzmann equation (DVBE) without the force term,
	\begin{equation}\label{eq2.0}
		\partial_tf_i+\bm c_i\cdot\nabla f_i=-\frac{1}{\lambda}(f_i-f_i^{eq}),
	\end{equation}
	where the $f_i(\bm x,t)$ is density distribution function for particle moving with velocity $\bm c_i$
	at position $\bm x$ and time $t$, and $\lambda$ is the relation time, $f_i^{eq}=f_i^{eq}(P,\bm u)$ is the equilibrium distribution function.
	Based on the previous FDLBM \cite{Chen2020,Chen2021}, the two evolution equations of IFDLBM can be written as
	\begin{equation}\label{eq2.1}
		\hat{f}_i(\bm x,t+\Delta t)=\hat{f}^+_i(\bm x,t)-\Delta t \bm c_i\cdot \nabla f_i(\bm x,t+\frac{1}{2}\Delta t),
	\end{equation}
	and
	\begin{equation}\label{eq2.2}
		\bar{f}_i(\bm x,t+\frac{1}{2}\Delta t)=\bar{f}^+_i(\bm x,t)-\frac{1}{2}\Delta t \bm c_i\cdot \nabla \bar{f}^+_i(\bm x,t),
	\end{equation}
	where
	\begin{eqnarray}\label{eq2.3}
		&\hat{f}_i(\bm x,t)=f_i(\bm x,t)-\frac{1}{2}\Delta t(-\lambda(f_i(\bm x,t)-f_i^{eq}(\bm x,t))),\\
		&\hat{f}^+_i(\bm x,t)=f_i(\bm x,t)+\frac{1}{2}\Delta t(-\lambda(f_i(\bm x,t)-f_i^{eq}(\bm x,t))),
	\end{eqnarray}
	and
	\begin{eqnarray}\label{eq2.4}
		&\bar{f}_i(\bm x,t)=f_i(\bm x,t)-\frac{1}{4}\Delta t(-\lambda(f_i(\bm x,t)-f_i^{eq}(\bm x,t))),\\
		&\bar{f}^+_i(\bm x,t)=f_i(\bm x,t)+\frac{1}{4}\Delta t(-\lambda(f_i(\bm x,t)-f_i^{eq}(\bm x,t))).
	\end{eqnarray}
	The gradient terms $\nabla f_i$ and $\nabla \bar{f}^+_i$ can be discretized by a mixed difference scheme,
	\begin{equation}
		\nabla \Pi_j^*=\frac{\partial \Pi_j^*}{\partial \chi_\alpha} \Bigg{|}_m=\eta \frac{\partial \Pi_j^*}{\partial \chi_\alpha} \Bigg{|}_c+(1-\eta)\frac{\partial \Pi_j^*}{\partial \chi_\alpha} \Bigg{|}_u ,
		\label{eq2.5}
	\end{equation}
	where $\Pi_j^*$ represents $f_j$ or $\bar{f}_j^+$, and the parameter $\eta
	\in [0,1]$. The terms $\dfrac{\partial \Pi_j^*}{\partial \chi_\alpha}
	\Bigg{|}_u$ and $\dfrac{\partial \Pi_j^*}{\partial \chi_\alpha} \Bigg{|}_c$
	represent second up-wind difference and central-difference schemes, which can be expressed as
	\begin{subequations}
		\begin{equation}
			\frac{\partial \Pi_j^*}{\partial \chi_\alpha} \Bigg{|}_c=\frac{\Pi_j^*(\chi_\alpha+\Delta \chi_\alpha,t)-\Pi_j^*(\chi_\alpha-\Delta \chi_\alpha,t)}{2\Delta \chi_\alpha},
			\label{eq2.6}
		\end{equation}
		\begin{equation}
			\frac{\partial \Pi_j^*}{\partial \chi_\alpha} \Bigg{|}_u=
			\begin{cases}
				\dfrac{3\Pi_j^*(\chi_\alpha,t)-4\Pi_j^*(\chi_\alpha-\Delta \chi_\alpha,t)+\Pi_j^*(\chi_\alpha-2\Delta \chi_\alpha,t)}{2\Delta \chi_\alpha}, \quad &if \quad {c_{i\alpha} \geq 0},\\
				-\dfrac{3\Pi_j^*(\chi_\alpha,t)-4\Pi_j^*(\chi_\alpha+\Delta \chi_\alpha,t)+\Pi_j^*(\chi_\alpha+\Delta \chi_\alpha,t)}{2\Delta \chi_\alpha}, \quad &if \quad {c_{i\alpha}< 0}.
			\end{cases}
			\label{eq2.7}
		\end{equation}
	\end{subequations}
	Through the CE analysis in the Appendix, the equilibrium distribution function $f_i^{eq}(\bm x,t)$ can be designed as
	\begin{equation}\label{eq2.8}
		f_i^{eq}=\bar{\omega}_i\frac{P}{c_s^2}+\omega_i\rho_0+\omega_i[\frac{\bm c_i\cdot\bm u}{c_s^2}+\frac{\bm u\bm u:(\bm c_i\bm c_i-c_s^2\bm I)}{2c_s^2}],
	\end{equation}
	where $\bar{\omega}_i$ and $\omega_i$ are the weight coefficients determined by the discrete velocity model.
	And the calculation of macroscopic quantities $\bm u$ and $P$ are given by
	\begin{equation}\label{eq2.10}
		\bm u=\sum_i\bm c_if_i=\sum_i\bm c_i\bar{f}_i=\sum_i\bm c_i\hat{f}_i,
	\end{equation}
	and
	\begin{equation}\label{eq2.11}
		P=\frac{c_s^2}{\bar{\omega}_0}(-\sum_{i\neq0}\hat{f}_i+(1-\omega_0)\rho_0+\frac{\omega_0}{2c_s^2}\bm u\bm u)=\frac{c_s^2}{\bar{\omega}_0}(-\sum_{i\neq0}\bar{f}_i+(1-\omega_0)\rho_0+\frac{\omega_0}{2c_s^2}\bm u\bm u)
	\end{equation}
	
	In the present work, we take the D2Q9 lattice model for simulation.
	The discrete velocities in D2Q9 model can be expressed as
	\begin{subequations}
		\begin{equation}
			\textbf{c}_{j}=\left(
			\begin{matrix}
				0& 1 &0 &-1 &0 &1 &-1 &-1 &1\\
				0& 0& 1 &0 &-1  &1 &1 &-1 &-1\\
			\end{matrix}
			\right)c,\\
			\label{eq2.9}
		\end{equation}
		\begin{equation}
			\omega_0=4/9,\omega_{j=1-4}=1/9,\omega_{j=5-9}=1/36,
		\end{equation}
		\begin{equation}
			\bar{\omega}_0=-5/9,\bar{\omega}_{j=1-4}=1/9,\bar{\omega}_{j=5-9}=1/36.
		\end{equation}
	\end{subequations}
	
	For the Power-law fluid, the viscosity of non-Newtonian fluids depends on the strain rate tensor. Different from the traditional methods, the local nature of FDLBM makes it possible to calculate the strain rate tensor locally at each grid point, rather than estimate the velocity gradients. According to the Eq. \eqref{A11} in the Appendix, the
	strain rate tensor can be calculated by the secondary moments of $f^{ne}$,
	\begin{equation}\label{eq2.12}
		\nabla \bm u+\nabla\bm u^T=-\frac{1}{\lambda c_s^2}\sum \bm c_i\bm c_i f_i^{ne},
	\end{equation}
	where
	\begin{equation}\label{eq2.13}
		f^{ne}_i=f_i-f_i^{eq}.
	\end{equation}
	To simulate the power-law fluids, the relaxation time of FDLBM is related with the viscosity. Because the apparent viscosity of power-law fluid varies with position, the modified relaxation time $\lambda$ can be rewritten as
	\begin{equation}\label{eq2.14}
		\lambda=\mu/c_s^2,
	\end{equation}
	where the $\mu$ can be obtained by Eqs. \eqref{eq2.12}, \eqref{eq1.4} and \eqref{eq1.4.1}. Combing Eq. \eqref{eq1.4}, Eq. \eqref{eq2.12} and Eq. \eqref{eq2.14}, it can be conclude that $\lambda=\mu(\lambda)/c_s^2$, which means that the calculation of $\lambda$ is an implicit form. It can not be solved by analytical techniques, so $\mu$ is approximated with the $\lambda$ at the last time step in the simulation. For steady-state problem or practical applications without sacrificing sufficient accuracy, the equation becomes an explicit form and it can be applied efficiently.
	For the power-law fluid flows in the lid-driven cavity, Eq. \eqref{eq1.4} can be non-dimensionalized to produce the following dimensionless number analogous to the $Re$ number:
	\begin{equation}\label{eq2.15.1}
		Re=\frac{U^{2-n}L^n}{\mu},
	\end{equation}
	where $U$ is the maximum velocity in the cavity of height $L$.
	
	\section{Coordinate conversion and boundary processing}\label{LBMflow}
	One feature of the IFDLBM is geometric flexibility. First, the physical and computational domain are denoted by $(x,y)$ and $(\xi,\eta)$, respectively. And they satisfy the following relationship,
	\begin{equation}\label{eq3.1}
		\xi=\xi(x,y),\quad \eta=\eta(x,y).
	\end{equation}
	Then Eqs. \eqref{eq2.1} and \eqref{eq2.2} can be transformed into the generalized curvilinear coordinate. For the physical and computational domains, the following condition should be satisfied
	\begin{equation}\label{eq3.2}
		\left[
		\begin{matrix}
			\xi_x& \xi_y \\
			\eta_x& \eta_y\\
		\end{matrix}
		\right]=\frac{1}{J}\left[
		\begin{matrix}
			y_\eta& -x_\eta \\
			-y_\xi & x_\xi\\
		\end{matrix}
		\right],\\
	\end{equation}
	where $J$ is the transformation Jacobian matrix, it can be obtained by
	\begin{equation}\label{eq3.3}
		J=x_\xi y_\eta-x_\eta y_\xi.
	\end{equation}
	According to the chain rule, the gradient term of Eqs. \eqref{eq2.1} and \eqref{eq2.2} can be rewritten as
	\begin{eqnarray}\label{eq3.4}
		\bm c_i\cdot \nabla f_i & &=c_{ix}\frac{\partial f_i}{\partial x}+c_{iy}\frac{\partial f_i}{\partial y}\nonumber\\
		& &=c_{ix}\left(\frac{\partial f_i}{\partial \xi}\xi_x +\frac{\partial f_i}{\partial \eta}\eta_x\right)+c_{iy}\left(\frac{\partial f_i}{\partial\xi}\xi_y+\frac{\partial f_i}{\partial \eta}\eta_y\right)\nonumber\\
		& &=(c_{ix}\xi_x+c_{iy}\xi_y)\frac{\partial f_i}{\partial\xi}+(c_{ix}\eta_x+c_{iy}\eta_y)\frac{\partial f_i}{\partial\eta}\nonumber\\
		& &=c_{i\xi}\frac{\partial f_i}{\partial\xi}+c_{i\eta}\frac{\partial f_i}{\partial \eta},
	\end{eqnarray}
	\begin{equation}\label{eq3.5}
		\bm c_i\cdot \nabla \bar{f}^+_i=c_{i\xi}\frac{\partial \bar{f}^+_i}{\partial\xi}+c_{i\eta}\frac{\partial \bar{f}^+_i}{\partial \eta}.
	\end{equation}
	In the computational domain, $\bm c_i=(c_{i\xi},c_{i\eta})=(c_{ix}\xi_x+c_{iy}\xi_y,c_{ix}\eta_x+c_{iy}\eta_y)$ is
	the microscopic contravariant velocity vectors. Then, the evolution equations \eqref{eq2.1}
	and \eqref{eq2.2} can be rewritten as
	\begin{equation}\label{eq3.6}
		\hat{f}_i(\bm x,t+\Delta t)=\hat{f}^+_i(\bm x,t)-\Delta t ( c_{i\xi}\frac{\partial f_i}{\partial\xi}+c_{i\eta}\frac{\partial f_i}{\partial\eta})(\bm x,t+\frac{1}{2}\Delta t),
	\end{equation}
	\begin{equation}\label{eq3.7}
		\bar{f}_i(\bm x,t+\frac{1}{2}\Delta t)=\bar{f}^+_i(\bm x,t)-\frac{1}{2}\Delta t ( c_{i\xi}\frac{\partial \bar{f}^+_i}{\partial\xi}+c_{i\eta}\frac{\partial \bar{f}^+_i}{\partial\eta})(\bm x,t).
	\end{equation}
	It should be noted that the velocity vectors $(c_{ix},c_{iy})$ are constant on the physical domain, however the transformed velocity $(c_{i\xi},c_{i\eta})$ are not constant in the computational plane, they change with the position.
	
	For the isosceles trapezoidal cavity, the general relationship between the $(\xi,\eta)$ with $(x,y)$ can be given by,
	\begin{equation}\label{eq3.8.1}
		x=\frac{2}{\tan \theta}\xi\eta+\xi-\frac{1}{\tan \theta}\eta+\frac{1}{\tan \theta}, \quad y=\eta,
	\end{equation}
	and the derivative can be expressed as,
	\begin{eqnarray}\label{eq3.8.2}
		&x_\xi=\frac{2}{\tan \theta}\eta+1, \quad x_\eta=\frac{2}{\tan \theta}\xi-\frac{1}{\tan \theta}.\nonumber\\
		&y_\xi=0,\quad y_\eta=1.
	\end{eqnarray}
	
\begin{figure}[htbp]
		\centering
		\includegraphics[scale=0.4]{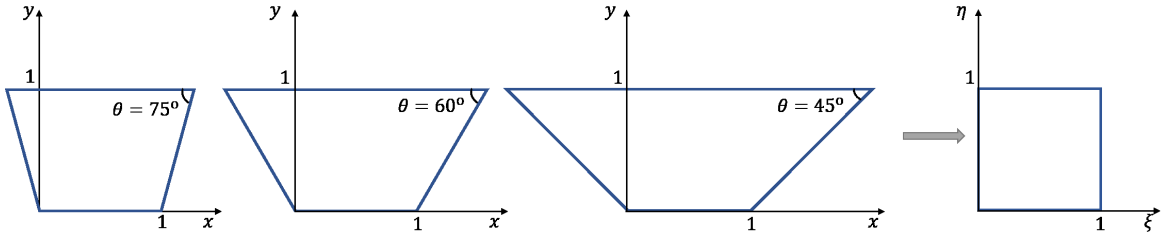}
		\caption{The physical domain transport to the computational domain.}\label{fig:transform}
	\end{figure}

	In the simulation, three physical conditions are taken into account. The bottom and height of the three types of trapezoidal cavities are equal to 1, and the top angles of the trapezoid are (i)case 1: $\theta=75^\circ$, $\tan\theta=2+\sqrt{3}$,
	(ii)case 2: $\theta=60^\circ$, $\tan\theta=\sqrt{3}$,
	(iii)case 3: $\theta=45^\circ$, $\tan\theta=1$. Three cases of isosceles trapezoid physical domains can be transported into the square computational domain by Eq. \eqref{eq3.8.1}, as shown in Fig. \ref{fig:transform}.

	In any numerical technique, the boundary information plays an important role. Thanks to the coordinate conversion, the standard boundary scheme can be applied for numerical simulation. In this work, the boundary conditions are treated by the non-equilibrium extrapolation scheme, which could keep the accuracy of the IFDLBM.

\begin{figure}[htbp]
		\centering
		\includegraphics[scale=0.35]{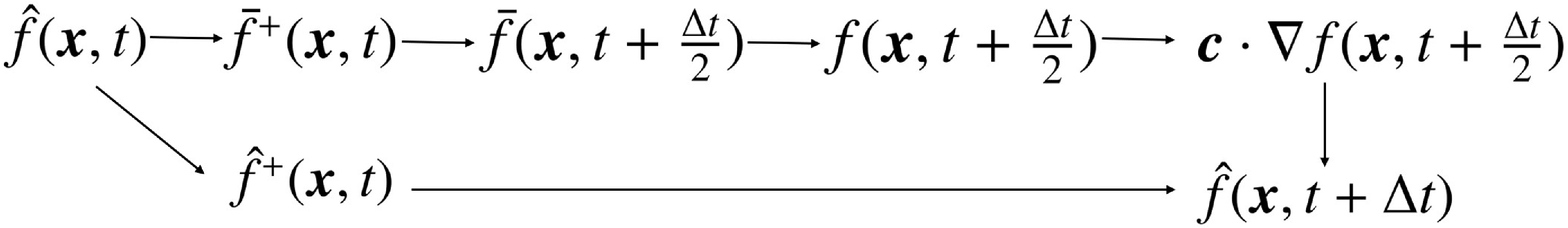}
		\caption{The evolution process of FDLBM} \label{fig:process}
	\end{figure}

	Now, we present the evolution process in Fig. \ref{fig:process}, and list the calculational process as follows:
	
	Step(1): initialize the fluid velocity and pressure, initialize the distribution function $f_i(\bm x,t)$ and $\hat{f}_i(\bm x, t)$ by $f_i^{eq}(\bm x, t)$, and calculate the transformed velocity $(c_{i\xi},c_{i\eta})$.
	
	Step(2): calculate the relaxation time $\lambda$ by $f_i^{(1)}(\bm x,t)$.
	
	Step(3): calculate $\hat{f}^+_i(\bm x,t)$ by Eq. \eqref{eq2.15},
	\begin{equation}\label{eq2.15}
		\hat{f}^+_i=\frac{2\lambda-\Delta t}{2\lambda+\Delta t}\hat{f}_i+\frac{2\Delta t}{2\lambda+\Delta t}f_i^{eq}.
	\end{equation}
	
	Step(4): calculate $\bar{f}^+_i(\bm x,t)$ by Eq. \eqref{eq2.16}, obtain the $\bar{f}_i(\bm x, t+\frac{\Delta t}{2})$ through the evolution equation \eqref{eq2.2}. Then calculate the distribution function $f_i(\bm x, t+\frac{\Delta t}{2})$ by Eq. \eqref{eq2.17}, and evaluate spatial term $\bm c_i\cdot \nabla f_i(\bm x, t+\frac{\Delta t}{2})$ by Eq. \eqref{eq2.5},
	\begin{equation}\label{eq2.16}
		\bar{f}_i^+=\frac{4\lambda-\Delta t}{4\lambda+2\Delta t}\hat{f}_i+\frac{3\Delta t}{4\lambda+2\Delta t}f_i^{eq},
	\end{equation}
	\begin{equation}\label{eq2.17}
		f_i=\frac{4\lambda}{4\lambda+\Delta t}\bar{f}_i+\frac{\Delta t}{4\lambda+\Delta t}f_i^{eq},
	\end{equation}
	
	Step(5): update the distribution function $\hat{f}_i(\bm x, t+\Delta t)$ by Eq. \eqref{eq2.1}.
	
	Step(6): calculate the macroscopic quantities $\bm u,P$ of fluid, and compute the boundary populations.
	
	Step(7): calculate the distribution function $f_i(\bm x,t+\Delta t)$ by Eq.  \eqref{eq2.18}.
	\begin{equation}\label{eq2.18}
		f_i=\frac{2\lambda}{2\lambda+\Delta t}\hat{f}_i+\frac{\Delta t}{2\lambda+\Delta t}f_i^{eq}.
	\end{equation}
	
	Step(8): calculate the global relative error (GRE) of $\bm u$ and $P$ by Eq. \eqref{eq2.19}, if GRE is more than $10^{-9}$, increment the time step and go back to step(2).
	
	\begin{equation}\label{eq2.19}
		E(\bm u)=\frac{\sqrt{\sum_{ij}|u_{i,j}(t_n)-u_{i,j}(t_{n-1})|^2}}{\sqrt{\sum_{ij}|u_{i,j}(t_n)|^2}},\quad E(P)=\frac{\sqrt{\sum_{ij}(P_{i,j}(t_n)-P_{i,j}(t_{n-1}))^2}}{\sqrt{\sum_{ij}P^2_{i,j}(t_n)}}.
	\end{equation}
	
	\section{Code validation and grid independence}\label{Numerical}
	
	\subsection{Lid-driven of power-law fluid in the square enclosure}\label{cylinder}
\begin{figure}[htbp]\centering
		\subfigure[]{ \label{fig:square-v_x}
			\includegraphics[scale=0.45]{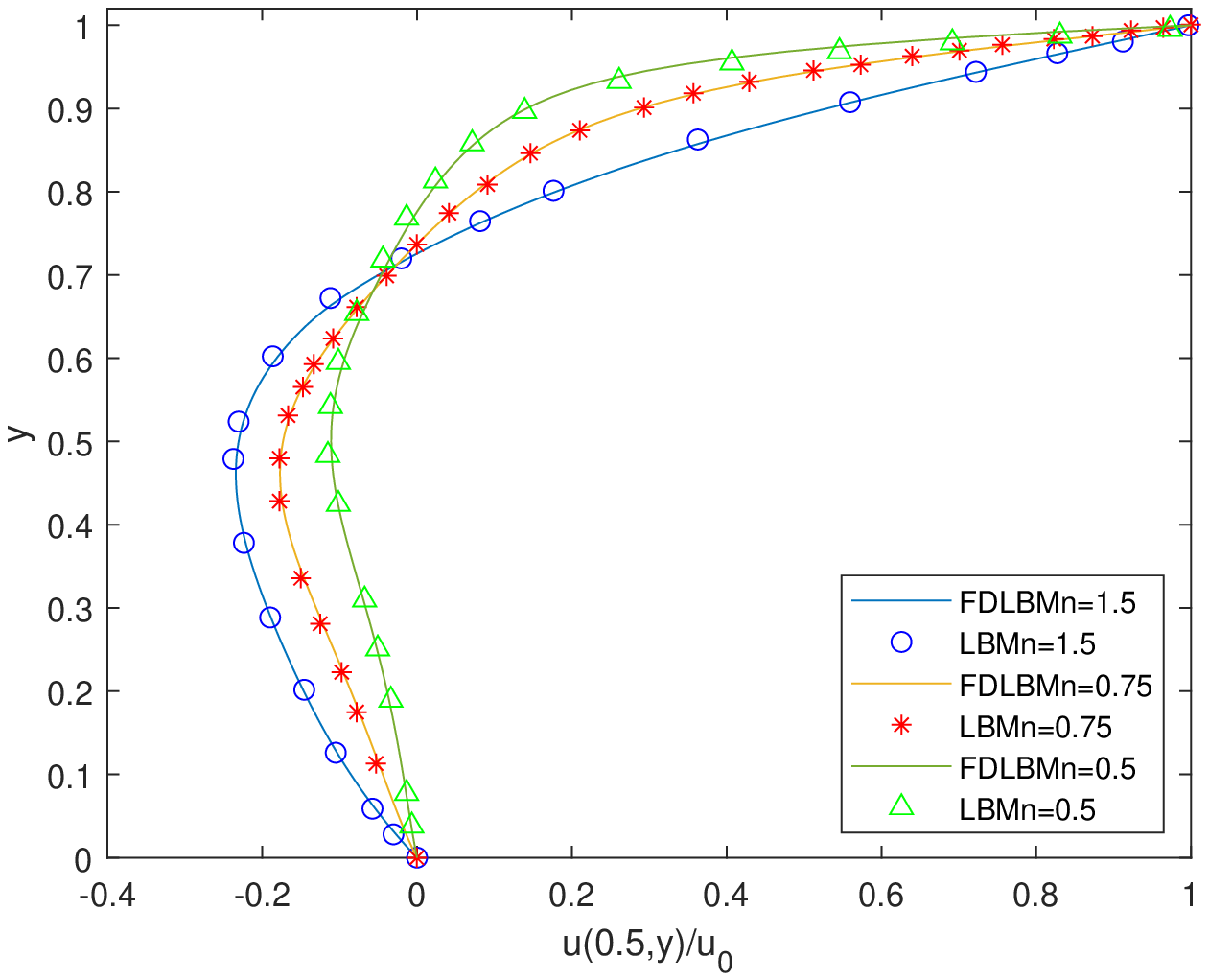}}
		\subfigure[]{ \label{fig:square-u_y}
			\includegraphics[scale=0.45]{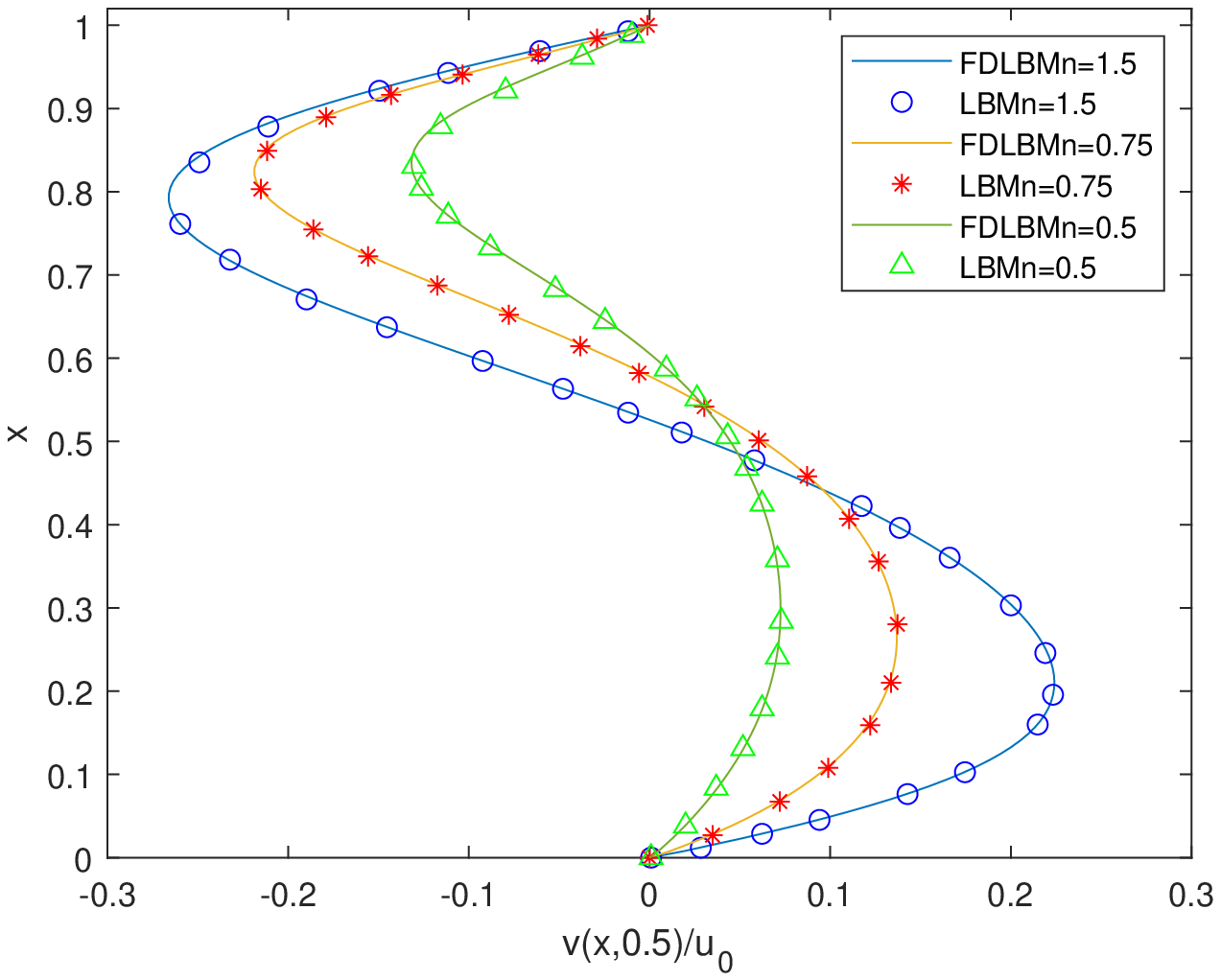}}
		\caption{Comparison between present results and those reported in some previous
			works for Power-law fluid (Re = 100); (a) the velocity component u along the vertical
			centreline of the cavity, (b) the component velocity v along the horizontal
			centreline of the cavity.}
		\label{fig:square}
	\end{figure}

	To verify the accuracy of the IFDLBM,
	we simulate the lid-driven of power-law fluid in the square enclosure.
	In the simulation, we take the grid $(128\times 128)$ for $Re=100$, and $c=1$. The Courant-Friedrichs-Lewy (CFL) condition number is set to be $0.5$. The lid velocity is set to be $0.1$ in initial conditions and boundary condition. Besides, the power-law index $n$ is taken as $0.5,0.75,1.5$. The numerical results are shown in the Fig. \ref{fig:square}. It is obvious that the numerical results of IFDLBM are in great agreement with the LBM \cite{chai2011non}. It can be indicated that the local computational scheme of viscosity is effective.
	
	\subsection{Lid-driven flow in the trapezoidal enclosure}\label{thermal}
\begin{figure}[htbp]\centering
		\subfigure[]{ \label{fig:tra-x-v-100}
			\includegraphics[scale=0.45]{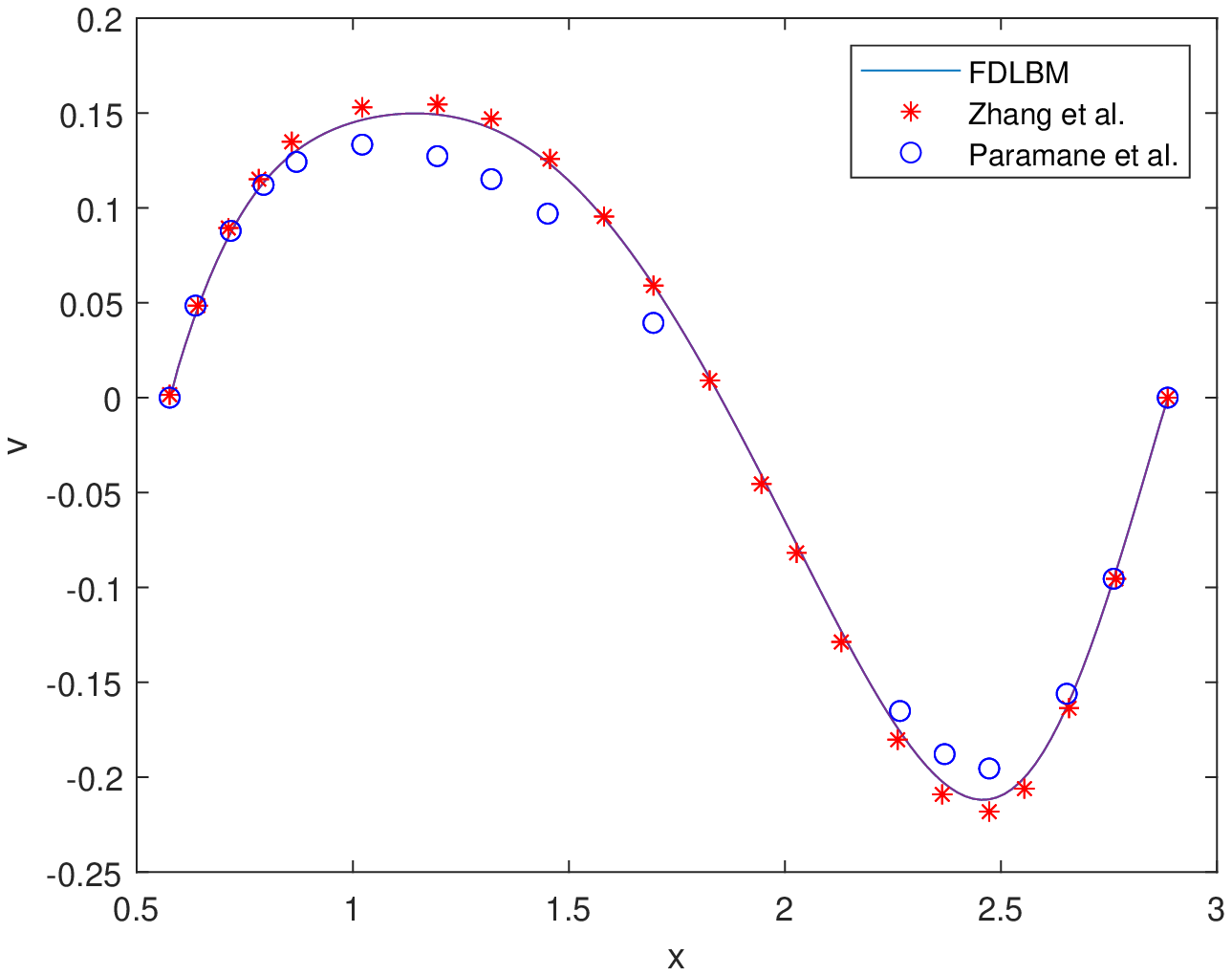}}
		\subfigure[]{ \label{fig:tra-y-u-100}
			\includegraphics[scale=0.45]{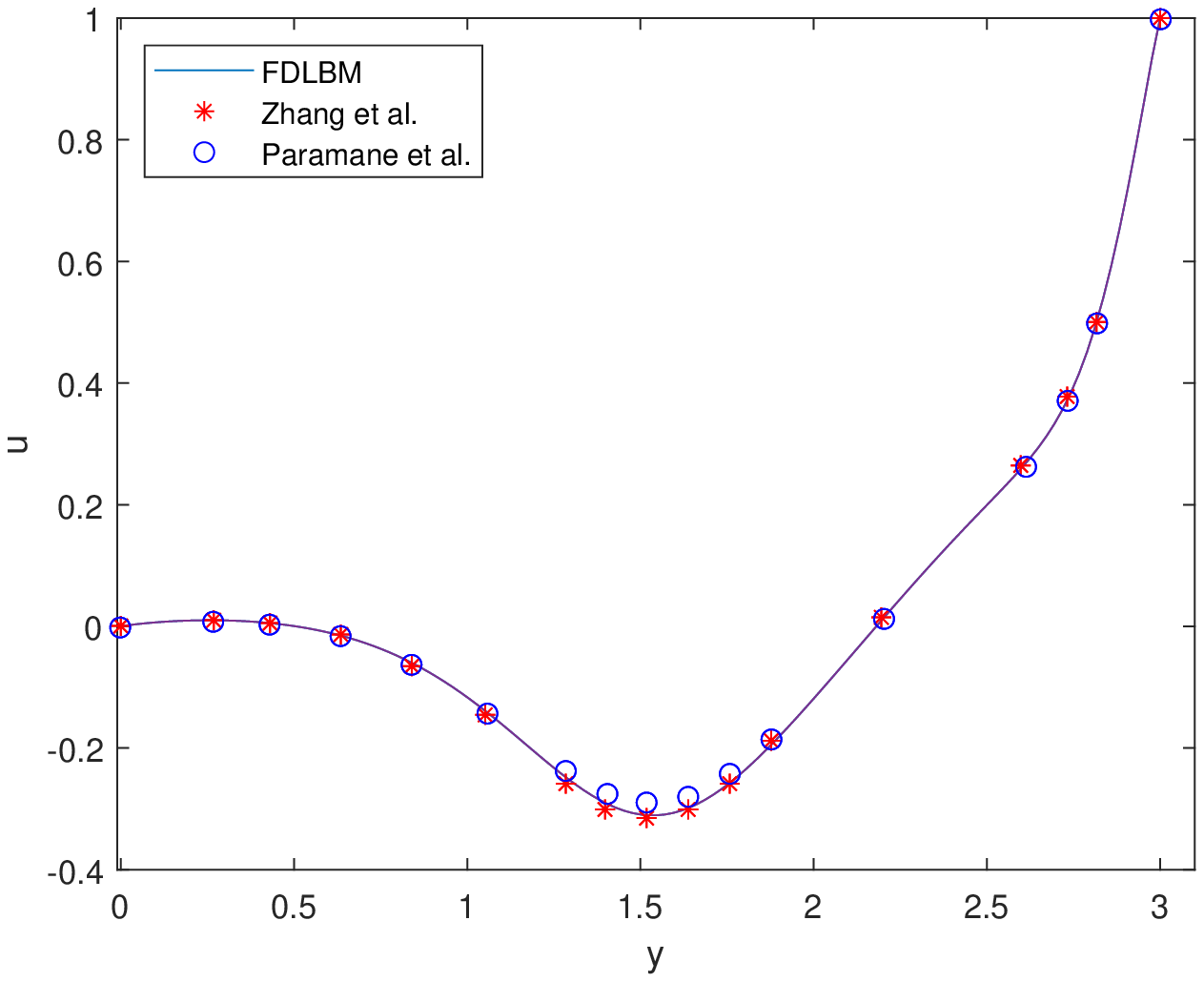}}
		\caption{Comparison between present results and those reported in some previous
			works for Power-law fluid (Re = 100); (a) the component velocity v along the horizontal centreline of the cavity, (b)the velocity component u along the vertical centreline of the cavity. }
		\label{fig:tra-100}
	\end{figure}

\begin{figure}[htbp]
		\subfigure[]{ \label{fig:tra-x-v-500}
			\includegraphics[scale=0.45]{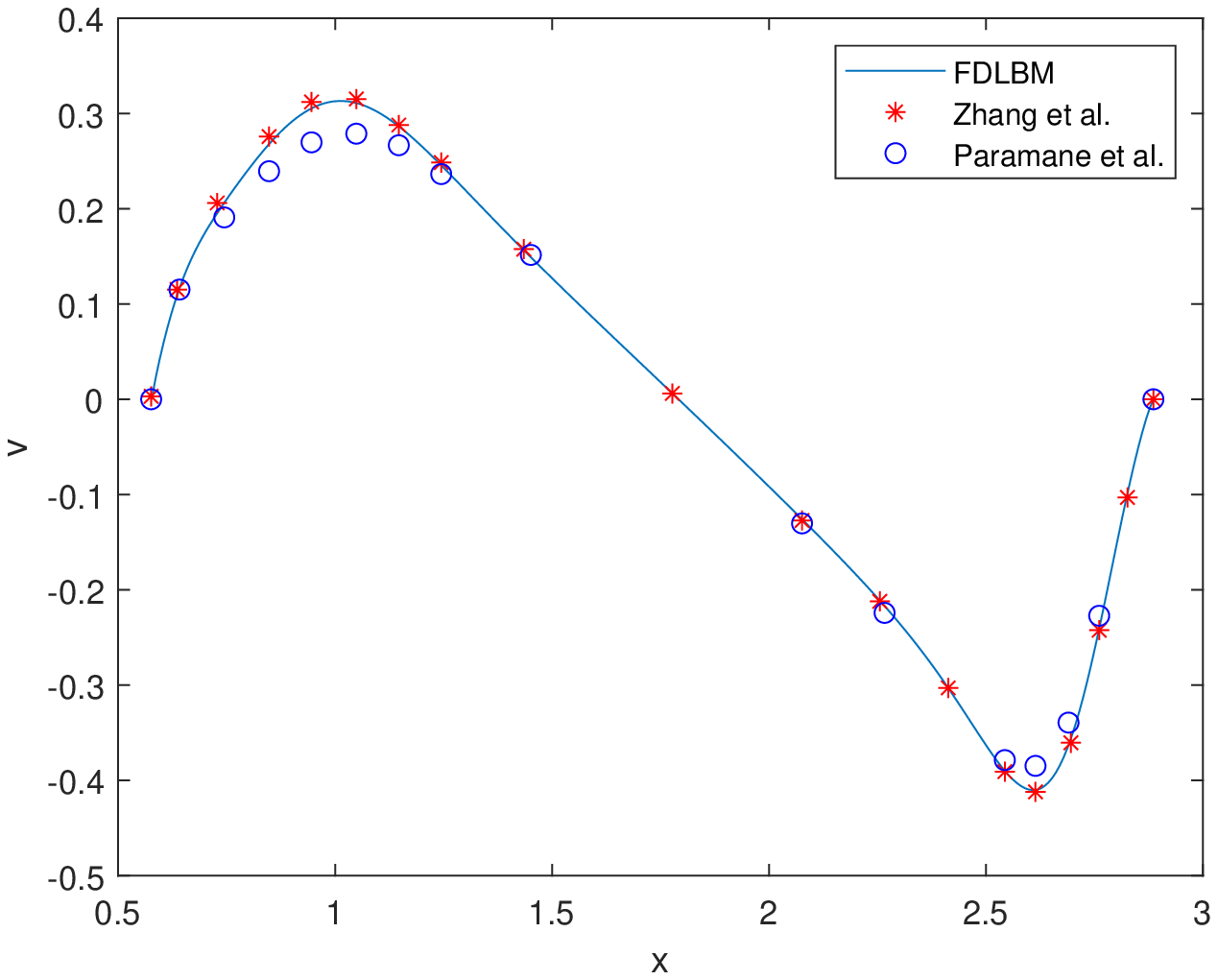}}
		\subfigure[]{ \label{fig:tra-y-u-500}
			\includegraphics[scale=0.45]{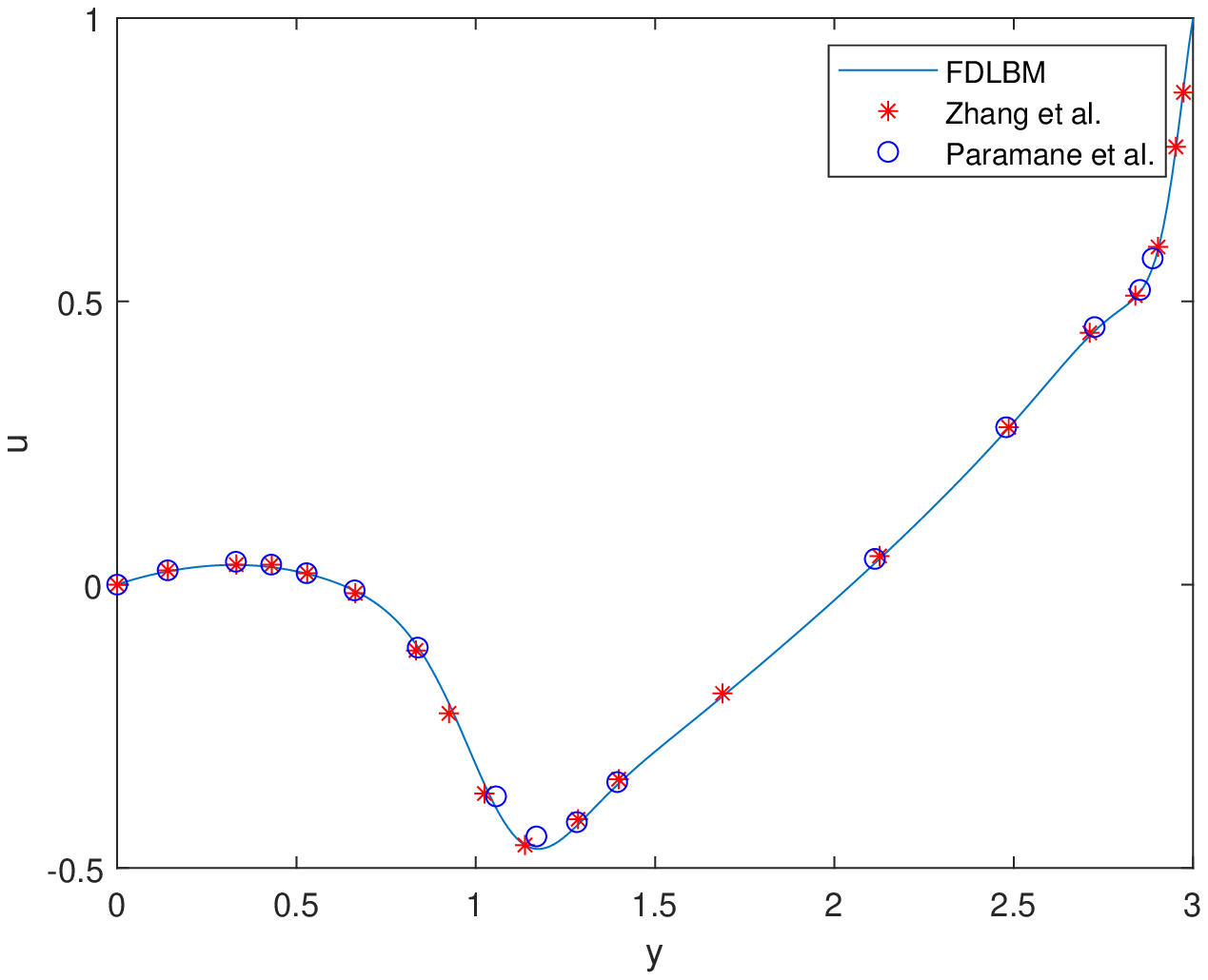}}
		\caption{Comparison between present results and those reported in some previous
			works for Power-law fluid (Re = 500); (a) the component velocity v along the horizontal centreline of the cavity, (b)the velocity component u along the vertical centreline of the cavity. }
		\label{fig:tra-500}
	\end{figure}

	In this subsection, we will adopt the coordinate conversion method to simulate the lid-driven flow in the trapezoidal enclosure. In order to verify the correctness of IFDLBM, the trapezoid size is consistent with that in the Ref.\cite{Zhang2010}. Different from Ref.\cite{Zhang2010}, we take the grid ($128\times128$) for $Re=100$ and the grids $(256\times256)$ for $Re=500$. And $u_0=0.1$, $c=1.0$ and $CFL=0.5$. The power-law index $n$ is taken as $1$, because only the Newtonian fluid has been studied in Ref.\cite{Zhang2010}. As shown in Fig. \ref{fig:tra-100} and \ref{fig:tra-500}, it can be found that the numerical results of IFDLBM agree well with previous work \cite{Zhang2010,PS2008}. It indicates the coordinate conversion method is valid and the IFDLBM can accurately simulate the flow in the trapezoidal enclosure.
	
\begin{figure}[htbp]
		\centering
		\includegraphics[scale=0.5]{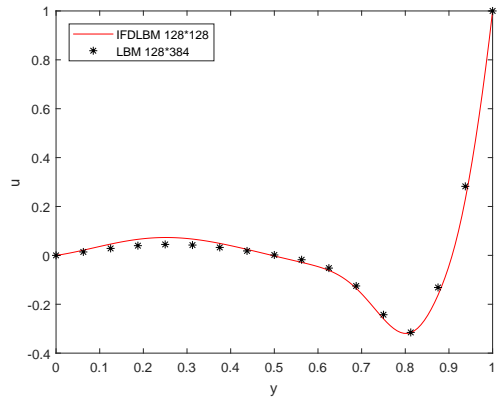}
		\caption{ Vertical component of velocity along $y$-axis for $Re=1200$ and $\theta=45^o$.} \label{fig:stable}
	\end{figure}

\begin{figure}[htbp]
		\centering
		\includegraphics[scale=0.6]{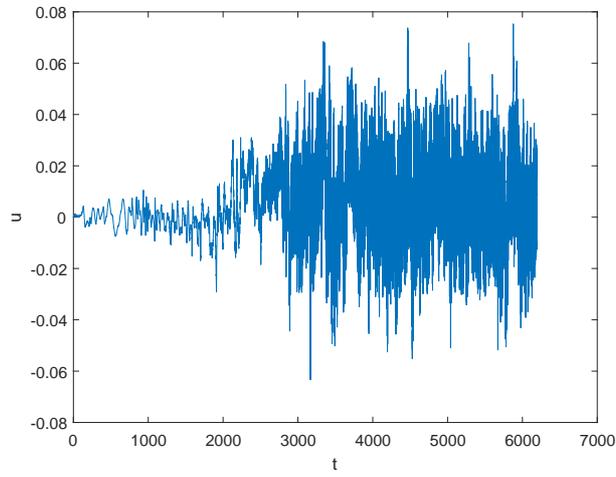}
		\caption{\textcolor{blue}{ The velocity $u$ at point $(1,0.5)$ for $Re=50000$ and $\theta=45^o$.}} \label{fig:ut_5w}
	\end{figure}

\begin{figure}[htbp]
		\centering
		\includegraphics[scale=0.5]{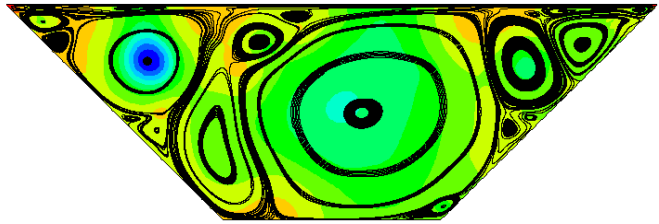}
		\caption{\textcolor{blue}{ Streamline plots at $Re=50000$ and $\theta=45^o$.}} \label{fig:5w}
	\end{figure}

	In order to verify the numerical stability of IFDLBM, some numerical simulations are carried out.
	Considering the case of $\theta=45^o$ in Fig. \ref{fig:transform} and $Re=1200$, the grid is fixed as $128*128$ for IFDLBM and $128*384$ for LBM \cite{Zhang2010}.
	The velocity of center line along $y$-axis are shown in Fig. \ref{fig:stable}.
	We can find the simulation results of the two numerical methods are in good agreement.
	\textcolor{blue}{However, if the $Re$ number increases, the LBM will be divergent when $Re>9000$, but the IFDLBM can still work well even if $Re=50000$.
		The simulation results for $Re=50000$ are shown in Figs. \ref{fig:ut_5w} and \ref{fig:5w}. The change of velocity $u$ at point $(1,0.5)$ is displayed in Fig. \ref{fig:ut_5w}. The streamline diagram of the fluid is shown in Fig. \ref{fig:5w}. It can be observed that the state of fluid transforms into turbulence. There are $20$ vortices in the trapezoidal cavity, which is much greater than in the case of $Re=500$, and the top vortex is severely squeezed by the main vortex.}
	These results indicate that IFDLBM has better numerical stability.
	
	\subsection{Grid independence test}
\begin{figure}[htbp]\centering
		\subfigure[Re=100]{ \label{fig:uy-100}
			\includegraphics[scale=0.3]{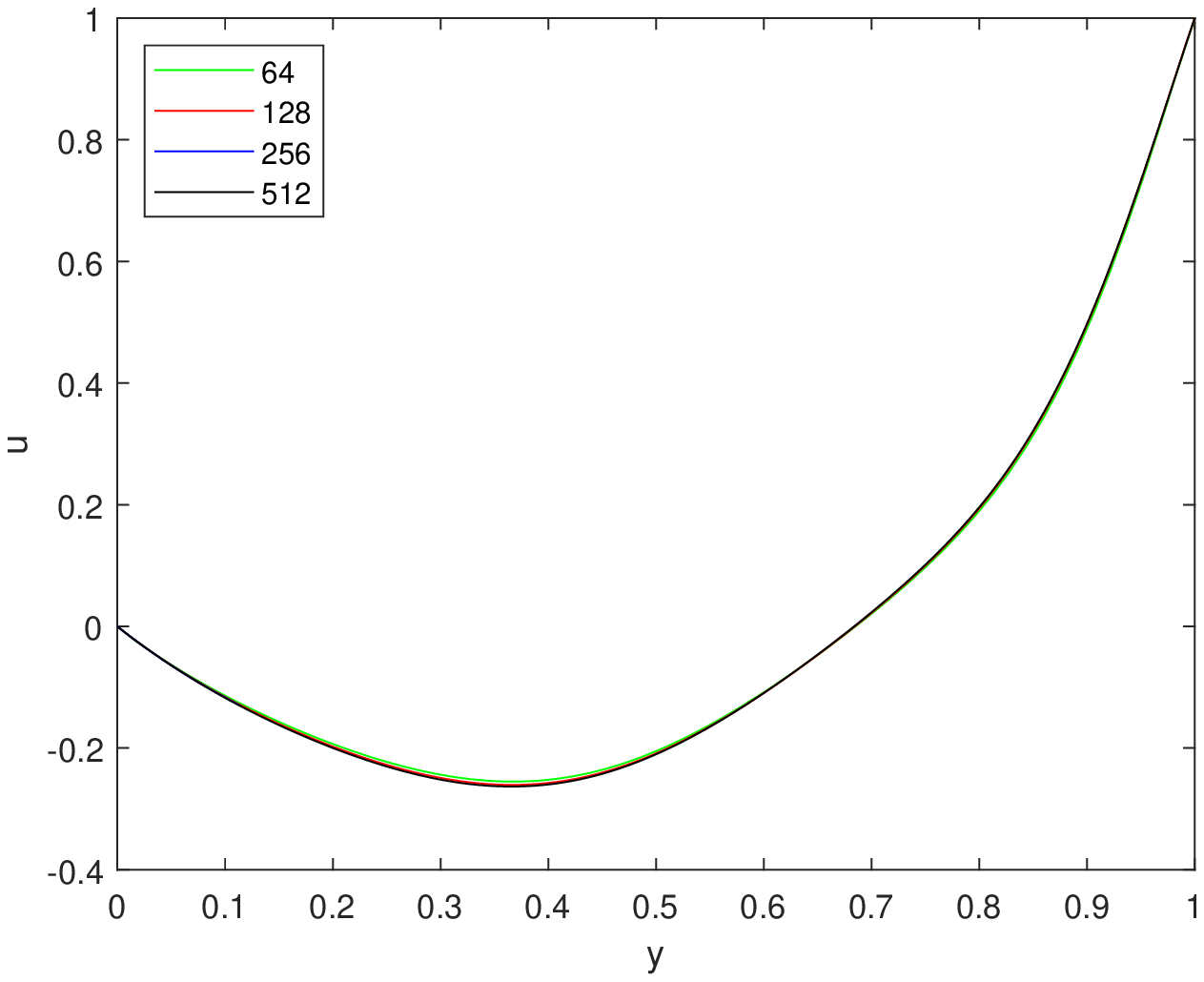}}
		\subfigure[Re=500]{ \label{fig:uy-500}
			\includegraphics[scale=0.3]{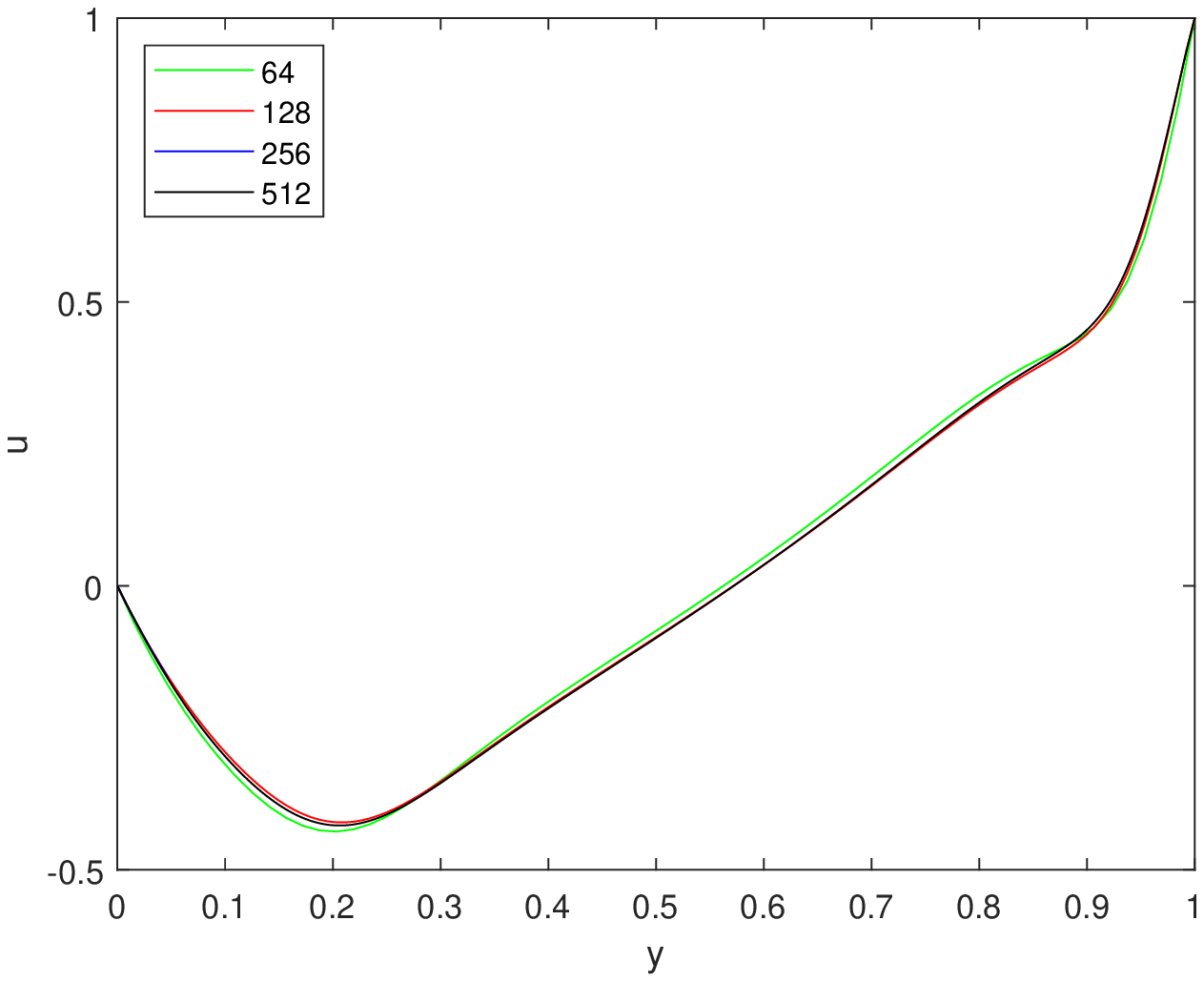}}
		\subfigure[Re=1000]{ \label{fig:uy-1000}
			\includegraphics[scale=0.3]{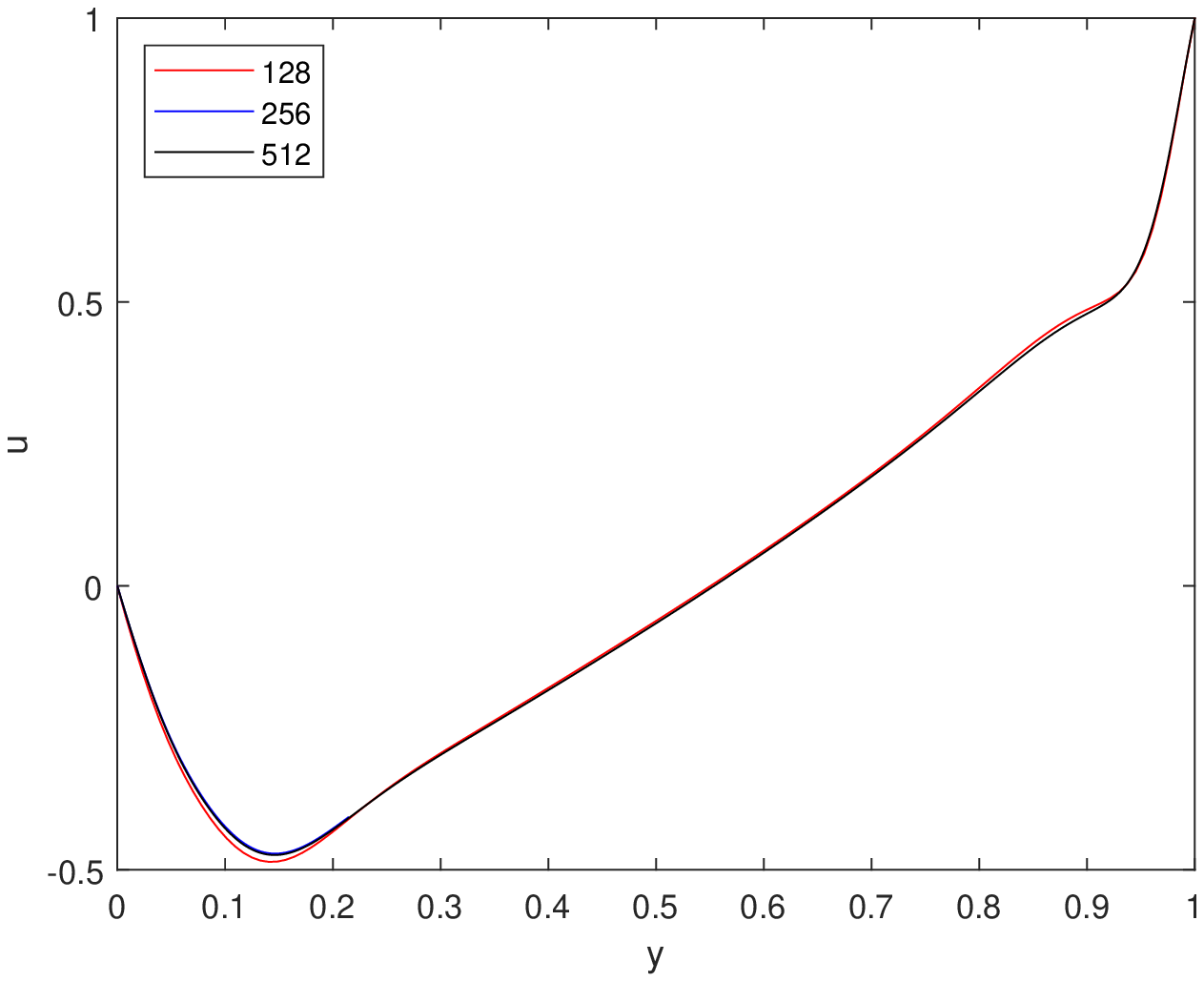}}
		\caption{ Vertical component of velocity along $y$-axis for different grid sizes; (a) Re=100, (b)Re=500, (c)Re=1000. }
		\label{fig:grid-size}
	\end{figure}

\begin{figure}[htbp]\centering
		\subfigure[]{ \label{fig:uy-local-100}
			\includegraphics[scale=0.3]{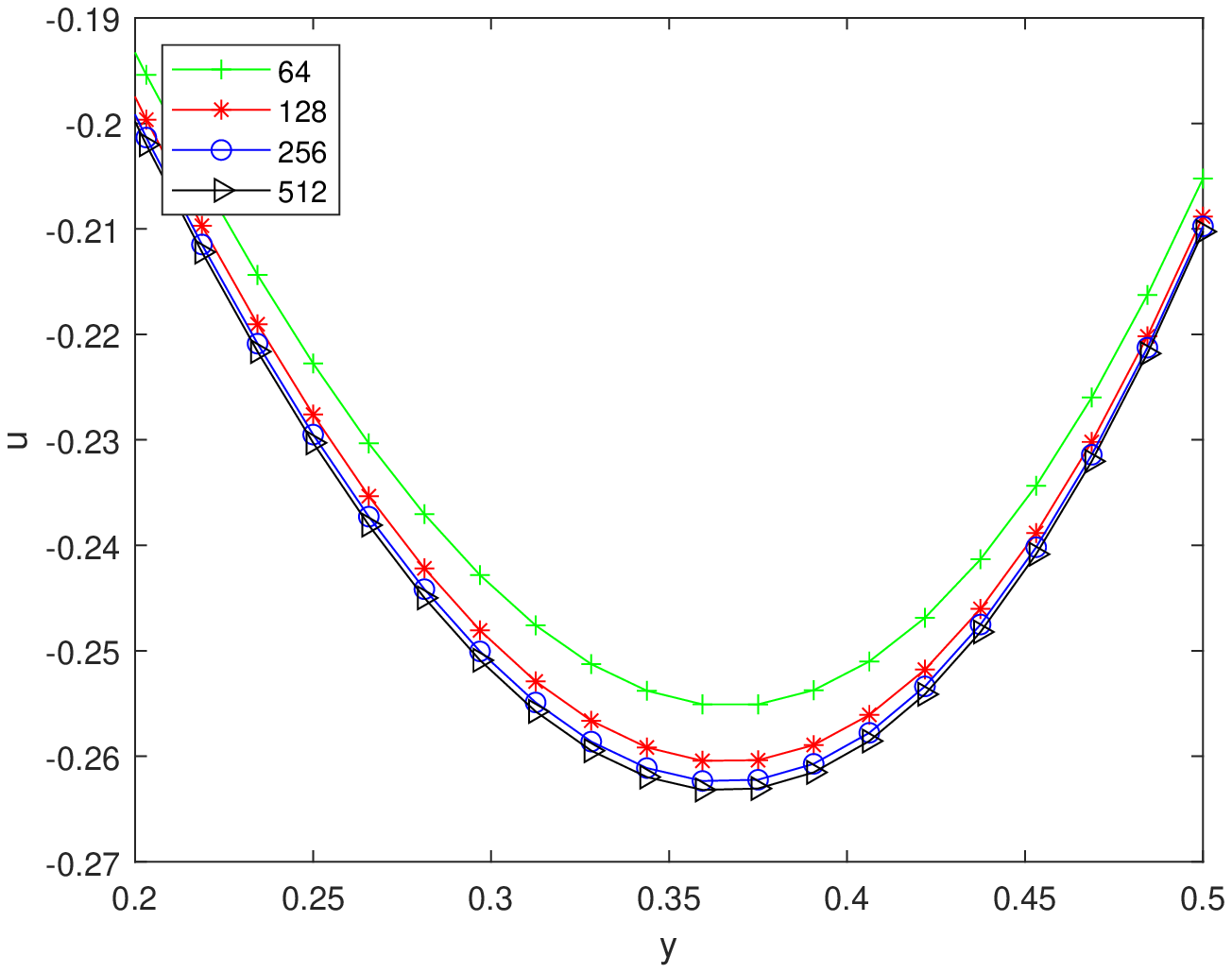}}
		\subfigure[]{ \label{fig:uy-local-500}
			\includegraphics[scale=0.3]{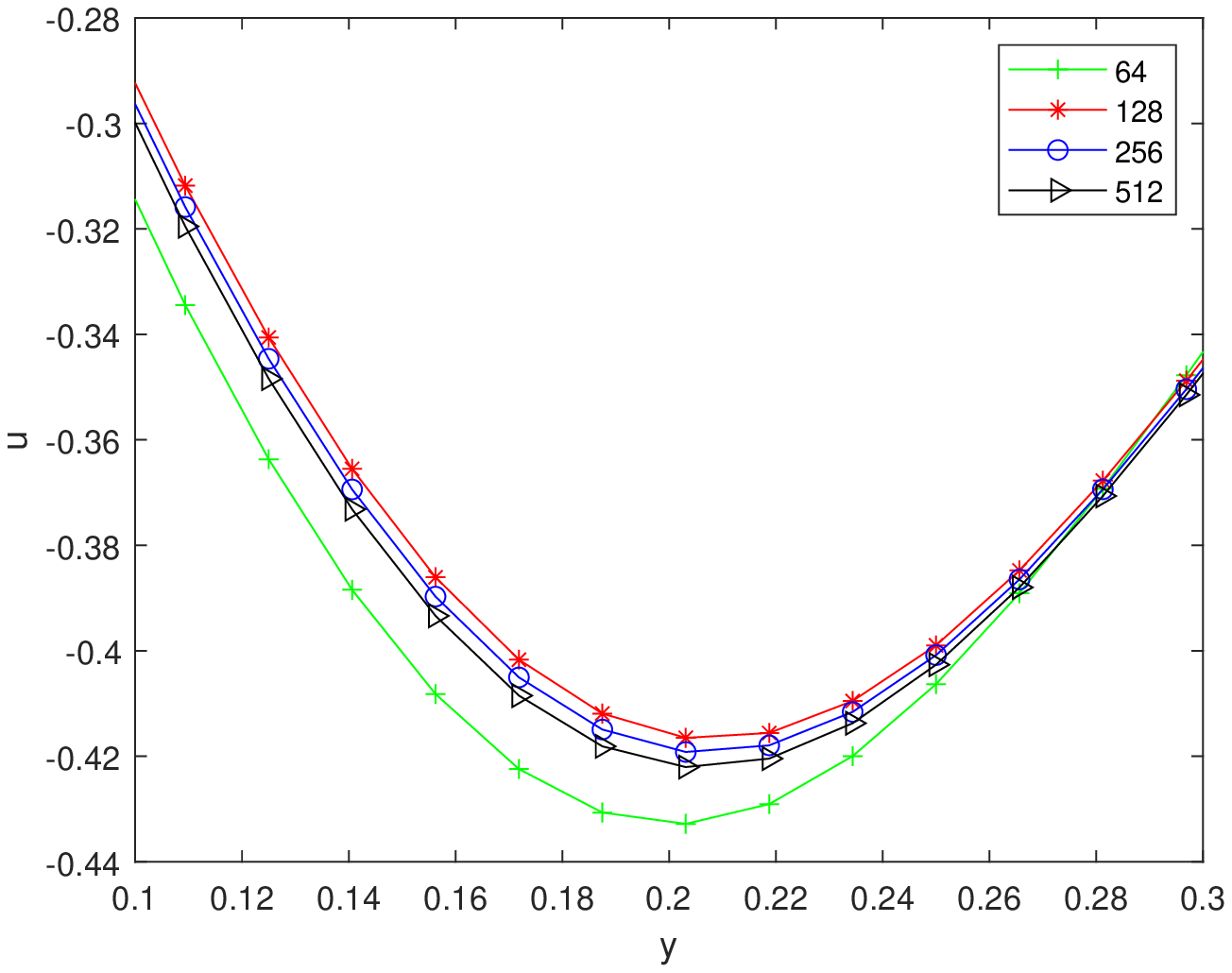}}
		\subfigure[]{ \label{fig:uy-local-1000}
			\includegraphics[scale=0.3]{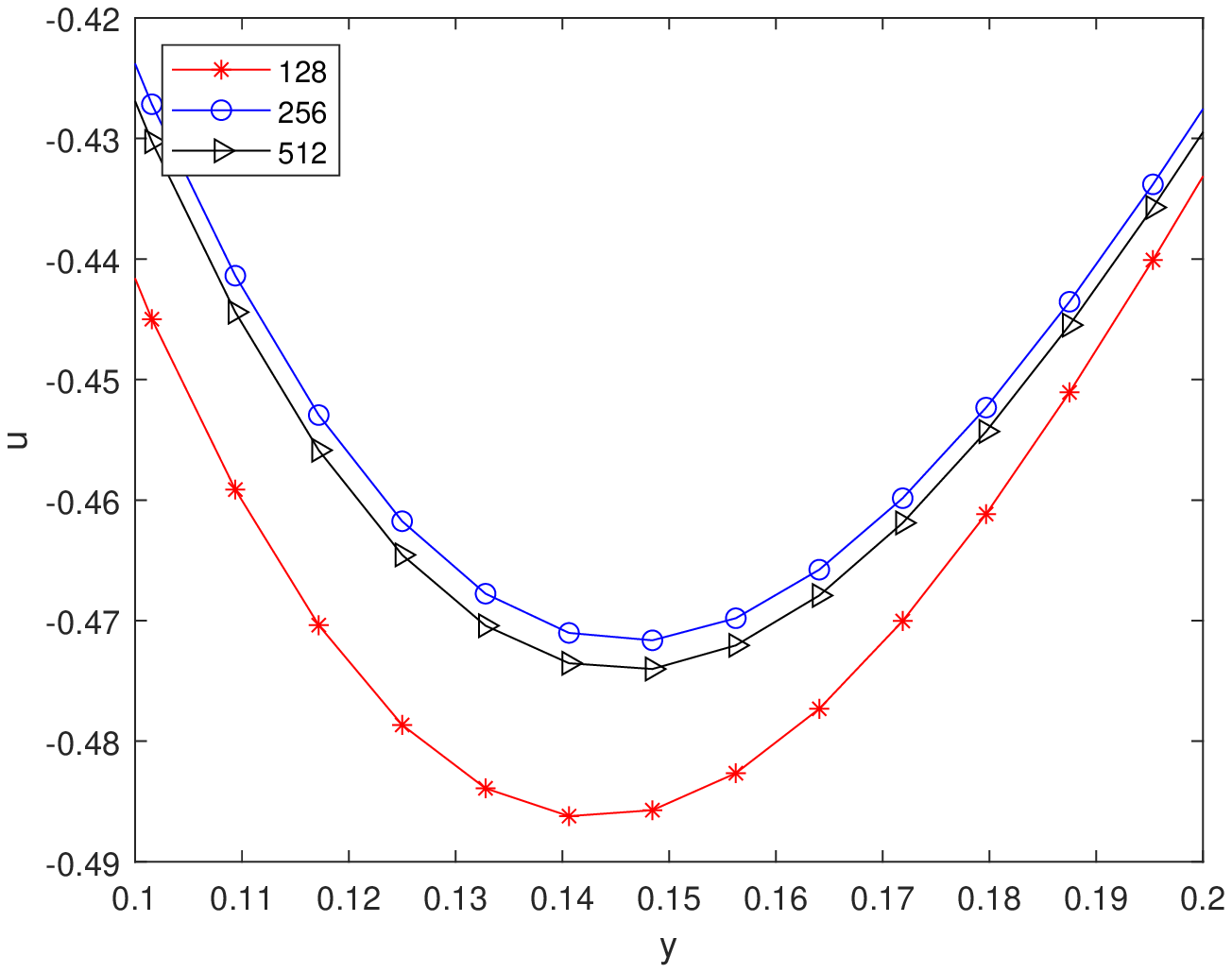}}
		\caption{ Vertical component of velocity along local $y$-axis for different grid sizes; (a) Re=100, (b)Re=500, (c)Re=1000. }
		\label{fig:grid-siz-locale}
	\end{figure}

	It is important to implement the grid independence analysis to confirm that the numerical technique is independent on the grid. To assess the effectiveness of the grid size on the method, we adopt four grid systems for $Re=100,500,1000$, i.e. $64\times 64$, $128\times 128$, $256\times 256$, $512\times 512$, respectively. The effects of different grids have been displayed in the Figs. \ref{fig:grid-size}. To select the suitable grid size, we magnify the local velocity $u$ in Fig. \ref{fig:grid-siz-locale}. It can been seen that the numerical results converges rapidly toward the results of grid $512\times512$ as the number of grid nodes increases. Since the numerical results with $256\times 256$ are nearest to the results with $512\times 512$.
	Considering the calculation efficiency and accuracy, the gird of $256\times256$ is adequate to simulate the problem.
	
	\section{Results and discussion}
	In this section, we are going to analyze the relationship between the behavior of power-law fluid with $Re$ number, angle $\theta$ and power-law index $n$. In order to study the rheological behavior effectively, two cases are taken into account, they are low $Re$ number condition ($Re=100$) and high $Re$ number condition $Re\geq 500$. To seek the law of TC flow, the power-law index $n$ will be changed from $0.5$ to $1.5$ and the angle $\theta$ will be varied from $45^o$ to $75^o$ for the two cases.
	
	\subsection{Low $Re$ number}
	
	(i) Effect of power-law index $n$ on the development of flow for low $Re$ number
	
\begin{figure}[htbp]\centering
		\subfigure[]{ \label{fig:75-100-1.5}
			\includegraphics[scale=0.45]{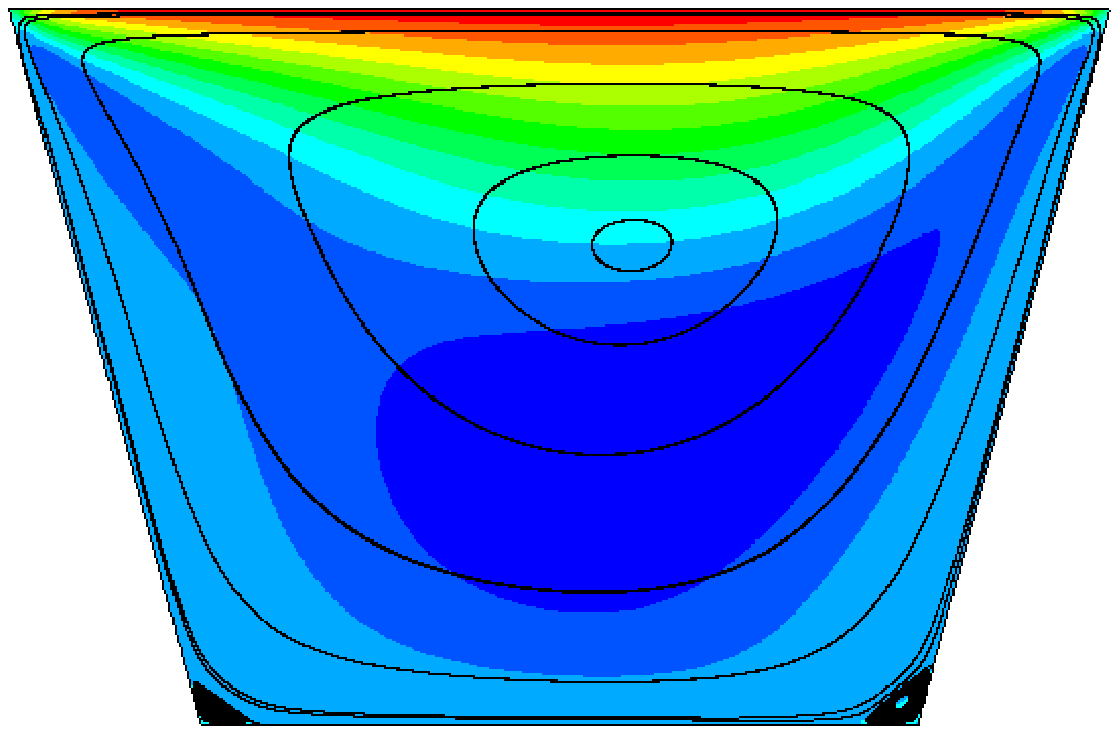}}
		\subfigure[]{ \label{fig:75-100-1.0}
			\includegraphics[scale=0.45]{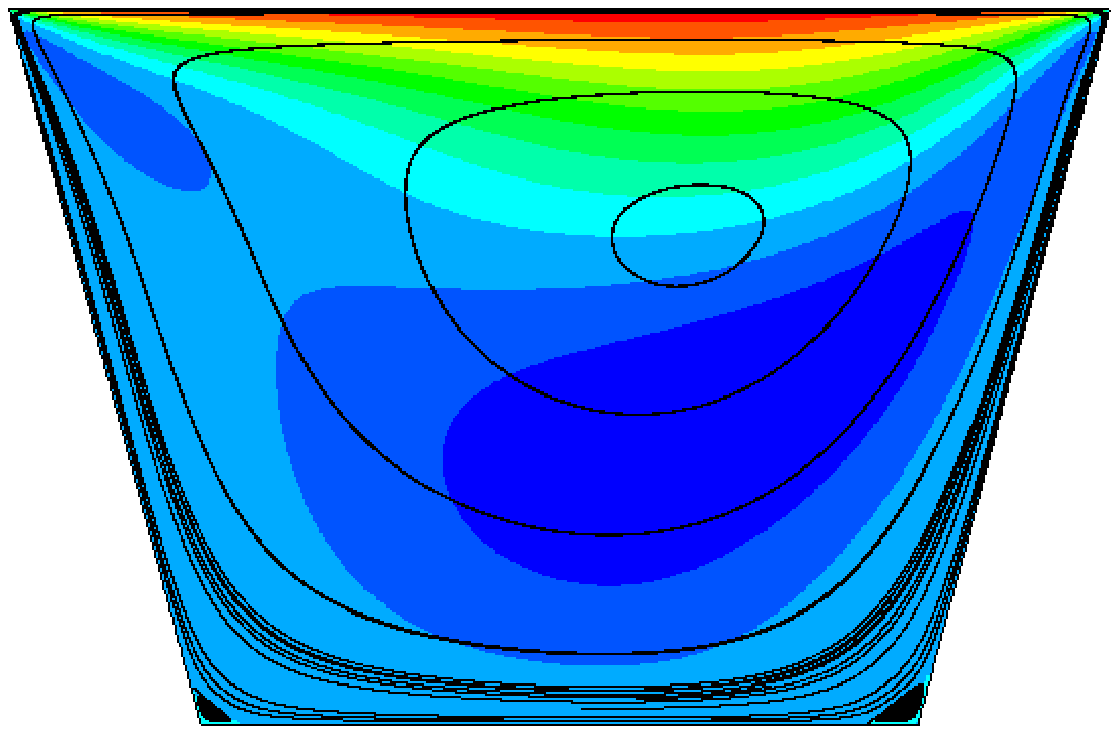}}
		\subfigure[]{ \label{fig:75-100-0.75}
			\includegraphics[scale=0.45]{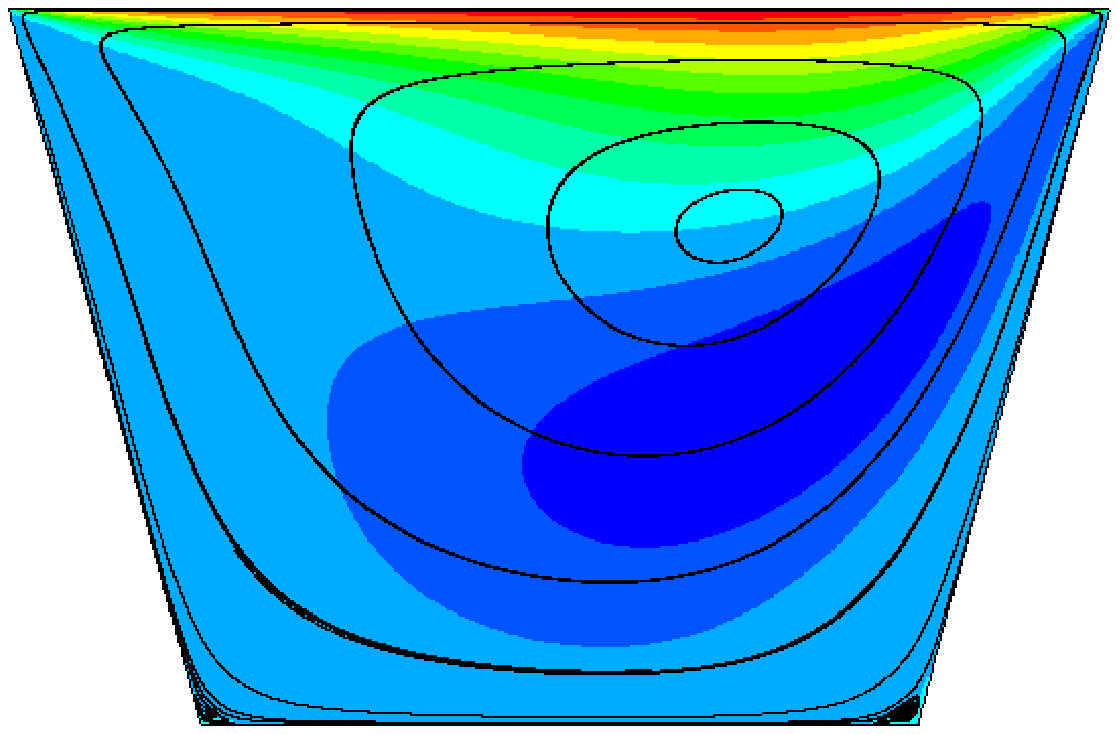}}
		\subfigure[]{ \label{fig:75-100-0.5}
			\includegraphics[scale=0.35]{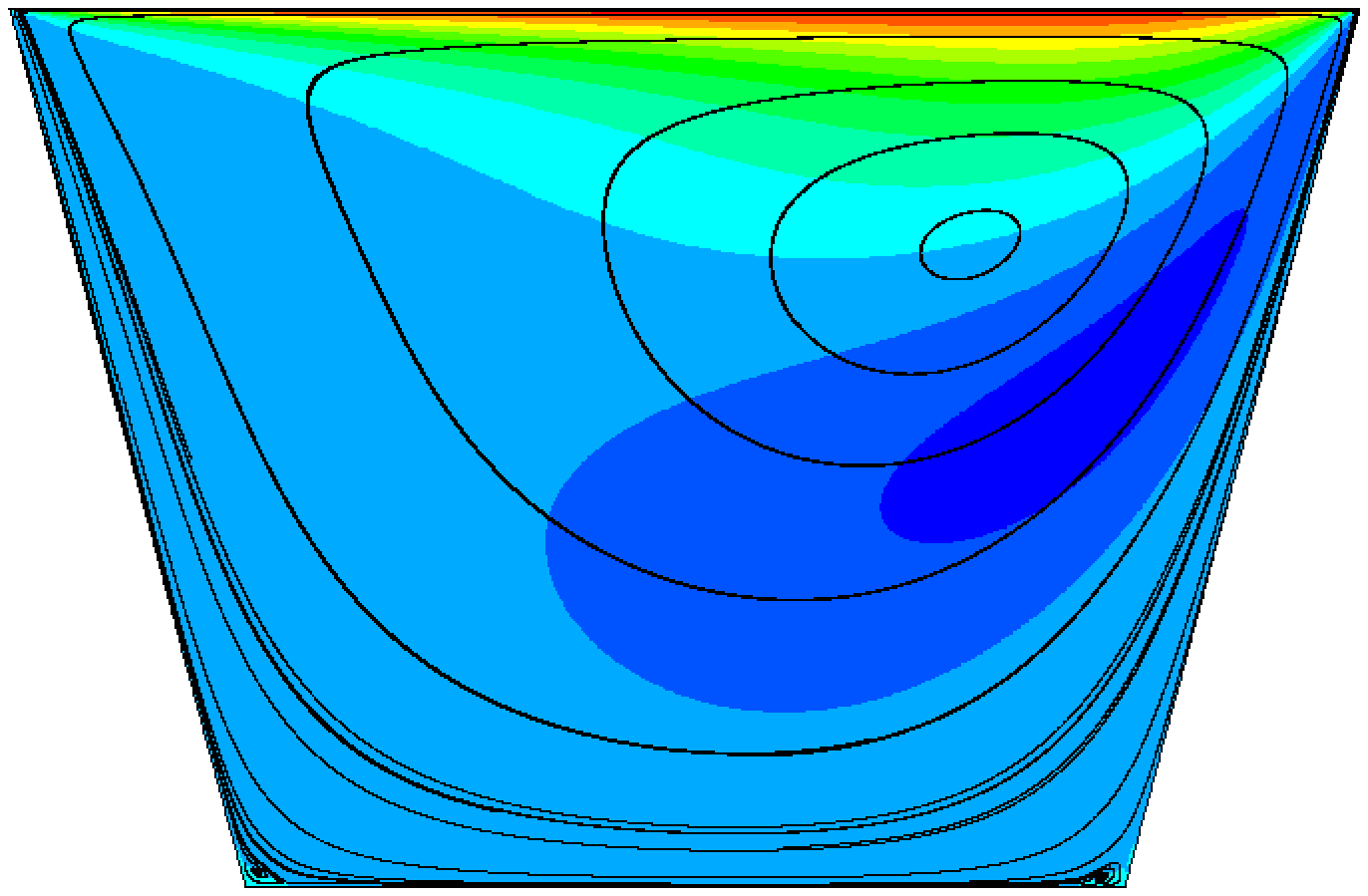}}
		\caption{ Streamline plots at $Re=100$ and $\theta=75^o$; (a) n=1.5, (b) n=1.0, (c) n=0.75, (d) n=0.5.}
		\label{fig:75-100}
	\end{figure}

	It is excepted that the behavior of power-law fluid will have an effect on the flow field. Because the power-law index $n$ determines the viscosity of the fluid.
	In this simulation, we fixed $\theta=75^o$ and $Re=100$. And the power-law index $n$ is varied from $0.5$ to $1.5$ to investigated the behavior of power-law fluid.
	We have presented some numerical results in Fig. \ref{fig:75-100}.

	According to the simulation results, it can be found that the TC flow will be in a stable state eventually when $Re=100$. However, the structure of vortices is different for various $n$. A first-order vortex appears in the central region of the cavity. At the same time, two secondary vortexes appear in the lower left and right corners of the cavity respectively, and the secondary vortex in the lower right corner is obviously larger than that in the lower left corner. As the power-law index $n$ increases, the center of the first-order vortex will gradually move closer to the center of the cavity. In addition, the range of secondary vortexes increases gradually as the power-law index $n$ increases.
	
\begin{figure}[htbp]\centering
		\subfigure[]{ \label{fig:xv-75-100}
			\includegraphics[scale=0.45]{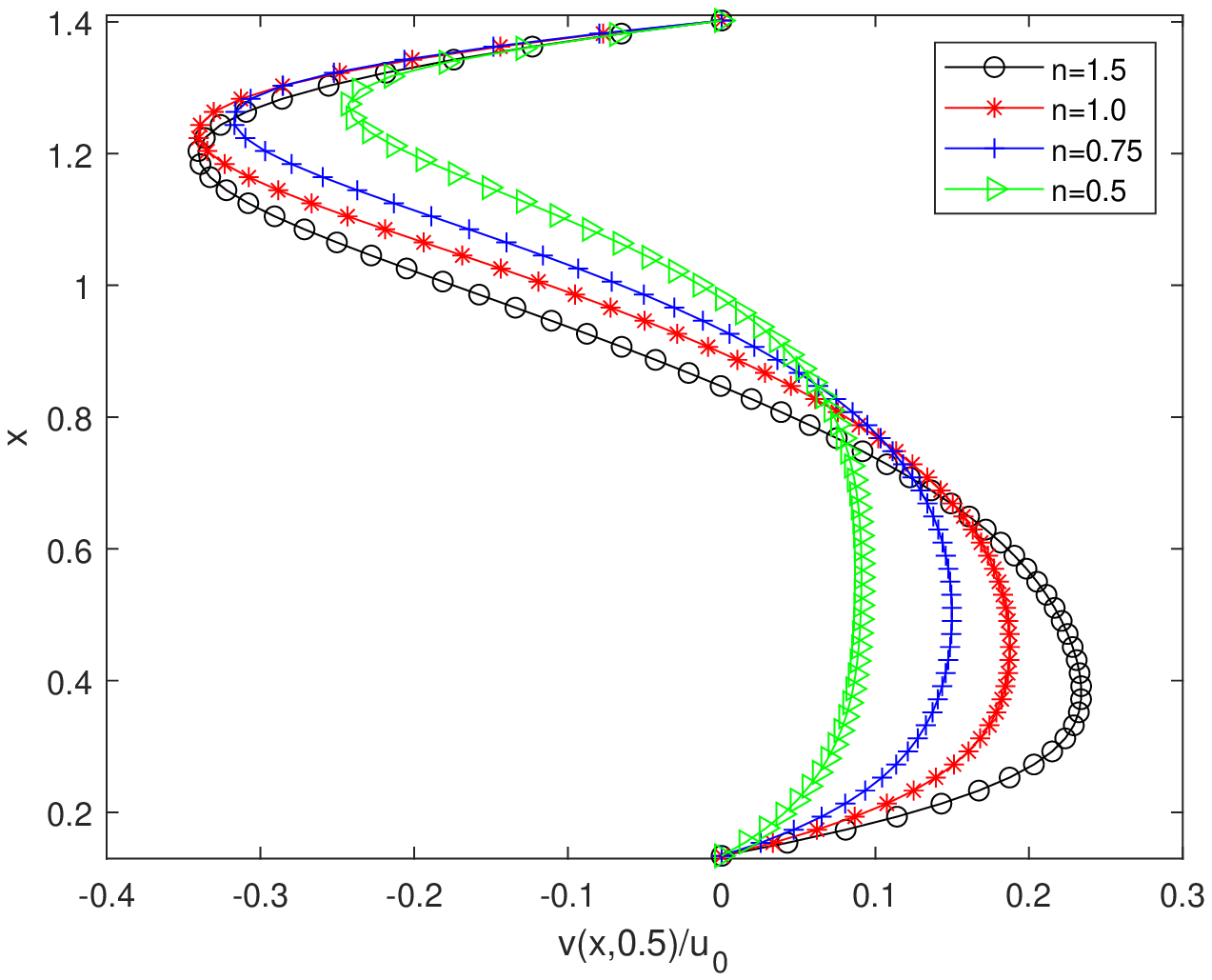}}
		\subfigure[]{ \label{fig:uy-75-100}
			\includegraphics[scale=0.45]{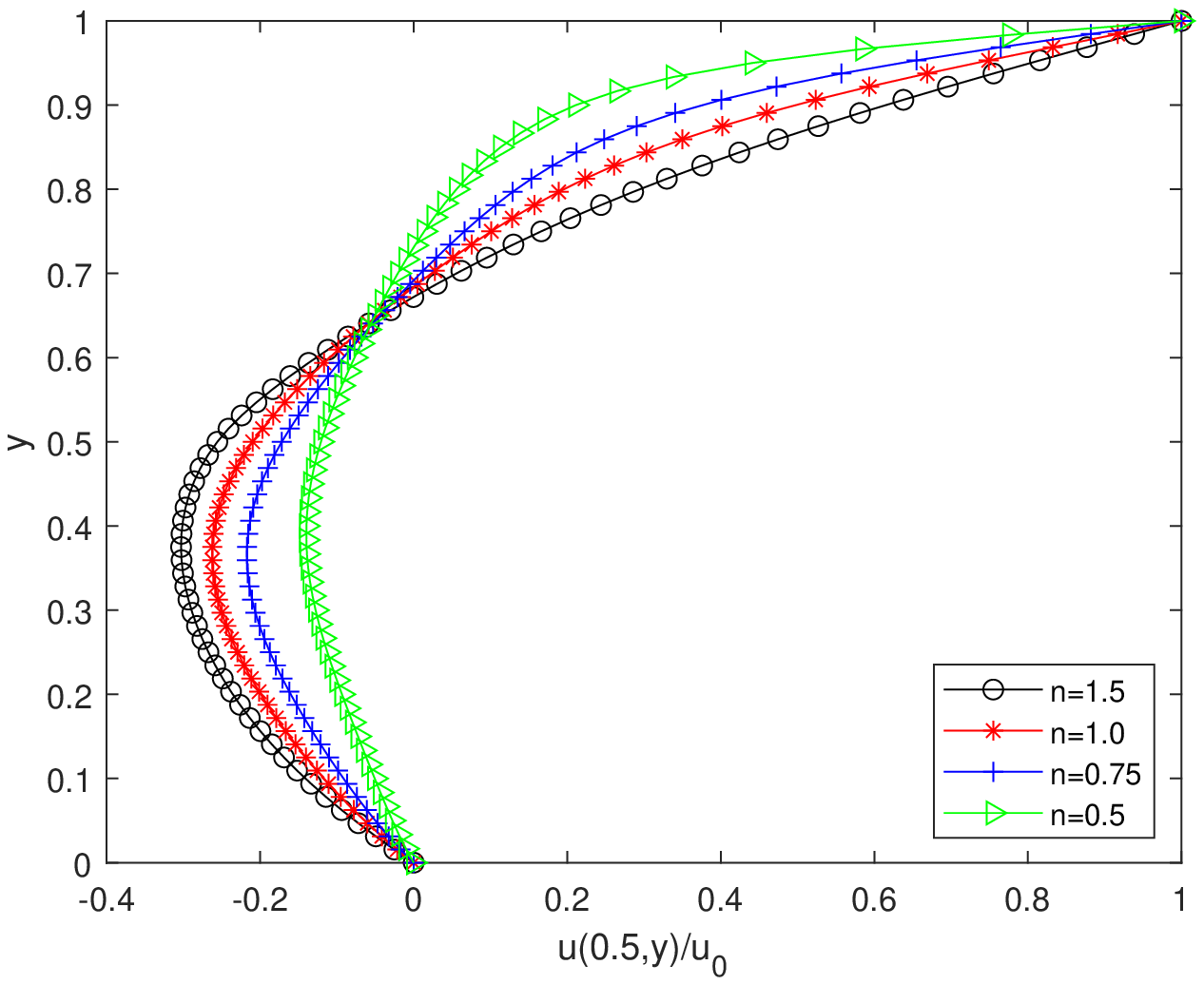}}
		\caption{ Vertical component of velocity for different power-law index $n$ with $Re=100$ and $\theta=75^o$; (a) velocity $v$ through $y/H=0.5$ along $x$-axis, (b)velocity $u$ through $x/L=0.5$ along $y$-axis.}
		\label{fig:75-100-uv}
	\end{figure}

	The numerical velocity in the central line through $x$-axis and $y$-axis are presented in the Fig. \ref{fig:75-100-uv}. As we can see, the velocity changes more dramatically when $n$ increases, and both the maximum and minimum values of velocity on the center line increase.
	
	\setlength{\tabcolsep}{1.2mm}{
		\begin{table}
			\caption{The location of eddies at different $n$ for isosceles trapezoidal cavity flow ($\theta=75^o$).} \label{table:75-eddy}
			\centering
			\begin{tabular}{ccccccc}
				\hline\hline
				n      & \multicolumn{2}{c}{First-primary eddy} & \multicolumn{2}{c}{Second-primary eddy (left)} & \multicolumn{2}{c}{Second-primary eddy (right)}  \\
				\cline{2-7}
				& x      & y                             & x      & y                                     & x      & y                                       \\
				\hline
				n=1.5  & 0.8697 & 0.6674                        & 0.2881 & 0.0249                                & 1.2452 & 0.0317                                  \\
				n=1.0  & 0.9509 & 0.6852                        & 0.2823 & 0.0194                                & 1.2478 & 0.0259                                  \\
				n=0.75 & 1.0067 & 0.6978                        & 0.2795 & 0.0158                                & 1.2480 & 0.0187                                  \\
				n=0.5  & 1.0937 & 0.7323                        & 0.2821 & 0.0168                                & 1.2409 & 0.0110                                  \\
				\hline\hline
			\end{tabular}
	\end{table}}
	
	(ii) Effect of angle $\theta$ on the development of flow for low $Re$ number

\begin{figure}[htbp]\centering
		\subfigure[]{ \label{fig:60-100-1.5}
			\includegraphics[scale=0.45]{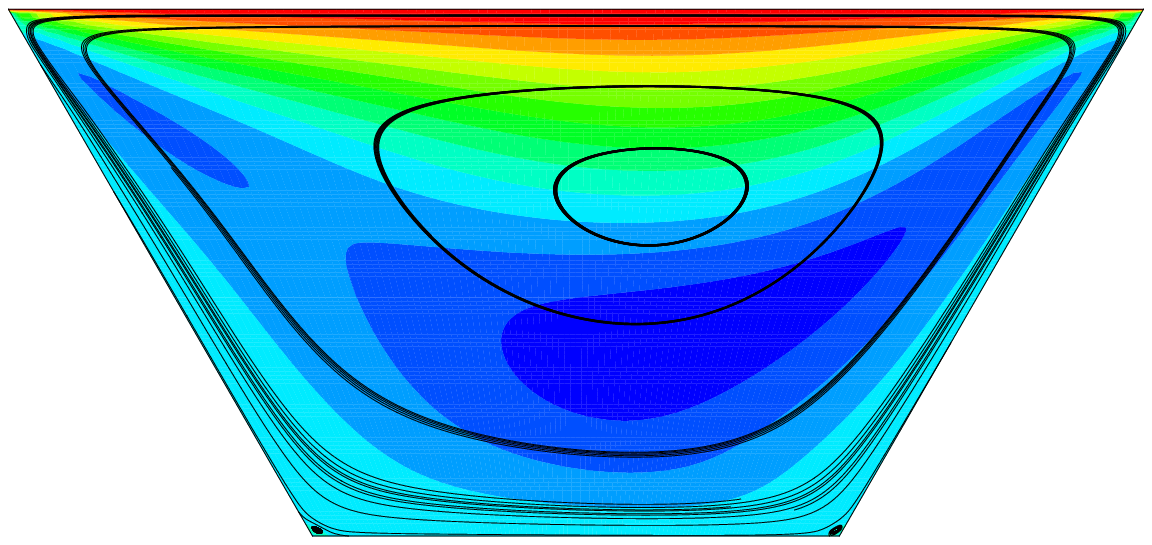}}
		\subfigure[]{ \label{fig:60-100-1.0}
			\includegraphics[scale=0.45]{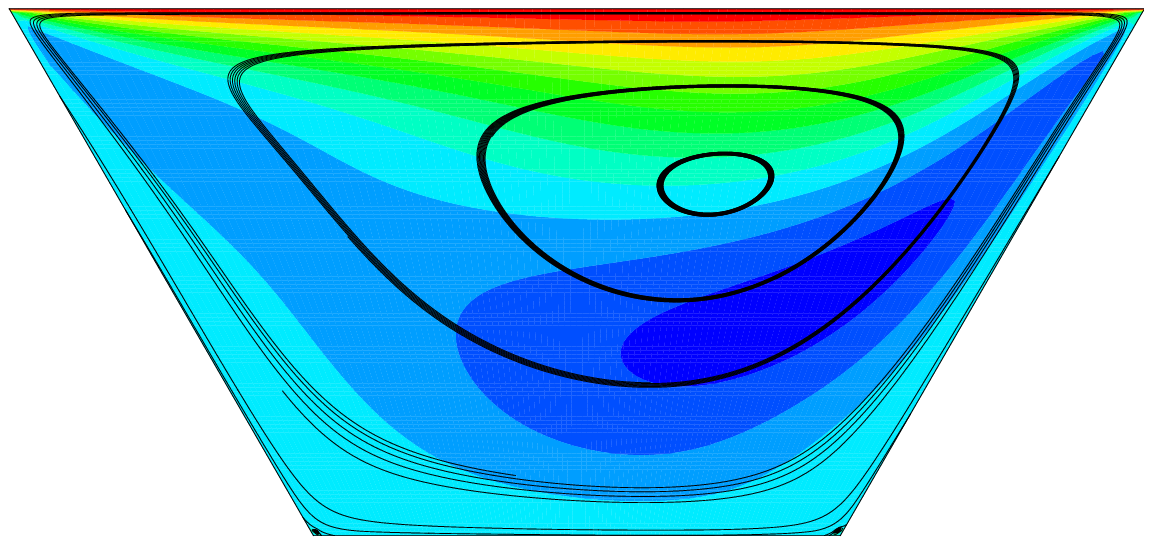}}
		\subfigure[]{ \label{fig:60-100-0.75}
			\includegraphics[scale=0.45]{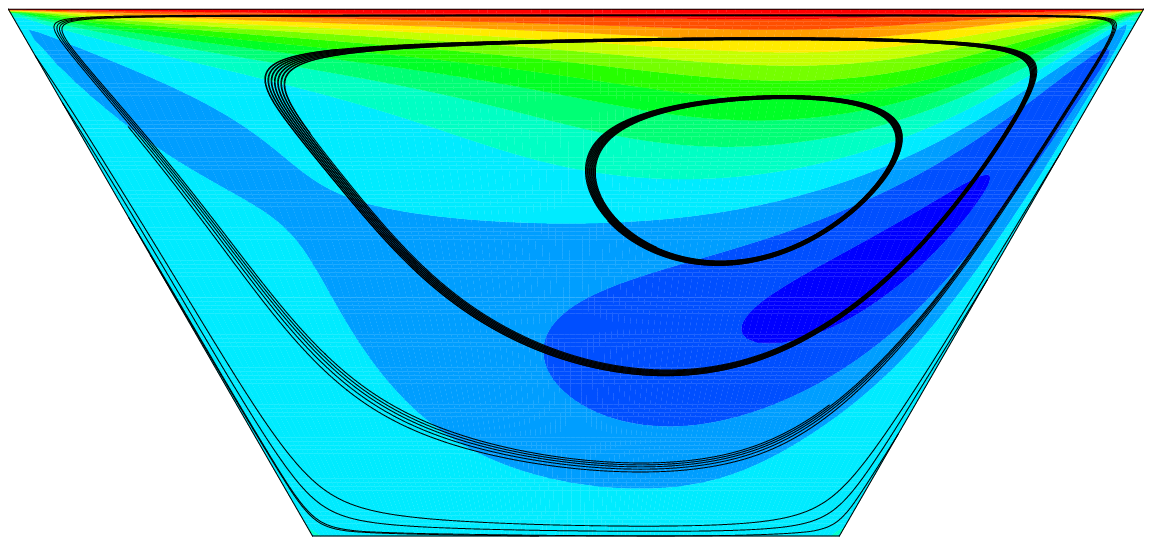}}
		\subfigure[]{ \label{fig:60-100-0.5}
			\includegraphics[scale=0.45]{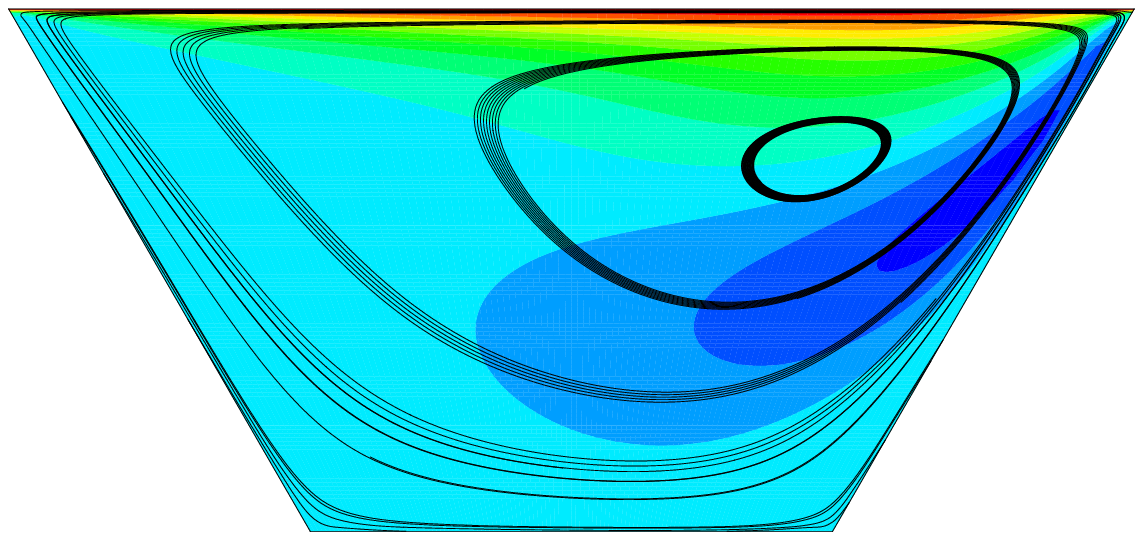}}
		\caption{ Streamline plots at $Re=100$ and $\theta=60^o$; (a) n=1.5, (b) n=1.0, (c) n=0.75, (d) n=0.5.}
		\label{fig:60-100}
	\end{figure}

	We also study the rheological behavior of power-law fluid with $\theta=60^o$ and $\theta=45^o$. As the $\theta$ decreases, the area of the cavity increases correspondingly. We show some numerical results with $\theta=60^o$ in Fig. \ref{fig:60-100}. It can be seen that the secondary vortexes in the lower left and right corner of the cavity are smaller than that in $\theta=75^o$. As the $n$ decreases, the secondary vortexes fade away and the first-order vortex gradually moves closer to the upper right corner of trapezoidal cavity. When $\theta=45^o$, there are no secondary vortexes in the lower left and right corner of the cavity, and the first-order vortex gradually moves closer to the center of trapezoidal cavity as the power-law index $n$ increases.
	
\begin{figure}[htbp]\centering
		\subfigure[]{ \label{fig:xv-60-100}
			\includegraphics[scale=0.45]{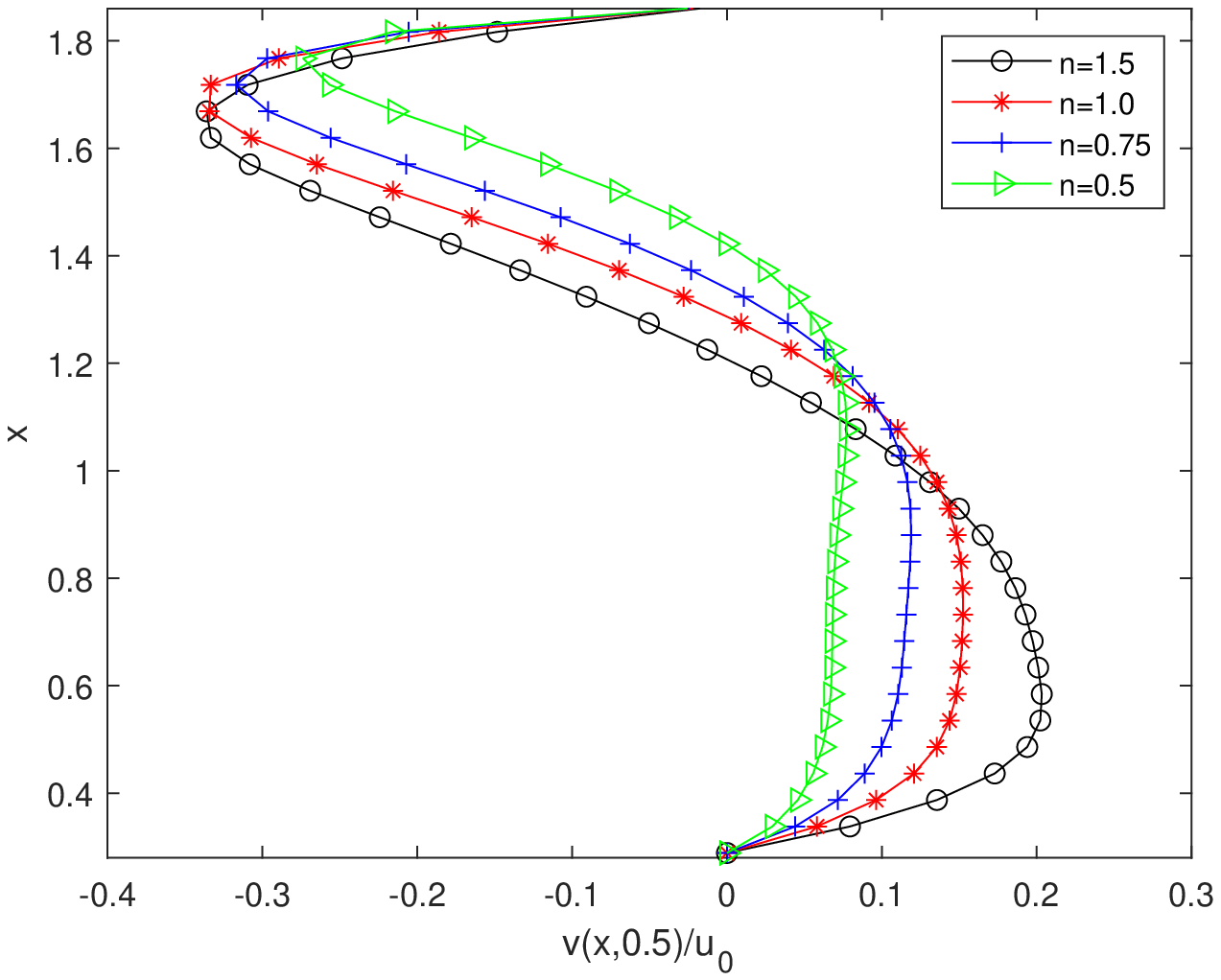}}
		\subfigure[]{ \label{fig:uy-60-100}
			\includegraphics[scale=0.45]{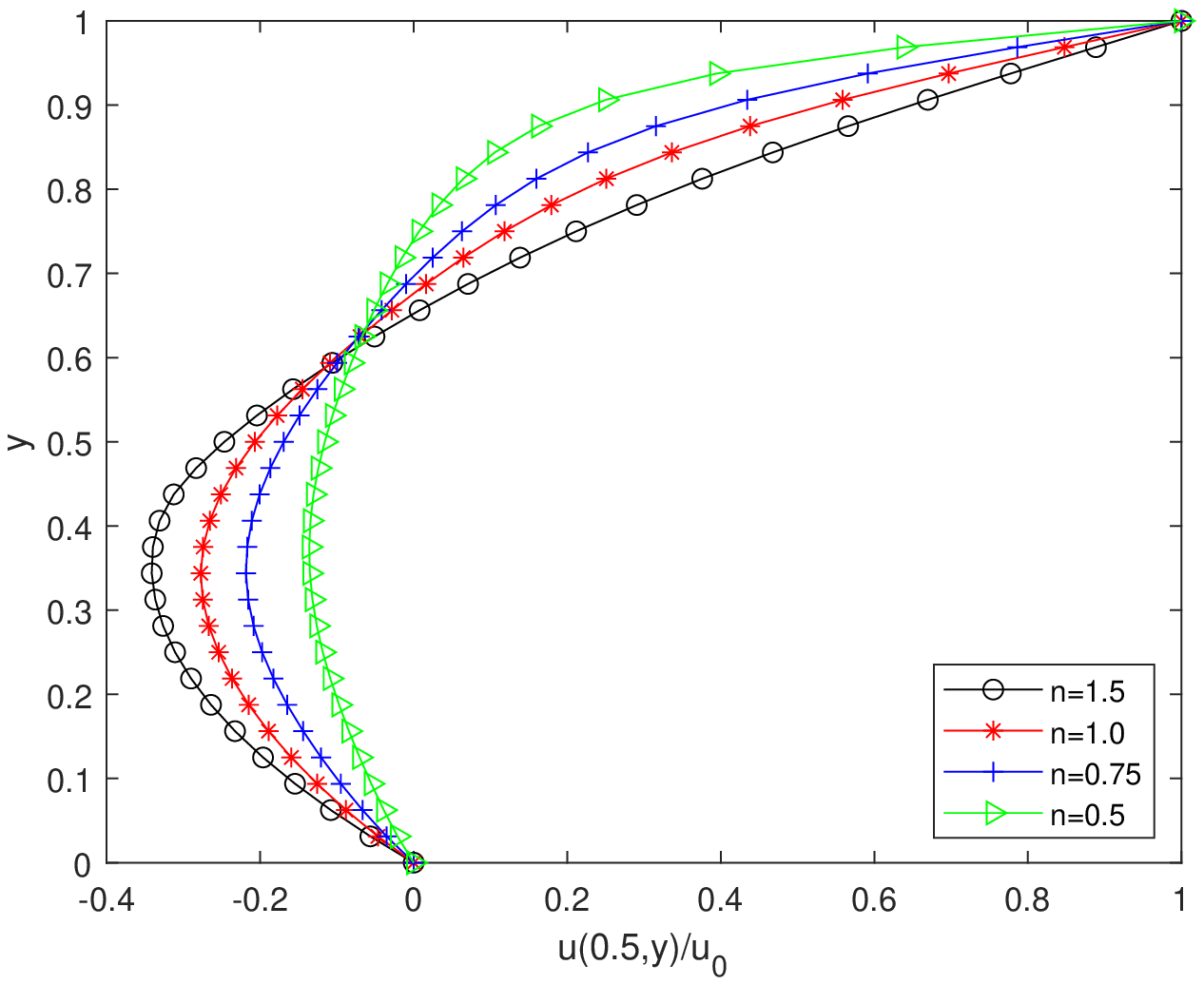}}
		\caption{ Vertical component of velocity for different power-law index $n$ with $Re=100$ and $\theta=60^o$; (a) velocity $v$ through $y/H=0.5$ along $x$-axis, (b)velocity $u$ through $x/L=0.5$ along $y$-axis.}
		\label{fig:60-100-uv}
	\end{figure}

\begin{figure}[htbp]\centering
		\subfigure[]{ \label{fig:xv-45-100}
			\includegraphics[scale=0.45]{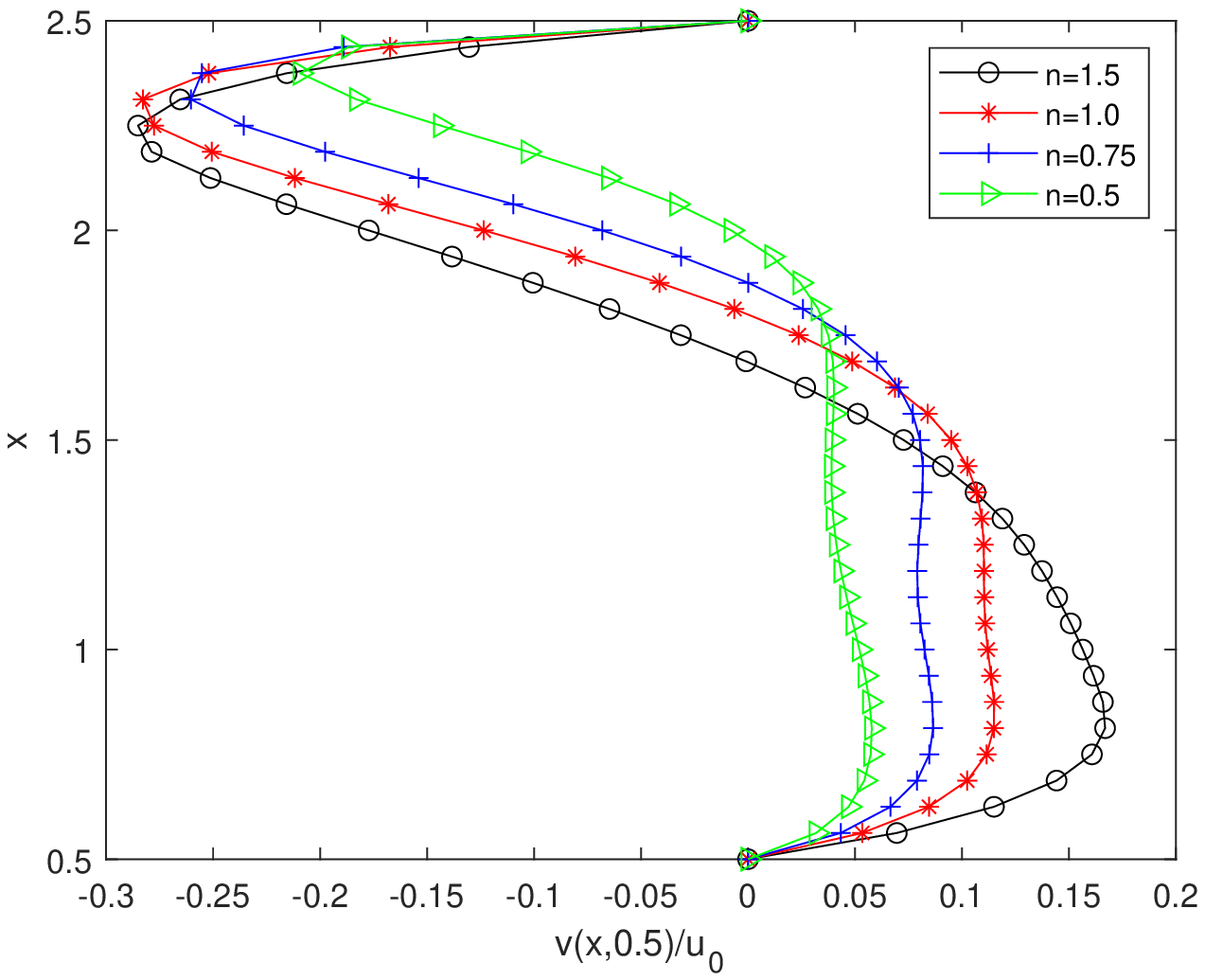}}
		\subfigure[]{ \label{fig:uy-45-100}
			\includegraphics[scale=0.45]{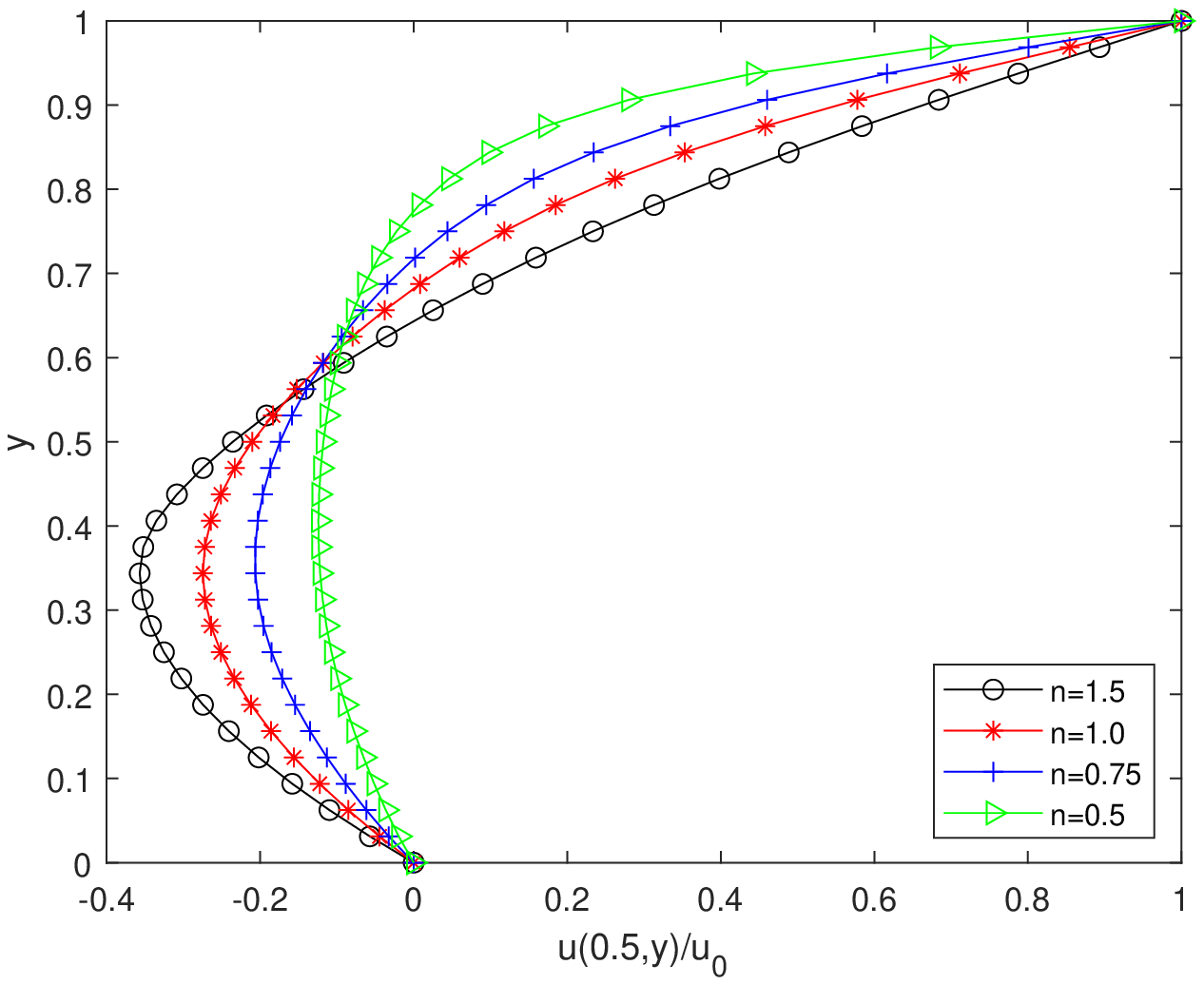}}
		\caption{ Vertical component of velocity for different power-law index $n$ with $Re=100$ and $\theta=45^o$; (a) velocity $v$ through $y/H=0.5$ along $x$-axis, (b)velocity $u$ through $x/L=0.5$ along $y$-axis.}
		\label{fig:45-100-uv}
	\end{figure}

	Figs. \ref{fig:60-100-uv} and \ref{fig:45-100-uv} show the numerical results of velocity in the central line through $x$-axis and $y$-axis. It is clear that the maximum and minimum values of velocity on the center line increase as the power-law index $n$ increases. And the velocity profiles are similar to the results with $\theta=75^o$.
	
	\setlength{\tabcolsep}{1.2mm}{
		\begin{table}
			\caption{The location of eddies at different $n$ for isosceles trapezoidal cavity flow ($\theta=60^o$).} \label{table:60-eddy}
			\centering
			\begin{tabular}{ccccccc}
				\hline\hline
				n      & \multicolumn{2}{c}{First-primary eddy} & \multicolumn{2}{c}{Second-primary eddy (left)} & \multicolumn{2}{c}{Second-primary eddy (right)}  \\
				\cline{2-7}
				& x      & y                             & x      & y                                     & x      & y                                       \\
				\hline
				n=1.5  & 1.2264 & 0.6465                        & 0.5852 & 0.0120                                & 1.5709 & 0.0119                                  \\
				n=1.0  & 1.3441 & 0.6684                        & 0.5831 & 0.0083                                & 1.5725 & 0.0095                                  \\
				n=0.75 & 1.4292 & 0.6872                        & --     & --                                    & --     & --                                      \\
				n=0.5  & 1.5544 & 0.7150                        & --     & --                                    & --     & --                                      \\
				\hline\hline
			\end{tabular}
	\end{table}}

\begin{figure}[htbp]
		\centering
		\includegraphics[scale=0.5]{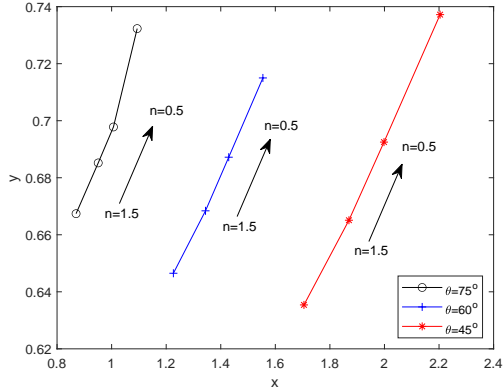}
		\caption{Variation of the central location of the first-order vortex with $\theta$ and $n$.} \label{fig:vort-local}
	\end{figure}

	In addition, the locations of the first-order vortex with different $\theta$ and $n$ are presented in Fig. \ref{fig:vort-local}. From the figure, with the decrease of $n$, the vortex center moves to the upper right corner of the trapezoidal cavity. As the angle decreases, the vortex center position moves toward the right side of the trapezoidal cavity. This phenomenon is consistent with the above results.
	
	\begin{figure}[htbp]
		\subfigure[]{ \label{fig:45-100-1.5}
			\includegraphics[scale=0.4]{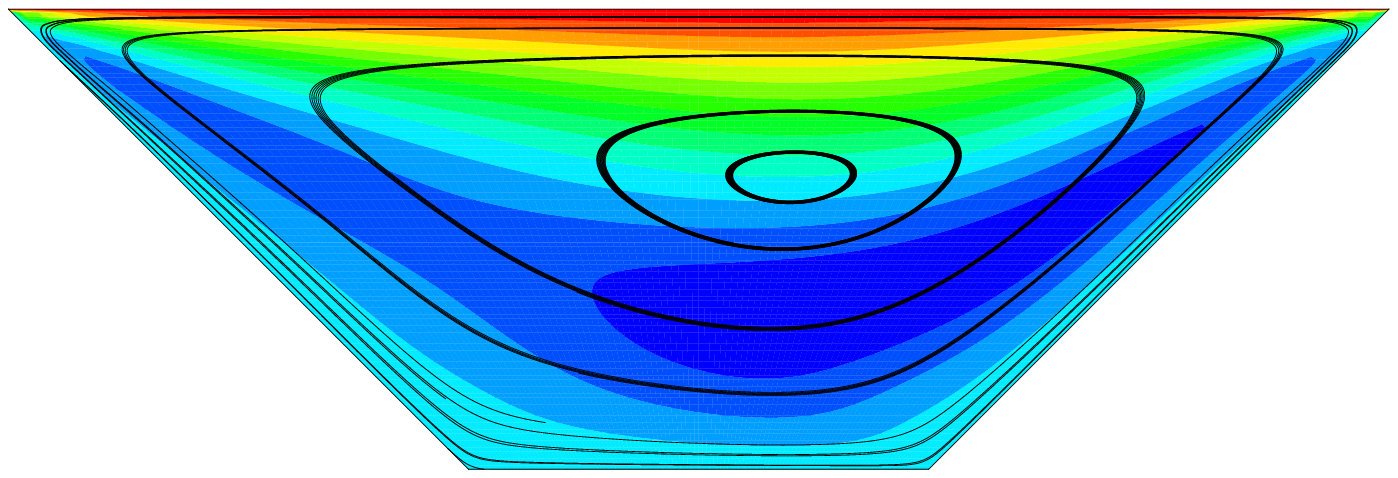}}
		\subfigure[]{ \label{fig:45-100-1.0}
			\includegraphics[scale=0.4]{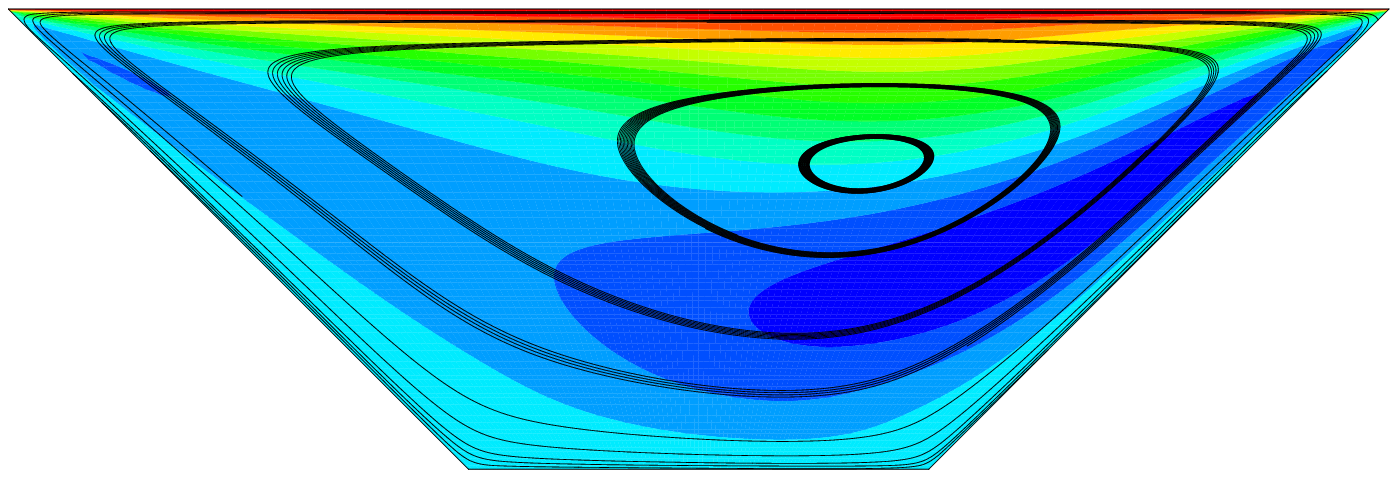}}
		\subfigure[]{ \label{fig:45-100-0.75}
			\includegraphics[scale=0.4]{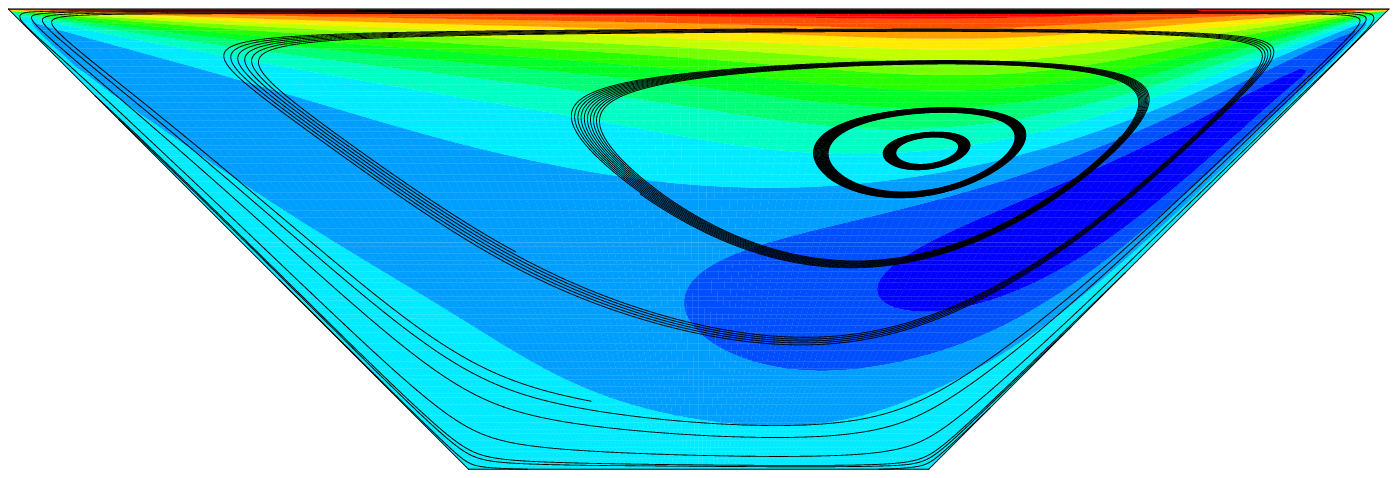}}
		\subfigure[]{ \label{fig:45-100-0.5}
			\includegraphics[scale=0.4]{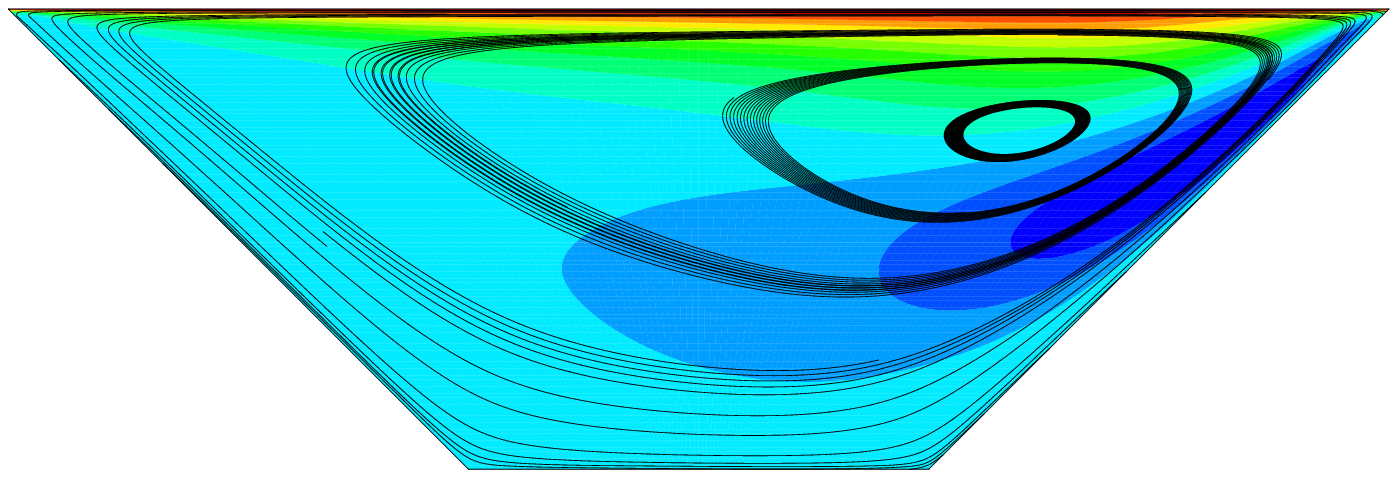}}
		\caption{ Streamline plots at $Re=100$ and $\theta=45^o$; (a) n=1.5, (b) n=1.0, (c) n=0.75, (d) n=0.5.}
		\label{fig:45-100}
	\end{figure}
	
	\begin{table}
		\caption{The location of eddies at different $n$ for isosceles trapezoidal cavity flow ($\theta=45^o$).} \label{table:45-eddy}
		\centering
		\begin{tabular}{ccc}
			\hline
			n      & \multicolumn{2}{c}{First-primary eddy}  \\
			\cline{2-3}
			& x      & y                              \\
			\hline
			n=1.5  & 1.7049 & 0.6354                         \\
			n=1.0  & 1.8698 & 0.6651                         \\
			n=0.75 & 1.9983 & 0.6925                         \\
			n=0.5  & 2.2039 & 0.7372                         \\
			\hline
		\end{tabular}
	\end{table}
	
	\subsection{High $Re$ number}
	
	(i) Effect of power-law index $n$ on the development of flow for high $Re$ number
	
	The discussion in section $5.2$ merely focuses on the low Re number, the trapezoid cavity flows are steady state when $Re=100$. As we all know, the flow state will change as $Re$ number increases. To study the influence of the $Re$ number on the behavior of power-law fluid, we take three cases of trapezoid cavity with different $Re$ numbers and power-law index $n$ into account.
	
\begin{figure}[htbp]\centering
		\subfigure[]{ \label{fig:60-1.5-500}
			\includegraphics[scale=0.3]{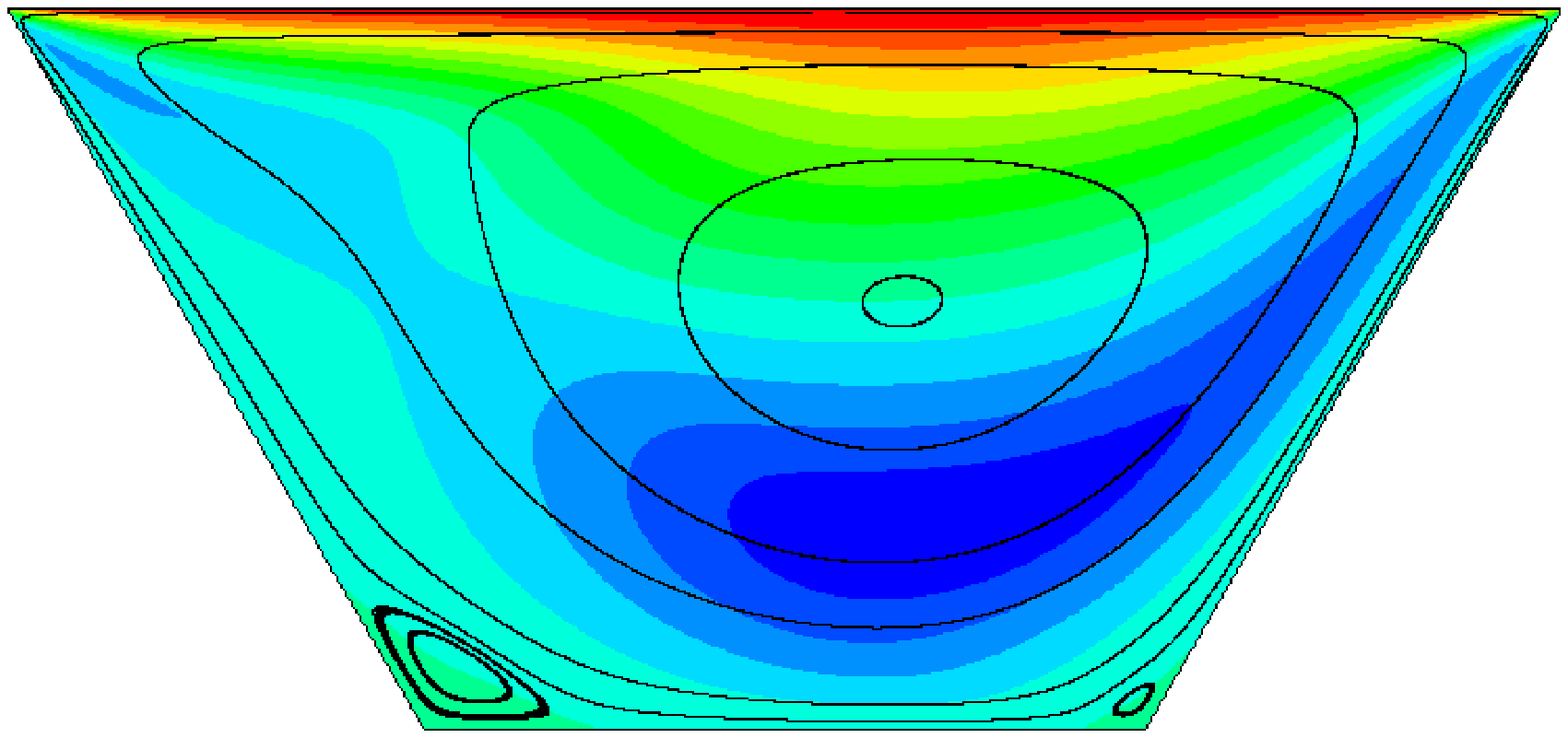}}
		\subfigure[]{ \label{fig:60-1.5-1000}
			\includegraphics[scale=0.3]{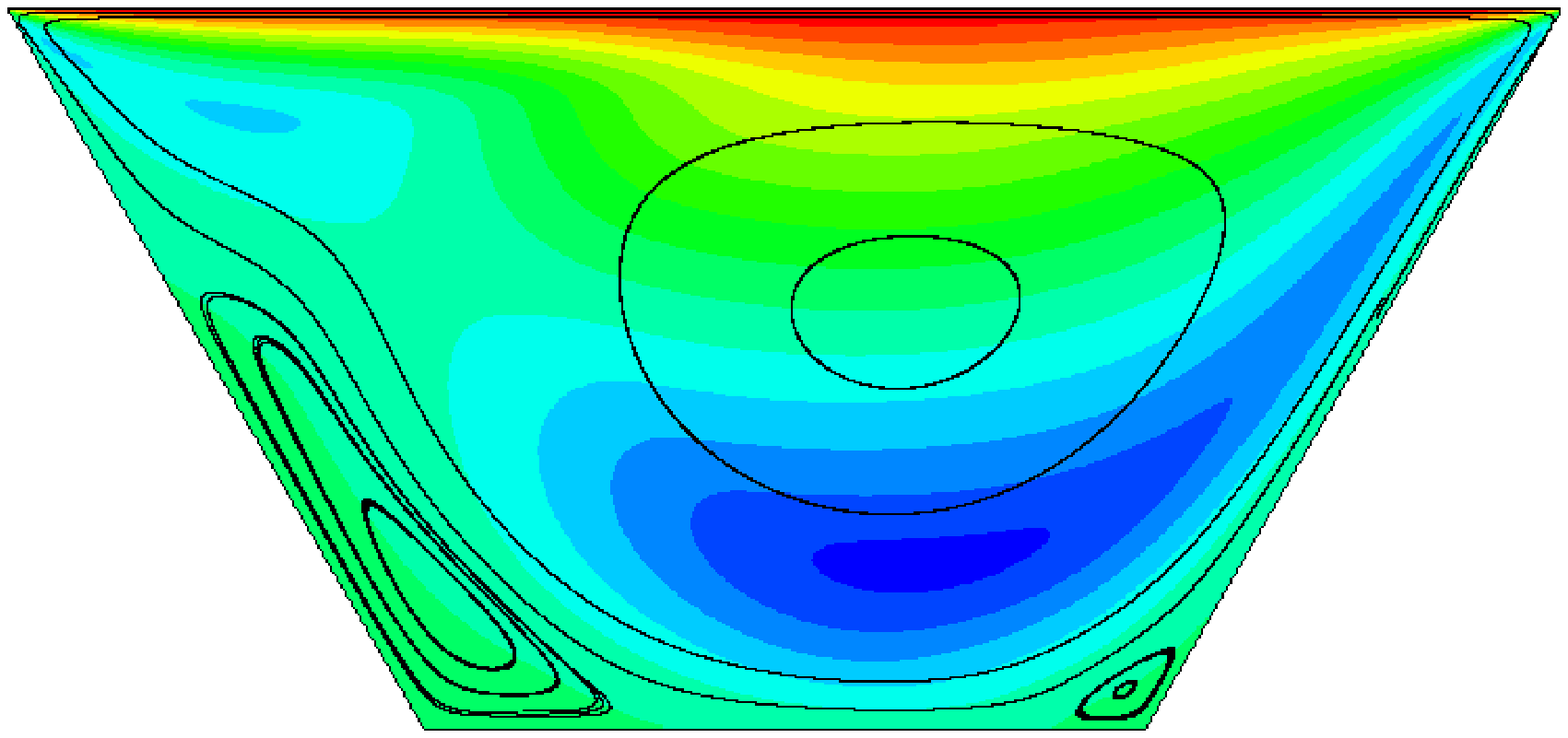}}
		\subfigure[]{ \label{fig:60-1.5-2000}
			\includegraphics[scale=0.3]{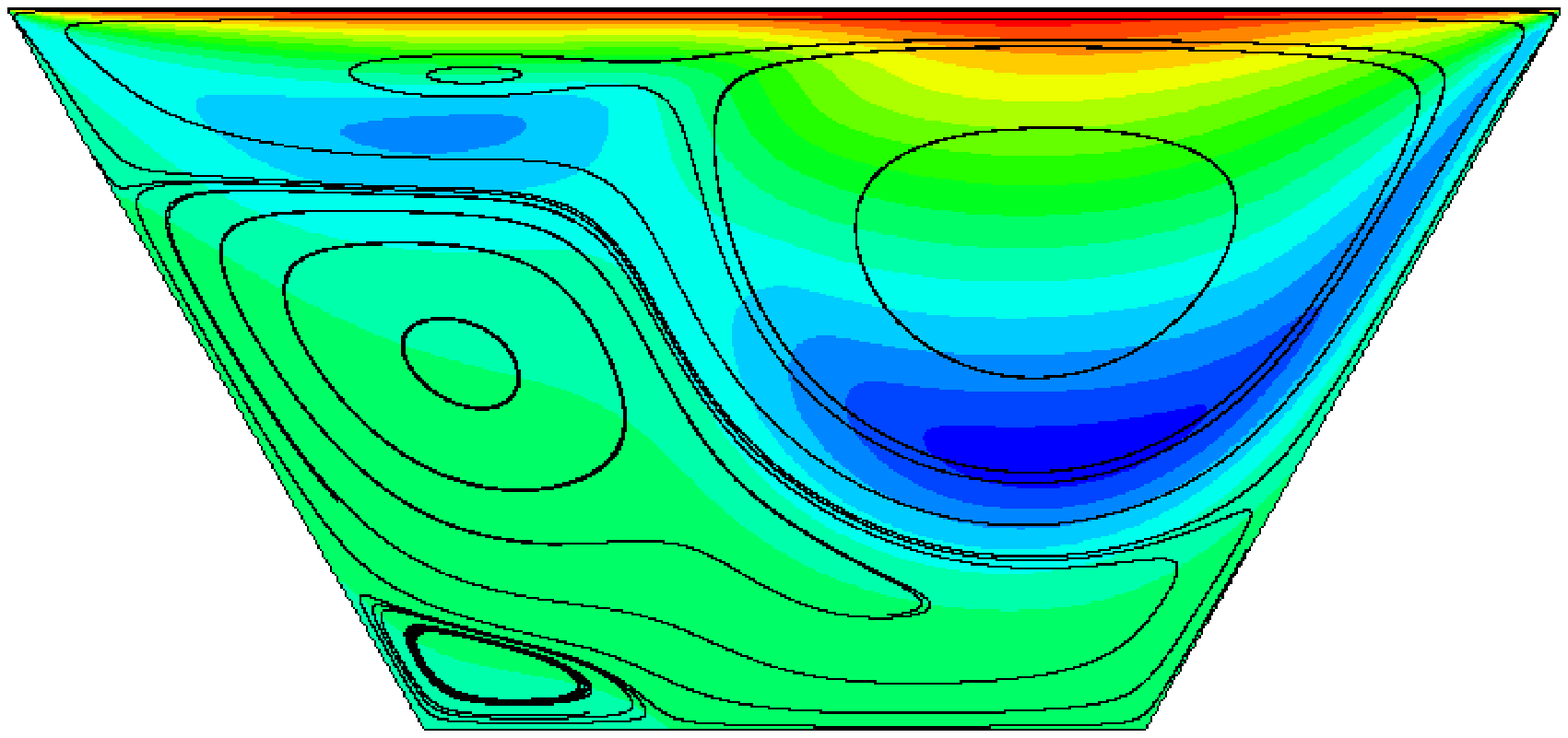}}
		\subfigure[]{ \label{fig:60-1.5-3000}
			\includegraphics[scale=0.3]{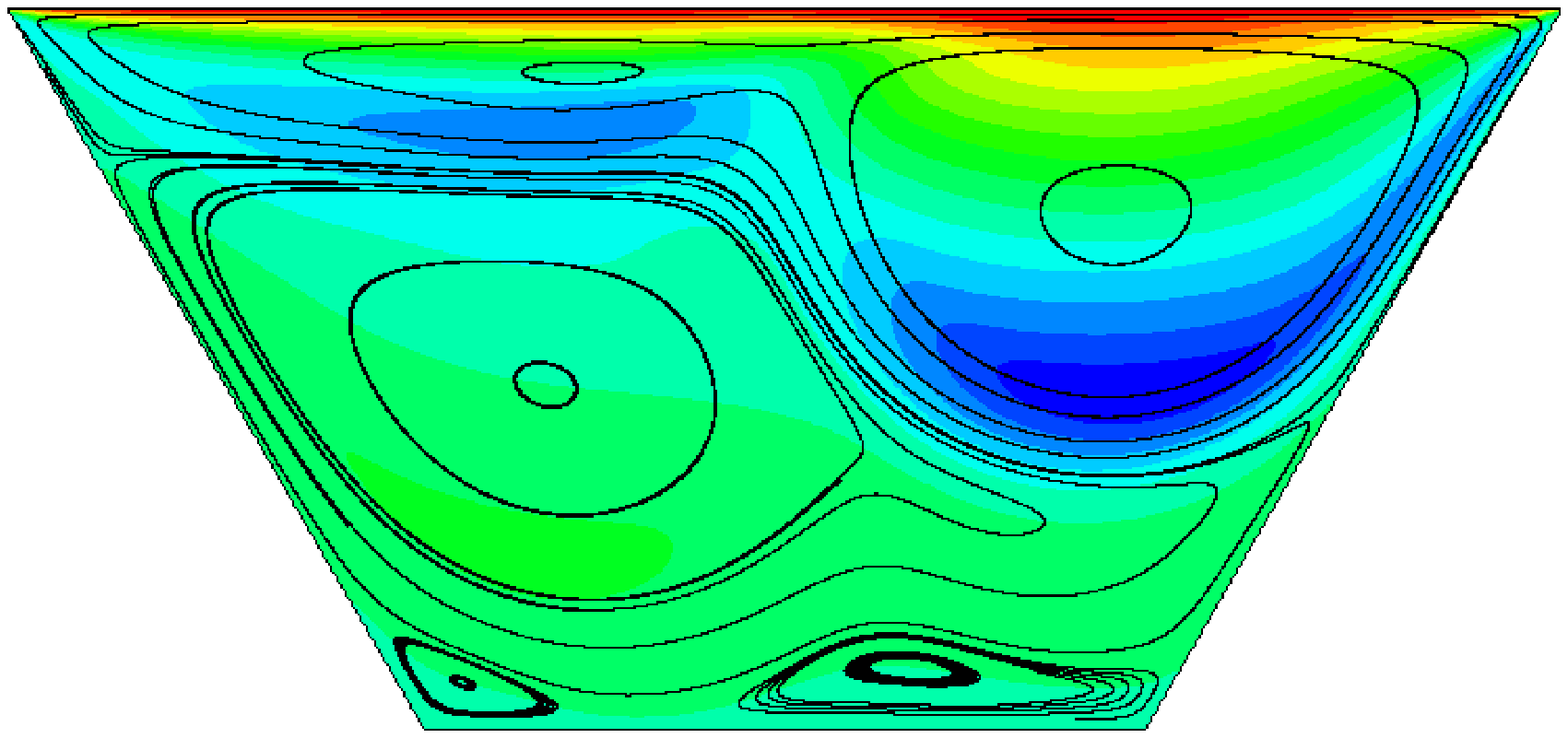}}
		\caption{ Streamline plots at $n=1.5$ and $\theta=60^o$; (a) Re=500, (b) Re=1000, (c) Re=2000, (d) Re=3000.}
		\label{fig:60-1.5}
	\end{figure}

	First of all, we consider the $\theta=60^o$ and $n=1.5$, and the computational grids is $256\times 256$. The typical study has been reported in Fig. \ref{fig:60-1.5} for different $Re$ numbers ranging from $500$ to $3000$. As we can see, the TC flows will eventually reach a stable state for any $Re$ number belonging to $[500,1000,2000,3000]$. The vortex structure in the cavity changes greatly as the $Re$ number changes. With the increase of $Re$ number, more and more vortices appear in the cavity. The center of the first-order large vortex is more and more away from the center line of the cavity and moves to the upper right of the cavity.
	
	A large vortex and two small angular vortices appear in the cavity when $Re$ reaches to $1000$. The vortex in the lower right corner is obviously larger than that in the lower left corner, and the flow function diagram in the trapezoidal cavity is similar to that in the square cavity. When the $Re$ number reaches $2000$, the flow phenomenon in the cavity is quite different from that in the square cavity.
	Four vortices appear in the trapezoidal cavity and are located at the top, middle and bottom of the cavity respectively.
	The angular vortex, which appears at the lower left corner gradually spread to the middle layer of the cavity as $Re$ increases to $2000$.
	Moreover, the scope and the intensity of the vortex increases significantly.
	The first-order vortex moves to the upper right of the cavity due to the squeeze of the vortex at the lower left corner, and the scope of the first-order vortex decreases significantly when $Re=2000$. As $Re$ increases to $3000$, it can be found that the range and intensity of the second-order vortex continue to increase, squeezing the upper first-order vortex, resulting in a decrease in the range and intensity of the first-order vortex. And there is a new third-order vortex appears at the lower right corner of the trapezoidal cavity. With its appearance, the range of the third-order vortex in the lower left corner decreased. All these phenomena indicate that the flow behavior in the cavity becomes more and more complicated with the increase of $Re$.

\begin{figure}[htbp]\centering
		\subfigure[]{ \label{fig:60-1.5-xu}
			\includegraphics[scale=0.45]{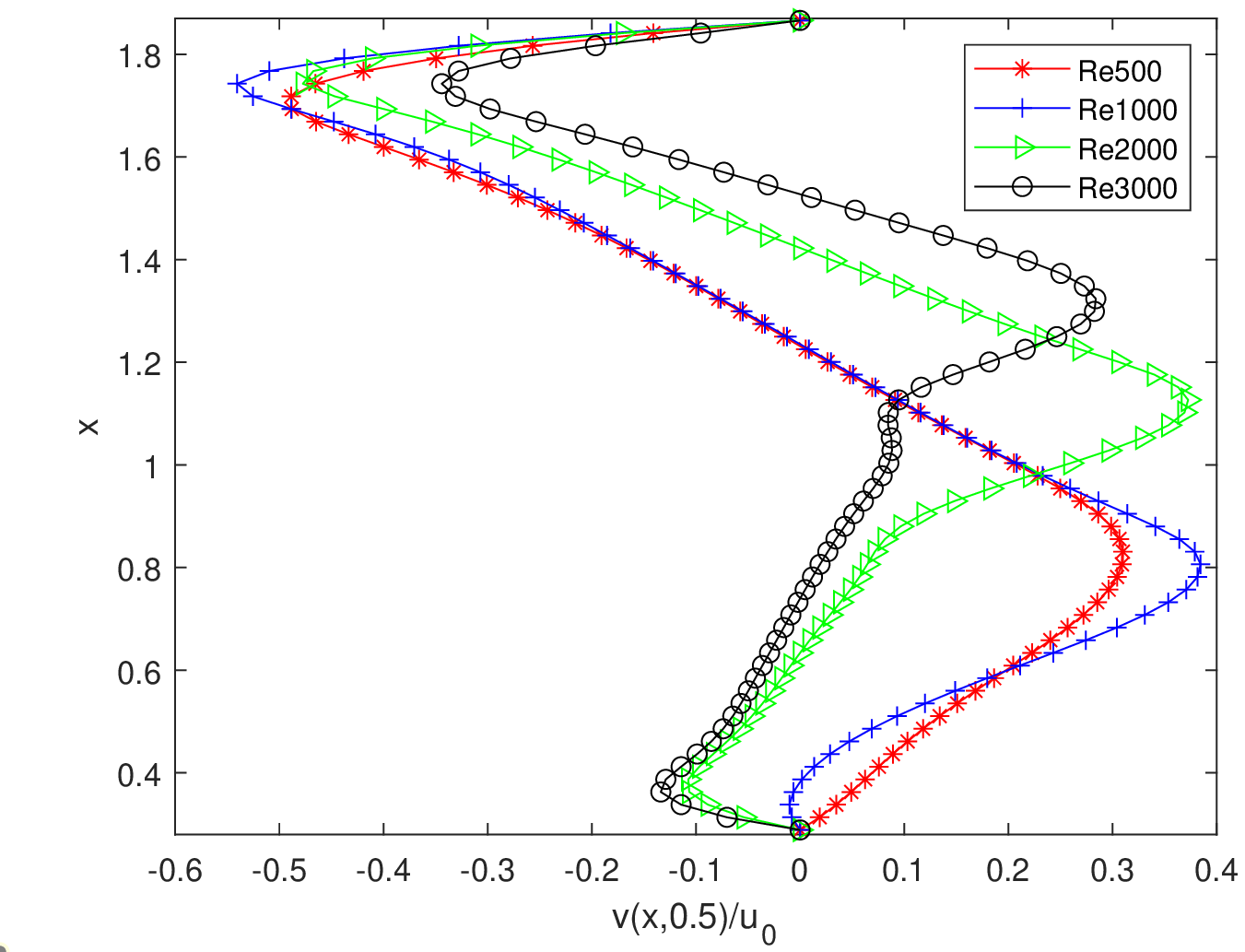}}
		\subfigure[]{ \label{fig:60-1.5-yu}
			\includegraphics[scale=0.45]{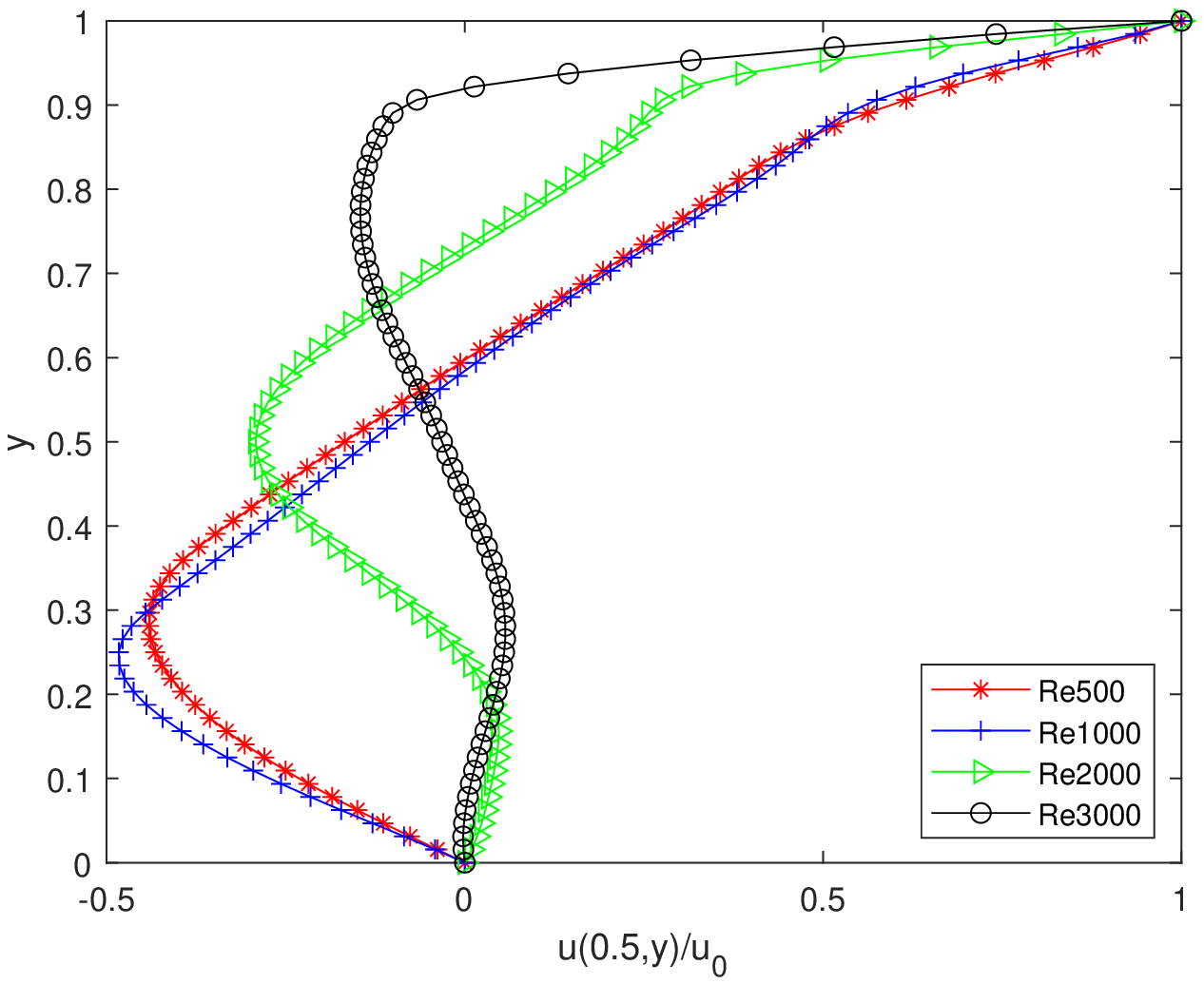}}
		\caption{ Vertical component of velocity for $\theta = 60^o$ and $n=1.5$; (a) velocity v through y/H = 0.5 along x-axis, (b)velocity u through x/L = 0.5 along y-axis.}
		\label{fig:60-1.5-uv}
	\end{figure}

	We also present the centerline velocity results in Fig. \ref{fig:60-1.5-uv}. It can be seen that the velocity profile changes significantly when $Re=2000,3000$. This is principal because the vortex shape in the trapezoidal cavity changes significantly with the increase of $Re$.
	
\begin{figure}[htbp]\centering
		\subfigure[]{ \label{fig:60-4000-uv}
			\includegraphics[scale=0.3]{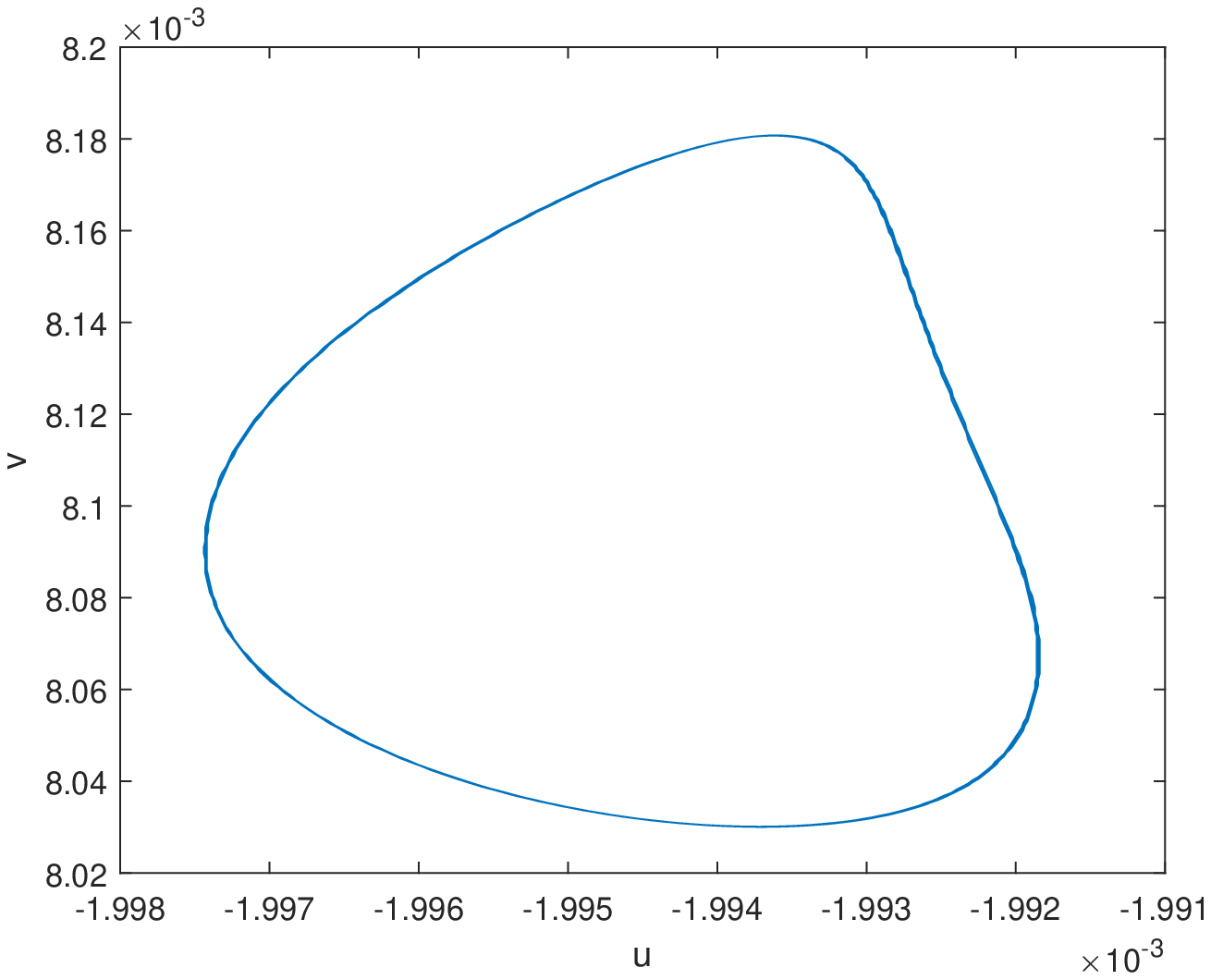}}
		\subfigure[]{ \label{fig:60-5000-uv}
			\includegraphics[scale=0.3]{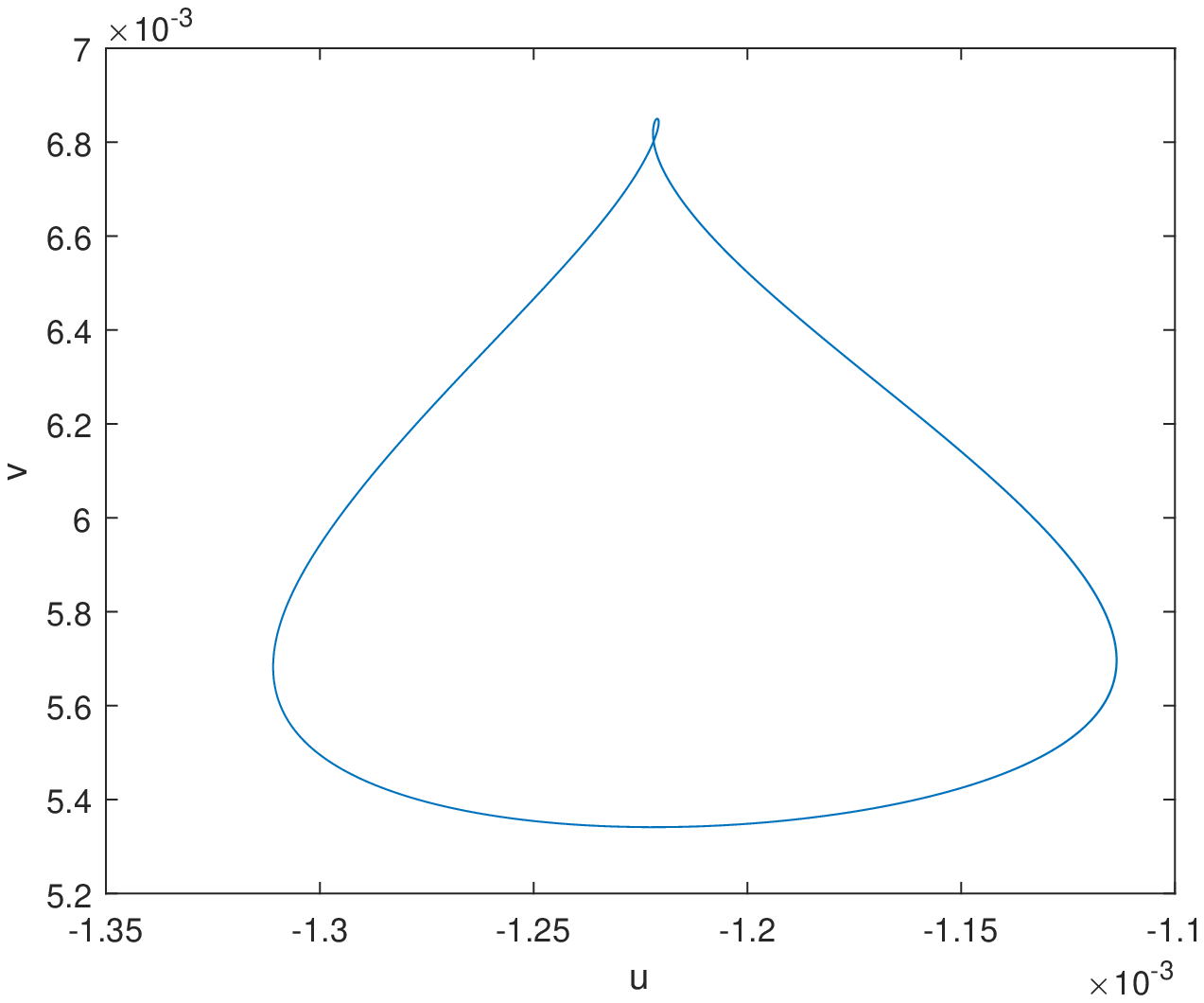}}
		
		\subfigure[]{ \label{fig:60-4000-u}
			\includegraphics[scale=0.3]{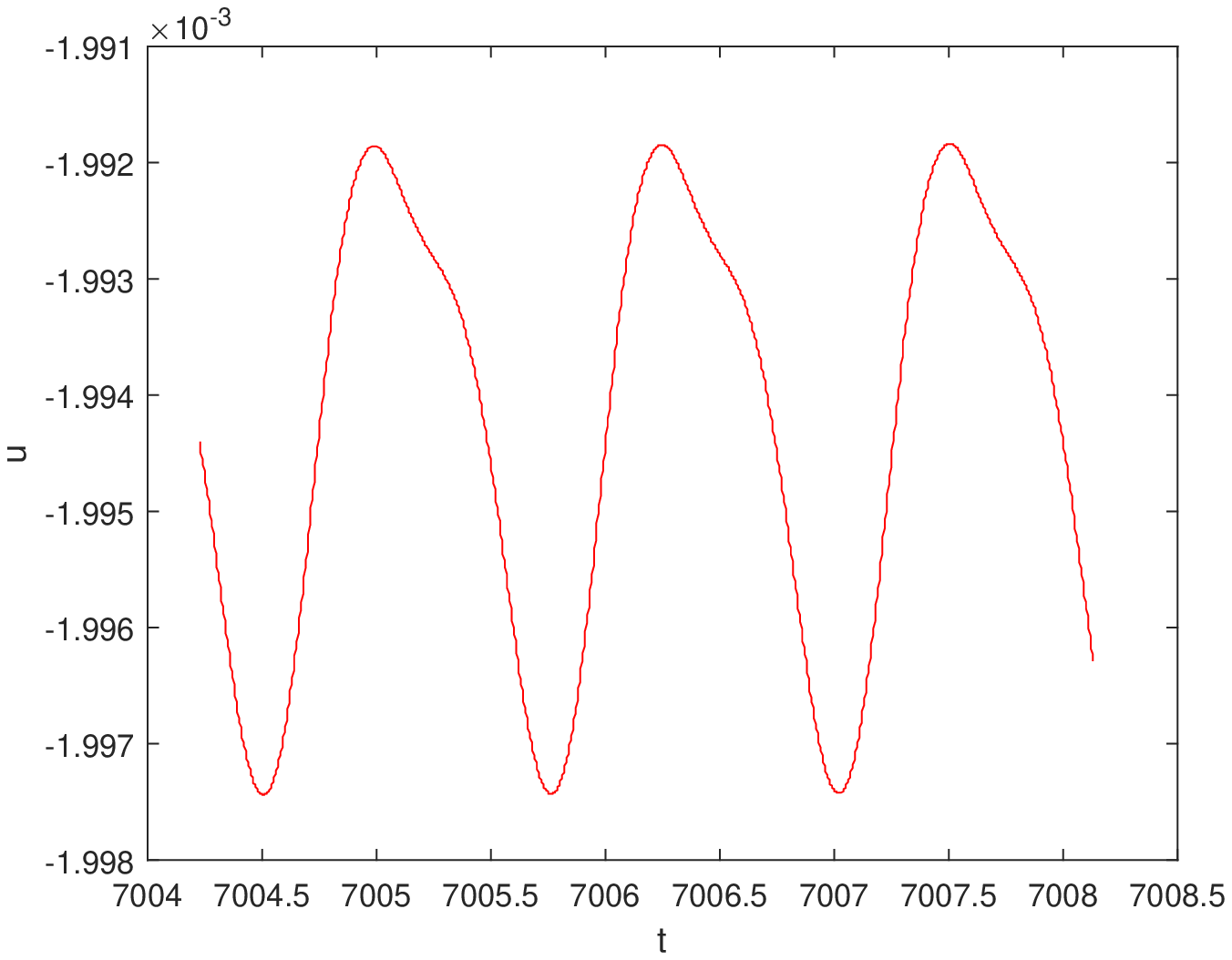}}
		\subfigure[]{ \label{fig:60-5000-u}
			\includegraphics[scale=0.3]{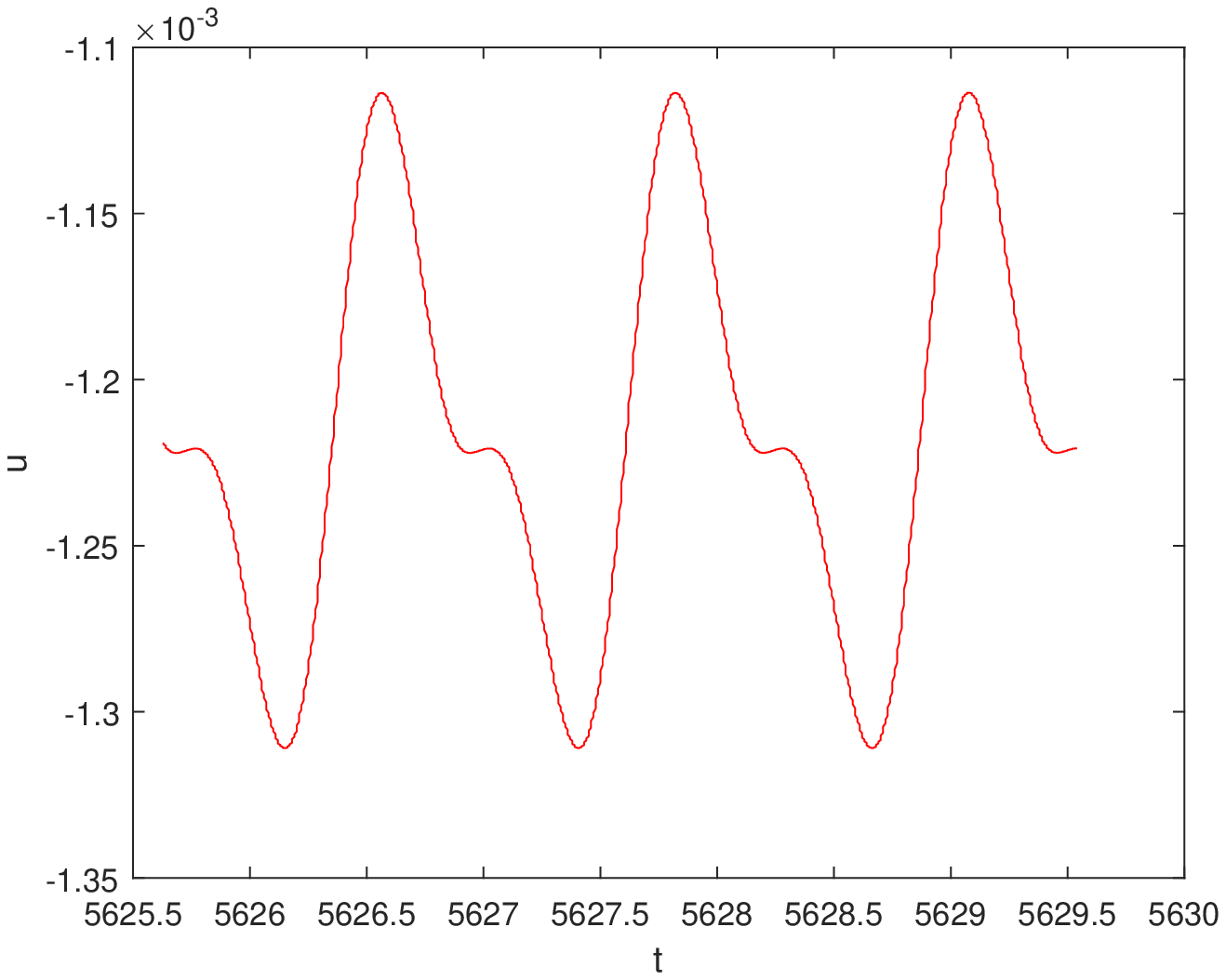}}
		
		\subfigure[]{ \label{fig:60-4000-v}
			\includegraphics[scale=0.3]{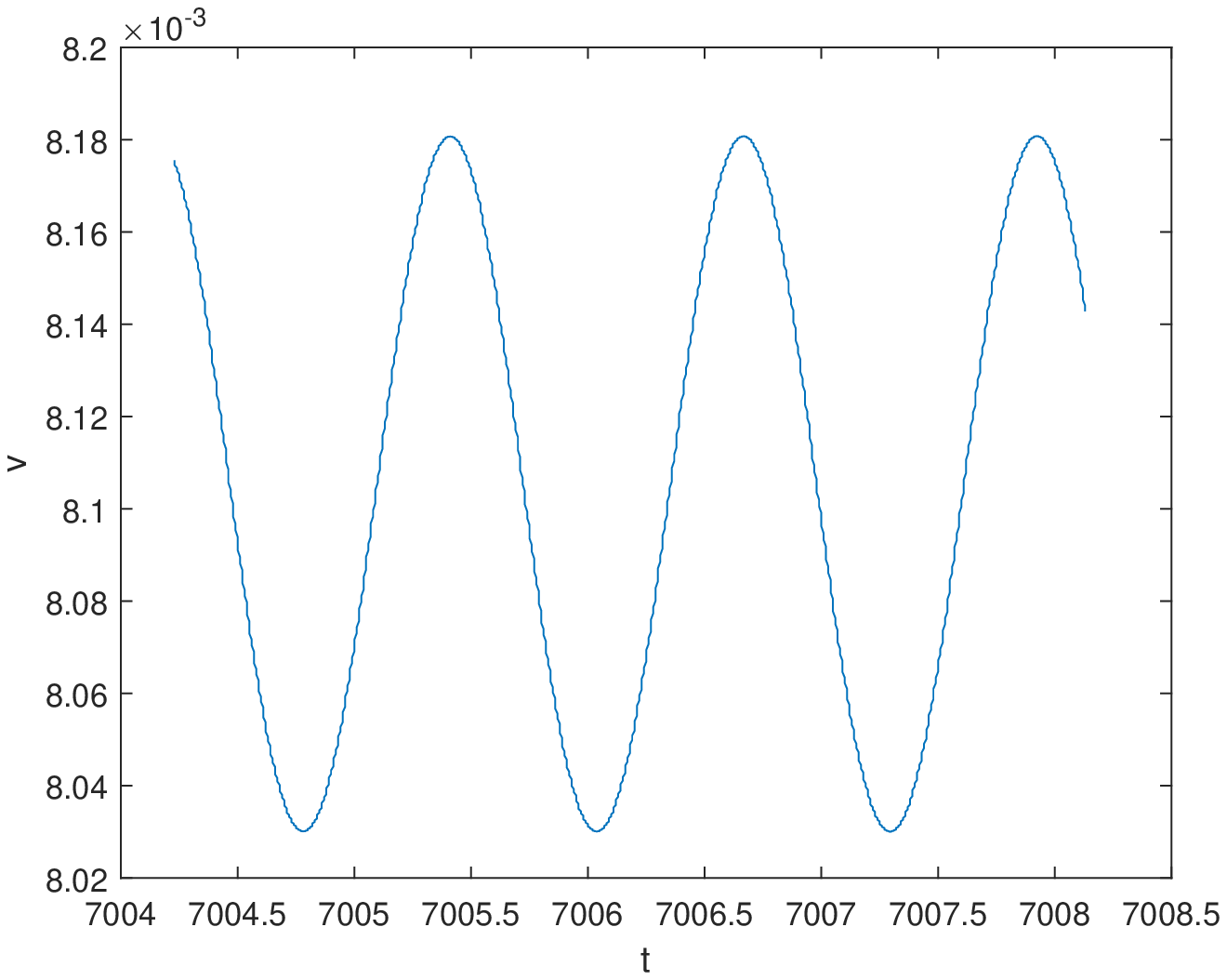}}
		\subfigure[]{ \label{fig:60-5000-v}
			\includegraphics[scale=0.3]{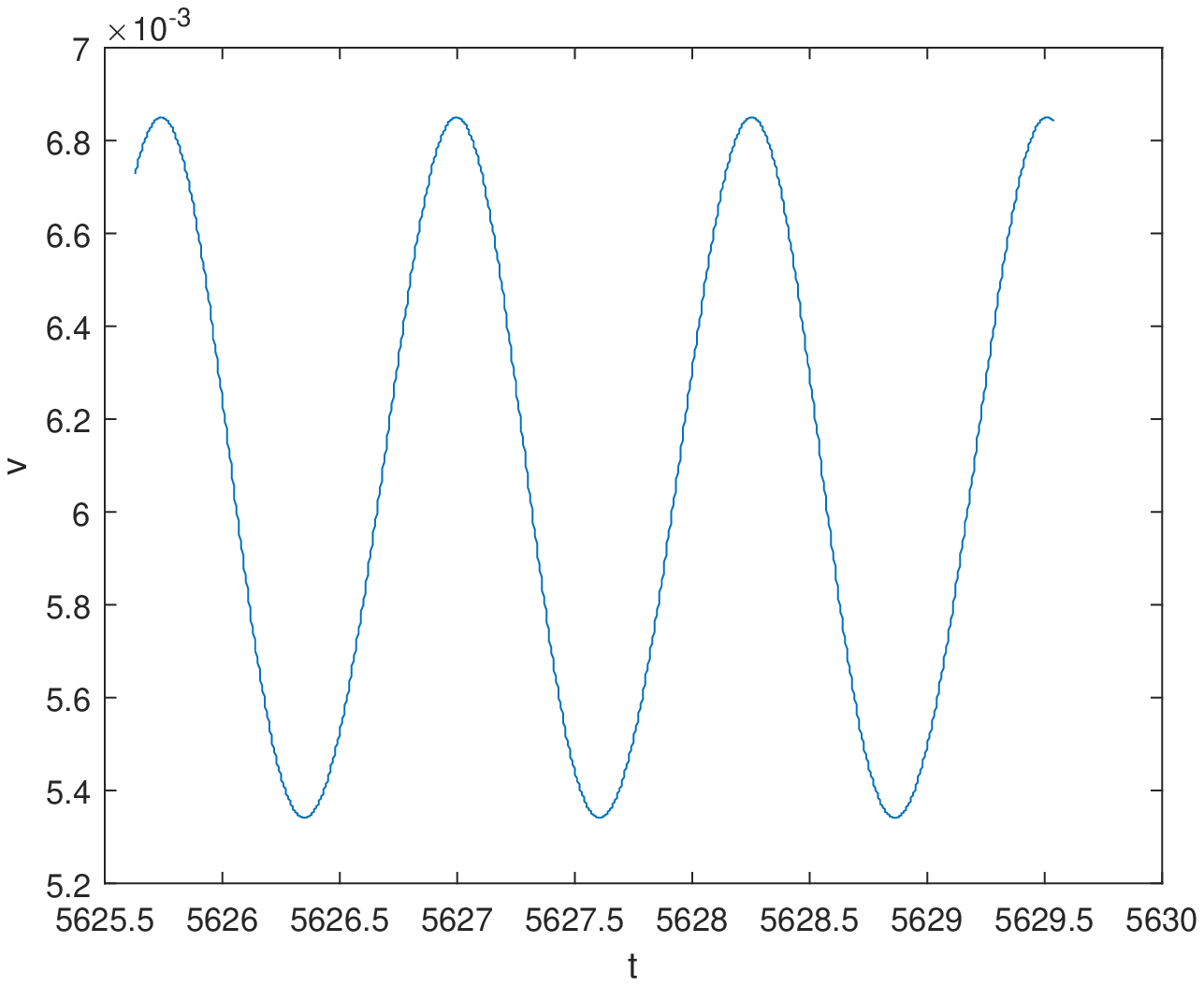}}
		
		\subfigure[]{ \label{fig:60-4000-E}
			\includegraphics[scale=0.3]{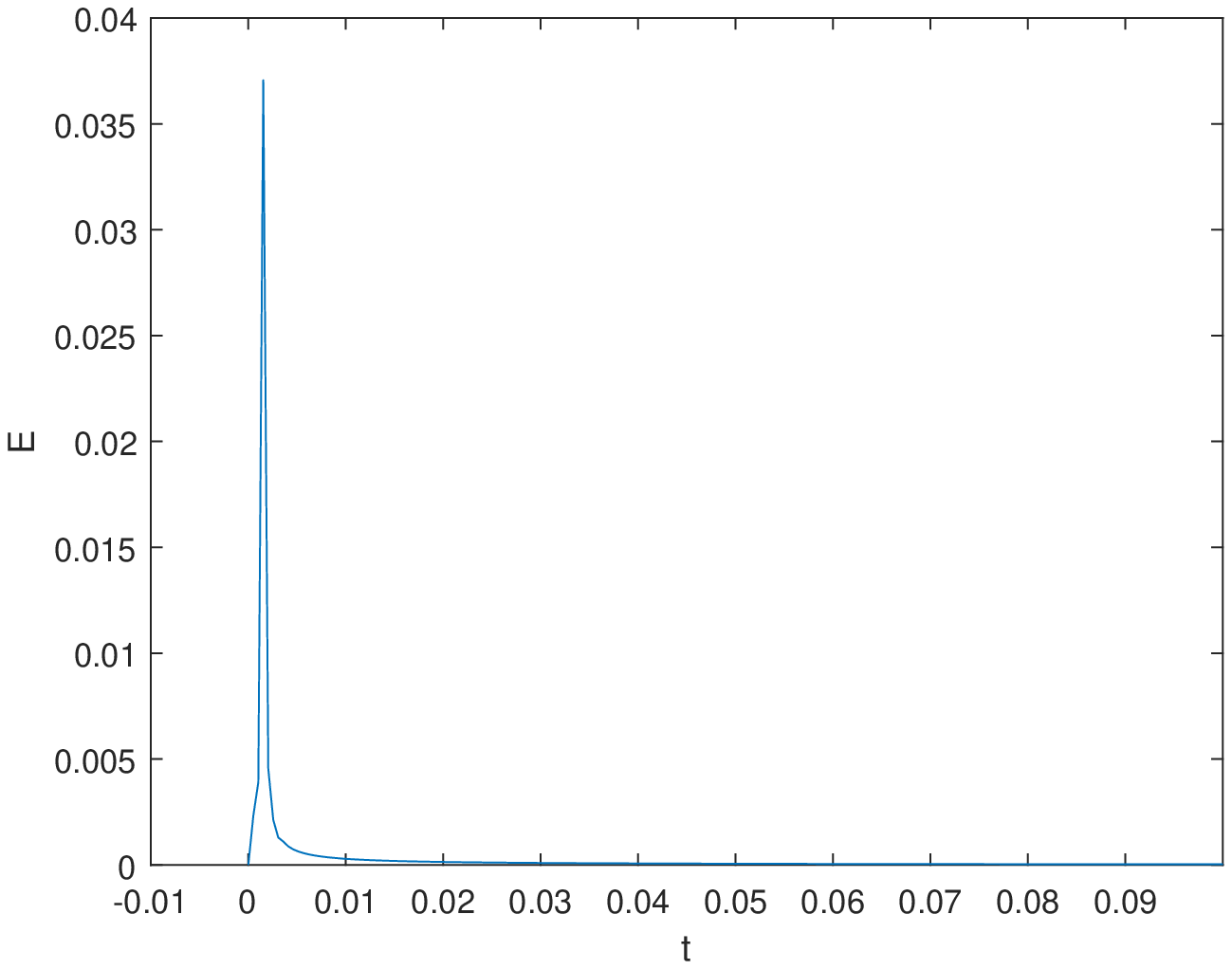}}
		\subfigure[]{ \label{fig:60-5000-E}
			\includegraphics[scale=0.3]{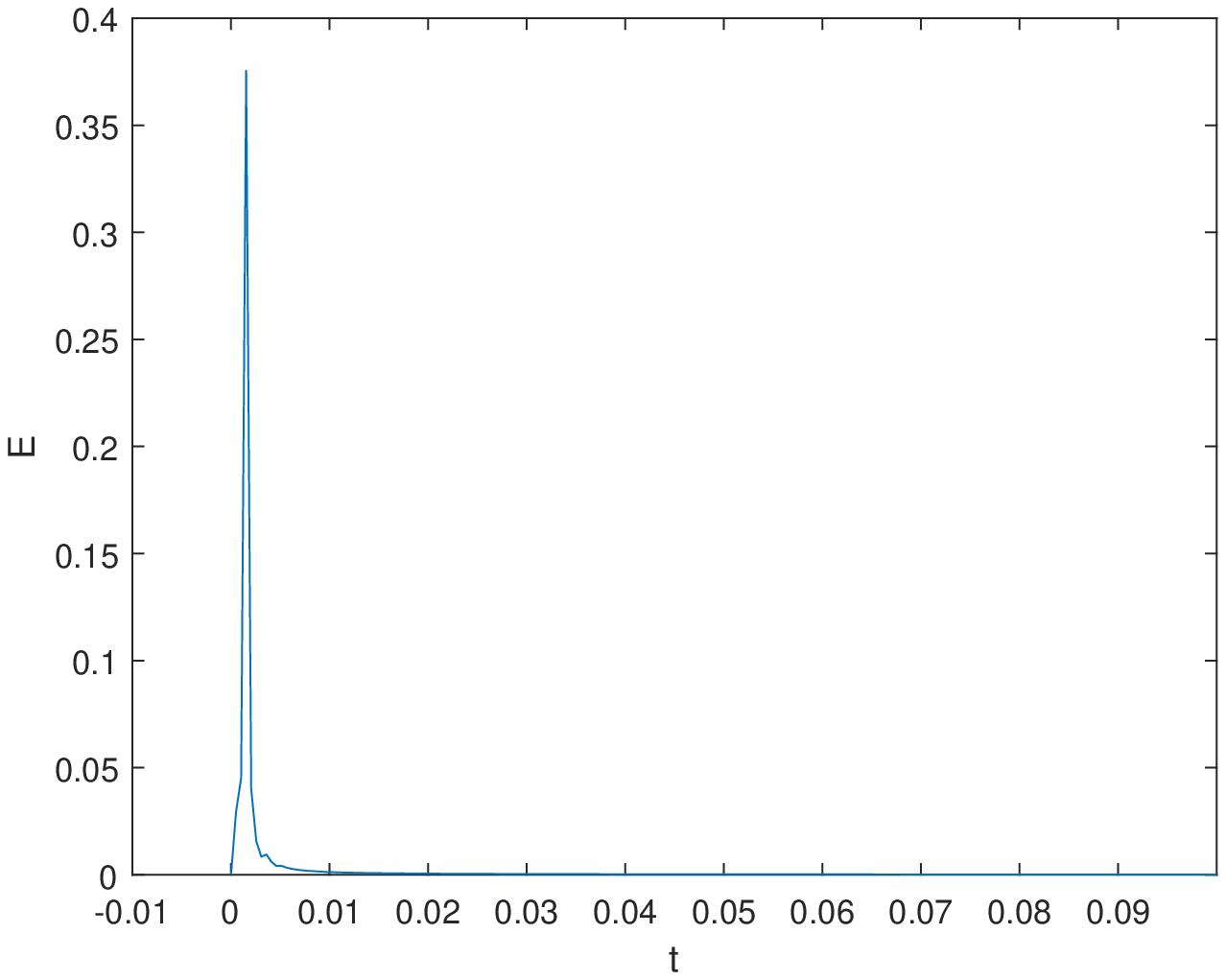}}
		\caption{ The information of TC flow with $n=1.5$ and $\theta=60^o$, the first column represents $Re=4000$ and the second column represents $Re=5000$; (a)(b) Phase-space trajectories of velocity, (c)(d) The evolution of velocity $u$, (e)(f) The evolution of velocity $v$, (g)(h) The Fourier power spectrum of kinetic energy.}
		\label{fig:60-4000}
	\end{figure}

	In addition, as the $Re$ number further increases, the flow in the cavity presents a periodic state. When $Re=4000,5000$, we also select a point in the cavity, whose coordinate is [1.07735,0.5], located at the center of the cavity, and track its velocity $(u,v)$. Figs. \ref{fig:60-4000-uv} and \ref{fig:60-5000-uv} show the phase diagrams of this point.
	As shown in Figs. \ref{fig:60-4000-u}, \ref{fig:60-4000-v}, \ref{fig:60-5000-u} and \ref{fig:60-5000-v}, the velocity changed periodically with time.
	We also track the energy at this point, which is defined as:
	\begin{equation}\label{eq5.1}
		E(t)=\frac{1}{2}\sum_\Omega||\bm u(\bm x,t)||^2d\bm x.
	\end{equation}
	where $\Omega$ represents the area of trapezoidal cavity. Then, by spectral analysis, we can get the principal frequency of periodic flow. According to Fig. \ref{fig:60-4000}, it can be concluded that when $Re=4000,5000$, the flow phenomenon presents a periodic state.
	
	\begin{figure}[htbp]\centering
		\subfigure[]{ \label{fig:60-1.0-500}
			\includegraphics[scale=0.3]{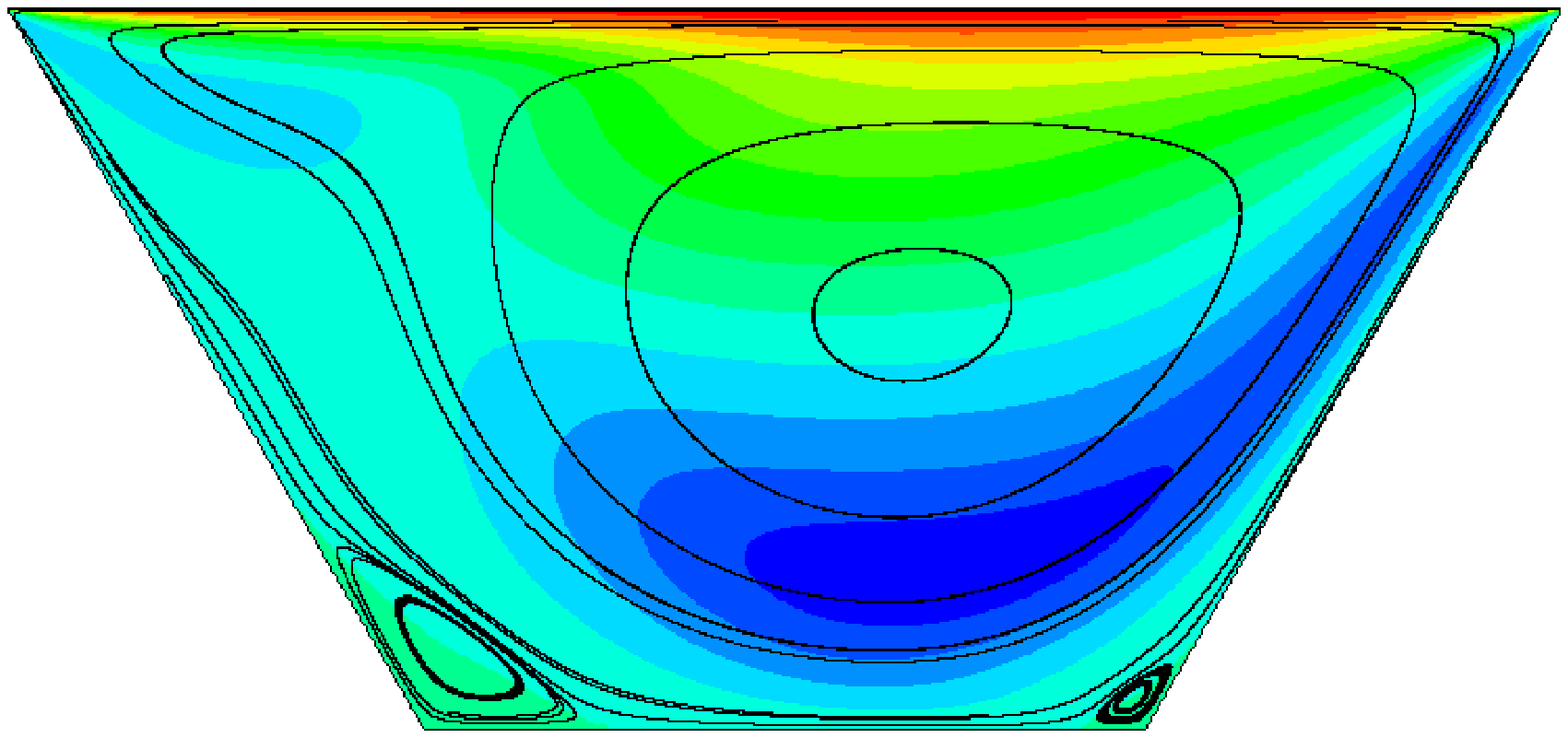}}
		\subfigure[]{ \label{fig:60-1.0-1000}
			\includegraphics[scale=0.3]{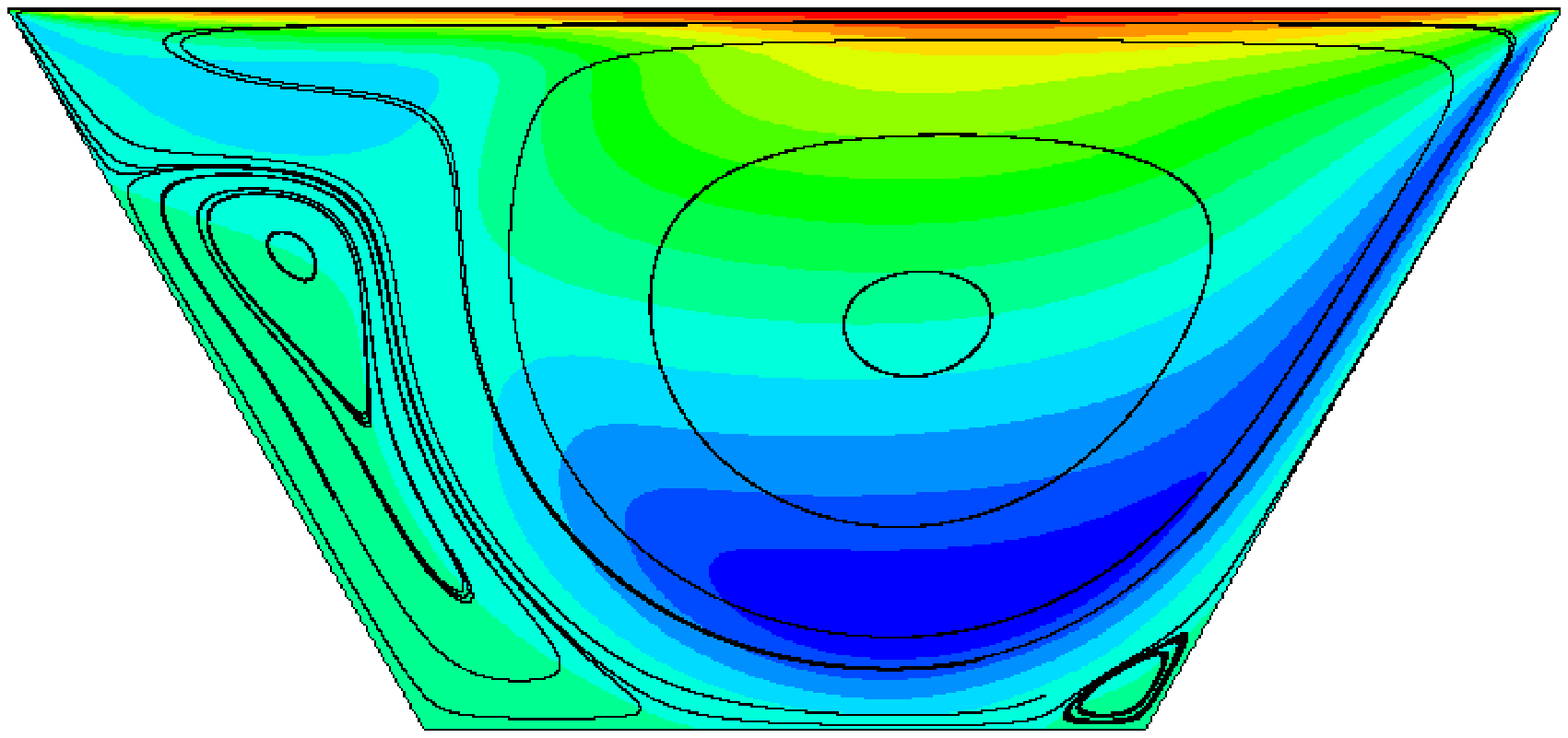}}
		\subfigure[]{ \label{fig:60-1.0-2000}
			\includegraphics[scale=0.3]{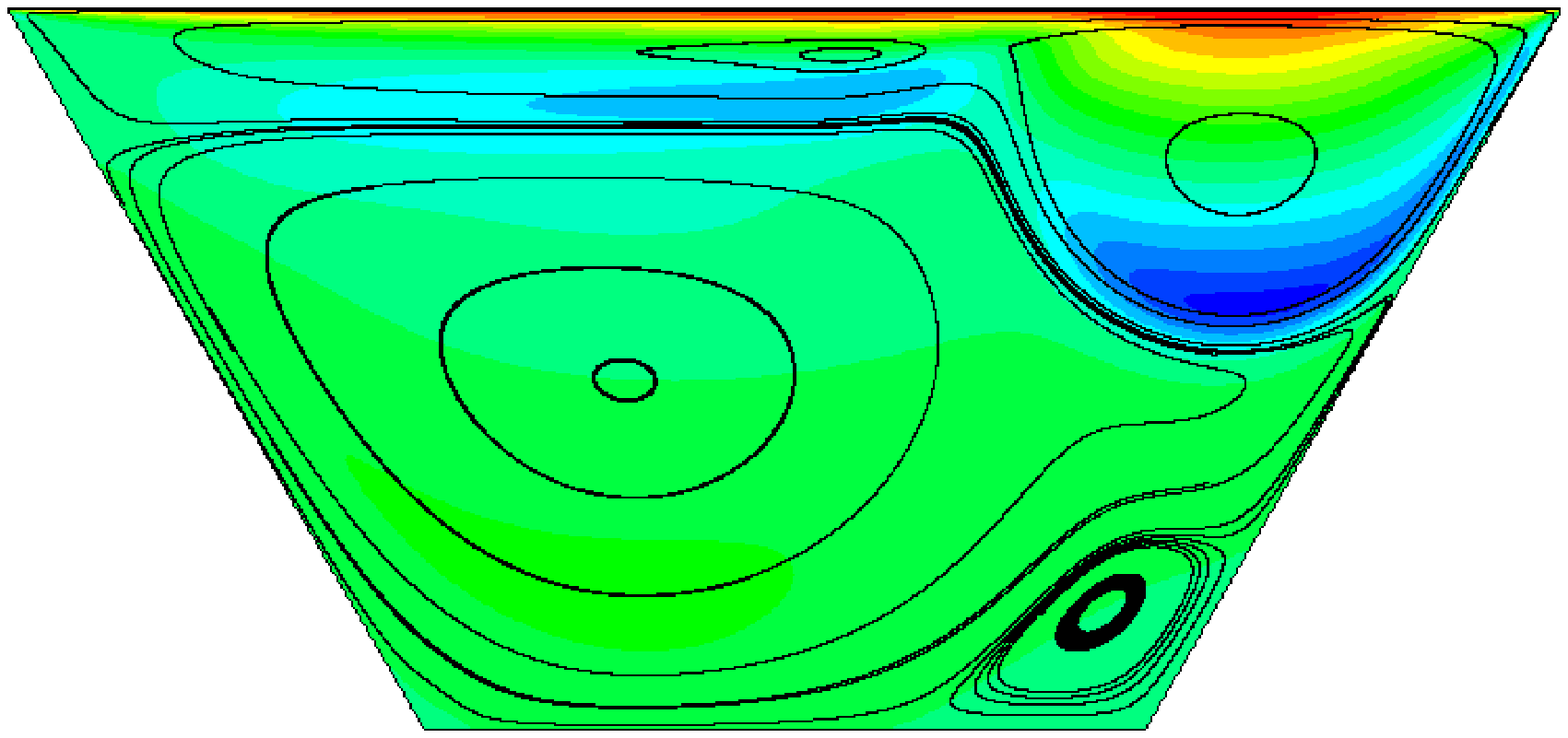}}
		\caption{ Streamline plots at $n=1.0$ and $\theta=60^o$; (a) Re=500, (b) Re=1000, (c) Re=2000.}
		\label{fig:60-1.0}
	\end{figure}
	
	\begin{figure}[htbp]\centering
		\subfigure[]{ \label{fig:60-1.0-xu}
			\includegraphics[scale=0.45]{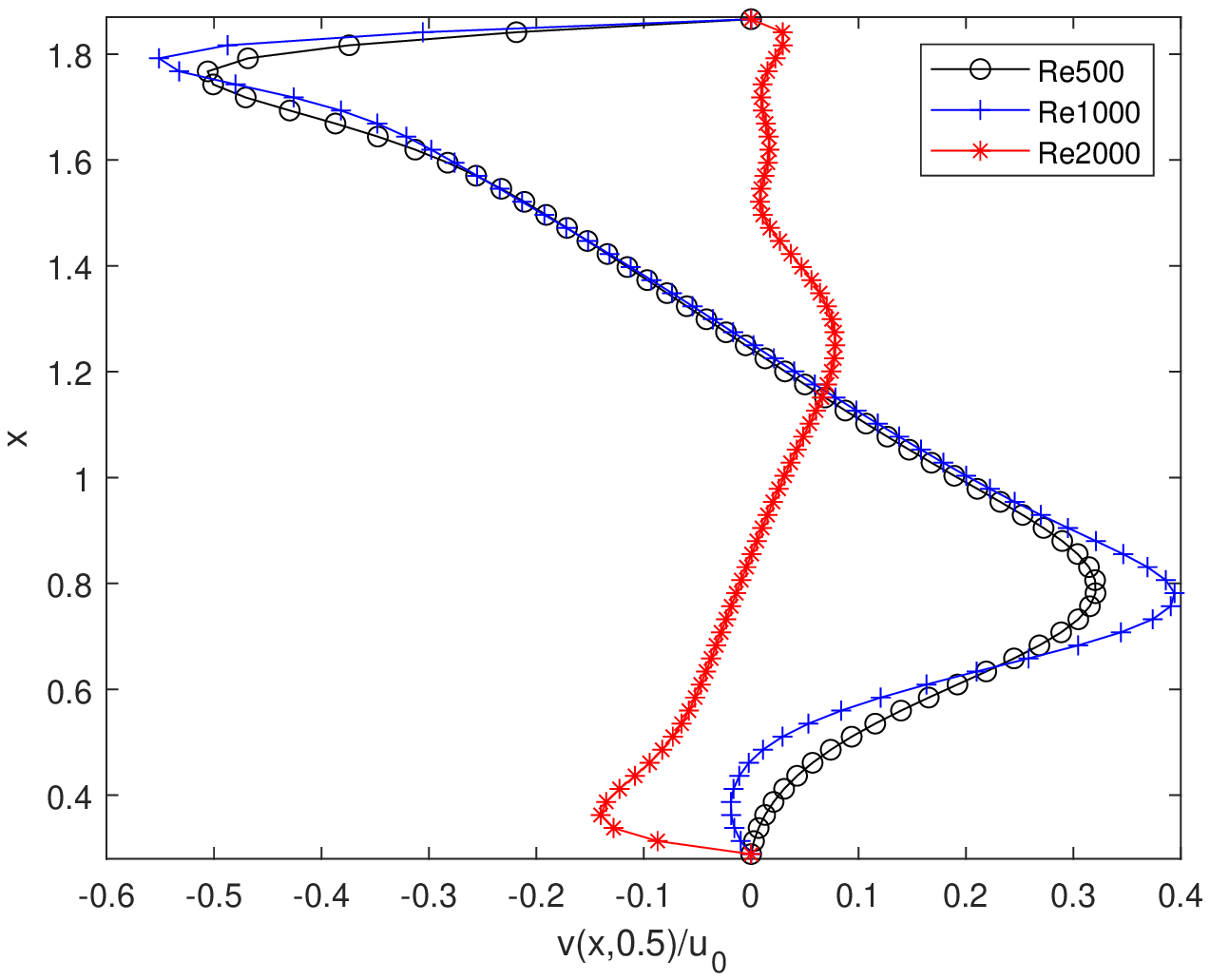}}
		\subfigure[]{ \label{fig:60-1.0-yu}
			\includegraphics[scale=0.45]{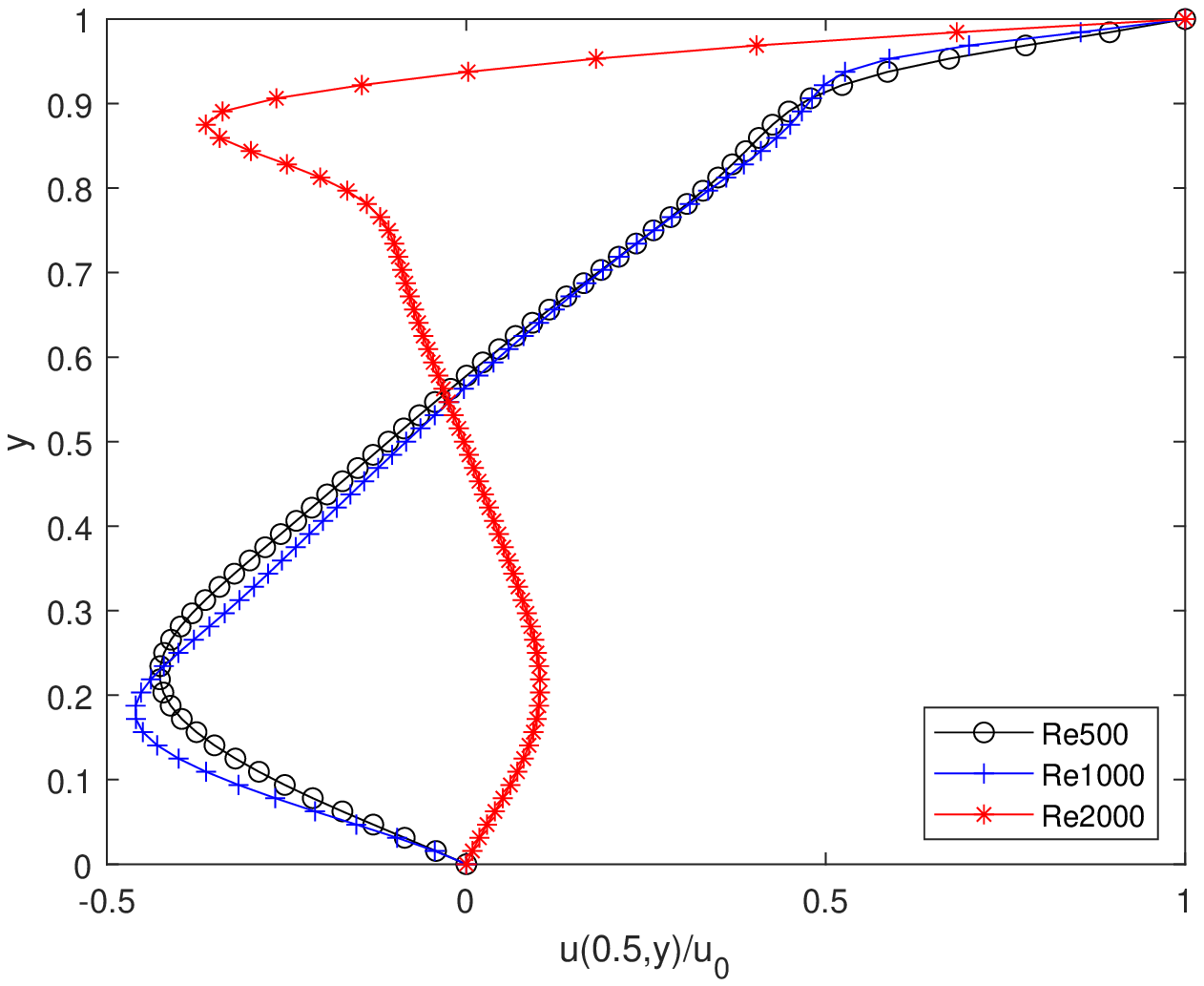}}
		\caption{ Vertical component of velocity for $\theta = 60^o$ and $n=1.0$; (a)	velocity v through y/H = 0.5 along x-axis, (b)velocity u through x/L = 0.5 along y-axis.}
		\label{fig:60-1.0-uv}
	\end{figure}
	
	Now, we consider the situation of $n=1.0$ and $\theta=60^o$. Fig. \ref{fig:60-1.0} presents the streamline plots with $Re$ changing from $500$ to $2000$. It is clear that the TC flow reaches a stable state when $Re$ number ranges from $500$ to $2000$. When $Re$ increases to $1000$, the range of secondary vortexes in the lower left and right corners begin to increase. Compared with $n=1.5$, the range of the second-order vortex grows larger and the squeezing of the first-order vortex is more obvious.
	As $Re$ number increases to $2000$, the shape of vortex in trapezoidal cavity changes obviously,
	and the distribution of vortexes are divided into three layers, which are similar to the case of $n=1.5$. However, the difference is that a three-stage vortex appears in the lower right corner, rather than the lower left.
	Compared with the case $n = 1.5$, due to the squeezing of the second-order vortex in the lower left corner, the range of the first-order vortex decreases more obviously, and the center of the first-order vortex is closer to the upper right corner of the cavity. Meanwhile, the range of second-order vortex grows bigger, the vortex in the upper left corner becomes more flattened after being squeezed, and the center of the third-order vortex in the lower right corner is closer to the right side of the trapezoidal cavity. This phenomenon is consistent with the change of vortexes shapes in Fig. \ref{fig:60-1.0}.
	
	The results of the centerline velocity with different $Re$ numbers are shown in the Fig. \ref{fig:60-1.0-uv}. As we can see, the velocity profile changes drastically when $Re=2000$. This is because the TC flow becomes more complex and the vortex morphology changes as $Re$ increases.
	
	\begin{figure}[htbp]\centering
		\subfigure[]{ \label{fig:60-1.0-2500-uv}
			\includegraphics[scale=0.3]{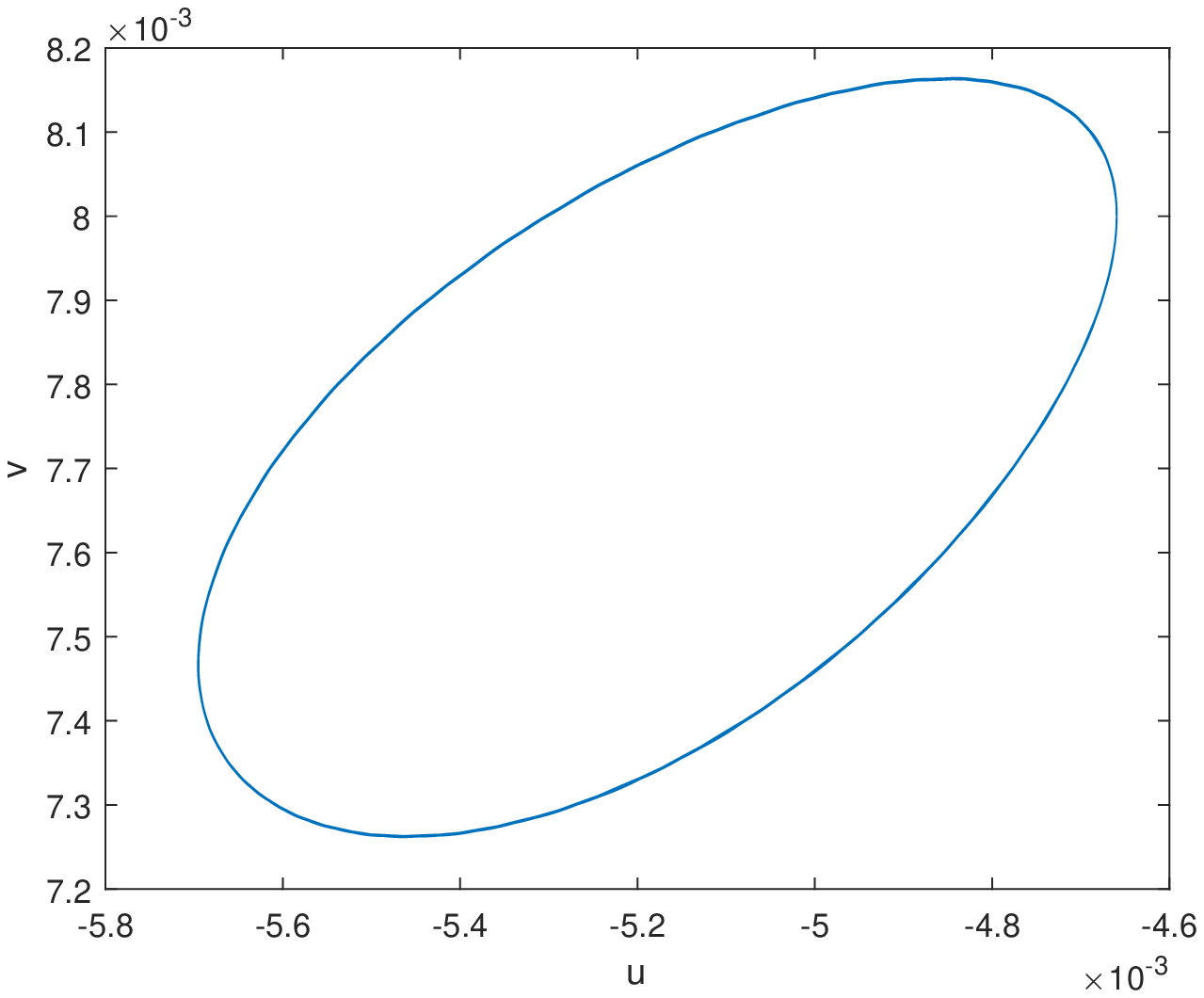}}
		\subfigure[]{ \label{fig:60-1.0-3000-uv}
			\includegraphics[scale=0.3]{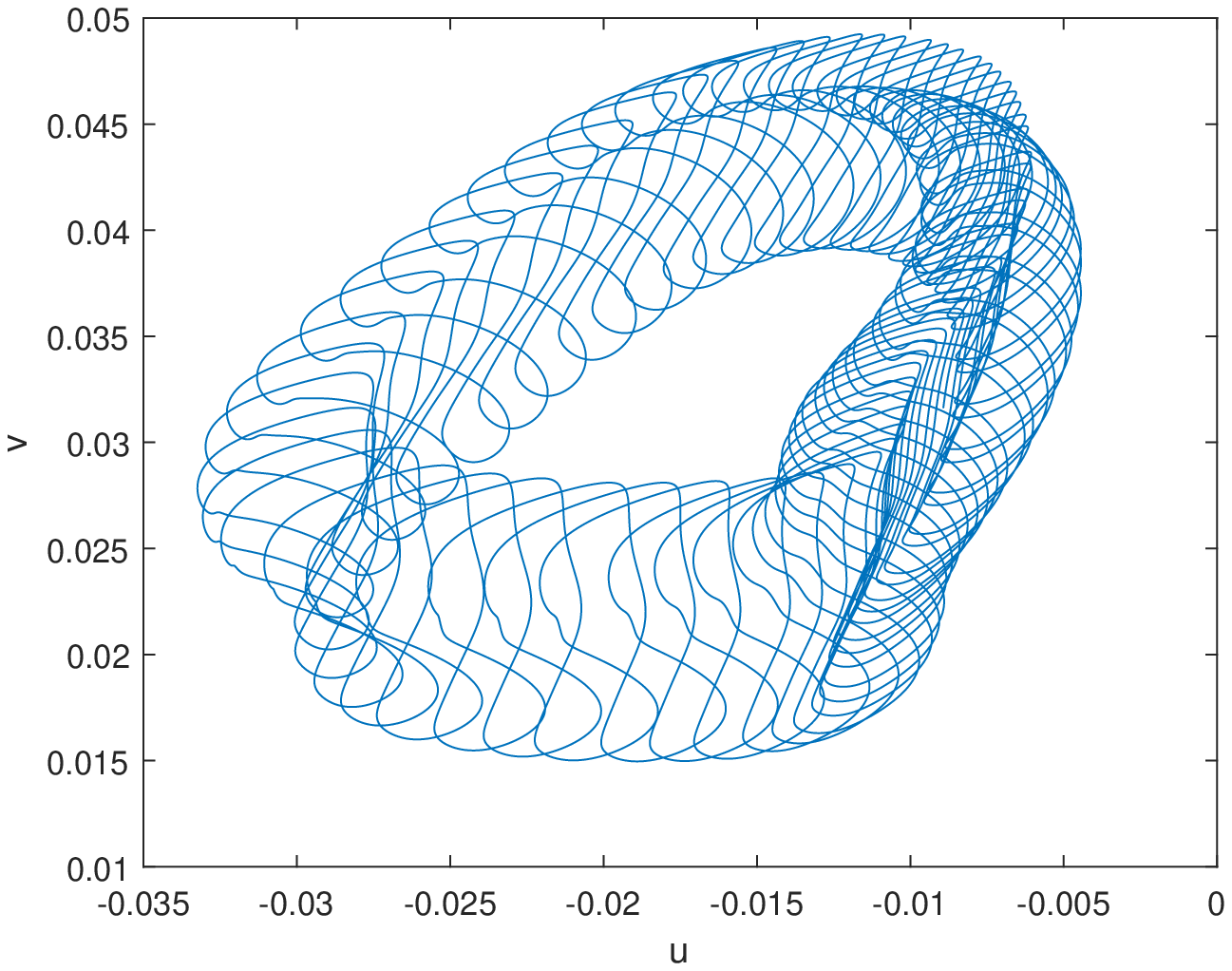}}
		\subfigure[]{ \label{fig:60-1.0-4000-uv}
			\includegraphics[scale=0.3]{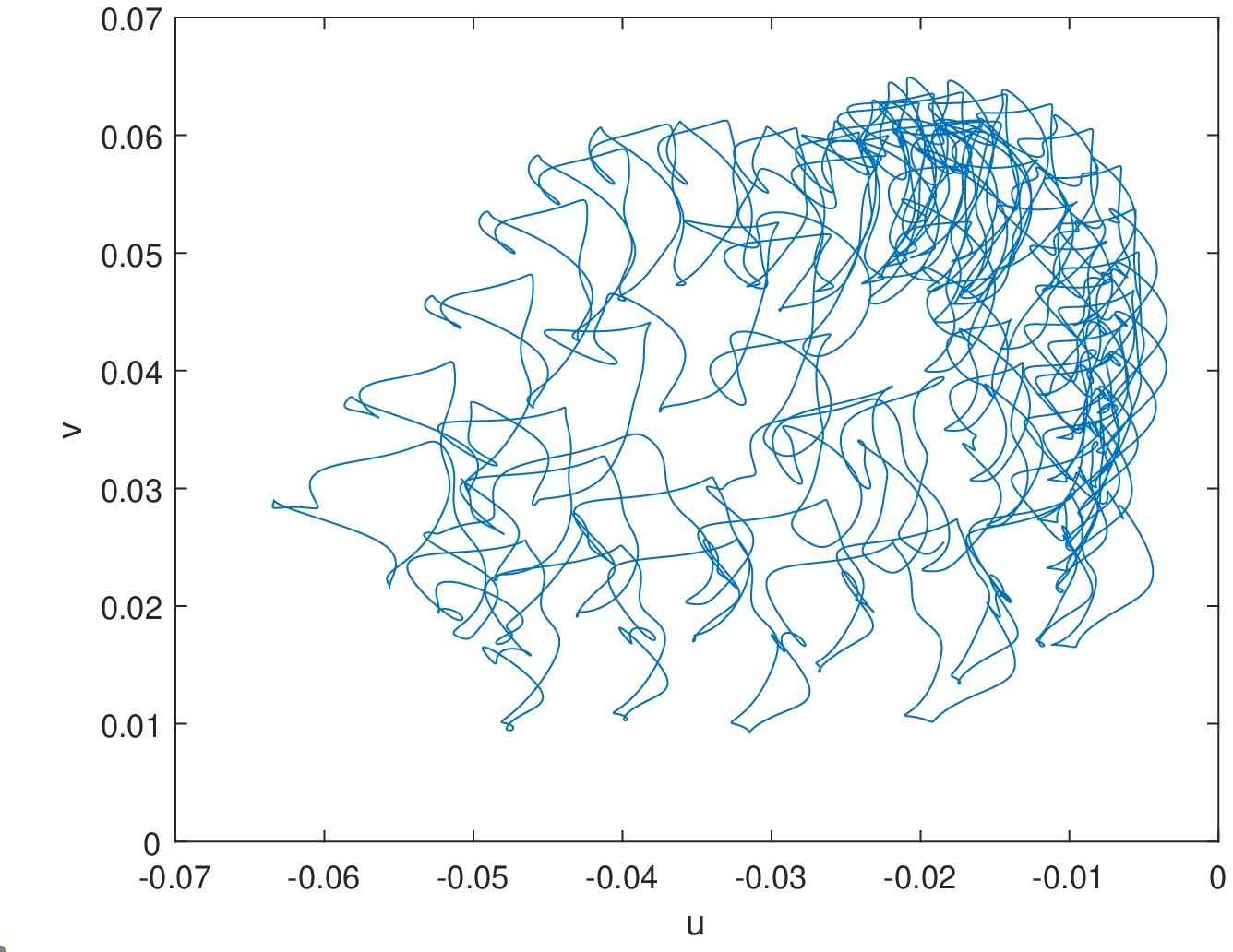}}
		
		\subfigure[]{ \label{fig:60-1.0-2500-u}
			\includegraphics[scale=0.3]{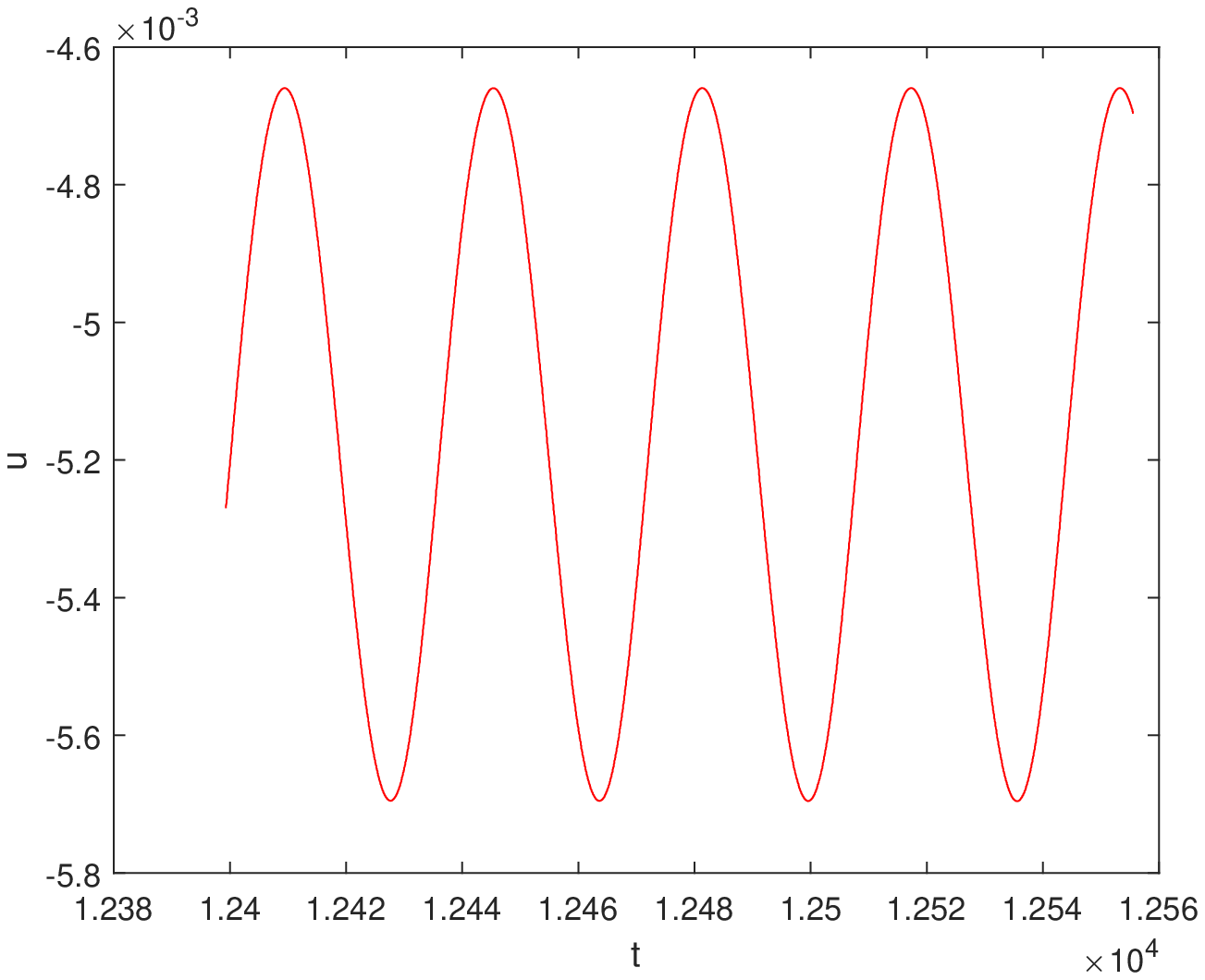}}
		\subfigure[]{ \label{fig:60-1.0-3000-u}
			\includegraphics[scale=0.3]{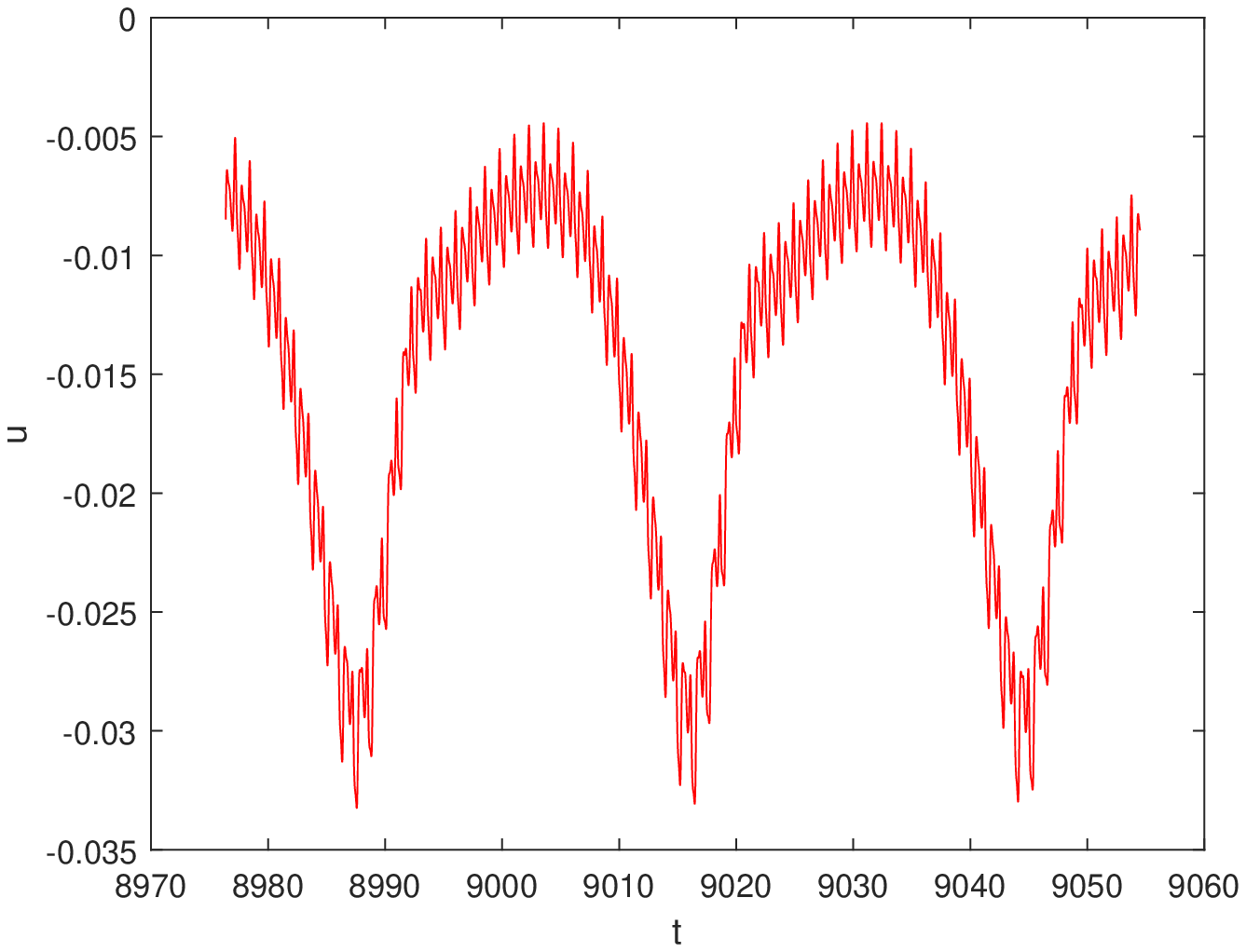}}
		\subfigure[]{ \label{fig:60-1.0-4000-u}
			\includegraphics[scale=0.3]{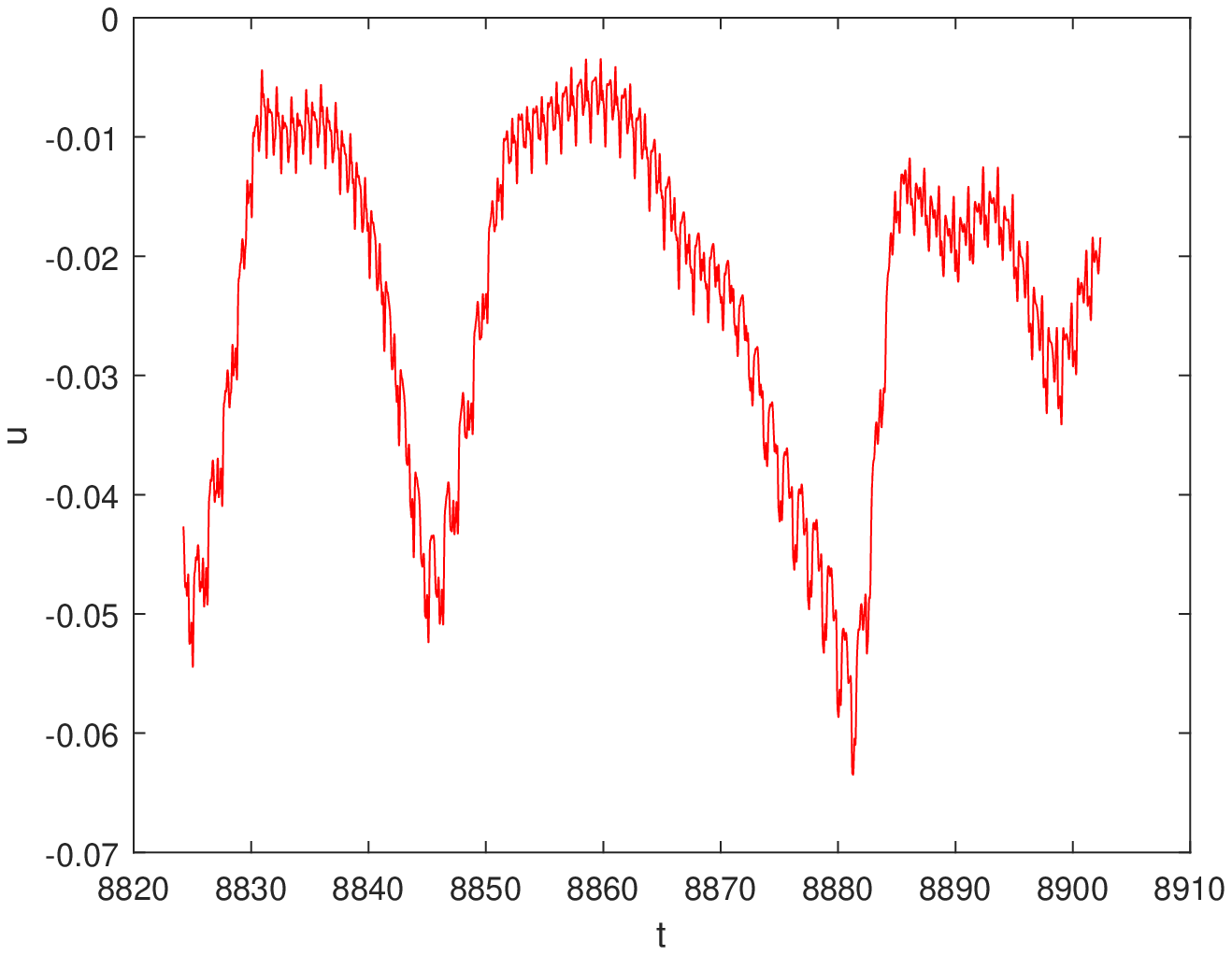}}
		
		\subfigure[]{ \label{fig:60-1.0-2500-v}
			\includegraphics[scale=0.3]{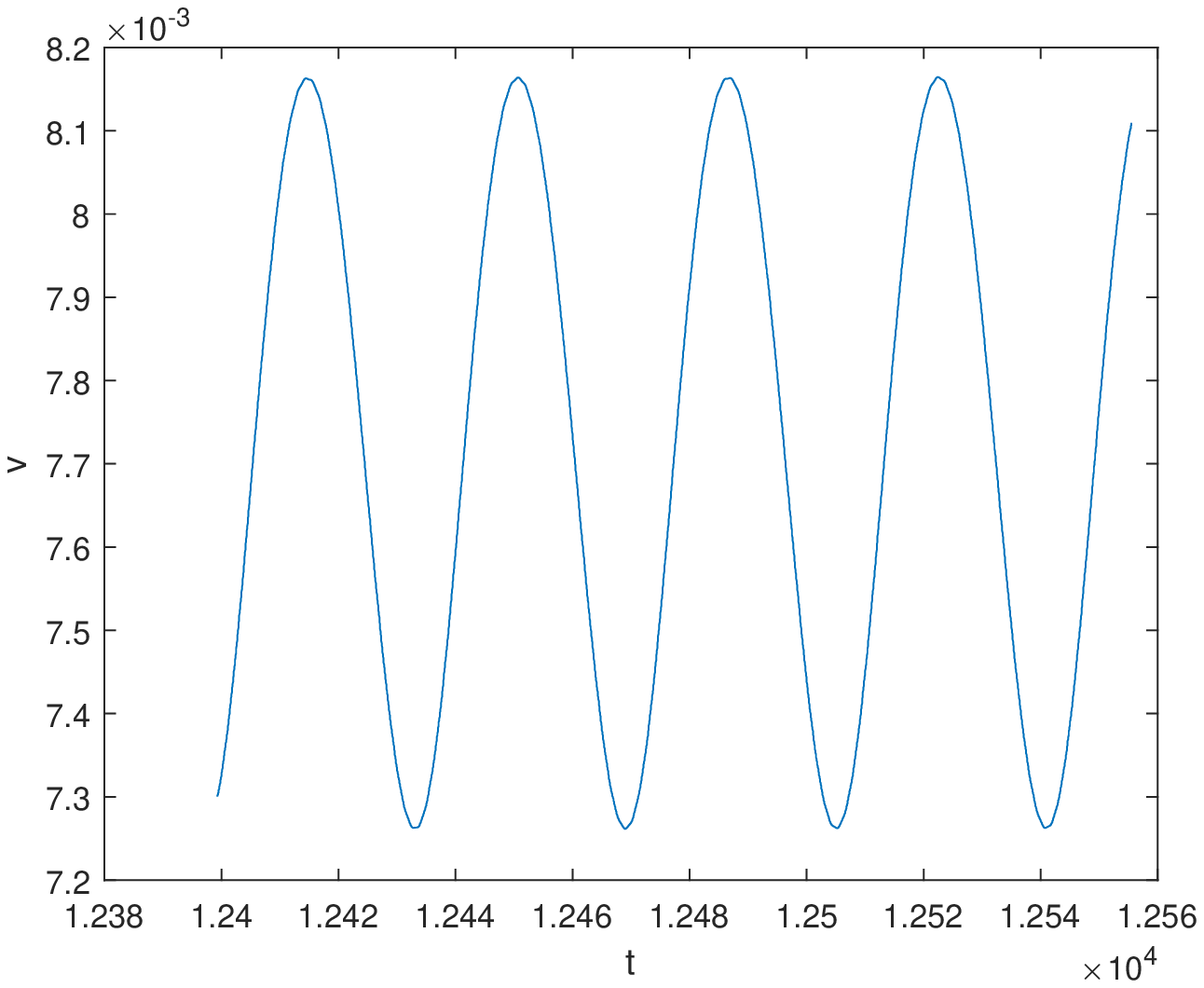}}
		\subfigure[]{ \label{fig:60-1.0-3000-v}
			\includegraphics[scale=0.3]{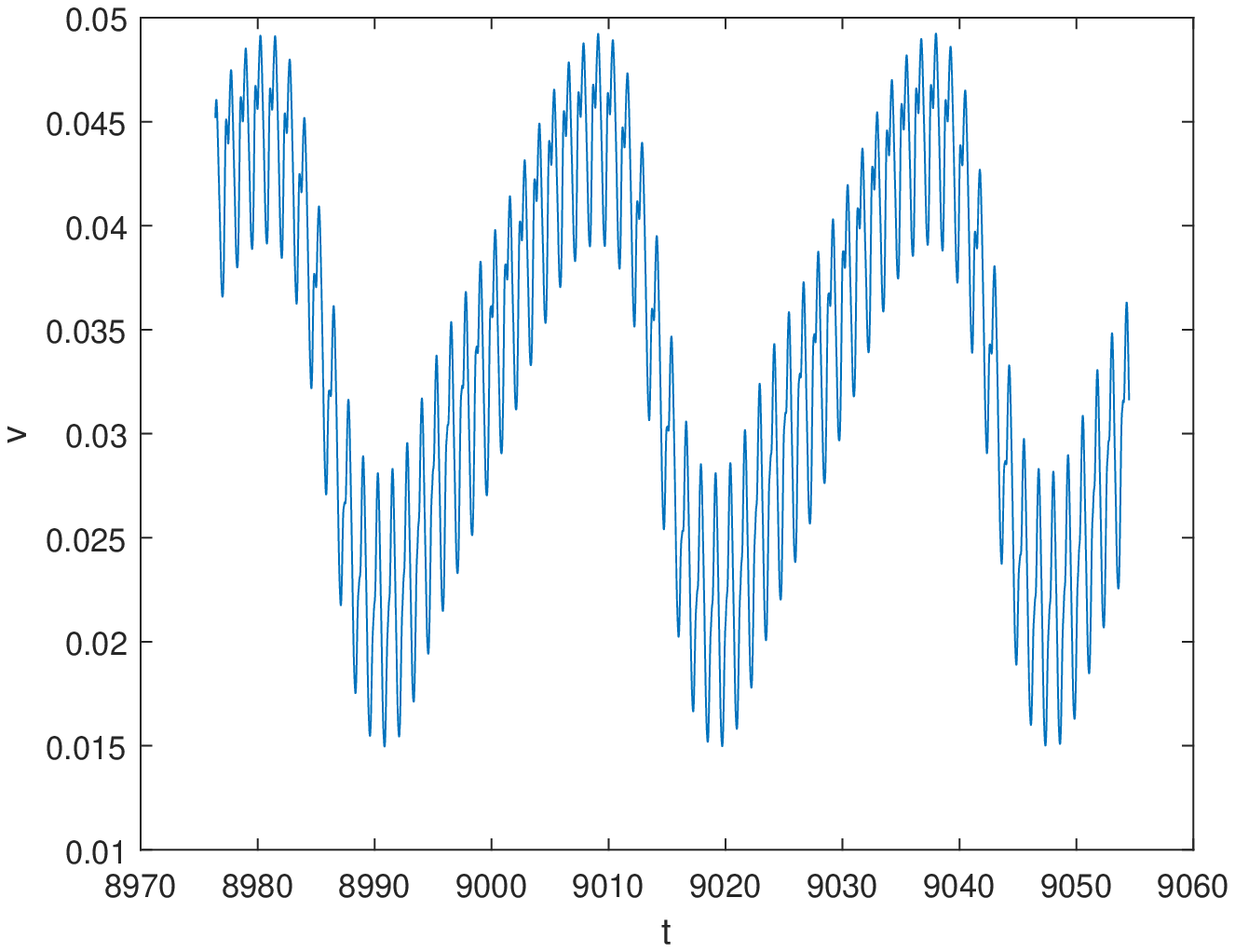}}
		\subfigure[]{ \label{fig:60-1.0-4000-v}
			\includegraphics[scale=0.3]{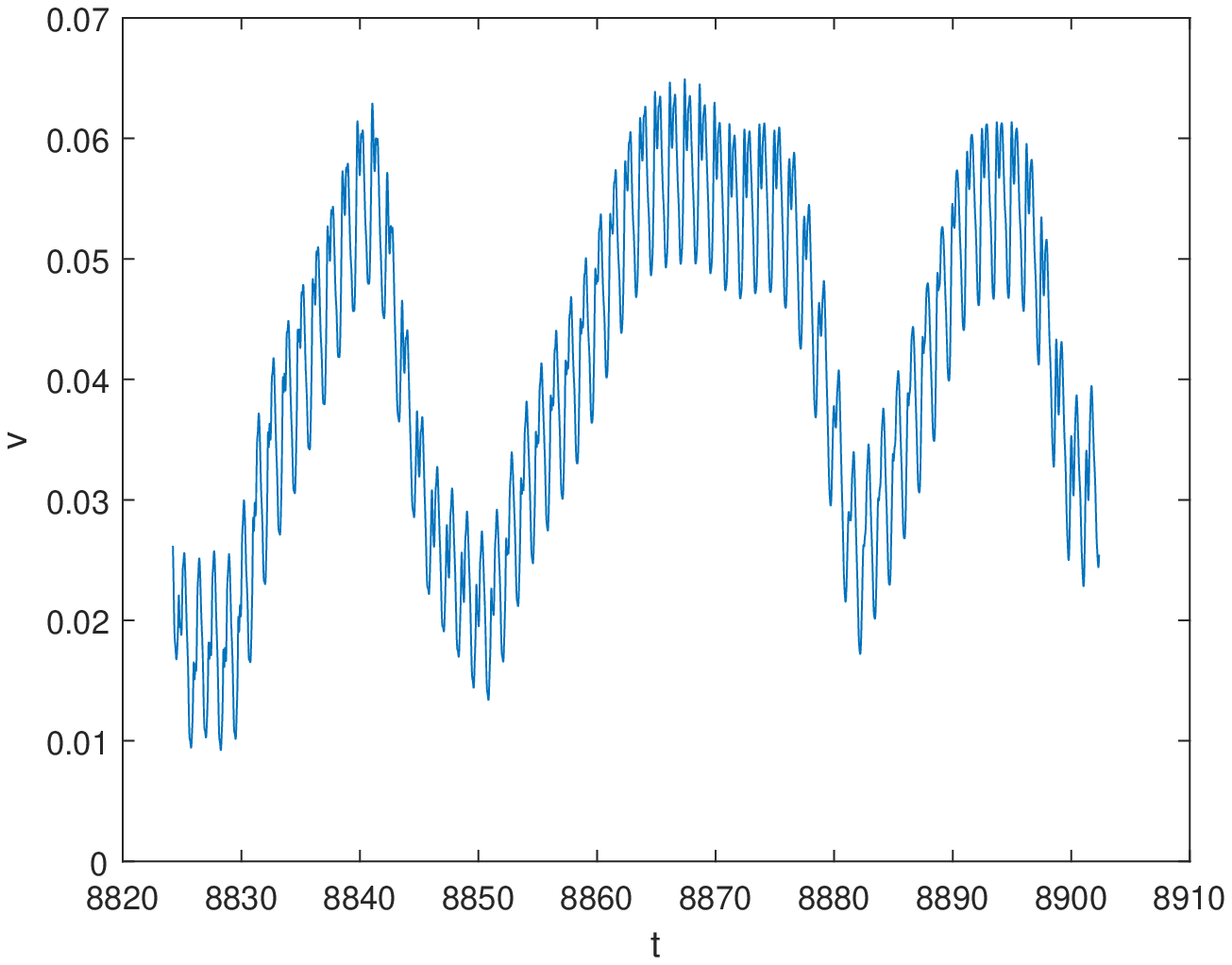}}
		
		\subfigure[]{ \label{fig:60-1.0-2500-E}
			\includegraphics[scale=0.3]{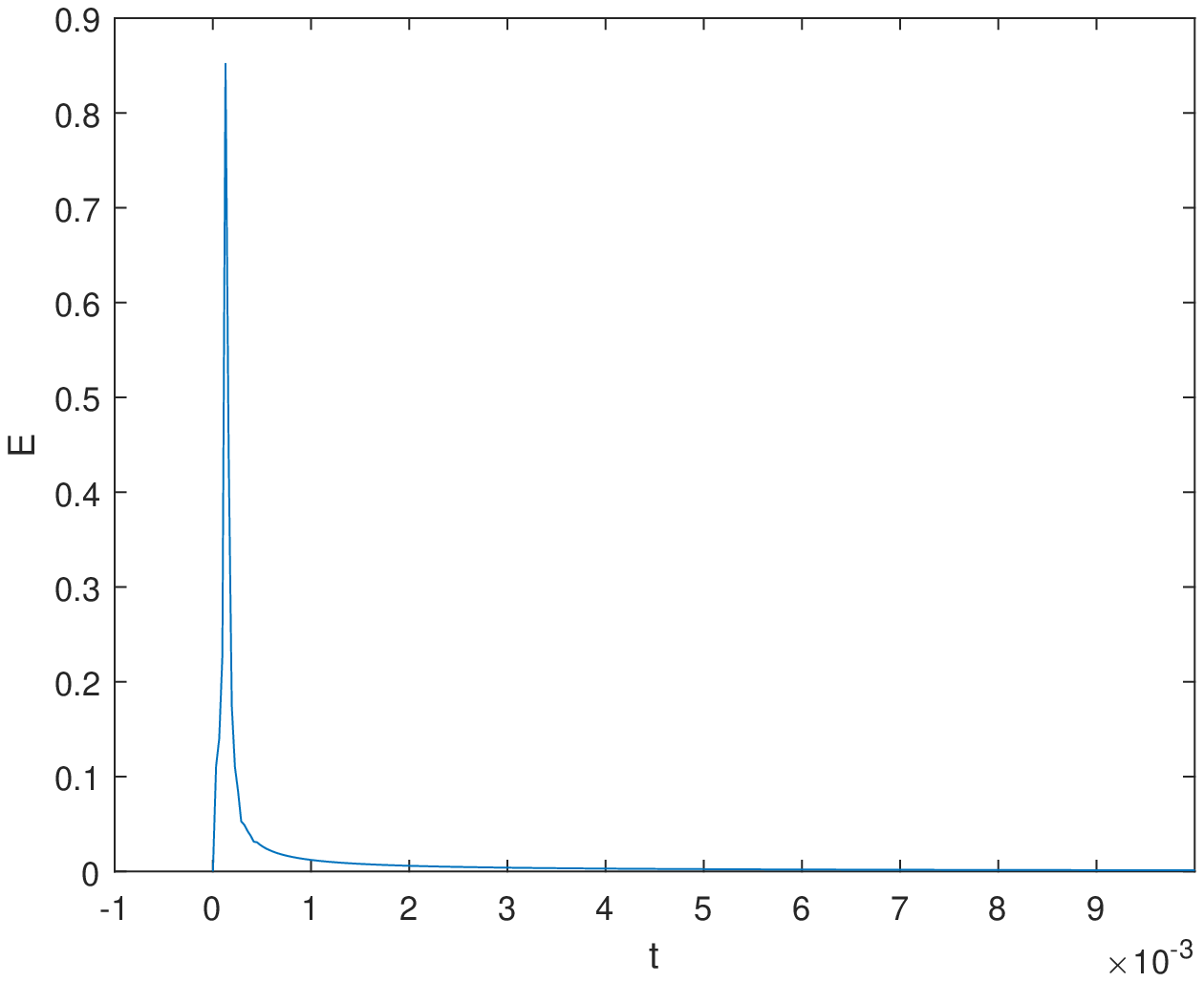}}
		\subfigure[]{ \label{fig:60-1.0-3000-E}
			\includegraphics[scale=0.3]{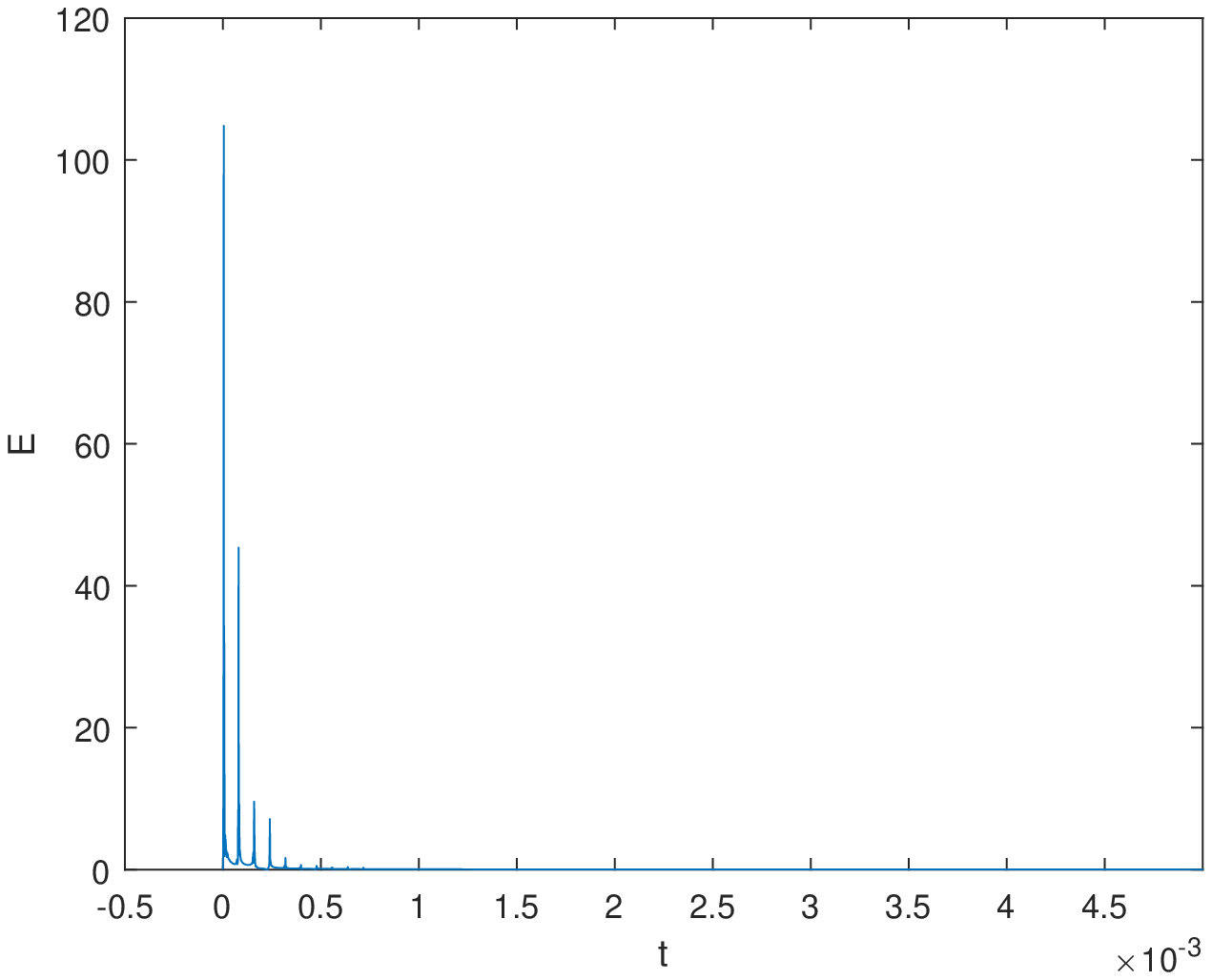}}
		\subfigure[]{ \label{fig:60-1.0-4000-E}
			\includegraphics[scale=0.3]{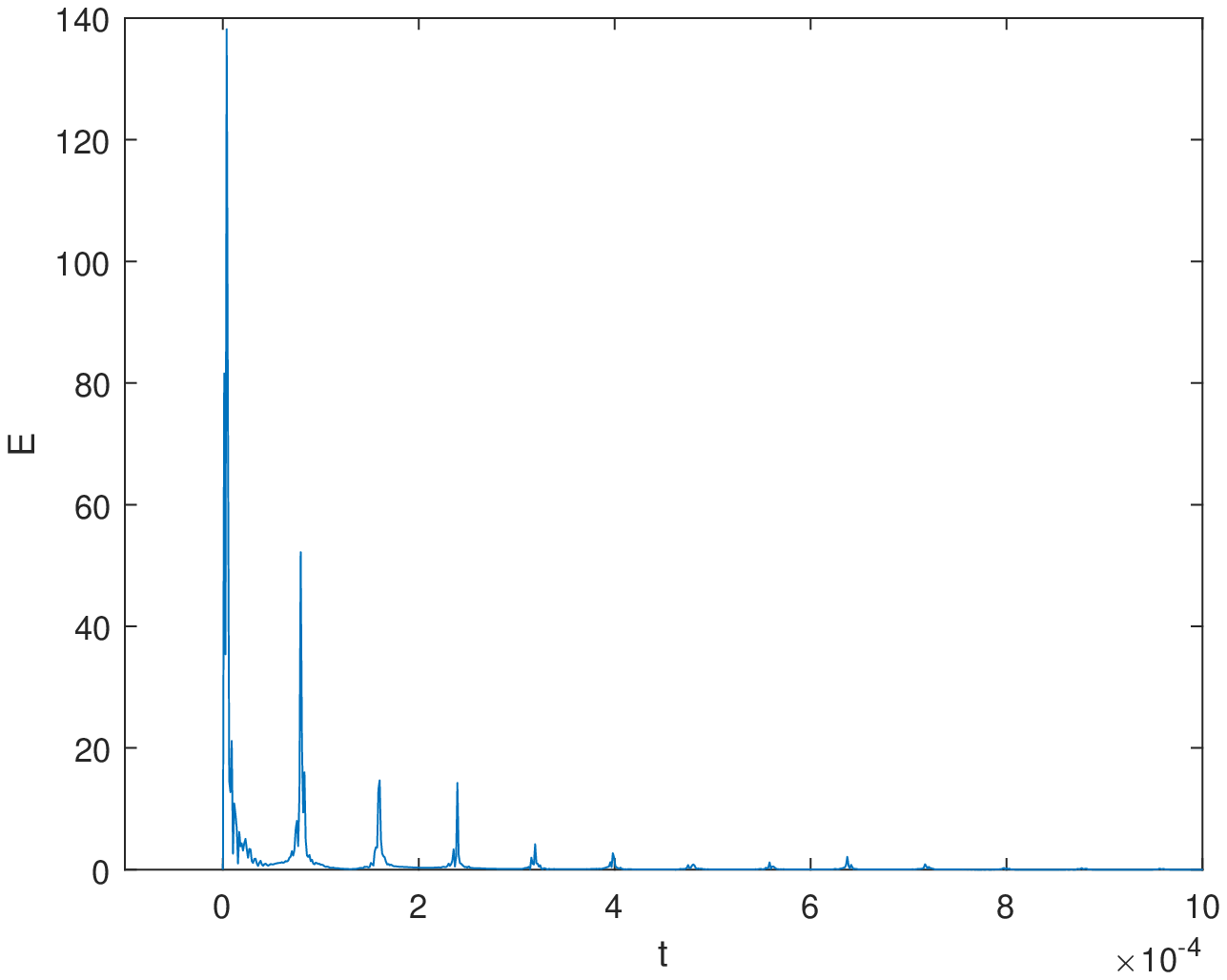}}
		\caption{ The information of TC flow with $n=1.0$ and $\theta=60^o$, the first column represents $Re=2500$, the second column represents $Re=3000$ and the third column represents $Re=4000$; (a)(b)(c) Phase-space trajectories of velocity, (d)(e)(f) The evolution of velocity $u$, (g)(h)(i) The evolution of velocity $v$, (j)(k)(l) The Fourier power spectrum of kinetic energy.}
		\label{fig:60-1.0-2500}
	\end{figure}
	
	When the $Re$ number increases to $2500$, the velocity $(u,v)$ of the point $[1.07735,0.5]$ at the center of cavity is tracked. Figs. \ref{fig:60-1.0-2500-uv}, \ref{fig:60-1.0-2500-u}, \ref{fig:60-1.0-2500-v} and \ref{fig:60-1.0-2500-E} show the phase diagrams, the evolutions of velocity $(u,v)$ and the Fourier power spectrum of kinetic energy, respectively. It is indicated that the TC flow is a periodic flow when $Re=2500$.
	
	Furthermore, the results on TC flow with $Re=3000$ are presented in Figs. \ref{fig:60-1.0-3000-uv}, \ref{fig:60-1.0-3000-u}, \ref{fig:60-1.0-3000-v} and \ref{fig:60-1.0-3000-E}. The flow phenomenons is close to a periodic state. But the results are somewhat different from the periodic state, where the spectrum of energy has more than one principal frequency and the phase diagram is not a simple closed ring. We define this state as quasi-periodic. Then, the study results on TC flow with $Re=4000$ are shown in Figs. \ref{fig:60-1.0-4000-uv}, \ref{fig:60-1.0-4000-u}, \ref{fig:60-1.0-4000-v} and \ref{fig:60-1.0-4000-E}. It is clear that the phase diagrams and velocity evolution diagrams become more complex and irregular, and the energy spectrum appears multiple local peaks. So the TC flow is turbulence when $Re=4000$. According to those flow phenomenons, we can find that the flow at low Reynolds numbers will develop into periodic flows and then it will turn to the turbulence flow as the Re number increases.
	
	\begin{figure}[htbp]\centering
		\subfigure[]{ \label{fig:60-0.5-500}
			\includegraphics[scale=0.3]{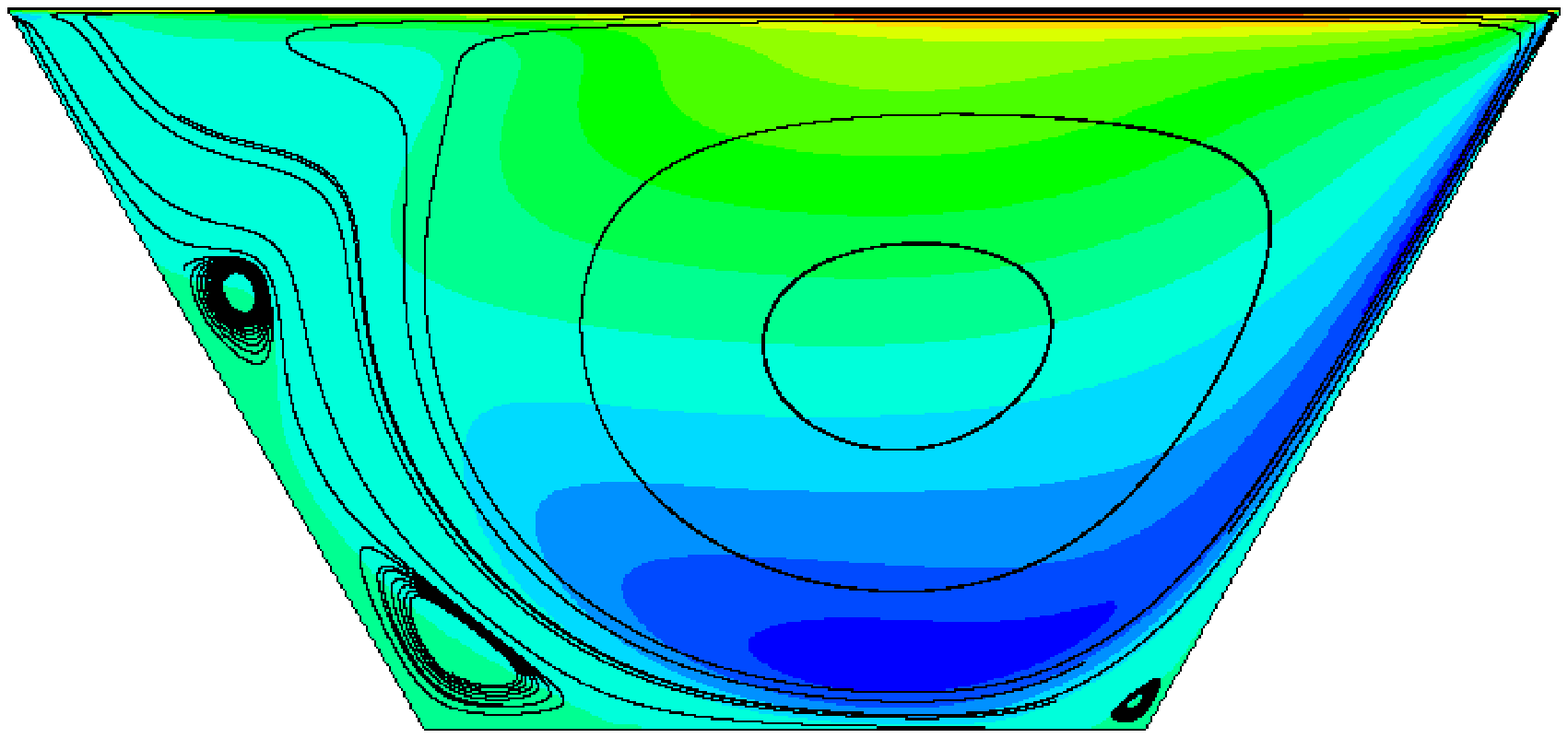}}
		\subfigure[]{ \label{fig:60-0.5-750}
			\includegraphics[scale=0.3]{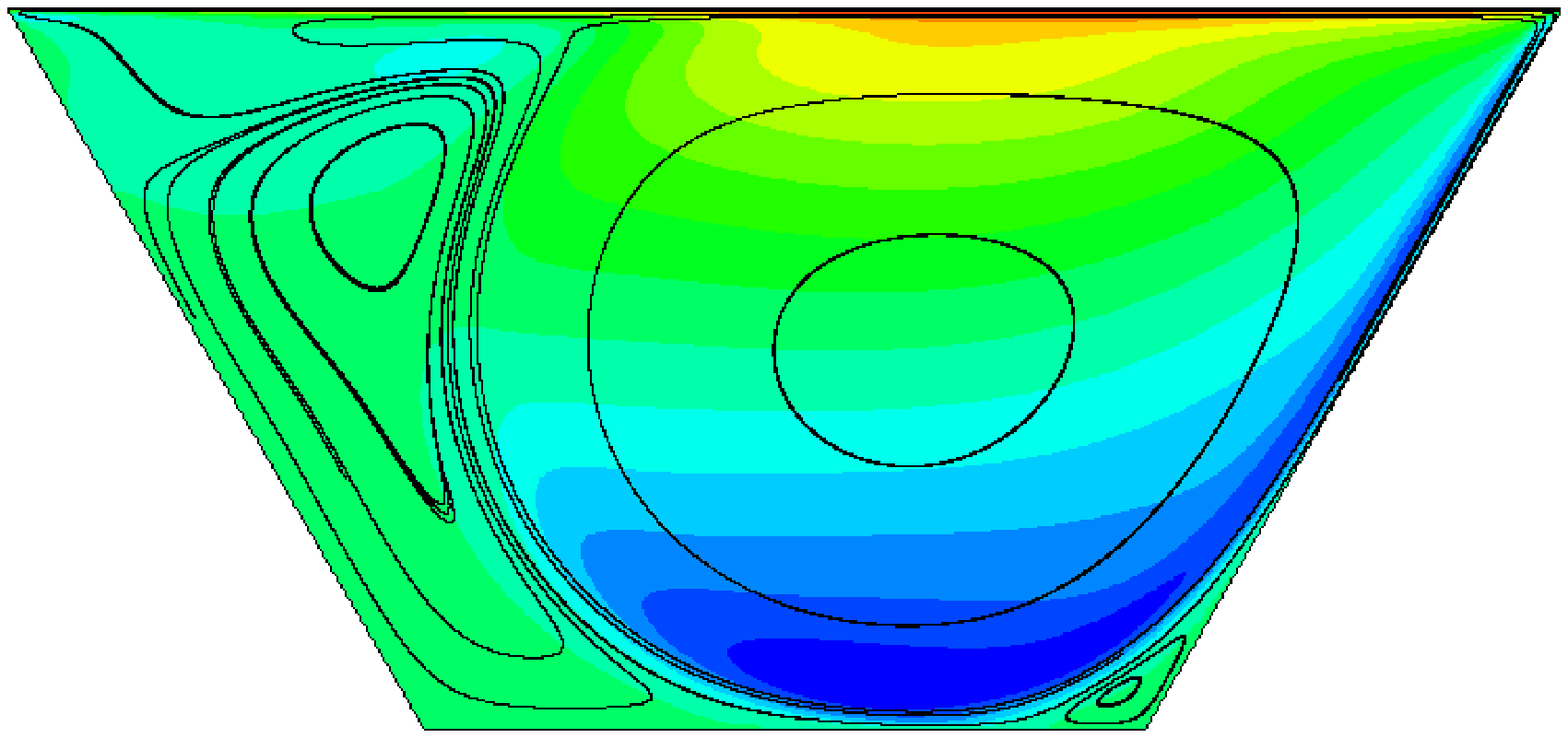}}
		\caption{ Streamline plots at $n=0.5$ and $\theta=60^o$; (a) Re=500, (b) Re=750.}
		\label{fig:60-0.5}
	\end{figure}
	
	\begin{figure}[htbp]\centering
		\subfigure[]{ \label{fig:60-0.5-xu}
			\includegraphics[scale=0.45]{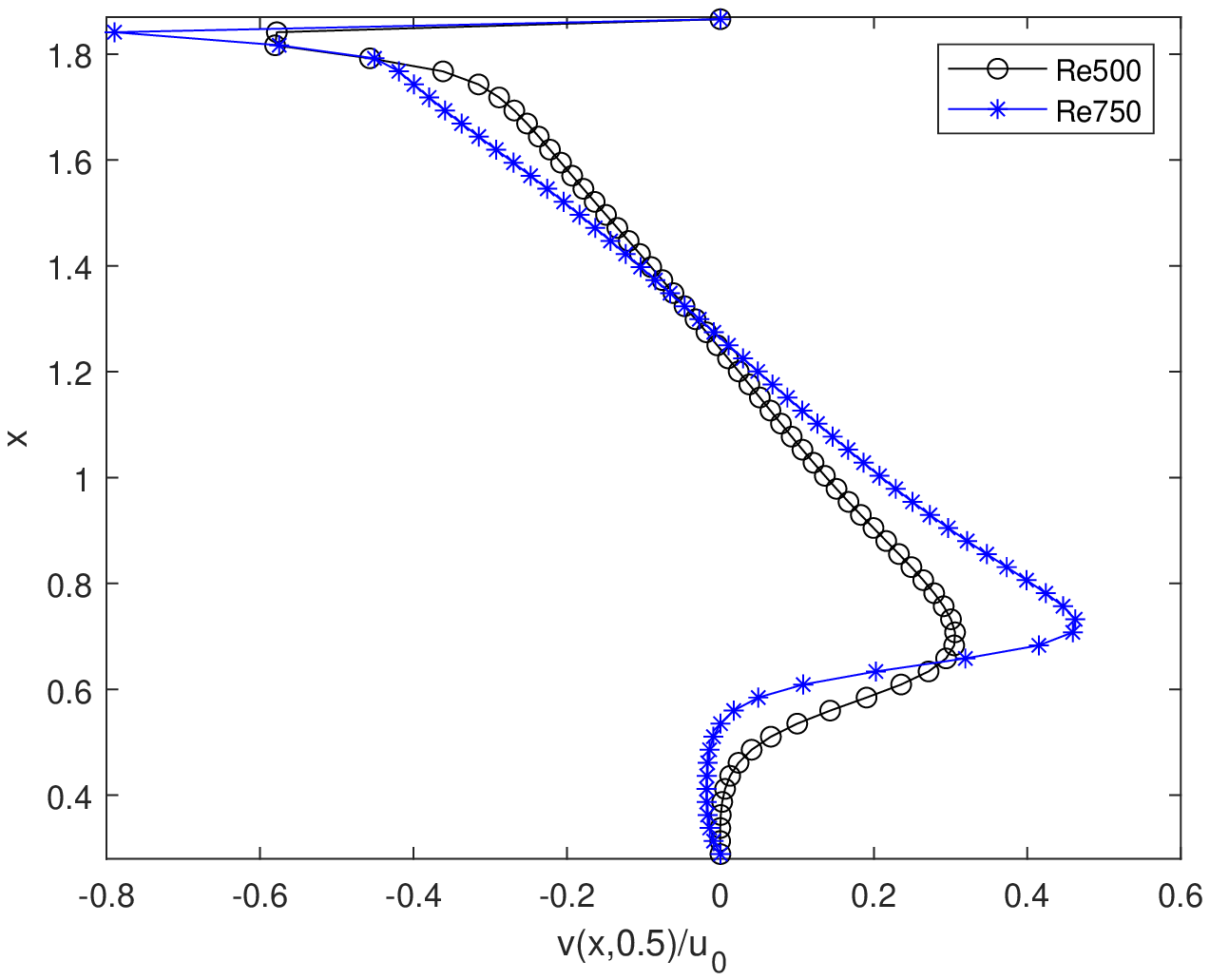}}
		\subfigure[]{ \label{fig:60-0.5-yu}
			\includegraphics[scale=0.45]{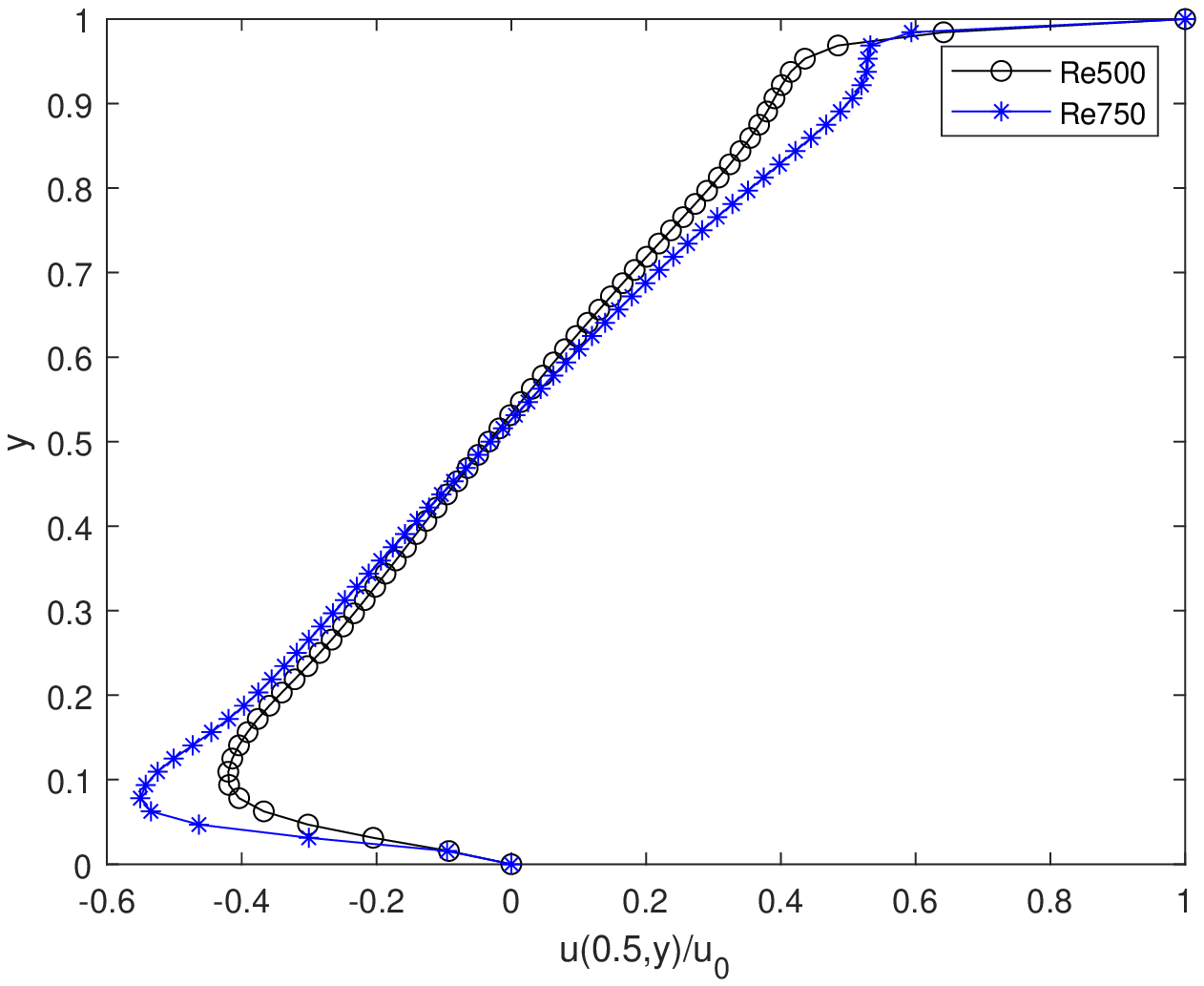}}
		\caption{ Vertical component of velocity for $\theta= 60^\circ$ and $n=0.5$; (a) velocity v through y/H = 0.5 along x-axis, (b)velocity u through x/L = 0.5 along y-axis.}
		\label{fig:60-0.5-uv}
	\end{figure}
	
	Next, we discuss the situation of $\theta=60^o$ and $n=0.5$. The Streamline plots are presented in Fig. \ref{fig:60-0.5}. When $Re=500$ and $Re=750$, the TC flow is a stable flow. There is a secondary vortex in the upper left corner of the trapezoidal cavity when $Re=500$, while this vortex is absent in the other two cases ($n=1.5$ and $n=1.0$). In addition, as the $Re$ number increases to $750$, the two second-order vortices on the left fuse into a larger vortex, and squeeze the first-order vortex. The results on centerline velocity are shown in Fig. \ref{fig:60-0.5-uv}.
	The development trend of velocity with $Re=500$ is similar to that with $Re=750$,
	because the structure of the first-order vortex does not change. However, when $Re=750$, the peak velocity is a little larger.

	\begin{figure}[htbp]\centering
		\subfigure[]{ \label{fig:60-0.5-1000-uv}
			\includegraphics[scale=0.3]{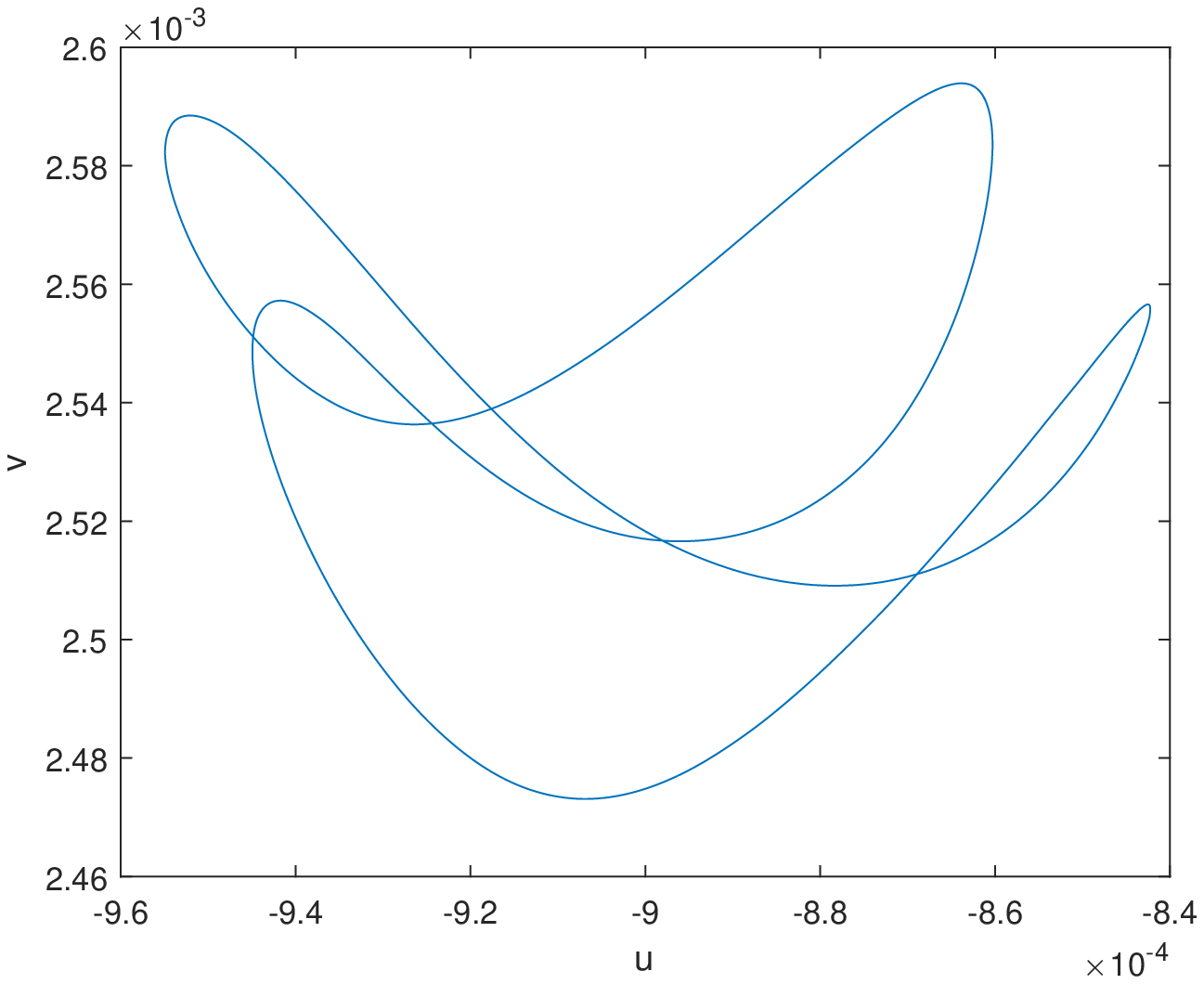}}
		\subfigure[]{ \label{fig:60-0.5-2000-uv}
			\includegraphics[scale=0.3]{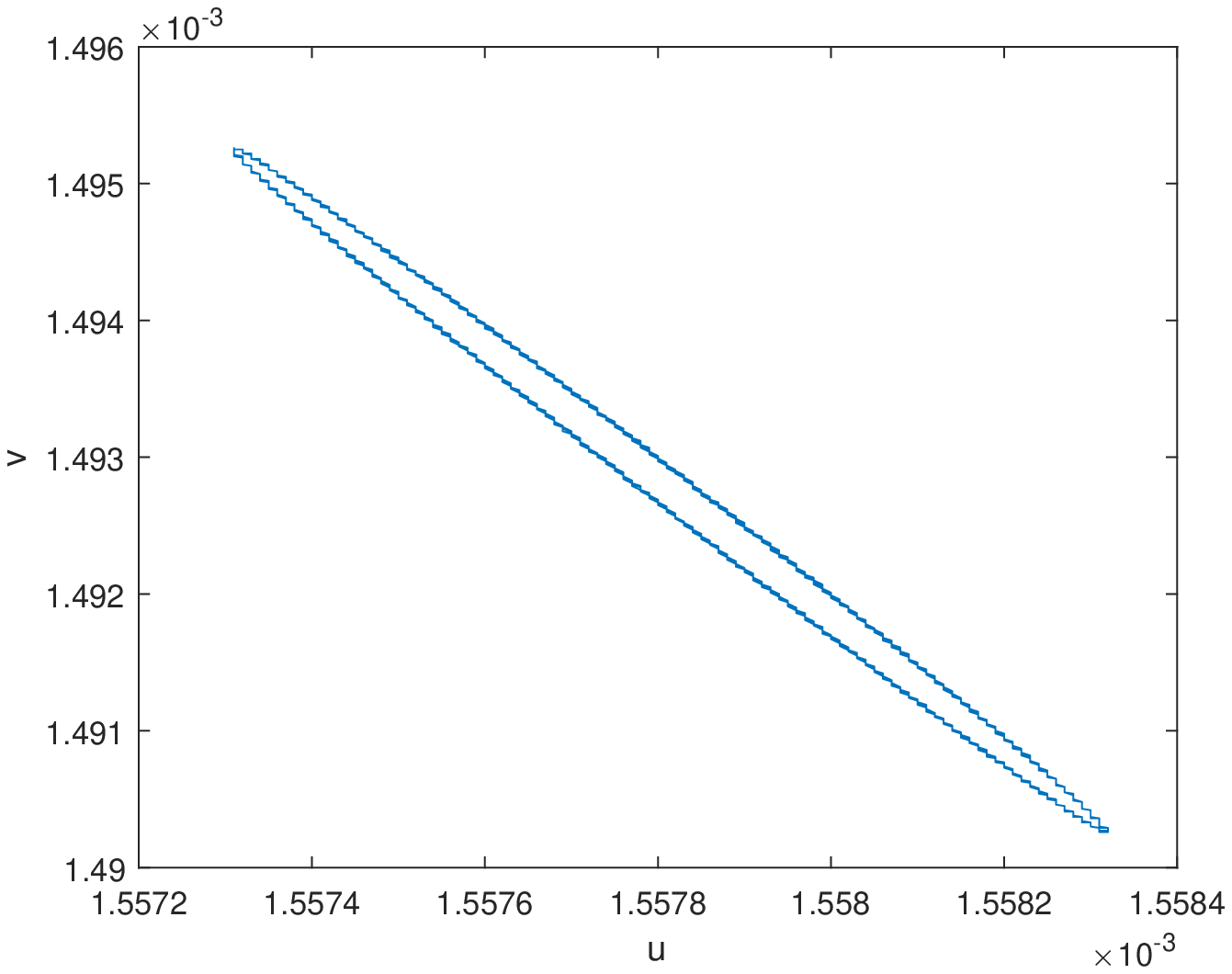}}
		
		\subfigure[]{ \label{fig:60-0.5-1000-u}
			\includegraphics[scale=0.3]{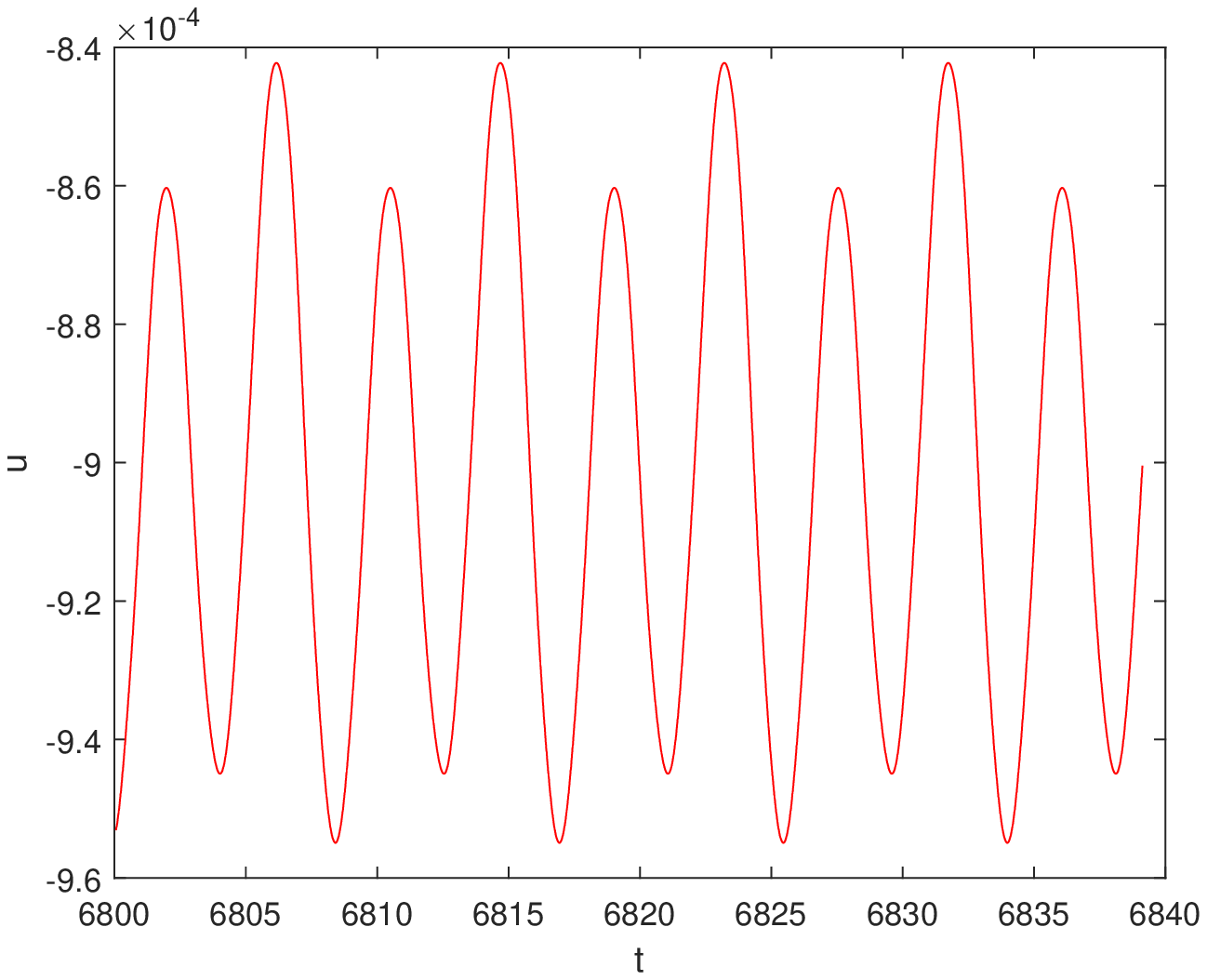}}
		\subfigure[]{ \label{fig:60-0.5-2000-u}
			\includegraphics[scale=0.3]{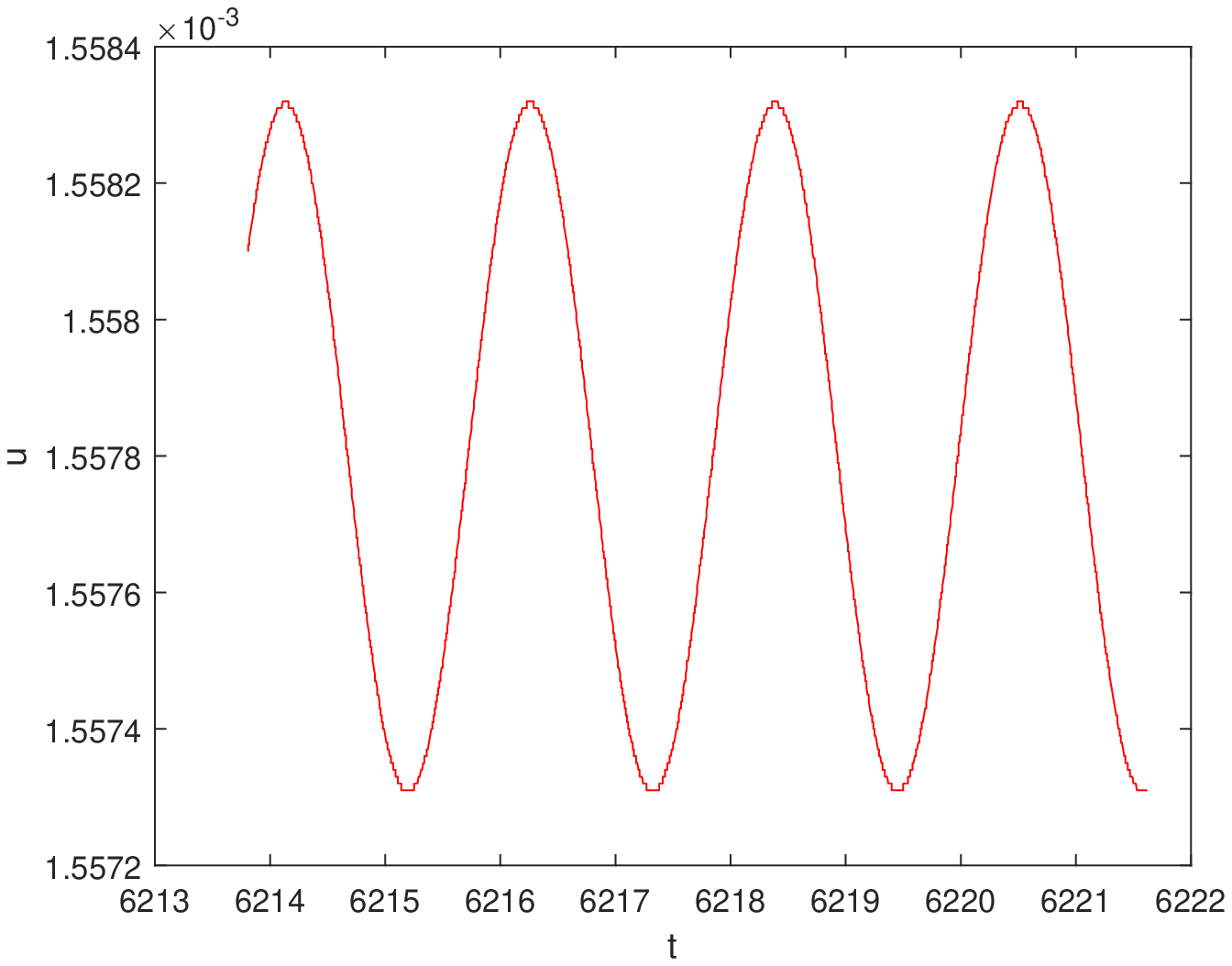}}
		
		\subfigure[]{ \label{fig:60-0.5-1000-v}
			\includegraphics[scale=0.3]{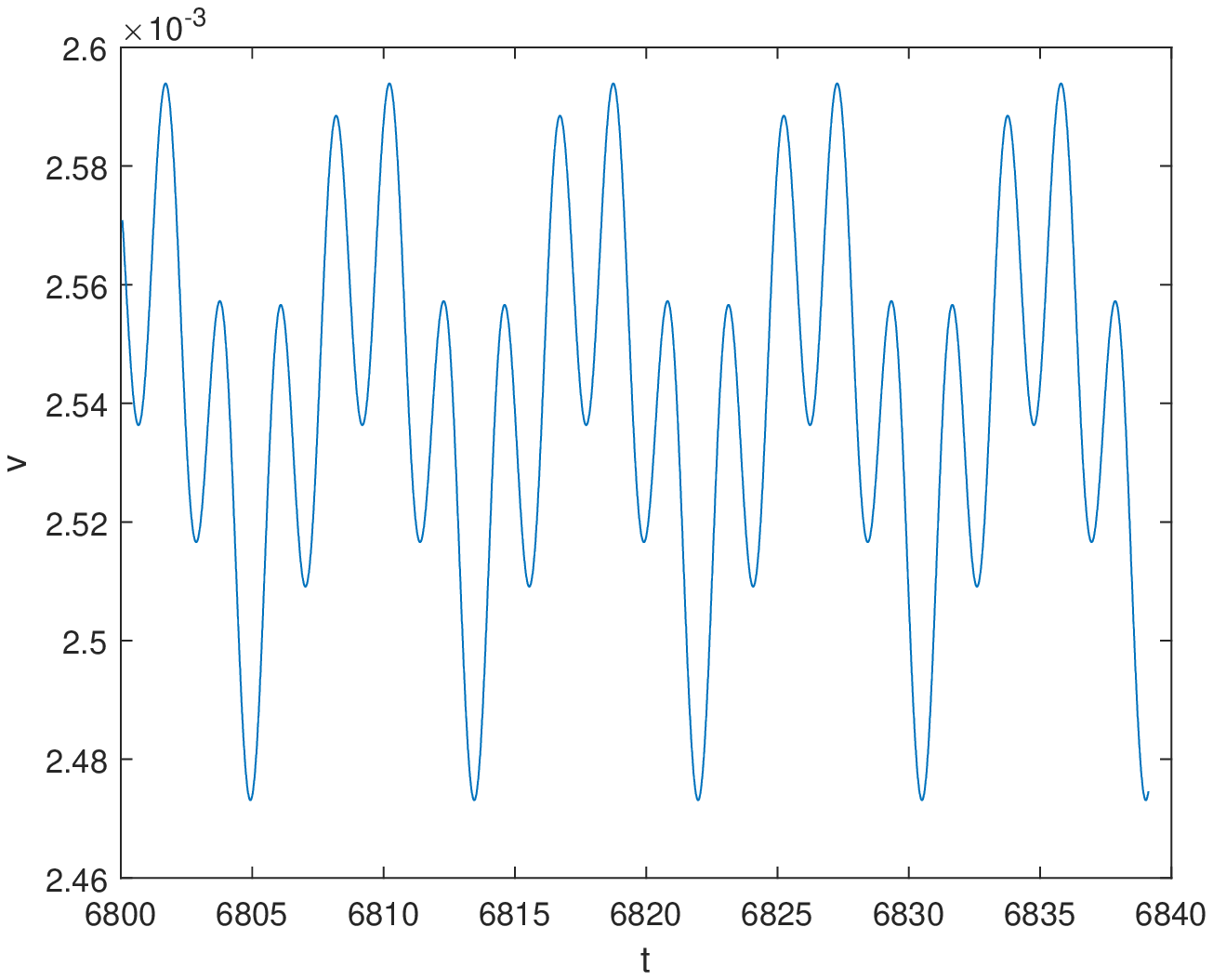}}
		\subfigure[]{ \label{fig:60-0.5-2000-v}
			\includegraphics[scale=0.3]{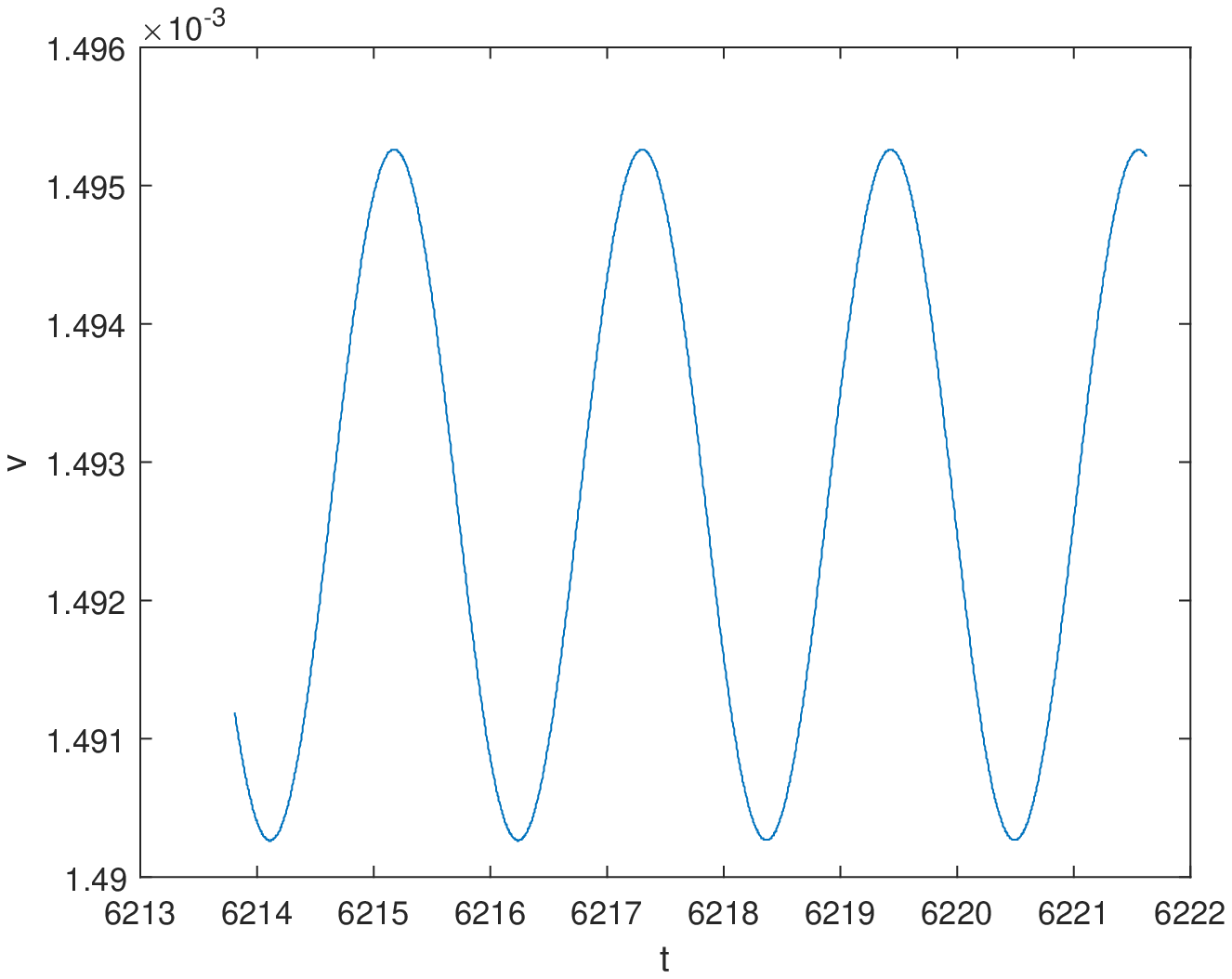}}
		
		\subfigure[]{ \label{fig:60-0.5-1000-E}
			\includegraphics[scale=0.3]{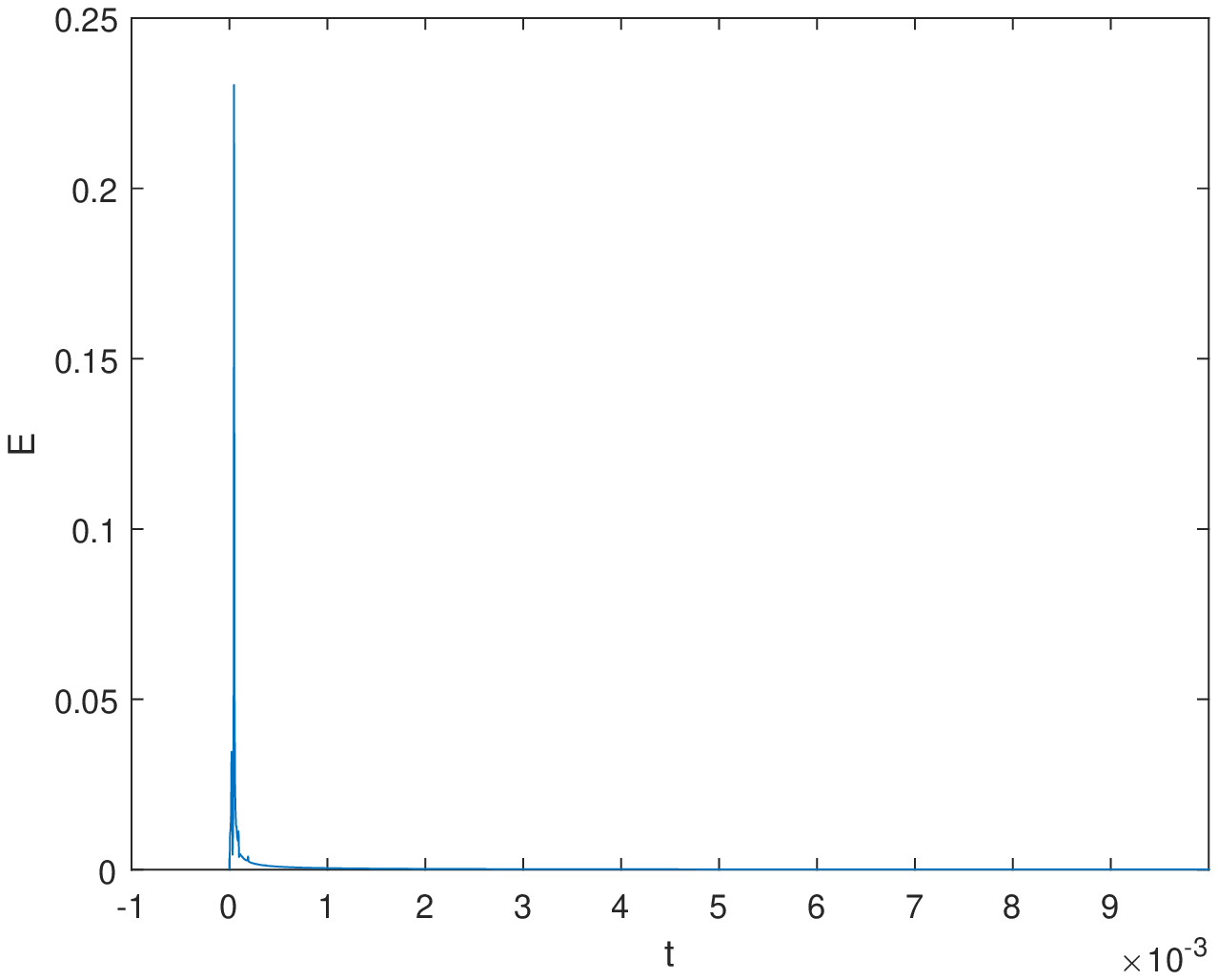}}
		\subfigure[]{ \label{fig:60-0.5-2000-E}
			\includegraphics[scale=0.3]{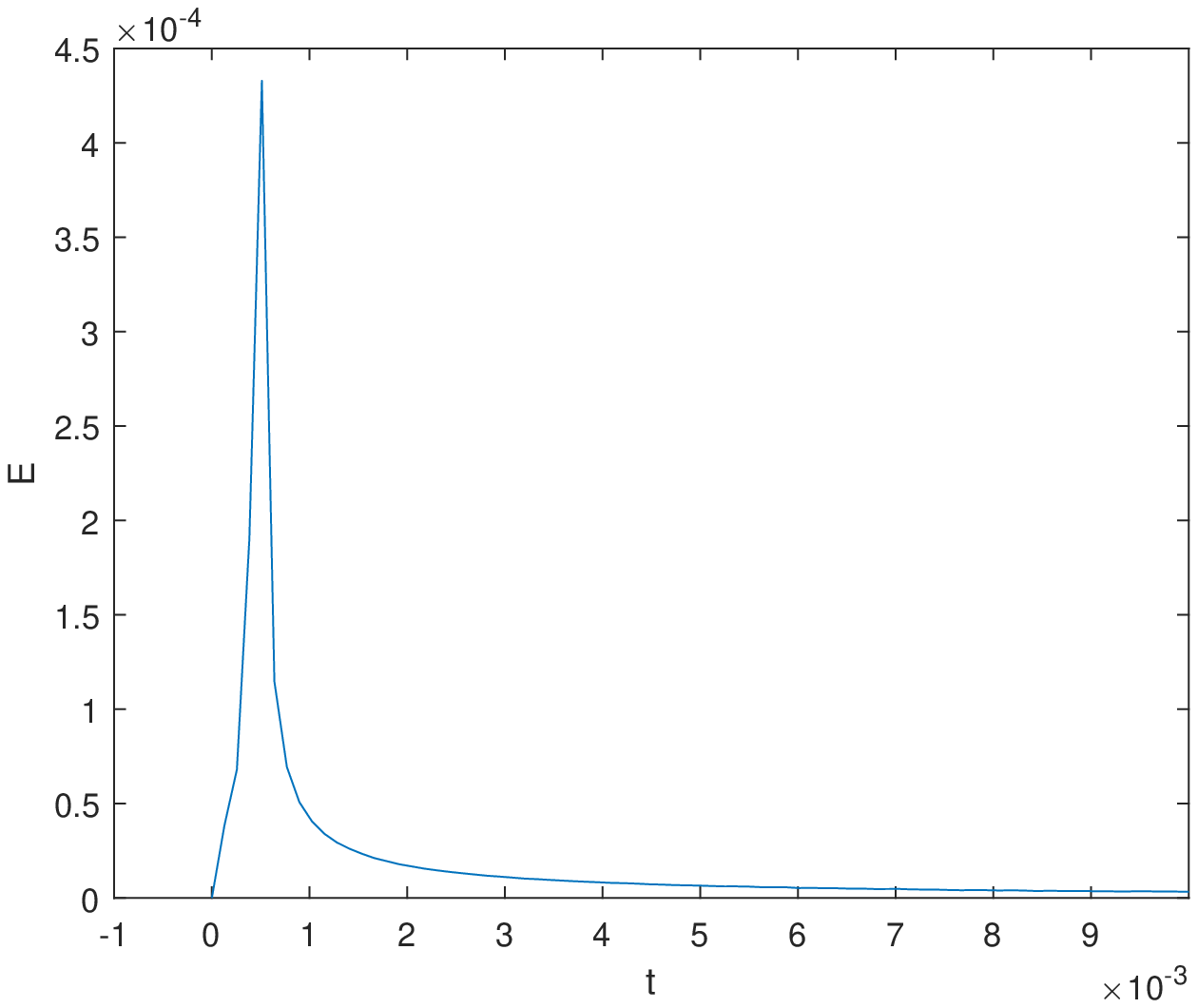}}
		\caption{ The information of TC flow with $n=0.5$ and $\theta=60^\circ$, the first column represents $Re=1000$ and the second column represents $Re=2000$; (a)(b) Phase-space trajectories of velocity, (c)(d) The evolution of velocity $u$, (e)(f) The evolution of velocity $v$, (g)(h) The Fourier power spectrum of kinetic energy.}
		\label{fig:60-0.5-1000}
	\end{figure}
	
	As $Re$ continues to increase, the TC flow exhibits a periodic state. The relevant results with $Re=1000$ and $Re=2000$ are displayed in Fig. \ref{fig:60-0.5-1000}. In conclusion, as the power-law index $n$ decreases, the critical Reynolds number of TC flow from steady state to periodic state also decreases.
	
	\begin{figure}[htbp]\centering
		\subfigure[]{ \label{fig:60-500-xu}
			\includegraphics[scale=0.45]{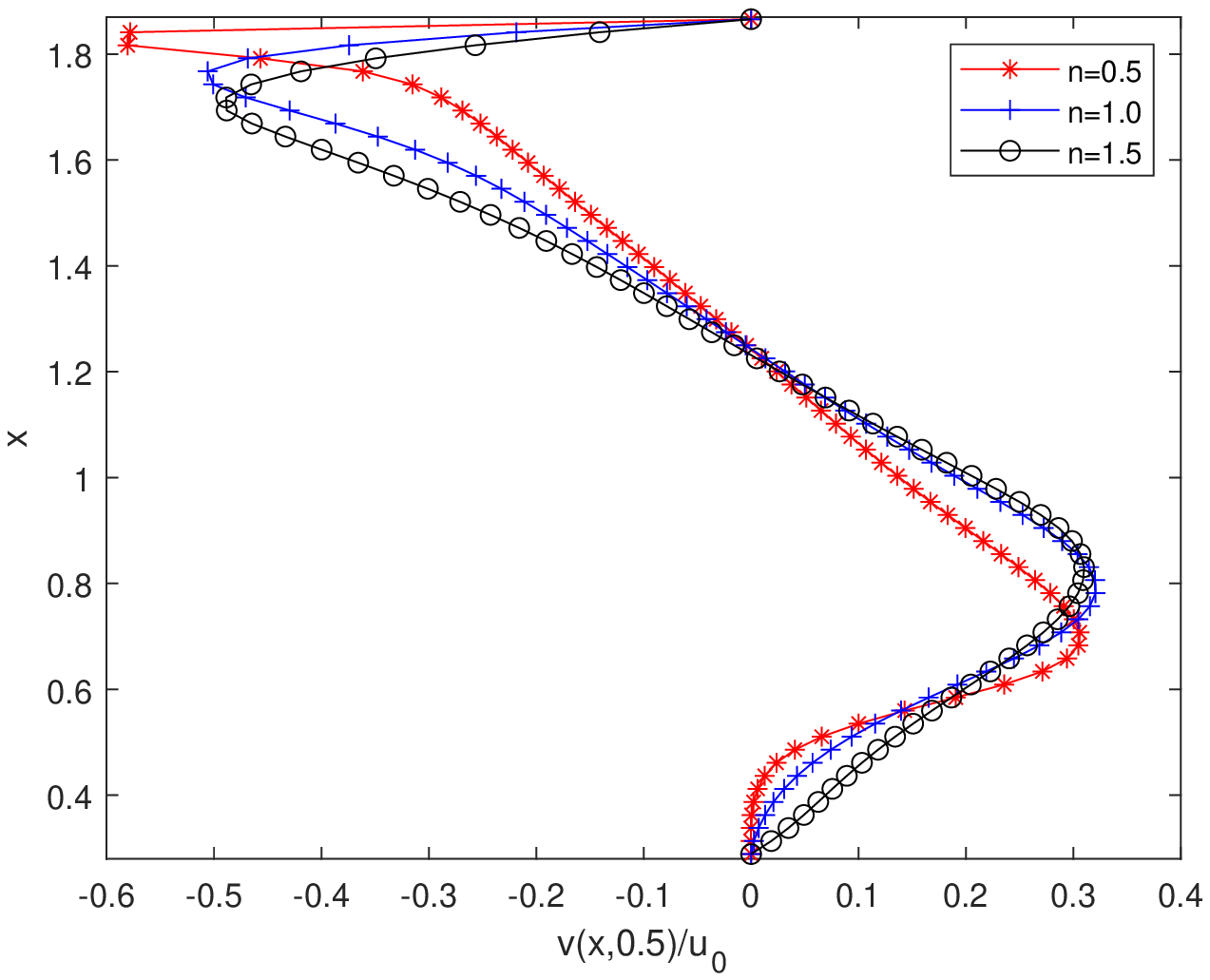}}
		\subfigure[]{ \label{fig:60-500-yu}
			\includegraphics[scale=0.45]{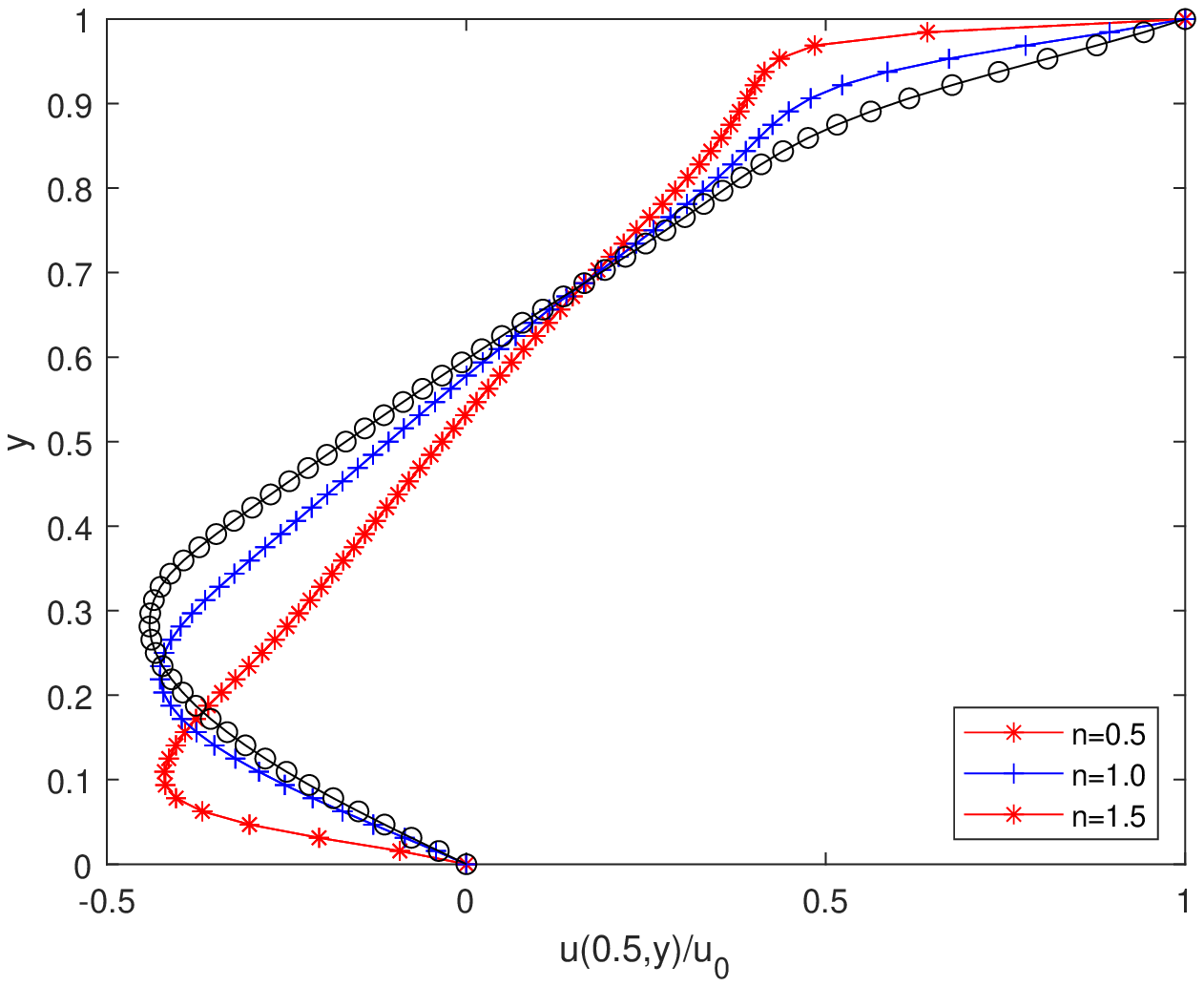}}
		\caption{ Vertical component of velocity for $\theta= 60^\circ$ and $Re=500$; (a) velocity v through y/H = 0.5 along x-axis, (b) velocity u through x/L = 0.5 along y-axis.}
		\label{fig:60-500}
	\end{figure}

	In addition, for $Re=500$ and $\theta=60^\circ$, we also compare the difference of centerline velocity under different power-law index.
	It can be seen that as $n$ decreases, the velocity changes more dramatically.
	Generally speaking, there are several reasons accounting for this phenomenon.
	When $n=1.5$, it is shear-thickening fluid, the higher the speed, the more viscous the fluid will be.
	This characteristic will hinder the flow of the fluid and make the velocity change to be flat.
	When n=0.5, it is a shear-thinning fluid, the higher the velocity, the lower the viscosity.
	Hence the flow of the fluid will be promoted, so the change of velocity will be more drastic.

	(ii) Effect of vertical angle $\theta$ on the development of flow for high $Re$ number
	
\begin{figure}[htbp]\centering
		\subfigure[]{ \label{fig:75-1.5-1000}
			\includegraphics[scale=0.32]{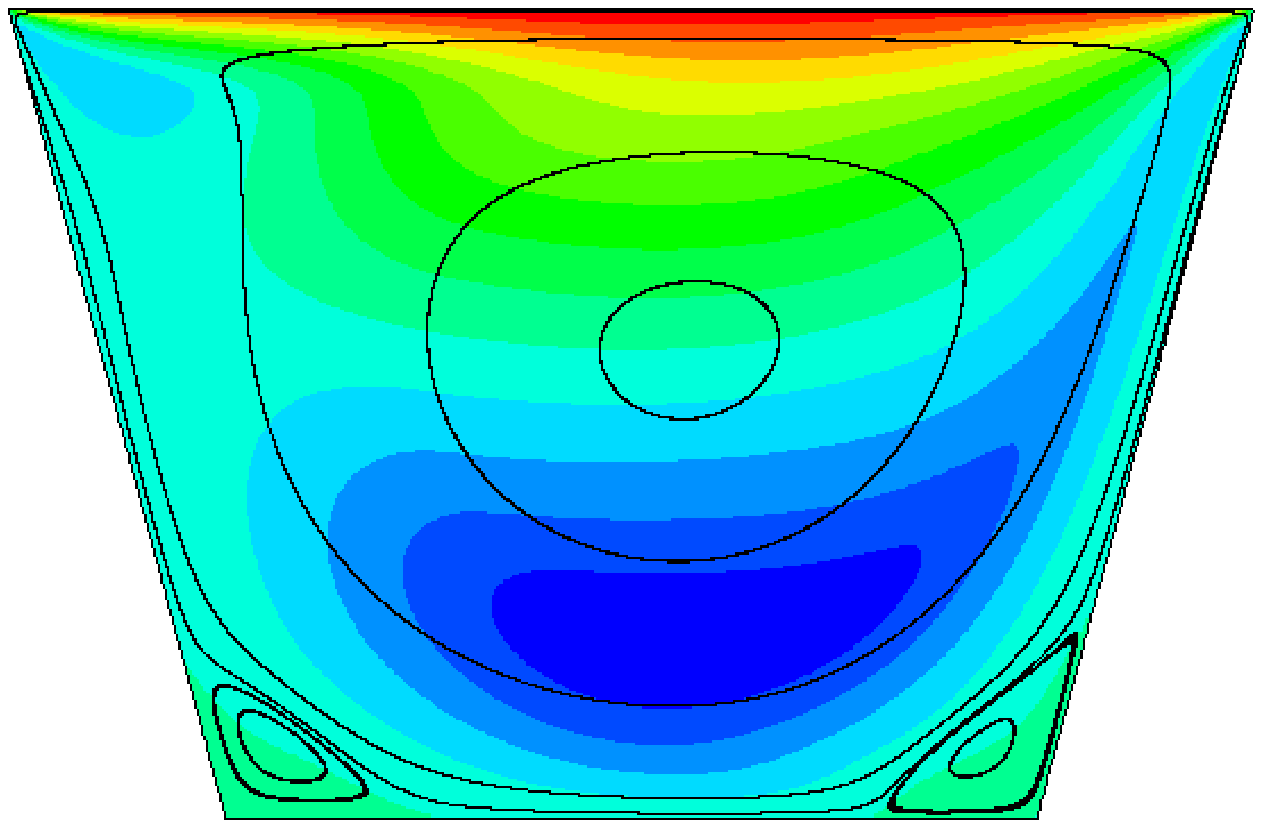}}
		\subfigure[]{ \label{fig:75-1.5-2000}
			\includegraphics[scale=0.32]{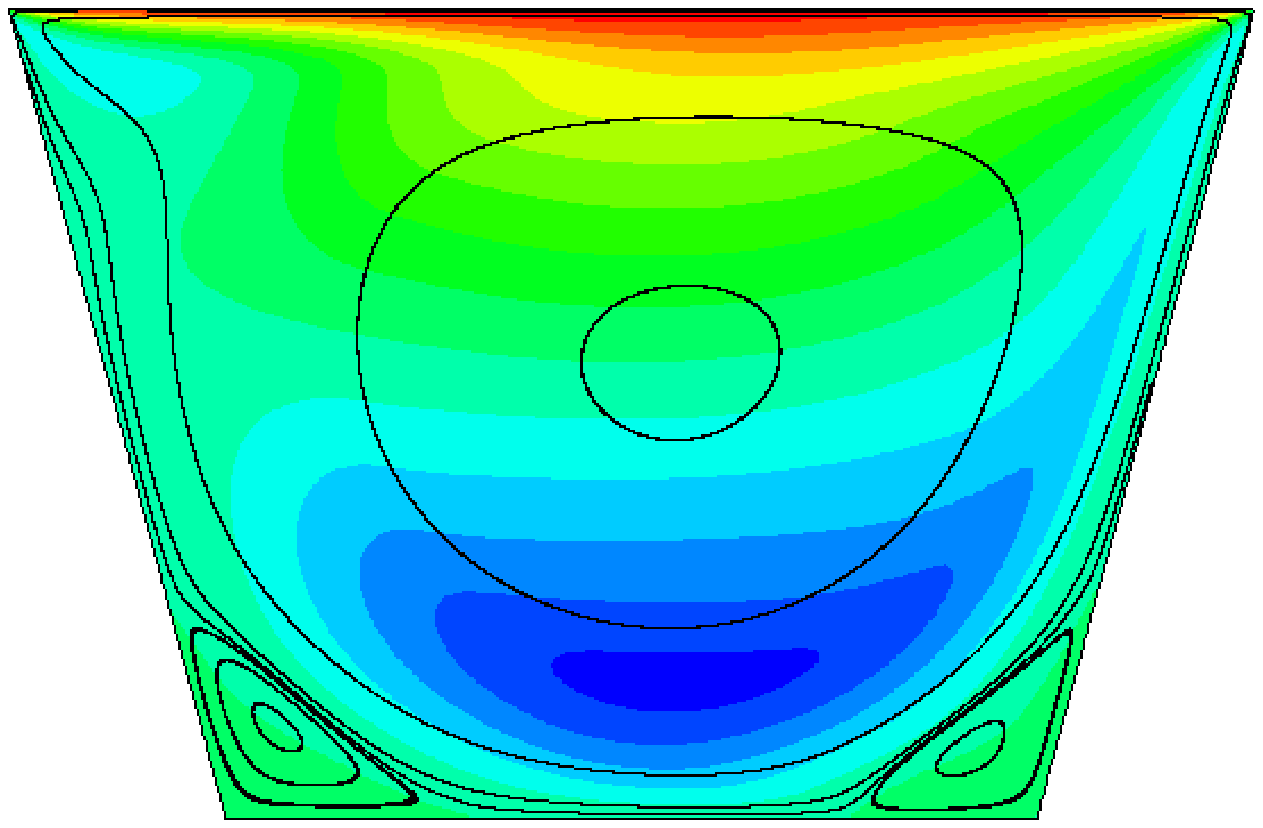}}
		
		\subfigure[]{ \label{fig:75-1.5-4000}
			\includegraphics[scale=0.32]{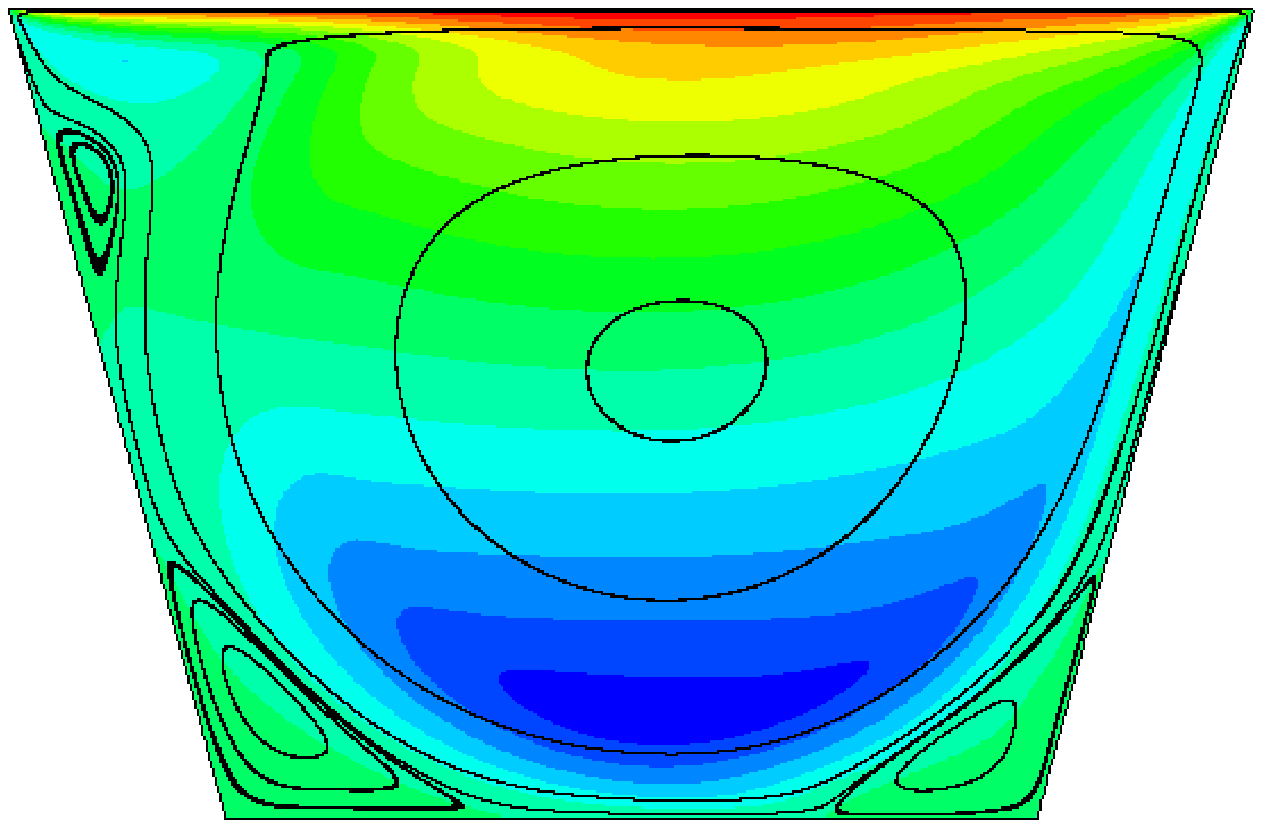}}
		\subfigure[]{ \label{fig:75-1.5-5000}
			\includegraphics[scale=0.32]{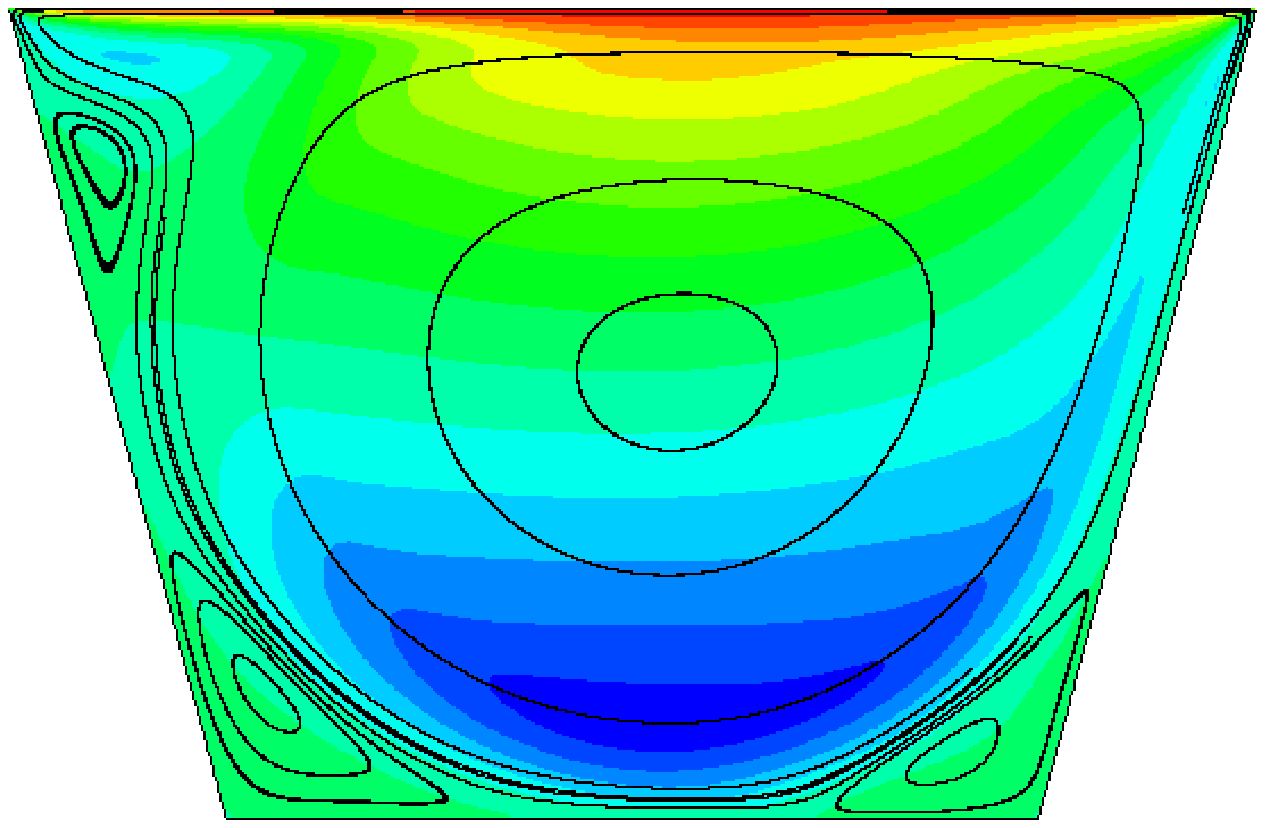}}
		\caption{ Streamline plots at $n=1.5$ and $\theta=75^\circ$; (a) Re=1000, (b) Re=2000, (c) Re=4000, (d) Re=5000.}
		\label{fig:75-1.5}
	\end{figure}

	In this section, we will fix the power-law index $n=1.5$, and adjust $\theta$ and $Re$ to observe the development of TC flow. First, we consider the case of $\theta=75^\circ$. We present some  streamline plots in Fig. \ref{fig:75-1.5}. As the Re varies between $1000$ and $5000$, the TC flow remains steady-state. When $Re=1000$, the first-order vortex occupies the central position, and there are two second-order vortexes in the lower left and right corners respectively. With the increase of $Re$ number, the range of the two secondary vortices in the lower left and right corner gradually increases. In addition, when $Re$ number increases to $4000$, a third-order vortex appears in the upper left corner, and the third-order vortex will also become larger with the increase of $Re$. These phenomena are very similar to the phenomena of lid-driven flow in a square cavity. But the phenomena are slightly different from those with $\theta=60^\circ$. The main reason is that with the increase of $\theta$, the length of the roof decreases, the physical area is closer to the square, and the flow is more similar to the flow of the square cavity. These results show that the TC flow tends to flatten as $\theta$ increases.
	
\begin{figure}[htbp]\centering
		\subfigure[]{ \label{fig:75-1.5-6000-uv}
			\includegraphics[scale=0.3]{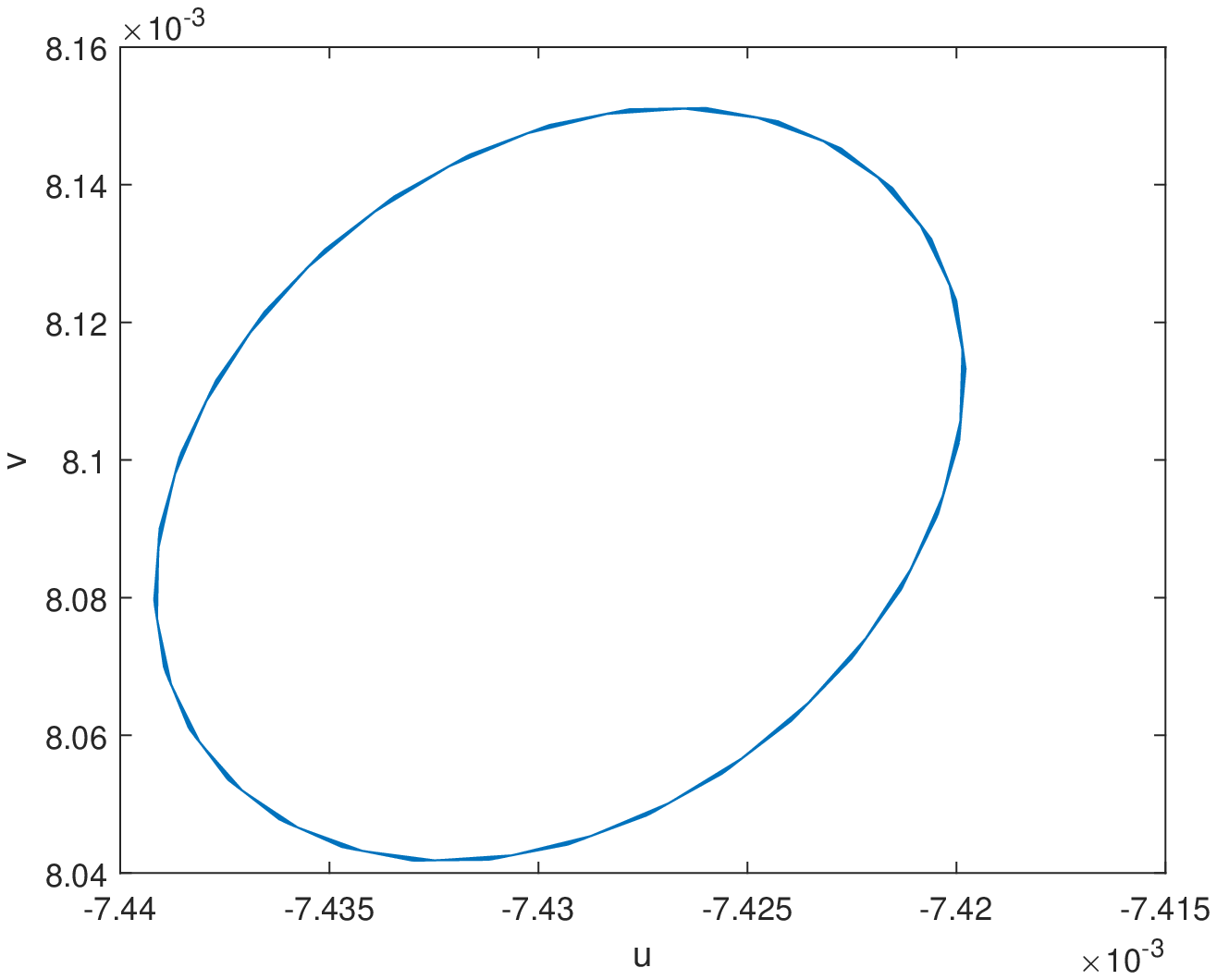}}
		\subfigure[]{ \label{fig:75-1.5-7000-uv}
			\includegraphics[scale=0.3]{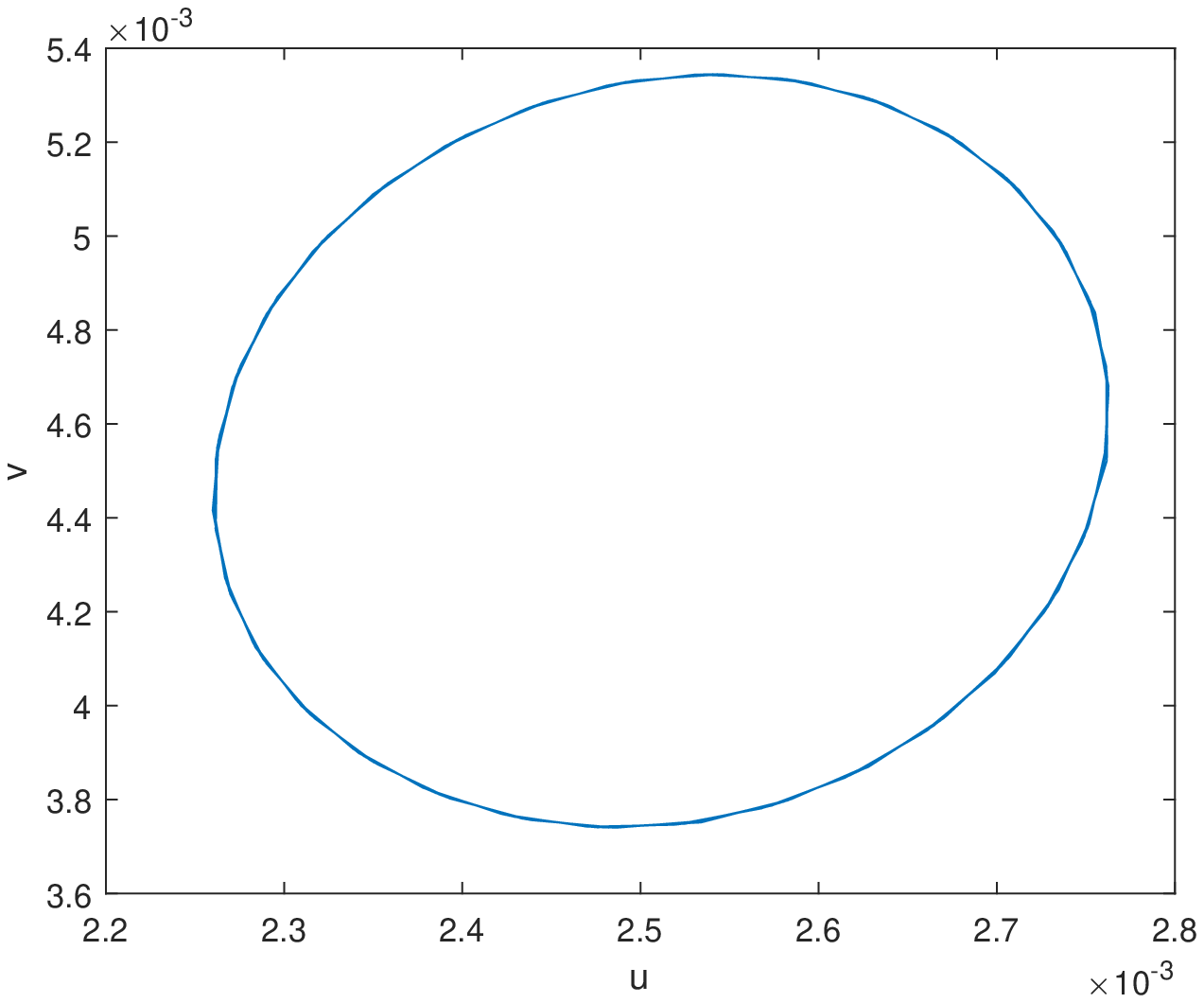}}
		
		\subfigure[]{ \label{fig:75-1.5-6000-u}
			\includegraphics[scale=0.3]{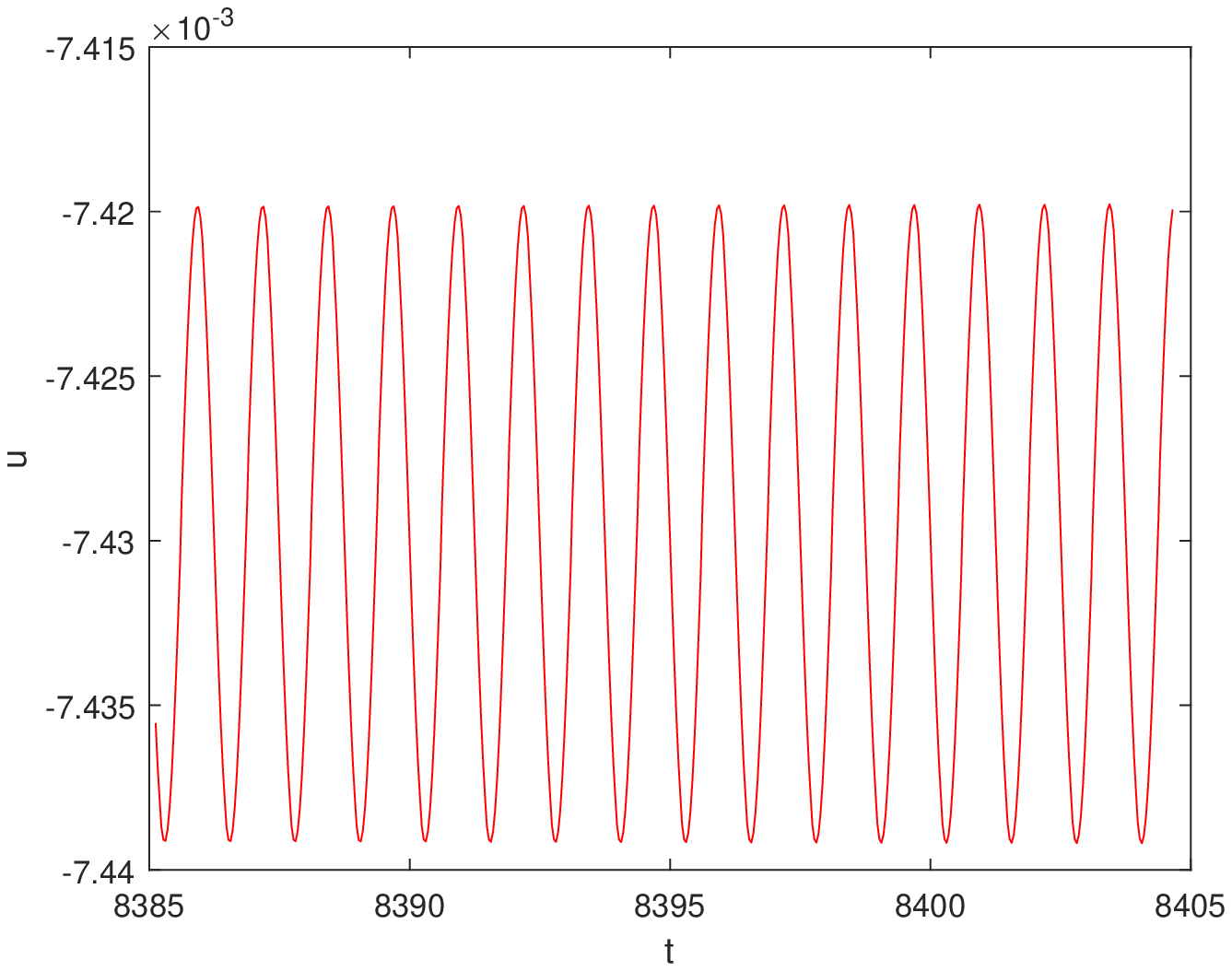}}
		\subfigure[]{ \label{fig:75-1.5-7000-u}
			\includegraphics[scale=0.3]{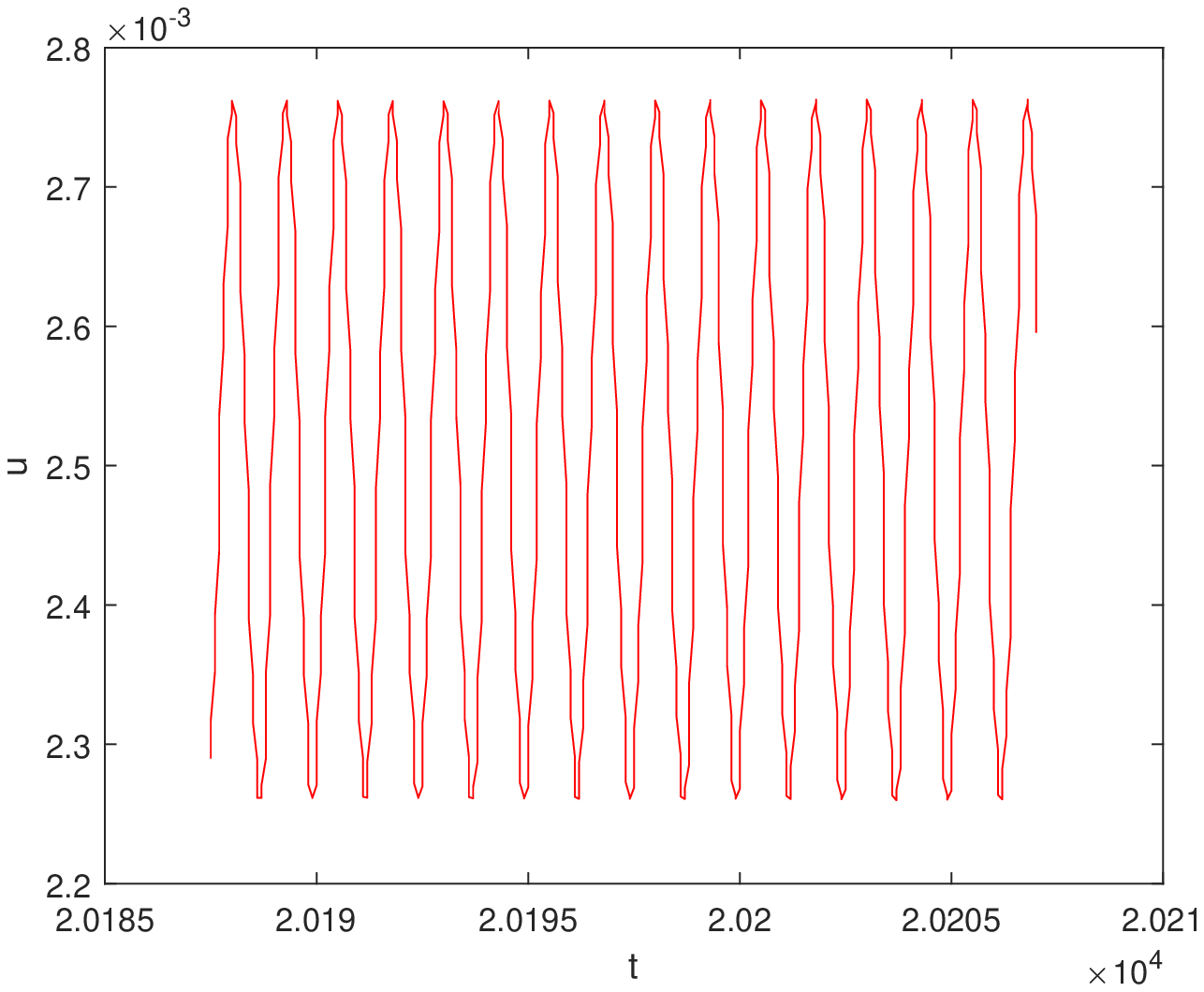}}
		
		\subfigure[]{ \label{fig:75-1.5-6000-v}
			\includegraphics[scale=0.3]{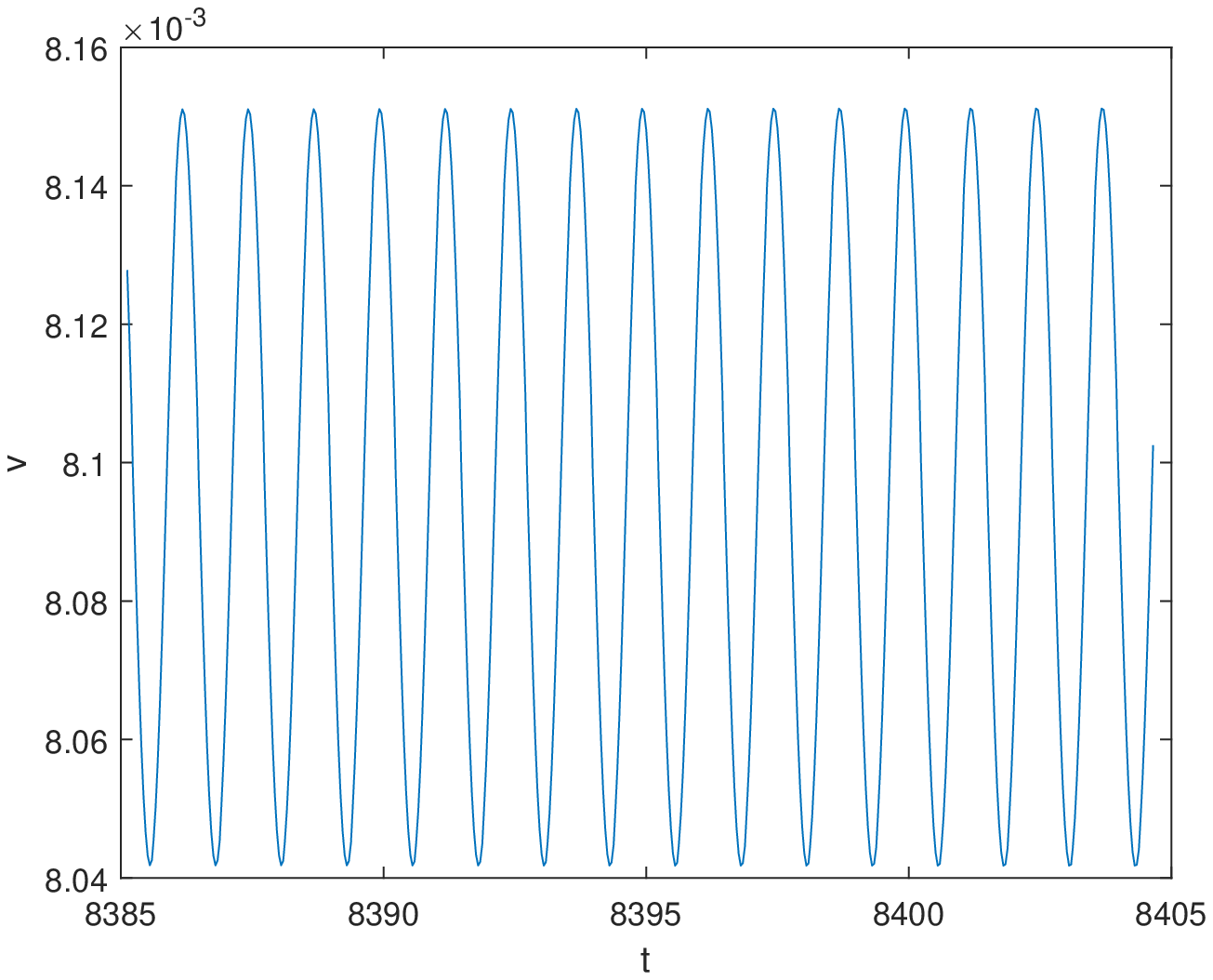}}
		\subfigure[]{ \label{fig:75-1.5-7000-v}
			\includegraphics[scale=0.3]{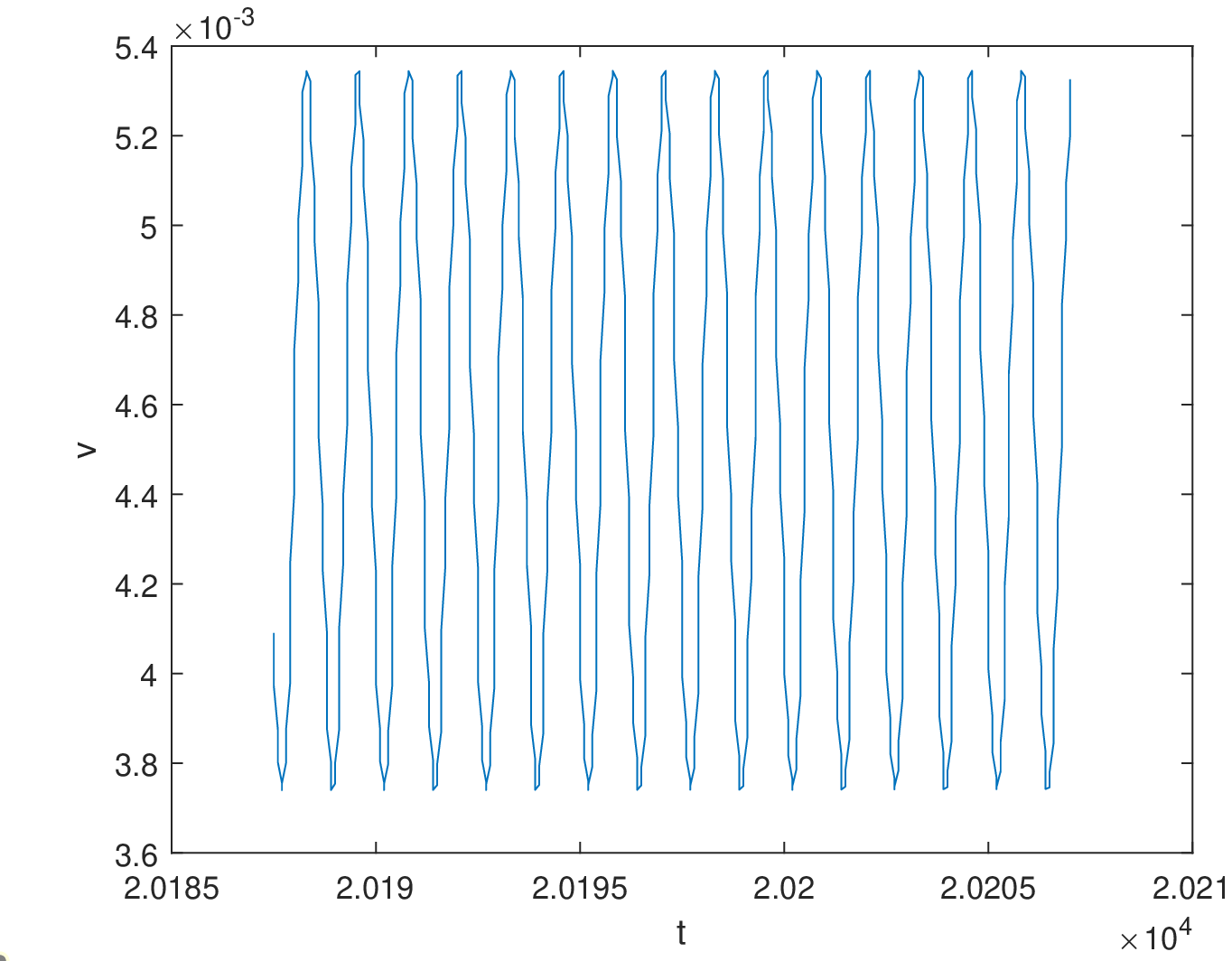}}
		
		\subfigure[]{ \label{fig:75-1.5-6000-E}
			\includegraphics[scale=0.3]{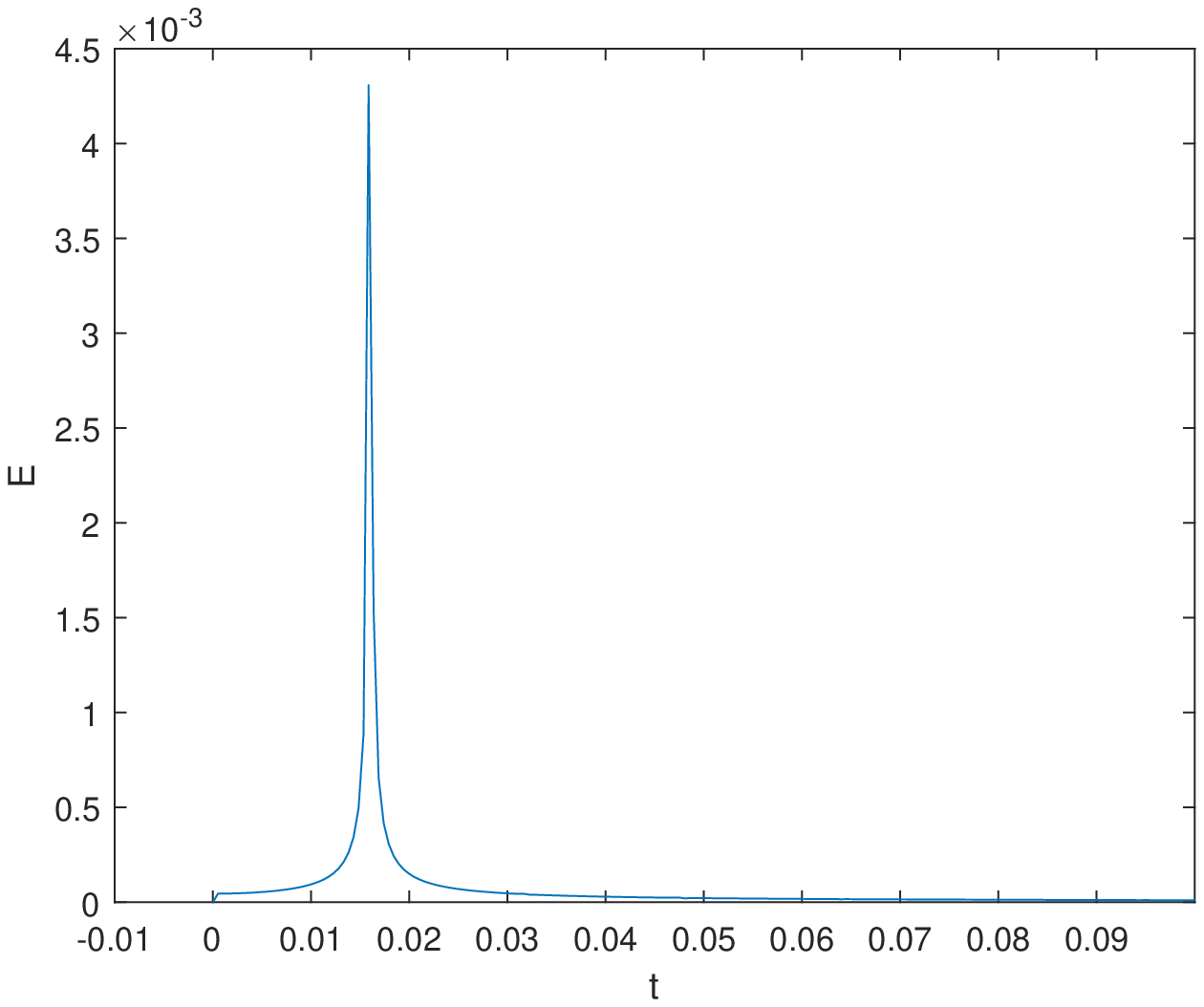}}
		\subfigure[]{ \label{fig:75-1.5-7000-E}
			\includegraphics[scale=0.3]{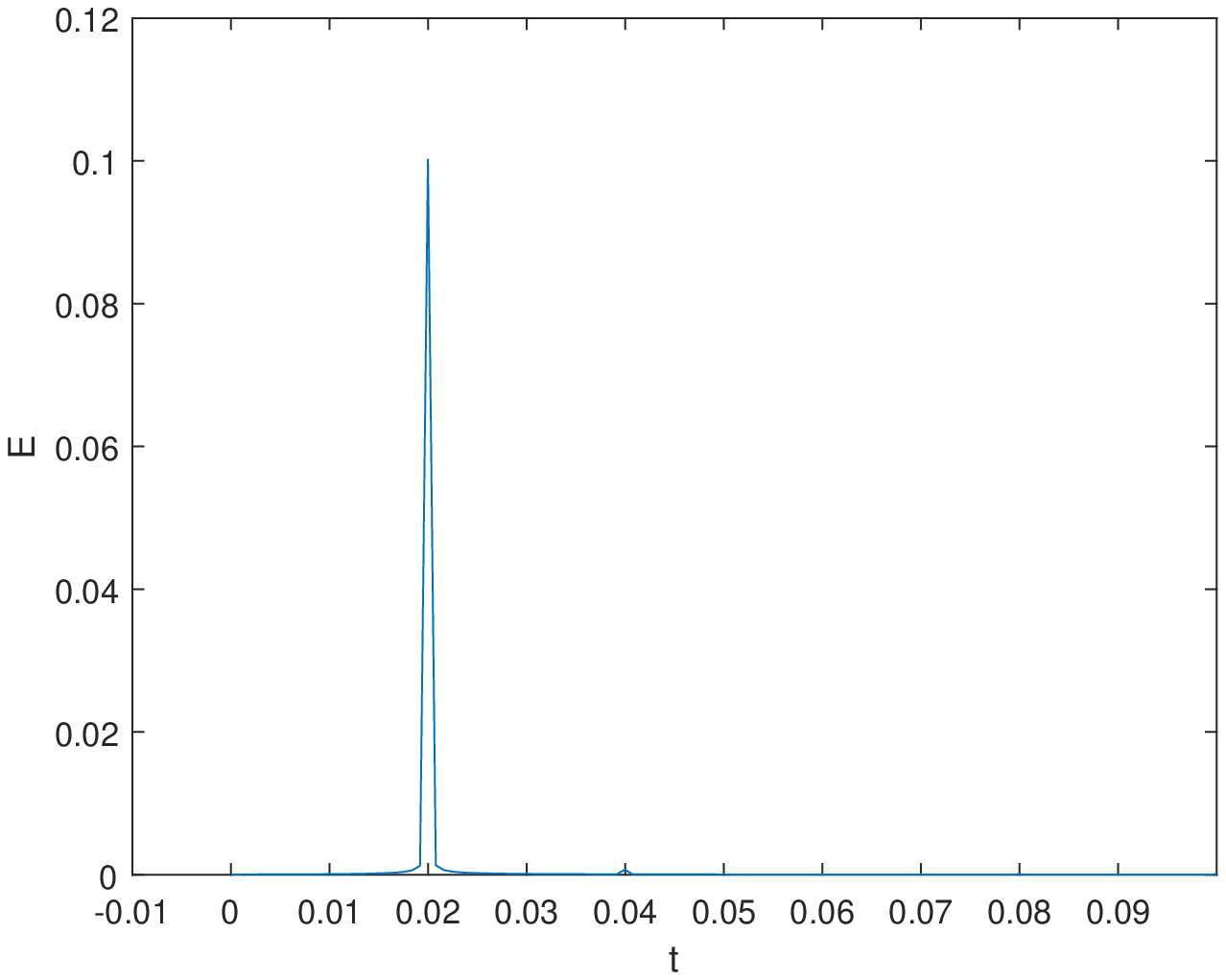}}
		\caption{ The information of TC flow with $n=1.5$ and $\theta=75^o$, the first column represent $Re=6000$ and the second column represent $Re=7000$; (a)(b) Phase-space trajectories of velocity, (c)(d) The evolution of velocity $u$, (e)(f) The evolution of velocity $v$, (g)(h) The Fourier power spectrum of kinetic energy.}
		\label{fig:75-1.5-6000}
	\end{figure}

	Some results with $Re=6000$ and $Re=7000$ are also shown in Fig. \ref{fig:75-1.5-6000}. It can be observed that the TC flow is a periodic flow.
	Compared with the cases of $\theta=60^o$, the closed loop in the phase diagram is simpler.
	This is mainly because the change of physical area makes the flow more gentle, and the vortex structure in the periodic flow will be simpler, and the shape of the closing ring will naturally become simpler.

\begin{figure}[htbp]
		\subfigure[]{ \label{fig:45-1.5-1000}
			\includegraphics[scale=0.3]{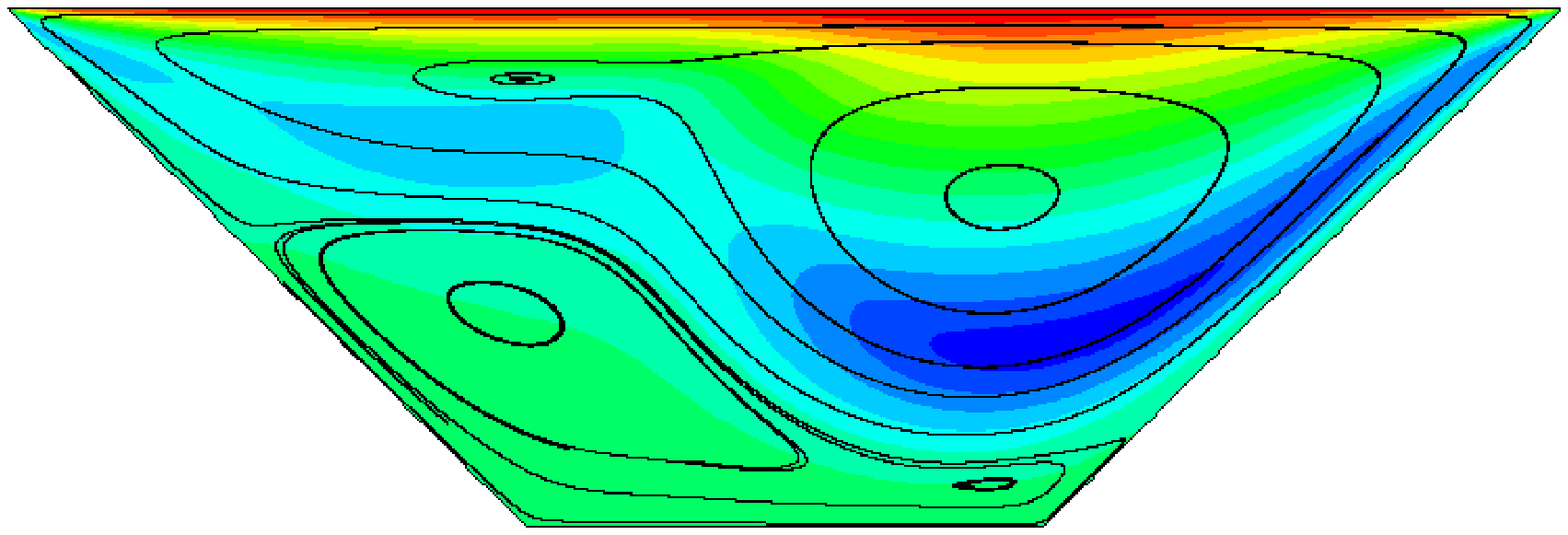}}
		\subfigure[]{ \label{fig:45-1.5-2000}
			\includegraphics[scale=0.3]{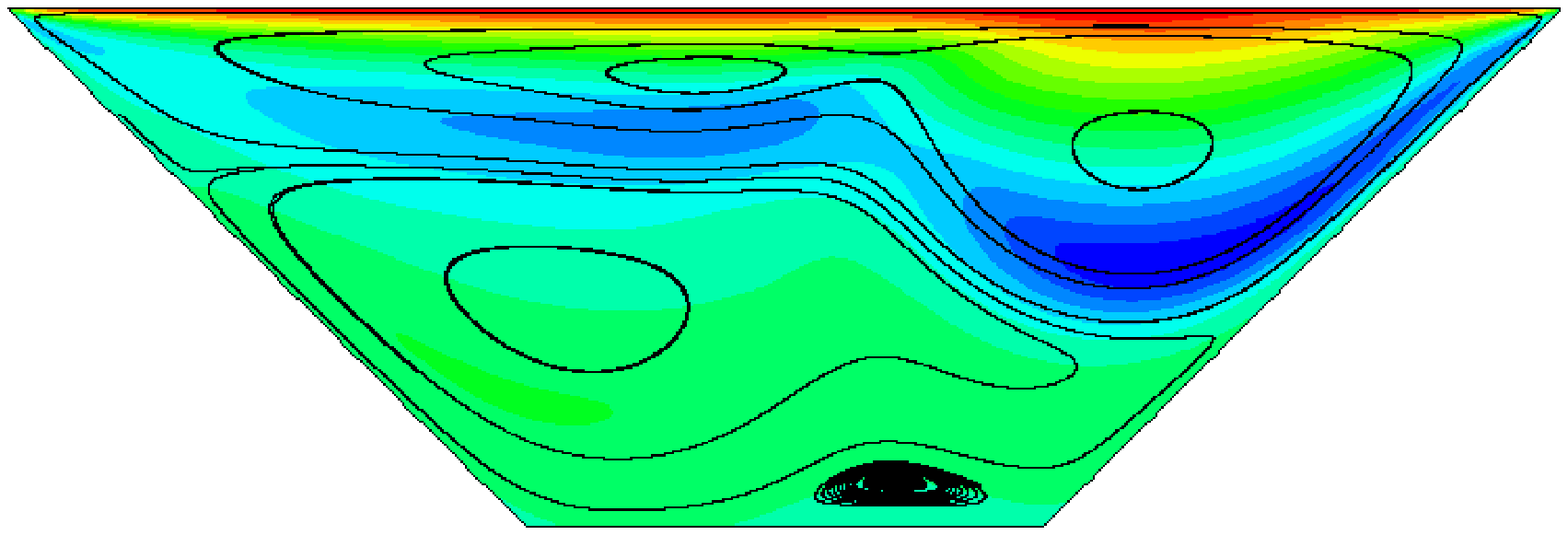}}
		\subfigure[]{ \label{fig:45-1.5-4000}
			\includegraphics[scale=0.3]{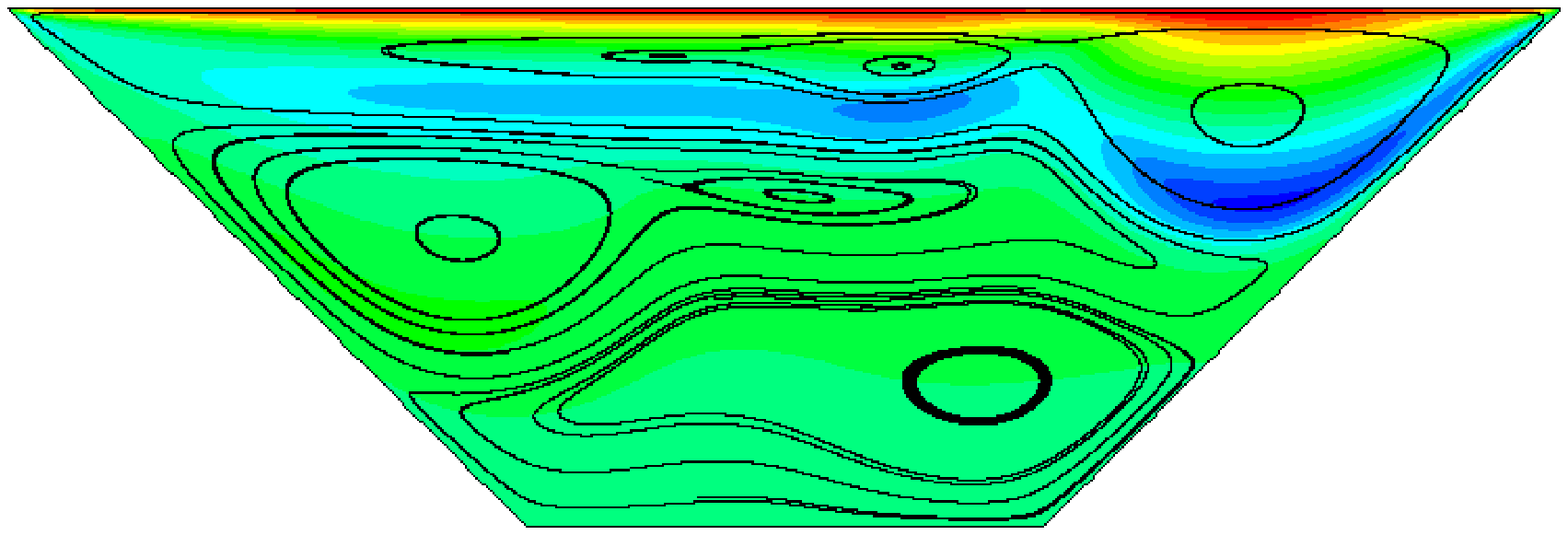}}
		\subfigure[]{ \label{fig:45-1.5-5000}
			\includegraphics[scale=0.3]{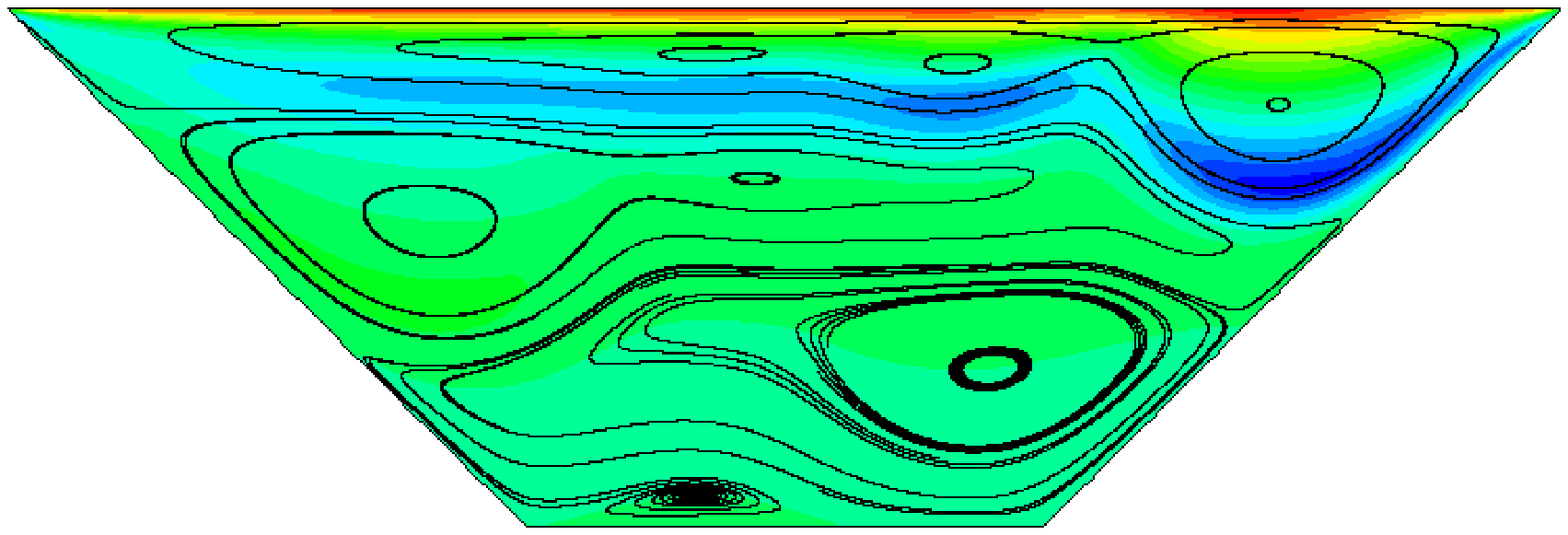}}
		\caption{ Streamline plots at $n=1.5$ and $\theta=45^o$; (a) Re=1000, (b) Re=2000, (c) Re=4000, (d) Re=5000.}
		\label{fig:45-1.5}
	\end{figure}

	Then, we study the development of TC flow with $\theta=45^o$ and $n=1.5$. The streamline plots with $Re$ ranging from $1000$ to $5000$ are displayed in the Fig. \ref{fig:45-1.5}. As the $\theta$ decreases, the shape of the vortexes in the trapezoidal cavity become more complex.
	
	When $Re=1000$, a new vortex has split away from the first-order vortex, which is located at the upper left corner of the cavity. The two secondary vortexes at the bottom are partially fused, and the secondary vortex at the lower left corner are obviously larger than that at the lower right corner.
	
	As $Re$ increases to $2000$, the range of second-order vortexes in the upper left corner and lower left corner gradually increases. Squeezed by these two vortices, the range of the first-order vortex becomes smaller and the center of the vortex moves to the upper right corner. At the same time, the two second-order vortices at the bottom are completely separated, and the second-order vortex at the lower right corner moves towards the bottom of the cavity. But on the whole, the number of vortices has not changed, and it is still four vortices.
	
	As $Re$ increases to $4000$, the number of vortices increases to 6, and the vortex structure is divided into three layers. The first layer consists of a first-order vortex, a second-order vortex and a third-order vortex separated from the first-order vortex. The second layer is made up of the second-order vortex on the left, and a new third-order vortex separated from the second-order vortex.
	The third layer is the secondary vortex in the lower right corner. Compared the vortexes with $Re=2000$, the scope of the secondary vortex at the third layer increases.
	
	When $Re=5000$, the shape of vortexes are similar to that of $Re= 4000$. However, the range of the first-order vortex is squeezed smaller, and the two vortices of the second layer are squeezed to the upper left of the cavity by the vortex in third layer. In addition, a fourth-order vortex appears at the bottom of the cavity, and it squeezes the third-order vortex.
	
	According to the above results, we conclude that the TC flow becomes more intense as $\theta$ decreases. The main reason is that when $\theta$ decreases, the length of roof increases and the drag distance lengthens, which will make the flow more complex and generate more small vortexes. The mutual extrusion of vortexes also makes the shape of vortexes more complicated.
	
	\begin{figure}[htbp]\centering
		\subfigure[]{ \label{fig:45-1.5-6000-uv}
			\includegraphics[scale=0.3]{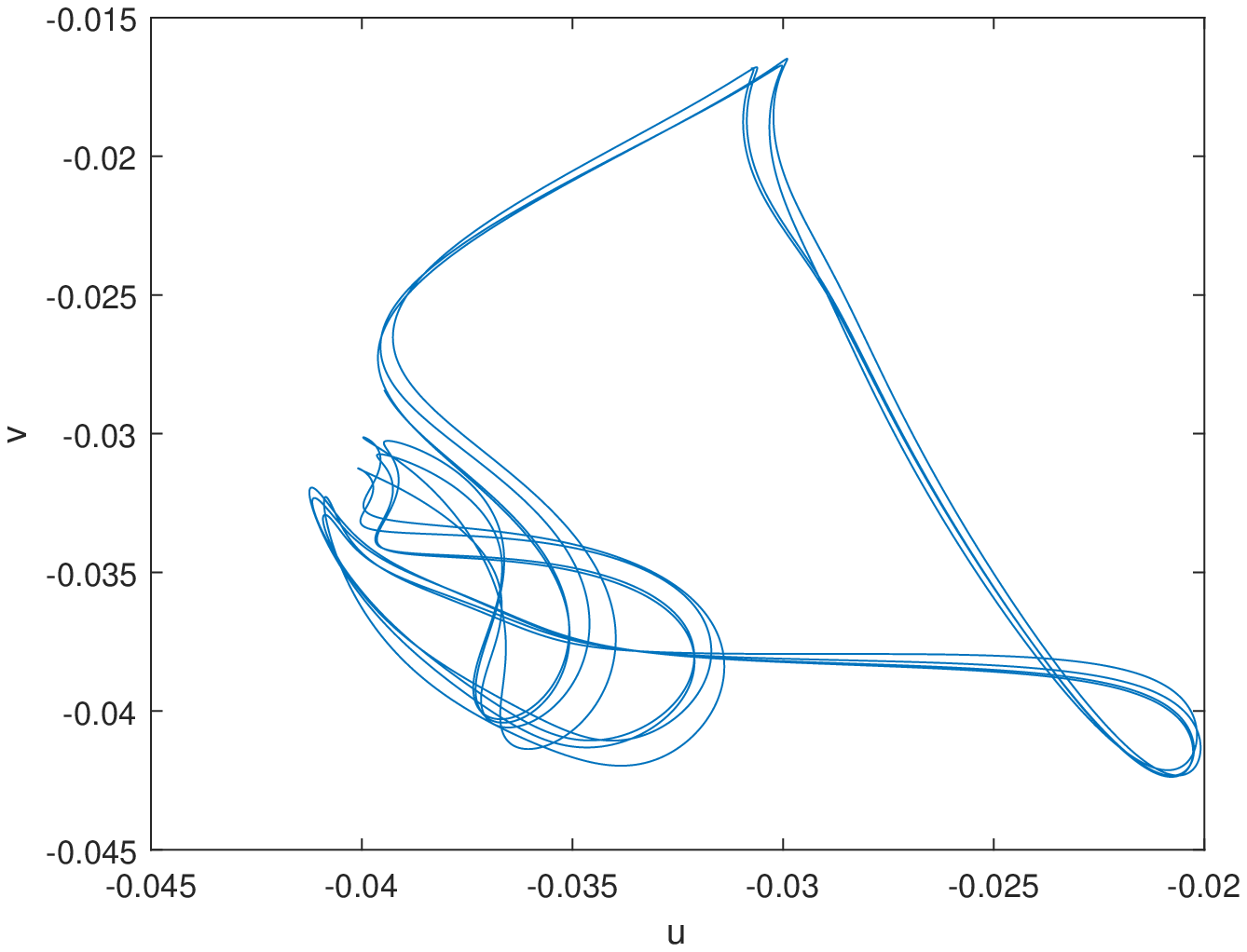}}
		\subfigure[]{ \label{fig:45-1.5-7000-uv}
			\includegraphics[scale=0.3]{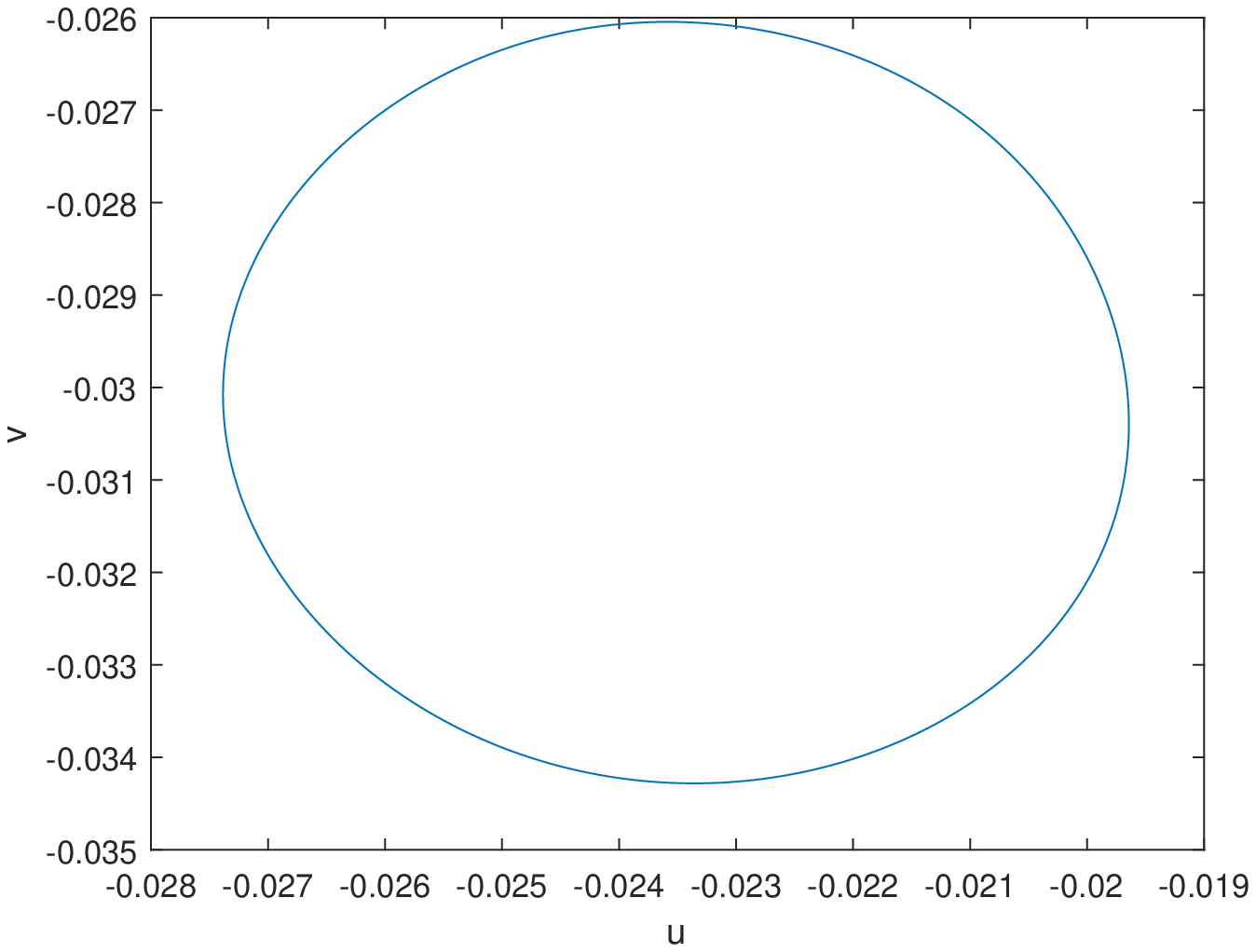}}
		
		\subfigure[]{ \label{fig:45-1.5-6000-u}
			\includegraphics[scale=0.3]{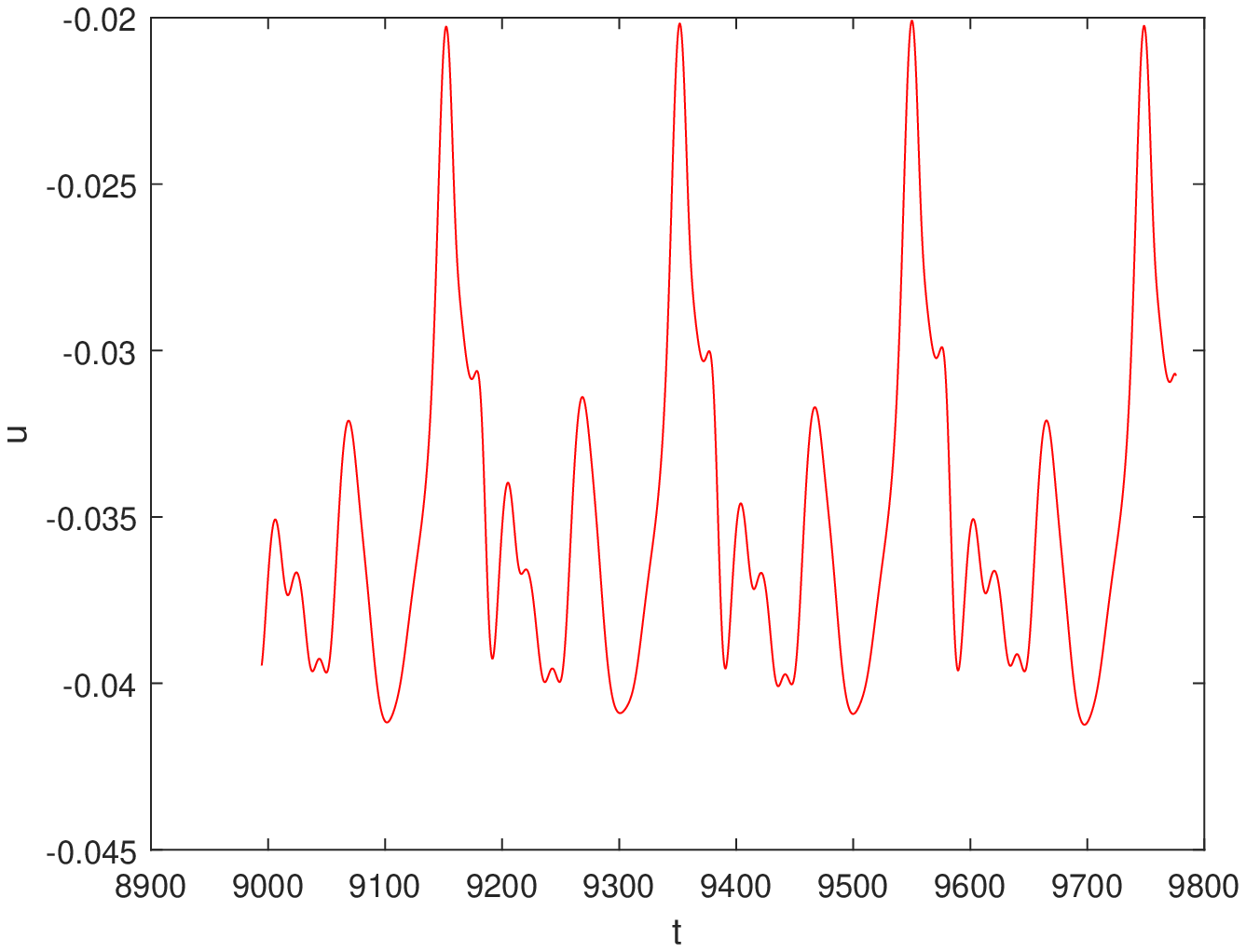}}
		\subfigure[]{ \label{fig:45-1.5-7000-u}
			\includegraphics[scale=0.3]{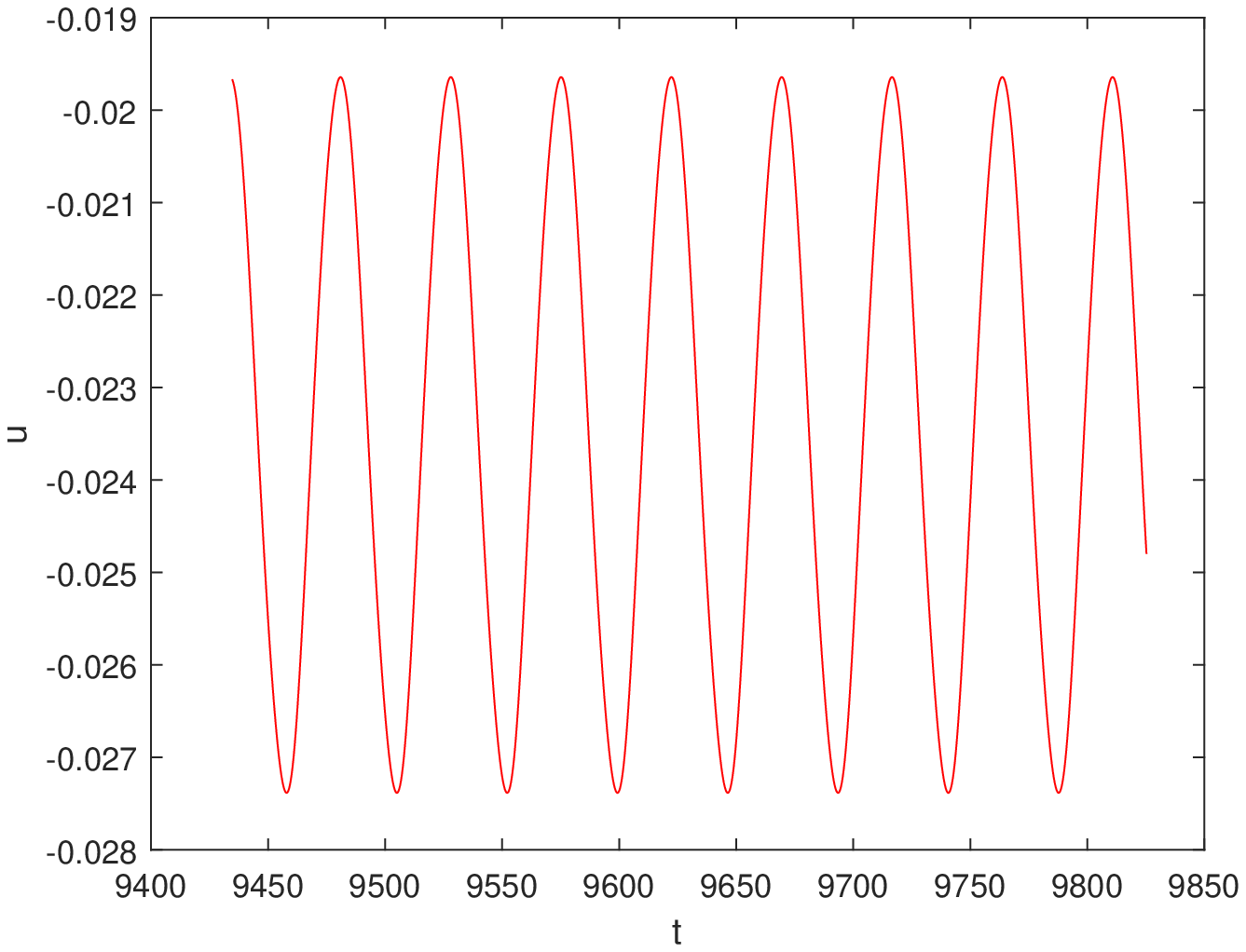}}
		
		\subfigure[]{ \label{fig:45-1.5-6000-v}
			\includegraphics[scale=0.3]{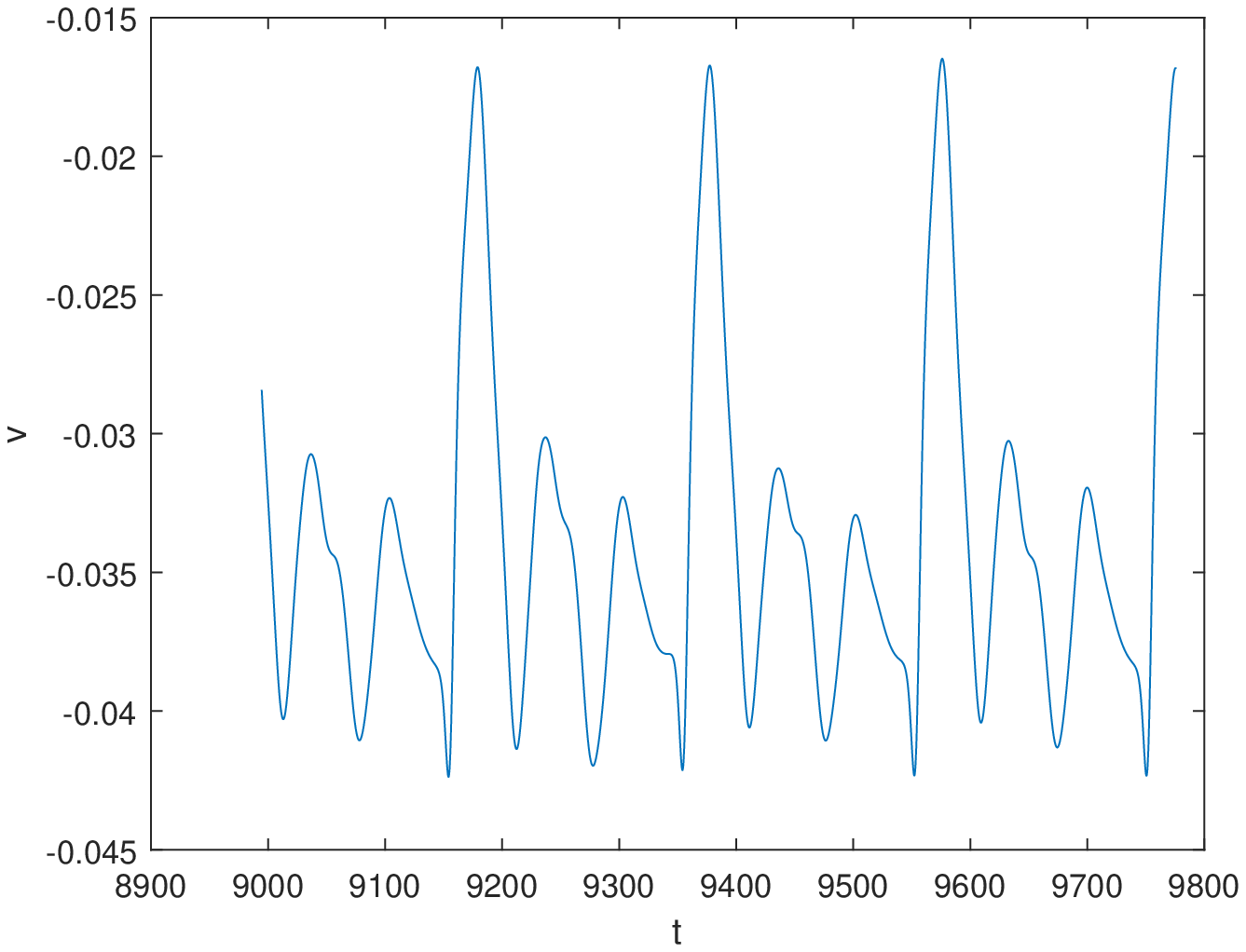}}
		\subfigure[]{ \label{fig:45-1.5-7000-v}
			\includegraphics[scale=0.3]{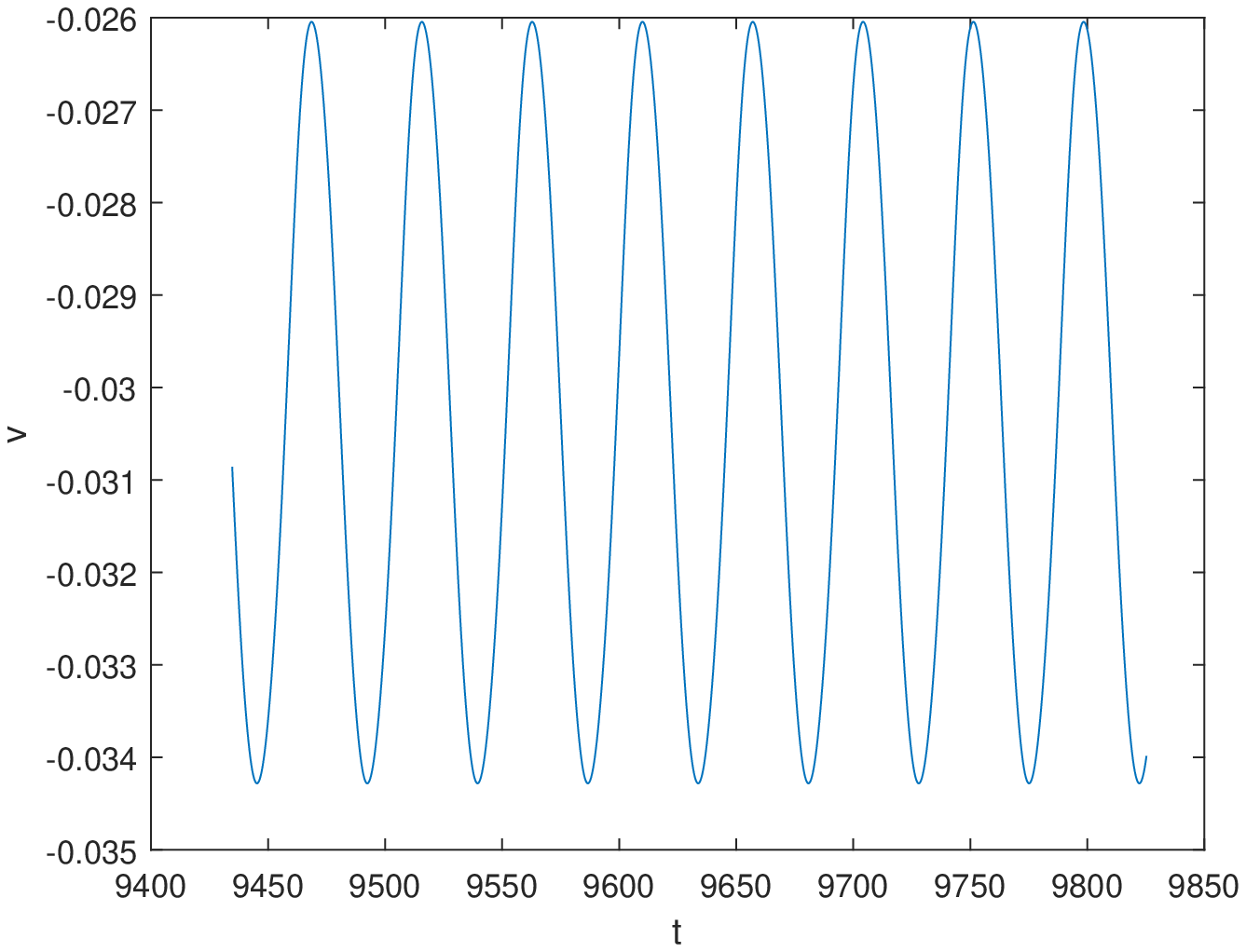}}
		
		\subfigure[]{ \label{fig:45-1.5-6000-E}
			\includegraphics[scale=0.3]{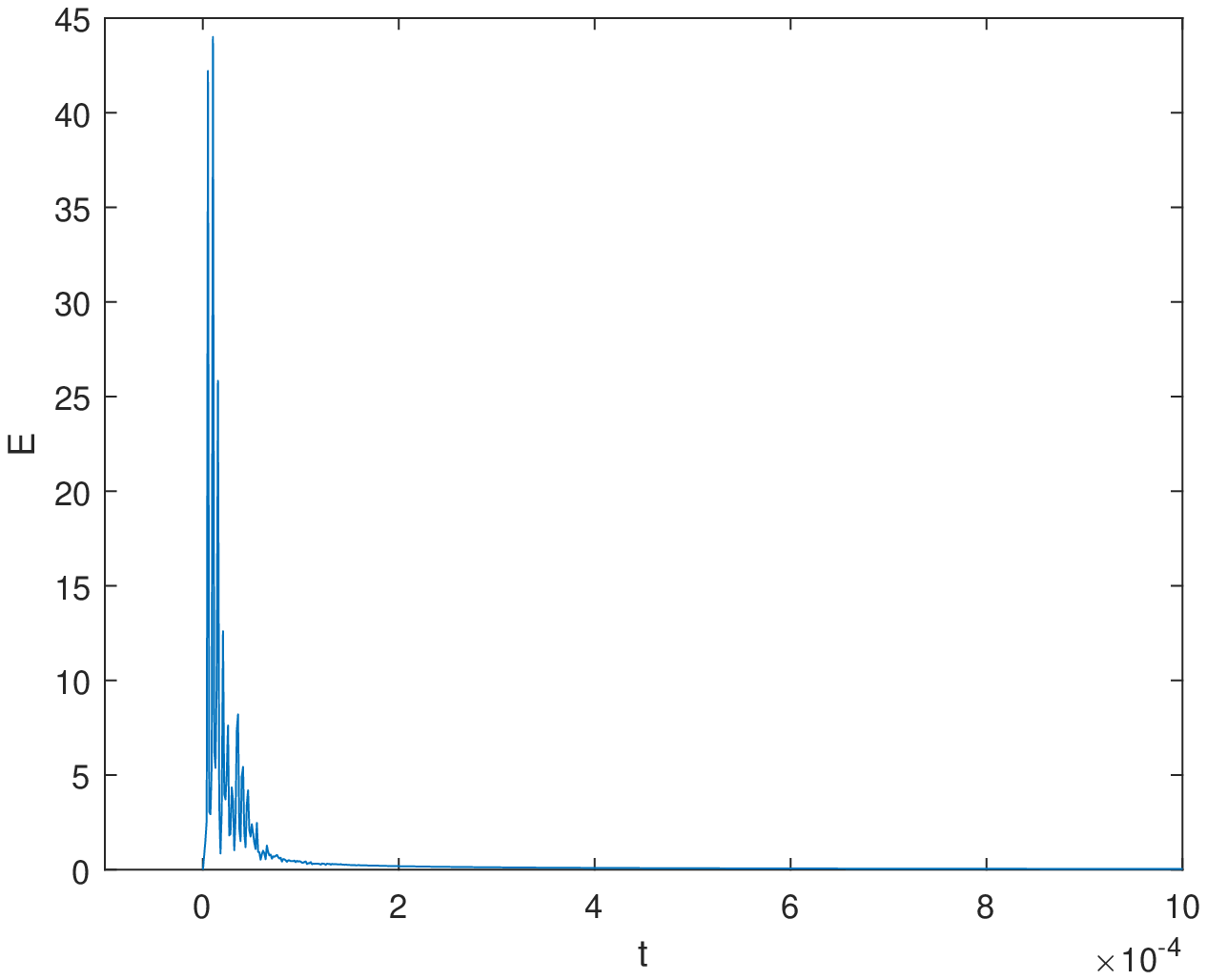}}
		\subfigure[]{ \label{fig:45-1.5-7000-E}
			\includegraphics[scale=0.3]{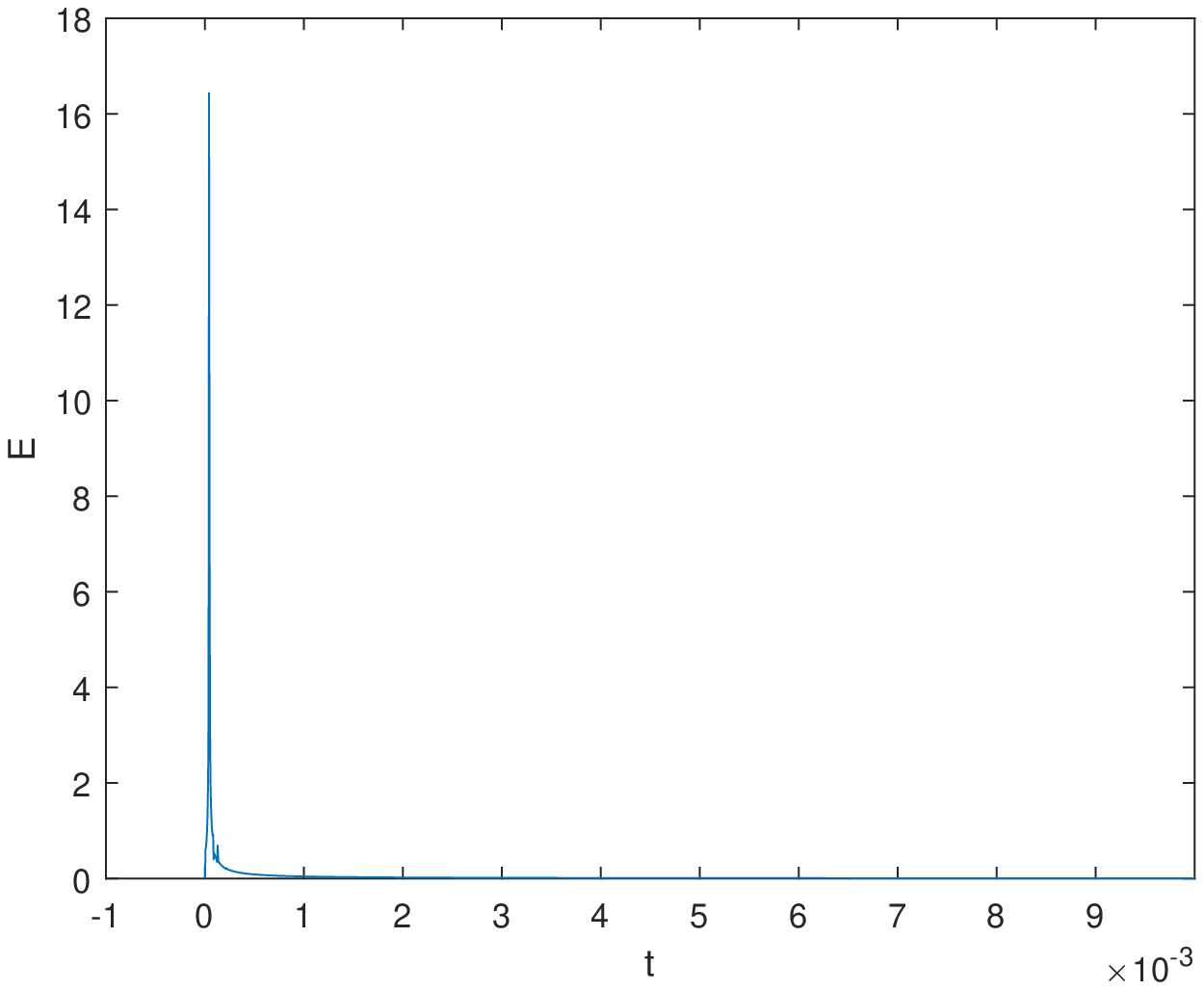}}
		\caption{The information of ITC flow with $n=1.5$ and $\theta=45^o$, the first column represent $Re=6000$ and the second column represent $Re=7000$; (a)(b) Phase-space trajectories of velocity, (c)(d) The evolution of velocity $u$, (e)(f) The evolution of velocity $v$, (g)(h) The Fourier power spectrum of kinetic energy.}
		\label{fig:45-1.5-6000}
	\end{figure}
	
	As we continue to increase the $Re$ number, the TC flow changes from a steady state to a periodic state. The relevant results are shown in Fig. \ref{fig:45-1.5-6000}. As shown in the figure, when $Re=6000$, it can be seen from the phase diagram and velocity curve presents an approximately periodic state, where the velocity in the phase diagram does not form a simple closed ring. Meanwhile, it can also be seen from the spectrum diagram of energy that multiple extreme points appear. Therefore, TC flow is a quasi-periodic flow when Re=6000. In addition, when $Re$ increases to $7000$, it can be seen from the figure that TC flow becomes a standard periodic flow.
	
	Combining the above results, we find that when $\theta=75^o$ and $45^o$, the TC flow reaches a periodic state at $Re=6000$, but when $\theta=60^o$, the TC stream reaches a periodic state at $Re=4000$. Therefore, the relationship between $\theta$ with critical $Re$ number from steady to periodic state is not monotonic.

	\begin{figure}[htbp]
		\centering
		\includegraphics[scale=0.5]{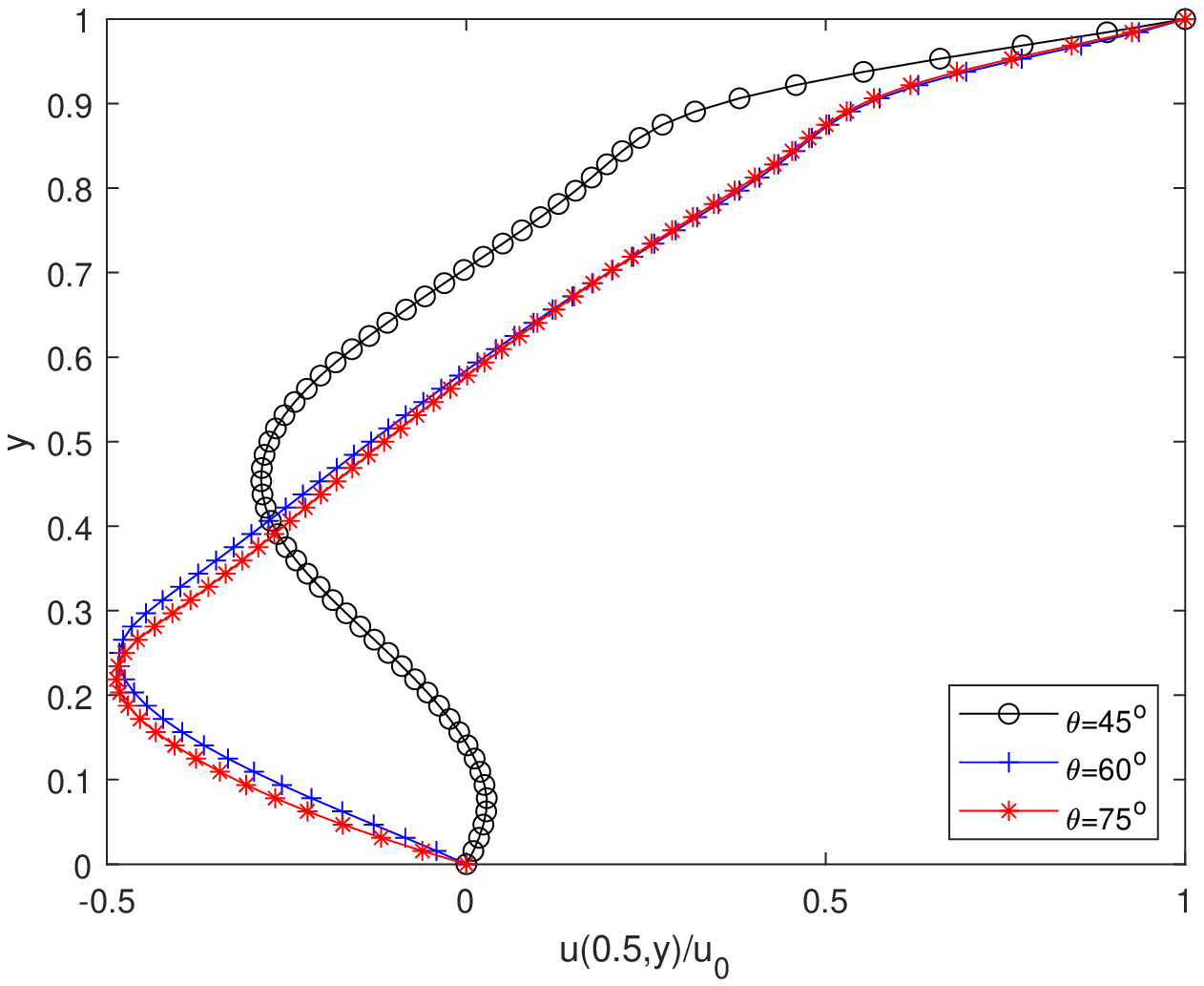}
		\caption{Vertical component of velocity for $n=1.5$ and $Re=500$ through x/L = 0.5 along y-axis.} \label{fig:1000-1.5-yu}
	\end{figure}
	
	We also present the centerline velocity with different $\theta$ in Fig. \ref{fig:1000-1.5-yu}. It can be found the velocity curves with $\theta=75^o$ is similar to that with $\theta=60^o$, but is different from that with $\theta=45^o$. It is indicated that the smaller the $\theta$ is, the greater the impact on the velocity. This is because the smaller the angle is, the greater the change degree of the vortex shape in the trapezoidal cavity is, and the greater the influence on the velocity is.
	
	\begin{figure}[htbp]\centering
		\subfigure[$\theta=75^o$]{ \label{fig:75}
			\includegraphics[scale=0.4]{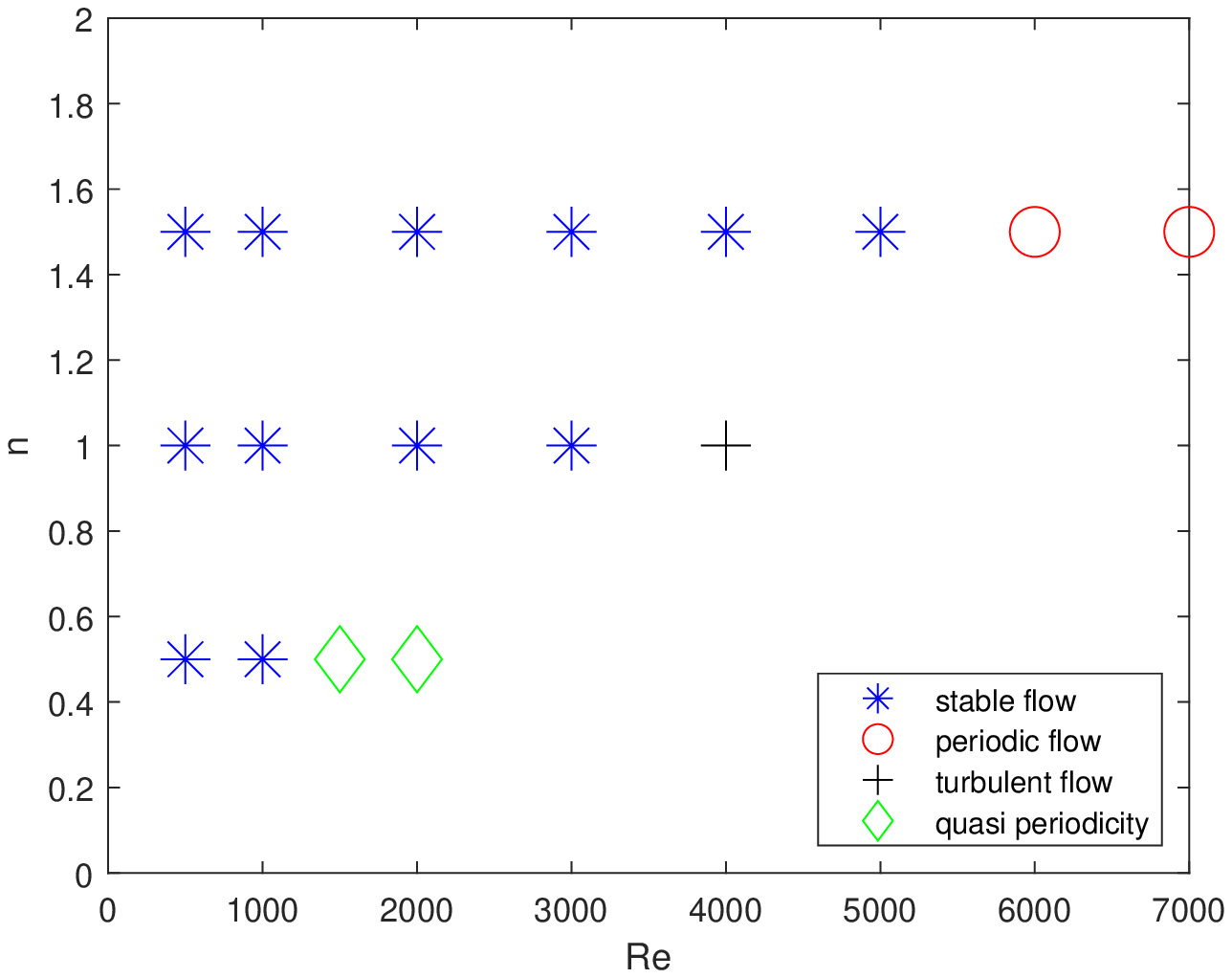}}
		\subfigure[$n=1.5$]{ \label{fig:1.5-3case}
			\includegraphics[scale=0.4]{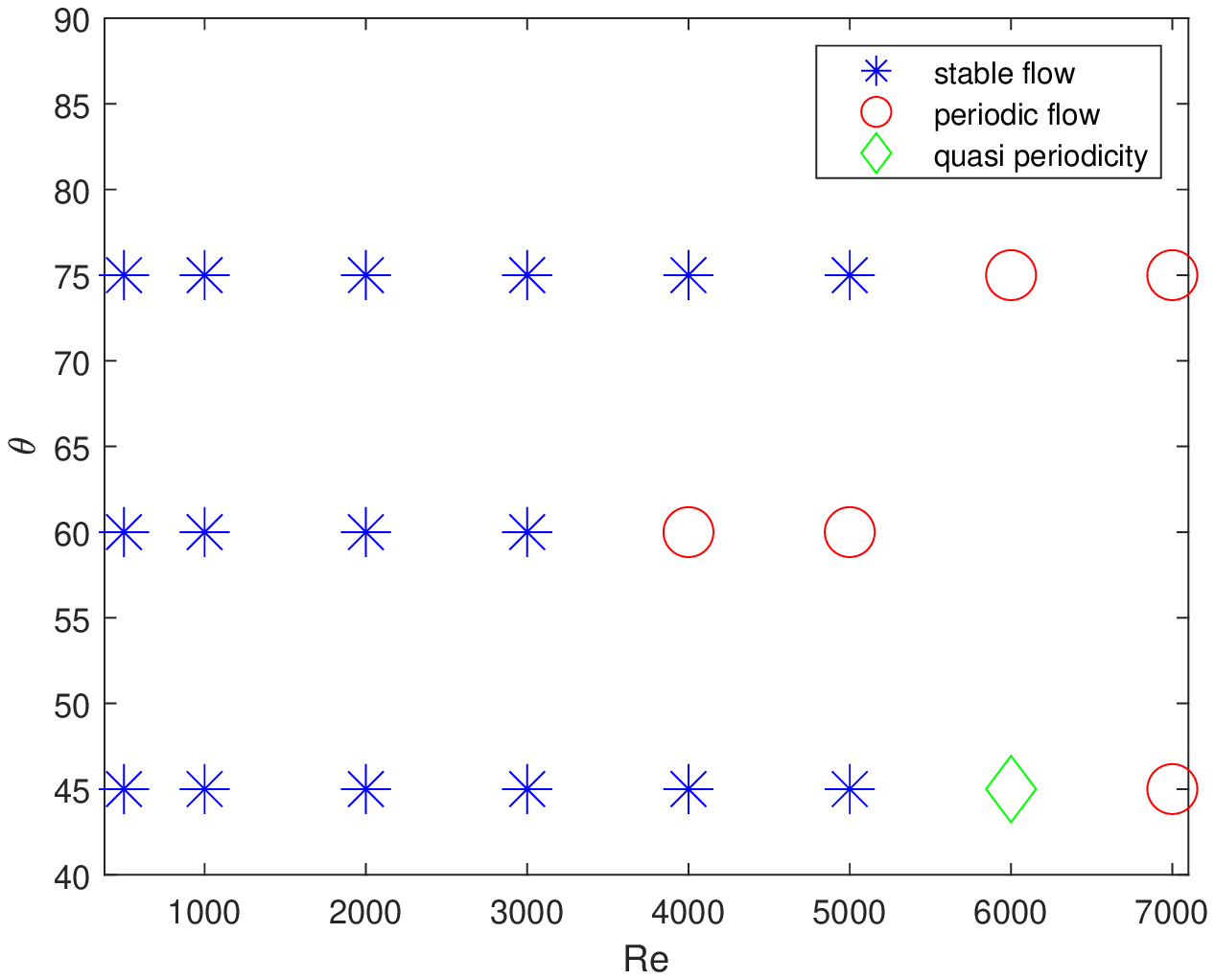}}
		
		\subfigure[$\theta=60^o$]{ \label{fig:60}
			\includegraphics[scale=0.4]{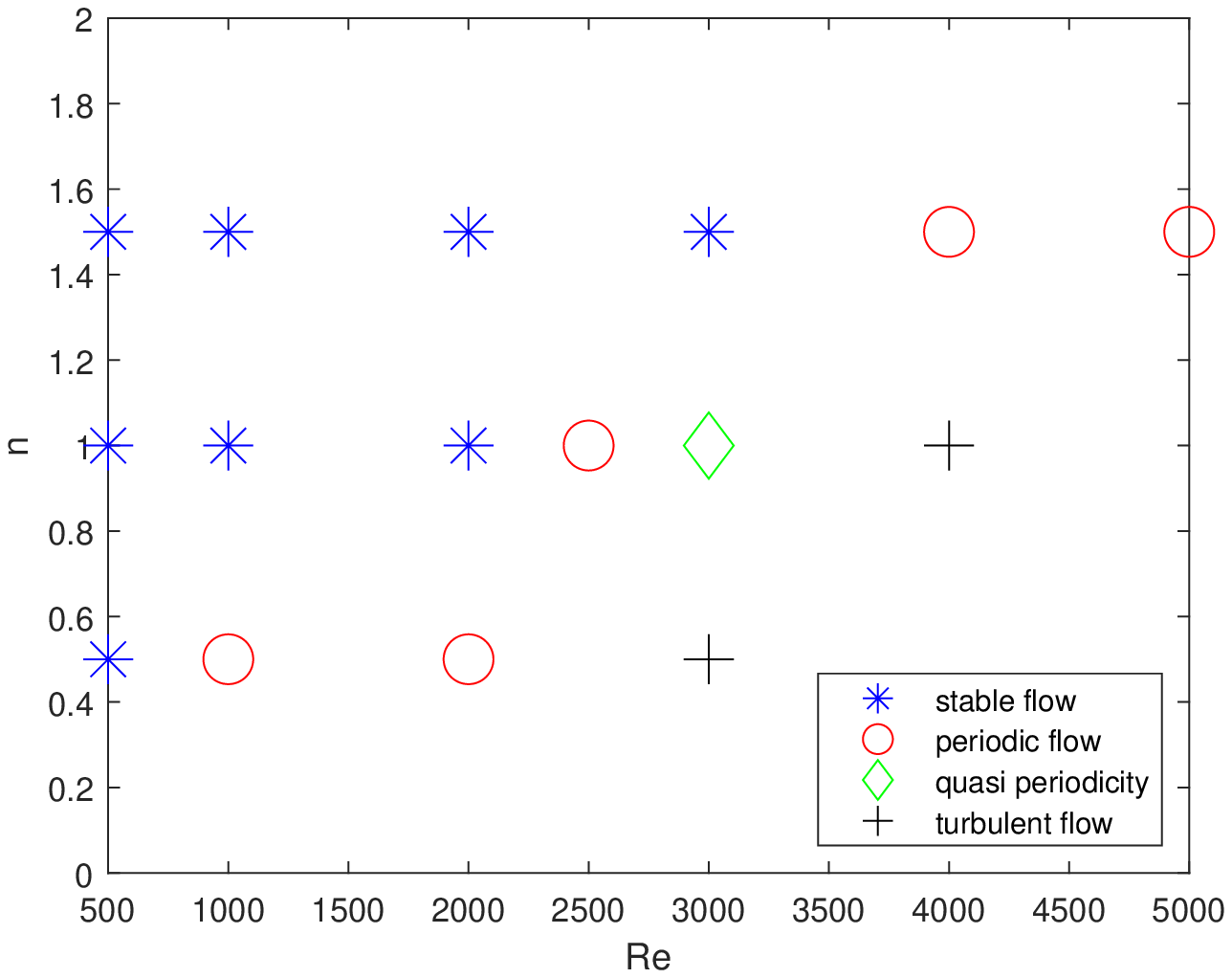}}
		\subfigure[$n=1.0$]{ \label{fig:1.0-3case}
			\includegraphics[scale=0.4]{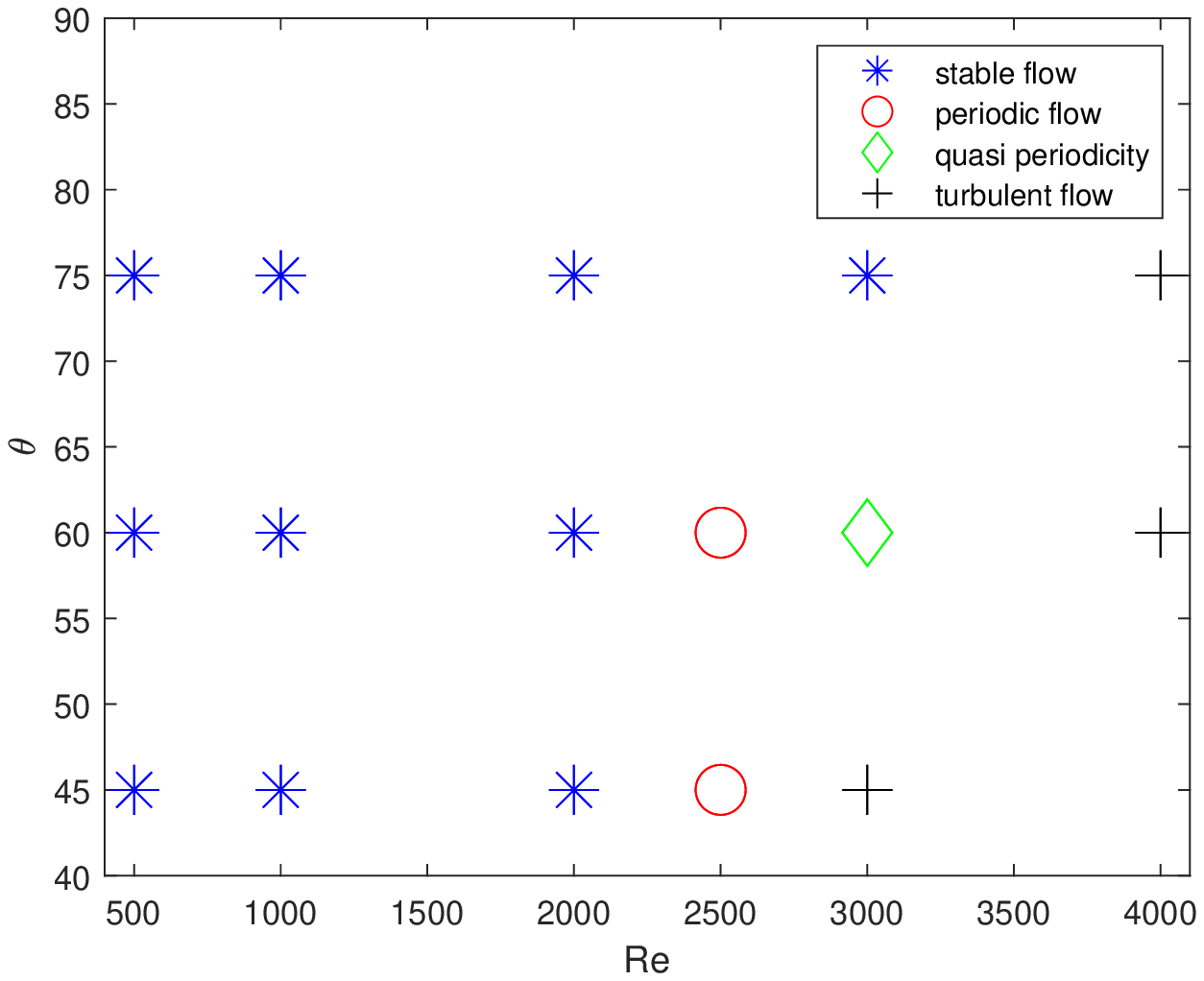}}
		
		\subfigure[$\theta=45^o$]{ \label{fig:45}
			\includegraphics[scale=0.4]{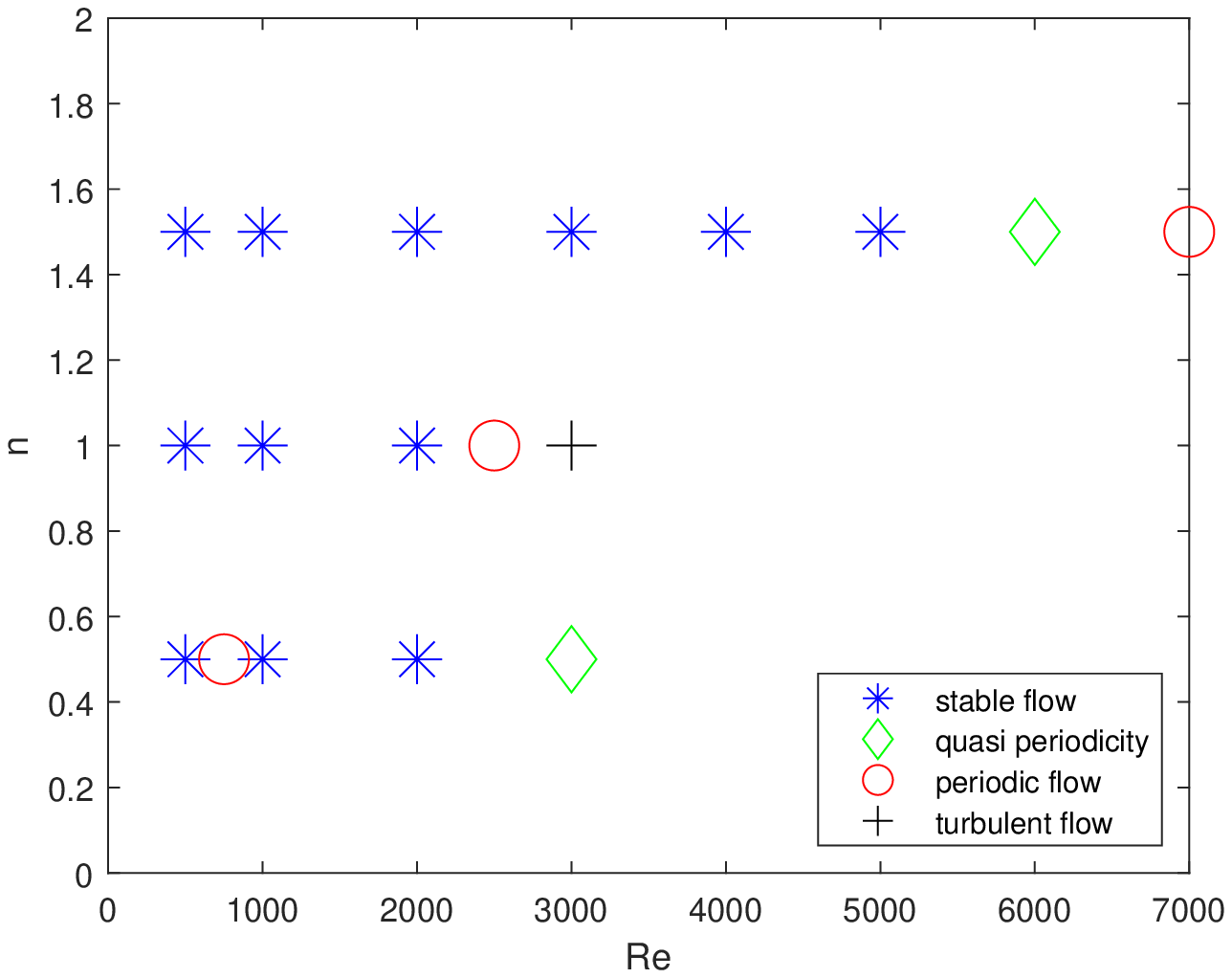}}
		\subfigure[$n=0.5$]{ \label{fig:0.5-3case}
			\includegraphics[scale=0.4]{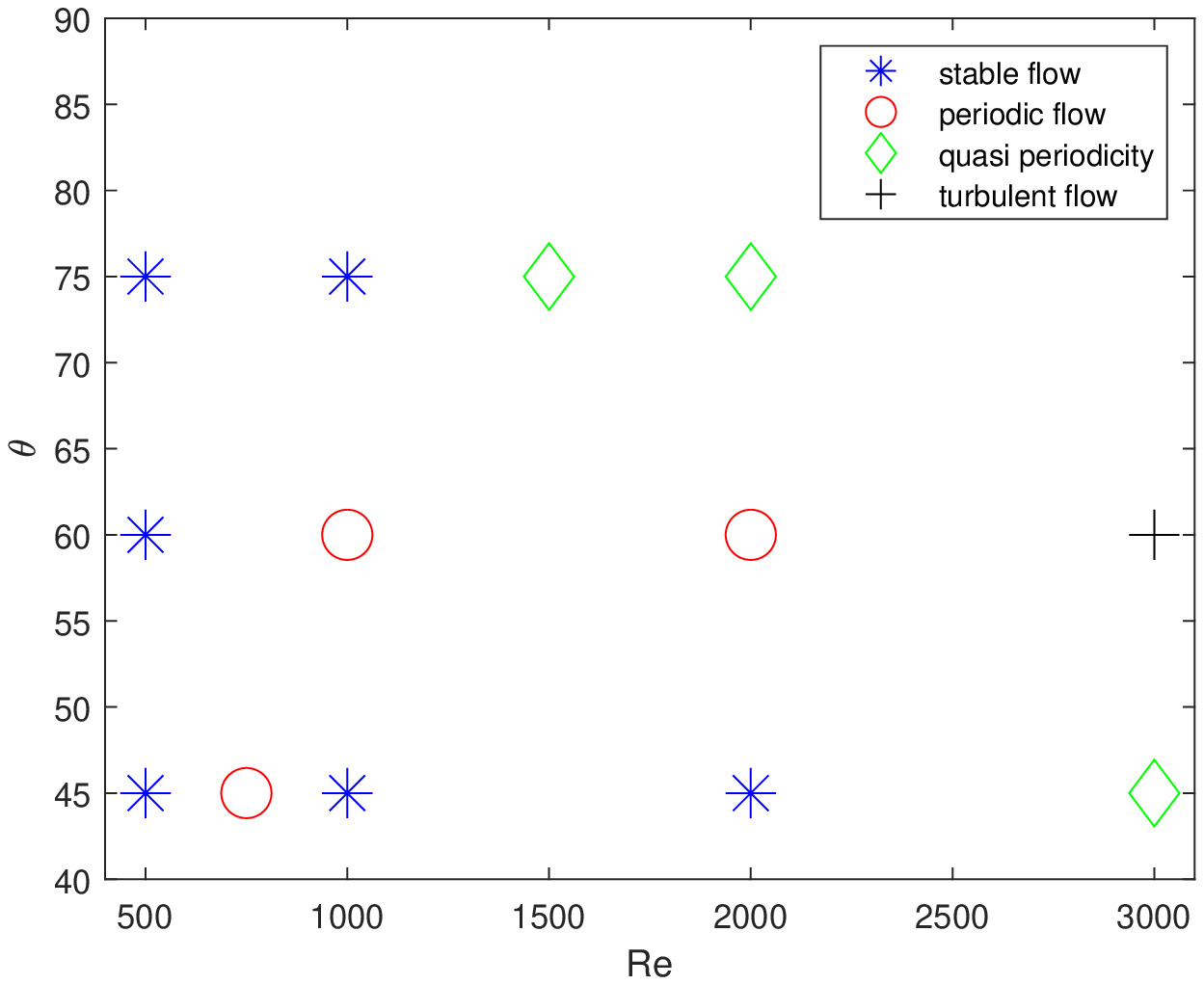}}
		\caption{ The distributions of flow state with different $\theta,n,Re$; (a) $\theta=75^o$, (c) $\theta=45^o$, (e) $\theta=60^o$, (b) $n=1.5$, (d) $n=1.0$, (f) $n=0.5$. }
		\label{fig:3case}
	\end{figure}
	
	Generally speaking, the flow state of lid-driven flow can be divided into three modes: steady flow, periodic flow and turbulence flow. According to the above simulation results, we summarize the flow states under different power-law index $n$ and different angles $\theta$, and show the results of flow states in Fig. \ref{fig:3case}. As shown in the figure, with the decrease of $n$, the critical $Re$ number from steady to periodic state decreases. The relationship between $\theta$  with the critical $Re$ number is not monotonic.

	\section{Conclusions}\label{Conclusion}
	The main work of this paper is to develop the IFDLBM and apply the IFDLBM to simulate power law fluid flow in a two-dimensional trapezoidal cavity. Due to the complex boundary of trapezoidal cavity, we use coordinate transformation method to transform the body-fitted gird of physical region into uniform grid of computational region. The effects of Re number, power-law index $n$ and angle $\theta$ on TC fluid are studied.
	
	It is found that when $Re$ is fixed at $100$, the cavity flow becomes more complicated with the increase of power-law index $n$. As $\theta$ decreases, the flow becomes gentle and the number of vortices decreases. When $\theta$ and $n$ are fixed, with the increase of Re, the development of the flow becomes more complex, the number and strength of vortices increase, and the TC flow gradually changes from steady flow to periodic flow and then to turbulent flow. In addition, the critical Re number from steady to periodic state decreases with the decrease of $n$. Finally, we study the effect of $\theta$ on TC flow at high Reynolds number. It can be found that the smaller the $\theta$ is, the more complicated the flow is. This is contrary to the conclusion at low $Re$ number.
	\section*{Acknowledgments}
	This work was financially supported by the National Natural Science Foundation of China (Grants No. 12072127 and No. 51836003) and the Fundamental Research Funds for the Central Universities, HUST (No. 2021JYCXJJ010). The computation was completed on the HPC Platform of Huazhong University of Science and Technology.
	\appendix
	
	\section{The Chapman Enskog analysis of the IFDLBM}\label{DTE}
	Inspired by the CE analysis of discrete unified gas kinetic scheme in Ref \cite{shang2021}, we will recover the NSE from the DVBE \eqref{eq2.0}.
	The CE analysis will be used to recover the incompressible NSE.
	The moment of distribution function $f_i^{eq}$ is designed as
	\begin{equation}\label{A1}
		\sum_if_i^{eq}=\rho_0 \quad \sum_i \bm c_if_i^{eq}=\bm u \quad \sum_i\bm c_i\bm c_if_i^{eq}=(P+c_s^2\rho_0)\bm I+\bm u\bm u \quad  \sum_i\bm c_i\bm c_i\bm c_if_i^{eq}=c_s^2\Delta\cdot\bm u,
	\end{equation}
	From the multi-scale technique, we can get
	\begin{equation}\label{A4}
		f_i=f_i^{(0)}+\epsilon f_i^{(1)}+\epsilon^2 f_i^{(2)},\quad \partial_t=\epsilon\partial_{t_1}+\epsilon\partial^2_{t_2},\quad \nabla=\epsilon\nabla_1.
	\end{equation}
	If we substitute Eq. \eqref{A4} to discrete Boltzmann equation, we can get
	\begin{subequations}
		\begin{equation}
			O(\varepsilon^0): -\frac{1}{\lambda}(f_i^{(0)}-f_i^{eq})=0\Leftrightarrow f_i^{(0)}=f_i^{eq}
			\label{eq:A6.1}
		\end{equation}
		\begin{equation}
			O(\varepsilon^1): \partial_{t_1}f_i^{(0)}+\bm c_i\cdot \nabla_1f_i^{(0)}=-\frac{1}{\lambda}f_i^{(1)},
			\label{eq:A6.2}
		\end{equation}
		\begin{eqnarray}
			O(\varepsilon^2): \partial_{t_2}f_i^{(0)}+\partial_{t_1}f_i^{(1)}+\bm c_i\cdot \nabla_1f_i^{(1)}=-\frac{1}{\lambda}f_i^{(2)}.
			\label{eq:A6.3}
		\end{eqnarray}
		\label{eq:A6}
	\end{subequations}
	Summing Eq. \eqref{eq:A6.2} yields
	\begin{equation}\label{A7}
		\nabla_1\cdot \bm u=0.
	\end{equation}
	Multiplying $\bm c_i$ to Eqs. \eqref{eq:A6.2} and \eqref{eq:A6.3}, and substituting them, one can obtain
	\begin{equation}\label{A8}
		\partial_{t_1}\bm u+\nabla_1\cdot[(P+c_s^2\rho_0)\bm I+\bm u\bm u]=0,
	\end{equation}
	\begin{equation}\label{A9}
		\partial_{t_2}\bm u+\nabla_1\cdot \sum_i\bm c_i\bm c_if_i^{(1)}=0,
	\end{equation}
	Multiplying $\bm c_i\bm c_i$ to Eq. \eqref{eq:A6.2} and substituting can deduce
	\begin{equation}\label{A10}
		\sum_i\bm c_i\bm c_if_i^{(1)}=-\lambda \left\{\partial_{t_1}[(P+c_s^2\rho_0)\bm I+\bm u\bm u]+\nabla_1\cdot c_s^2\Delta\cdot \bm u\right\},
	\end{equation}
	where
	under the low-Mach-number assumption, the term $\partial_{t_1}[(P+c_s^2\rho_0)\bm I+\bm u\bm u]$ can be ignored, then Eq. \eqref{A10} can be simplified as
	\begin{eqnarray}\label{A10.1}
		\sum_i\bm c_i\bm c_if_i^{(1)}&=&-\lambda c_s^2[(\nabla_1\cdot \bm u)\bm I+(\nabla_1\bm u+\nabla_1\bm u^T)],
	\end{eqnarray}
	if Eq. \eqref{A7} is substituted into Eq. \eqref{A10.1}, we can obtain
	\begin{eqnarray}\label{A11}
		\sum_i\bm c_i\bm c_if_i^{(1)}&=&-\lambda c_s^2(\nabla_1\bm u+\nabla_1\bm u^T).
	\end{eqnarray}
	with the help of Eq. \eqref{A11}, Eq. \eqref{A9} can be rewritten as
	\begin{equation}\label{A12}
		\partial_{t_2}\bm u=\nabla_1\cdot[\lambda c_s^2(\nabla_1\bm u+\nabla_1\bm u^T)].
	\end{equation}
	Through combining the results at $\varepsilon$ and $\varepsilon^2$ scales, i.e., Eqs. \eqref{A7}, \eqref{A8} and \eqref{A12}, the NSE \eqref{eq1.1} and \eqref{eq1.2} can be recovered correctly.

	\bibliographystyle{elsarticle-num}
	\bibliography{ref}
\end{document}